\documentclass[fleqn,usenatbib,useAMS]{mnras}

\usepackage{graphicx}    % Including Fig. files
\usepackage{amsmath}     % Advanced maths commands
\usepackage{amssymb}     % Extra maths symbols
\usepackage{multicol}    % Multi-column entries in tables
\usepackage{bm}          % Bold maths symbols, including upright Greek
\usepackage{pdflscape}   % Landscape pages
\usepackage[version=4]{mhchem} % Chemical formulas
\usepackage{wrapfig}     % Wrapping text around figures
\usepackage{cancel}      % Cancel text
\usepackage{threeparttable} % Notes in tables
\usepackage[normalem]{ulem}
\usepackage{subcaption}
\usepackage[switch]{lineno}
\usepackage{overpic}

\usepackage{changes} % Use [final] to apply changes, remove it to show changes
\definecolor{red}{RGB}{255, 0, 0}
\definecolor{green}{RGB}{0, 100, 230}
\setdeletedmarkup{\textcolor{red}{\sout{#1}}}
\setaddedmarkup{\textcolor{green}{\textbf{#1}}}

\newcommand{\kms}{\,km\,s$^{-1}$} % kilometres per second
\usepackage[T1]{fontenc}
\usepackage{ae,aecompl}

\usepackage{newtxtext,newtxmath}
\title[ATOMIUM SiO lines]{ATOMIUM: Inner circumstellar envelopes of oxygen-rich AGB stars as revealed by highly excited SiO lines}

\author[B. Pimpanuwat et al.]{B. Pimpanuwat$^{1,2}$\thanks{Contact e-mail: \href{mailto:bannawit@narit.or.th}{bannawit@narit.or.th, bpimpanuwat@gmail.com}},
S. Etoka$^{2}$,
M. D. Gray$^{1,2}$,
A. M. S. Richards$^{2}$,
A. Baudry$^{3}$,
F. Herpin$^{3}$,
\newauthor T. Danilovich$^{4,5}$,
L. Decin$^{5,6}$,
M. O. Lewis$^{7}$,
I. El Mellah$^{8}$,
C. A. Gottlieb$^{9}$,
Y. Mori$^{4}$,
H. S. P. M{\"u}ller$^{10}$,
\newauthor R. Sahai$^{11}$,
K. T. Wong$^{12}$,
J. A. Yates$^{13}$,
and A. Zijlstra$^{2}$
\\
$^{1}$National Astronomical Research Institute of Thailand (Public Organization), Chiang Mai 50180, Thailand\\
$^{2}$Jodrell Bank Centre for Astrophysics, The University of Manchester, Manchester M13 9PL, UK\\
$^{3}$Laboratoire d'Astrophysique de Bordeaux, Univ. Bordeaux, CNRS, B18N, allée Geoffroy Saint-Hilaire, 33615 Pessac, France\\
$^{4}$School of Physics \& Astronomy, Monash University, Wellington
Road, Clayton 3800, Victoria, Australia\\
$^{5}$Institute of Astronomy, KU Leuven, Celestijnenlaan 200D, 3001
Leuven, Belgium\\
$^{6}$School of Chemistry, University of Leeds, Leeds LS2 9JT, UK\\
$^{7}$Leiden Observatory, Leiden University, PO Box 9513, 2300 RA Leiden, the Netherlands\\
$^{8}$Departament de Física, EEBE, Universitat Politècnica de Catalunya, c/Eduard Maristany 16, 08019 Barcelona, Spain\\
$^{9}$Harvard-Smithsonian Center for Astrophysics, 60 Garden Street,
Cambridge, MA 02138, USA\\
$^{10}$Universität zu Köln, Astrophysik/I. Physikalisches Institut, 50937
Köln, Germany\\
$^{11}$California Institute of Technology, Jet Propulsion Laboratory,
Pasadena, CA, 91109, USA\\
$^{12}$Theoretical Astrophysics, Department of Physics and Astronomy,
Uppsala University, Box 516, SE-751 20 Uppsala, Sweden\\
$^{13}$Department of Computer Science, University College London, London, WC1E 6BT, UK
}

\date{Accepted 2026 May 05. Received 2026 May 05; in original form 2025 August 06 }

\pubyear{2025}

\begin{document}
\label{firstpage}
\pagerange{\pageref{firstpage}--\pageref{lastpage}}
\maketitle

% Abstract of the paper
\begin{abstract}
Silicon monoxide (SiO) traces the physical conditions and dynamics in the circumstellar envelopes (CSEs) of AGB stars. We present high-resolution ALMA Band 6 observations of highly excited SiO emission in 14 oxygen-rich AGB stars. We cover transitions from $\varv=0$ to $\varv=8$, including first detections of \ce{^{28}SiO} $\varv=3,4,8$, $J=6-5$, \ce{^{29}SiO} $\varv=6$, $J=6-5$, and \ce{^{30}SiO} $\varv=4,5$, $J=6-5$, some of which are masers. The $\varv=8$ transition is the highest $\varv$-state observed in an AGB star yet. Masers in $\varv=0$ are detected clearly in V PsA and IRC+10011 and tentatively in T Mic. R Hya exhibits the richest SiO spectrum. SiO $J=6-5$ absorption is seen in R Aql, R Hya, S Pav, and T Mic, with features indicative of both infalls and outflows, and tentative detection of \ce{^{28}SiO} $\varv=8$, $J=6-5$ absorption is found towards S Pav and R Aql. Highly excited SiO emission is often distributed in arcs or clumps with velocity gradients; components in R Hya and U Her align with predicted shock fronts. Detection rates show no significant difference between low and high mass-loss rate stars, although line overlap may affect some intensities. Maser detections appear uncorrelated with pulsation period or phase. The radius enclosing 90 per cent of compact SiO emission shows a tentative correlation with mass-loss rate. These results highlight the role of mass loss and CSE geometry in shaping high-excitation SiO emission.
\end{abstract}

% Select between one and six entries from the list of approved keywords.
% Don't make up new ones.
\begin{keywords}
masers -- stars: AGB and post-AGB -- stars: circumstellar matter -- stars: mass-loss
\end{keywords}

%%%%%%%%%%%%%%%%%%%%%%%%%%%%%%%%%%%%%%%%%%%%%%%%%%

%%%%%%%%%%%%%%%%% BODY OF PAPER %%%%%%%%%%%%%%%%%%
%\linenumbers

\section{Introduction}
\label{sec:intro}

Asymptotic giant branch (AGB) stars are evolved low- and intermediate-mass stars with initial masses of between 0.8 and 8 M$_{\sun}$ characterised by strong mass loss of $\sim$10$^{-7}$ to $\sim$10$^{-4}$ M$_{\sun}$ yr$^{-1}$ \citep{1996A&ARv...7...97H,2005ARA&A..43..435H,2018A&ARv..26....1H} that usually takes place over a span of $10^5$--$10^6$ yr \citep{2004agbs.book.....H}. High-resolution interferometric observations have shown that AGB winds may exhibit small-scale ($<$ arcsec) complex structures, such as arcs, shells, bipolar structures, clumps, spirals, tori, and rotating discs (e.g. \citealt{2004ASPC..313..141S,2006A&A...452..257M,2016A&A...596A..92K,2020Sci...369.1497D}), and are thought to be driven by stellar pulsations and radiation pressure on dust grains (see, for example, \citealt{1962MNRAS.124..417H}, \citealt{2018A&ARv..26....1H}, \citealt{2019MNRAS.484.4678M}, and \citealt{2019A&A...623A.158H}). They are responsible for approximately 85 per cent of gas and about 35 per cent of the dust that enrich the interstellar medium \citep{2005pcim.book.....T}, and are crucial for the production of over 90 molecules and 15 dust species (e.g. \citealt{1996A&ARv...7...97H,2004agbs.book.....H,2018A&ARv..26....1H}). Hydrodynamical calculations have also revealed that a nearby companion can cause gravitational perturbations that lead to a variety of the aforementioned morphologies (e.g. \citealt{1999ApJ...523..357M,2004ApJ...600..992G,2017MNRAS.468.4465C,2022MNRAS.513.4405A}).

SiO thermal emission in the ground vibrational state, $\varv=0$, originates from the cooler, more extended regions of the circumstellar envelope (CSE). These thermal emissions trace the bulk molecular gas and provide constraints on the distribution, temperature, and density of SiO in the circumstellar medium. SiO thermal lines are often observed in rotational, $J$, transitions in the millimetre and sub-millimetre wavelength ranges, making them valuable for characterising the global properties of the CSE. Studies of AGB stars using instruments like ALMA (Atacama Large Millimeter/submillimeter Array) and the IRAM (Institut de Radioastronomie Millimétrique) 30m telescope have provided high-resolution maps and multi-line surveys of SiO thermal emission, respectively, revealing intricate spatial and spectral structures and outflow dynamics (e.g. \citealt{2018A&A...617A..23B, 2018A&A...615A..28D, 2020A&A...642A..93H, 2021MNRAS.504.2687N, 2020A&A...641A..57M, 2024A&A...688A..16M}). Outside the dust formation zone, SiO abundance and distribution are influenced by the mass-loss rate and chemical processes occurring in the CSE (\citealt{2003A&A...411..123G}; Gottlieb et al., in prep.). Some observations suggest a possible trend of decreasing SiO abundance with increasing mass-loss rates \citep{2003A&A...411..123G,2020A&A...641A..57M}, which may be attributed to the molecule's incorporation into silicate grains in oxygen-rich (M-type, i.e. with C/O $<$ 1) AGB stars like IK Tau \citep{1993AJ....105..595S,2016A&A...585A...6G}. This interplay between gas-phase and dust-phase SiO highlights the role of SiO in dust nucleation and growth.

The CSE of an AGB star is often the production site of astrophysical masers. There have been many reports of detections of SiO maser lines in AGB stars to date, including those that originate from the less common \ce{^{29}SiO} and \ce{^{30}SiO} isotopologues \citep{1983ApJ...264L..65D,1993ApJ...407L..33C,2020ApJ...892...52L,2021ApJS..253...44R}. Some of the highest-frequency SiO maser transitions observed towards evolved stars known to date are \ce{^{29}SiO} $\varv=2$, $J=16-15$ in the red supergiant VY CMa (676 GHz, \citealt{2020ApJS..247...23A}) and \ce{^{28}SiO} $\varv=1$, $J=11-10$ in W Hya (474.18 GHz, \citealt{2020A&A...638A..19B}). The vibrational ground state lines of \ce{^{28}SiO} often have very few to no maser features \citep{2004ApJ...608..480B,2016A&A...589A..74D}, with the majority of the spectral profile dominated by thermal emission. 

Despite abundant data, the complex behaviour of SiO emission still requires further study, as the relative strengths of rotational transitions within a given vibrational state remain unexplained \citep{2004A&A...426..131S,2021ApJS..253...44R}. High-resolution Very Long Baseline Interferometry (VLBI) observations of SiO masers typically result in detections of apparent compact spots of a few milliarcseconds (mas) in diameter, where the apparent size of a spot results from maser beaming, usually much less than the physical extent of emission in each channel. Ring-like structures formed by low-$J$ SiO maser spots, commonly attributed to tangential amplification due to the geometry and kinematics of the masing region \citep{1985MNRAS.212..375C}, are typically observed at distances of a few optical photospheric radii, $R_*$, around RSG and AGB stars (see e.g. \citealt{1994ApJ...430L..61D,2004A&A...426..131S,2020MNRAS.495.1284Y}). This is because gas-phase SiO is mostly present here and emission from the gas accelerating or decelerating radially away from or toward the star maintains the highest velocity coherence along the line of sight when observed in tangential directions. Most previous studies have concentrated on the low-frequency, low-$J$ SiO transitions ($\varv=1,2$, $J=1-0$ and $2-1$ at 43 and 86 GHz), as high-frequency rovibrationally excited lines (e.g. $J=5-4$ and $6-5$ or high-$J$, hereafter) have recently become more accessible with improved observing facilities in high-resolution interferometer arrays and millimetre-wavelength VLBI. The Korean VLBI Network (KVN) has conducted long-term monitoring up to $J=3-2$ at 129 GHz \citep{2024IAUS..380..324Y,2024ApJS..275...20L} as a part of their Key Science Program, but there is currently no systematic monitoring of higher-$J$ SiO masers. The latest simultaneous four-band (22/43/86/129 GHz) KVN single-dish survey of 155 oxygen-rich AGB stars by \citet{2025AJ....170...84B} and \citet{2025AJ....170..266S} revealed distinct maser detection patterns across evolutionary types, with higher-$J$ SiO lines (especially $\varv=1$, $J=2-1$ and $J=3-2$) more frequently detected in semiregular and Mira variables.

A study by \citet{1997ApJ...487L.147B} suggests that there was evidence for an intimate tie between stellar periodicity and instabilities of the $\varv=1$, $J=1-0$ SiO maser ring structure in evolved stars. Comparisons between hydrodynamical models and observations by \citet{1996MNRAS.282.1359H} also suggest that a cloud of gas responsible for SiO masers around the star should experience alternate phases of outward acceleration caused by a shock wave and infalls under gravity once the shock is damped. There is a clear connection between the pulsation phase and SiO maser regions, and the associated shocks appear to support the correlations between SiO maser velocity, intensity, and stellar pulsations, as suggested by models (e.g. \citealt{2007A&A...470..191W}; \citealt{2009MNRAS.394...51G}).

Despite many extensive surveys (e.g. \citealt{1982ApJ...256L..55S}; \citealt{1996A&AS..115..117C}; \citealt{2002A&A...393..115M};  \citealt{2019ApJS..244...25S}), only a few studies have explored a potential correlation between SiO emission and mass-loss rates for low-$J$ transitions (43 and 86 GHz), such as the BAaDE survey \citep{2019ApJS..244...25S}. One notable finding involving low-$J$ SiO masers is detailed in \citet{2016ApJ...817..115C}. They found strong correlations between the mass-loss rate, absolute magnitude, kinematic luminosity, and the isotropic luminosity of the $\sim$43 GHz ($\varv=1, 2$, $J=1-0$) SiO masers, and proposed that stars with higher mass-loss rates or greater intensity of near-infrared radiation are likely to stimulate stronger SiO maser emission. No extensive studies of the effects of mass-loss rates on SiO emission originating from higher-$J$ states ($\sim$200–300 GHz) have been published thus far. Therefore, a large and representative set of (sub-)mm observations of these emission lines towards oxygen-rich stars is required to investigate the mechanisms behind the interplay of the transition between the pulsation-dominated and dust-driven winds in the inner CSEs.

This paper serves as an overview of a snapshot of high-$J$ SiO thermal and maser emission observed in oxygen-rich AGB stars of a wide range of mass-loss rates ($10^{-8}$--$10^{-5}$ M$_{\sun}$ yr$^{-1}$). Here, we present a summary of the observations obtained with ALMA and the data preparation process in Section \ref{sec:data}. Identification and characterisation of high-$J$ SiO line detections as well as absorption towards selected AGB sources are given in Section \ref{sec:SiO-id}, while a more in-depth discussion about the high-$J$ SiO absorption features is provided in Section \ref{sec:absorption}. Section~\ref{sec:dist} presents an overview of the SiO component distributions and their associated brightness temperatures. A tentative analysis of the effects of stellar pulsation period and phase on line detection rate and intensity are given in Section \ref{sec:compare}. We then discuss a potential link between angular positions at which compact, highly excited SiO emission is observed and mass-loss rates in Section \ref{sec:correlation}. Finally, we give our conclusions and remarks on future work in Section \ref{sec:conclusions}.

\begin{table*}
\centering
\caption{Summary of some circumstellar parameters of the 14 AGB sources in ATOMIUM, adapted from \citet{2022A&A...660A..94G}, \citet{2024A&A...681A..50W} and \citet{2025Danilovich}. Most values are those adopted by \citet{2025Danilovich} unless marked or stated otherwise. The original references are given in the notes below.}
%\resizebox{\linewidth}{!}{%
\begin{tabular}{|l|c|c|c|c|r|r|c|r|r|}
\hline \hline 
Name
& \multicolumn{1}{p{1.0cm}|}{\centering Variability \\ Type} 
& \multicolumn{1}{p{1.2cm}|}{\centering $\dot{M}$$^{(1)}$ \\ (M$_\odot$ yr$^{-1}$)}
& \multicolumn{1}{p{1.0cm}|}{\centering $P$$^{(2)}$ \\ (d)} 
& \multicolumn{1}{p{1.0cm}|}{\centering $D$ \\ (pc)}
& \multicolumn{1}{p{1.0cm}|}{\centering 2$R_*$$^{(3)}$ \\ (mas)}
& \multicolumn{1}{p{1.1cm}|}{\centering 2$R_*^{\rm{mm}}$$^{(4)}$ \\ (mas)}
& \multicolumn{1}{p{1.0cm}|}{\centering T$_{\rm{eff}}$$^{(5)}$ \\ (K)}
%& \multicolumn{1}{p{1.0cm}|}{\centering T$_{\rm{ex}}$ \\ (K)} 
& \multicolumn{1}{p{1.0cm}|}{\centering $V_*$$^{(6)}$ \\ (\kms)}
& \multicolumn{1}{p{1.0cm}|}{\centering Pulsation \\ Phase$^{(7)}$}\\
\hline
S Pav & SRa & 5.3E-8 $^{\rm{a}}$& 390 & 184 $^{\rm{i}}$& 16.6 & 20.42 (0.03) & 2752 $^{\rm{k}}$& $-$18.2 & 0.96\\
U Del & SRb & 6.0E-8 $^{\rm{a}}$& 120, 1163 & 333 $^{\rm{j}}$& 7.5 & 11.25 (0.18) & 3236 $^{\rm{k}}$& $-$6.8 & 0.01\\
V PsA & SRb & 1.6E-7 $^{\rm{a}}$& 148, 105 & 299 $^{\rm{j}}$& 10.5 & 11.86 (0.09) & 2360 $^{\rm{a}}$& $-$11.1 & $^\triangle$0.30\\
U Her & Mira & 1.9E-7 $^{\rm{b}}$& 406 & 271 $^{\rm{i}}$& 11.9 & 18.98 (2.95) & 2700 $^{\rm{k}}$& $-$14.9 & 0.76\\
SV Aqr & SRb & 3.3E-7 $^{\rm{a}}$& 93, 232 & 445 $^{\rm{i}}$& 8.8 & 6.65 (1.54) & 2180 $^{\rm{a}}$& 6.7 & $^\triangle$0.40\\
RW Sco & Mira & 3.4E-7 $^{\rm{c}}$& 389 & 560 $^{\rm{j}}$& 6.0 & $^\dag$1.06 (3.52) & 2500 $^{\rm{c}}$& $-$69.7 & 0.20\\
T Mic & SRb & 4.4E-7 $^{\rm{a}}$& 352, 178 & 175 $^{\rm{i}}$& 15.8 & 21.41 (0.04) & 2856 $^{\rm{k}}$& 25.5 & 0.84\\
$\pi^1$ Gru$^{(8)}$ & SRb & 7.7E-7 $^{\rm{d}}$& 196, 5750 & 164 $^{\rm{j}}$& 16.7 & 23.97 (3.39) & 3100 $^{\rm{n}}$& $-$11.7 & 0.01\\
R Hya$^{(8)}$ & Mira & 9.7E-7 $^{\rm{b}}$& 359 & 126 $^{\rm{i}}$& 26.0 & 31.87 (3.16) & 3100 $^{\rm{i}}$& $-$10.1 & 0.07\\
W Aql$^{(8)}$ & Mira & 2.8E-6 $^{\rm{e}}$& 488 & 380 $^{\rm{i}}$& 13.1 & 16.65 (0.06) & 2300 $^{\rm{m}}$& $-$23.0 & 0.16\\
R Aql & Mira & 5.0E-6 $^{\rm{b}}$& 269 & 266 $^{\rm{i}}$& 10.7 & 16.07 (0.04) & 3008 $^{\rm{j}}$& 47.2 & 0.21\\
IRC$-$10529 & Mira & 2.3E-5 $^{\rm{f}}$& 670 & 930 $^{\rm{i}}$& 8.9 & 17.68 (3.25) & 2000 $^{\rm{l}}$& $-$16.3 & *0.40\\
GY Aql & Mira & 3.2E-5 $^{\rm{g}}$& 464 & 410 $^{\rm{i}}$& 13.7 & 15.29 (0.17) & 2143 $^{\rm{k}}$& 34.0 & 0.12\\
IRC+10011 & Mira & 4.0E-5 $^{\rm{h}}$& 651 & 720 $^{\rm{i}}$& 14.0 & 17.93 (0.29) & 1800 $^{\rm{o}}$& 10.1 & *0.63\\
\hline
\end{tabular}%
%}
%\begin{tablenotes}
\\ \raggedright \textbf{Notes:}
$^{(1)}$Targets are listed in the order of increasing mass-loss rates, which have been scaled to the latest distance estimates. $^{(2)}$For $P$, the primary period is listed first, then followed by the secondary period (if any). $^{(3)}$Diameters calculated from stellar effective temperature $T_{\rm{eff}}$ and luminosity $L_*$ in milliarcseconds. $^{(4)}$Diameter of uniform disc (UD) fit at 241.75 GHz. $^\dag$RW Sco has low signal-to-noise, S/N, ratio and a small angular size, leading to large uncertainty. $^{(5)}$Stellar effective temperatures. $^{(6)}$Stellar systemic velocity according to \citet{2023A&A...674A.125B}. $^{(7)}$During extended-configuration observations, obtained by fitting a polynomial to the light curve from the AAVSO database. $^\triangle$Estimated from ASAS-SN light curves. *Extrapolated based on the data from \citet{1974ApJS...27..331H} and \citet{2016ApJ...822....3Y}, respectively. $^{(8)}$Known binary system. \textbf{References}: $^{\rm{a}}$\citet{2002A&A...391.1053O};
$^{\rm{b}}$\citet{2020A&A...638A..19B};
$^{\rm{c}}$\citet{1999A&AS..140..197G};
$^{\rm{d}}$\citet{2017A&A...605A..28D};
$^{\rm{e}}$\citet{2017A&A...605A.126R};
$^{\rm{f}}$\citet{2015A&A...581A..60D};
$^{\rm{g}}$\citet{1986ApJ...311..335W};
$^{\rm{h}}$\citet{2017A&A...606A.124D};
$^{\rm{i}}$\citet{2022A&A...667A..74A}; $^{\rm{j}}$\citet{2021AJ....161..147B}; $^{\rm{k}}$\citet{2012MNRAS.427..343M}; $^{\rm{l}}$\citet{2015A&A...581A..60D};
$^{\rm{m}}$\citet{2014A&A...569A..76D};
$^{\rm{n}}$\citet{2014A&A...570A.113M};
$^{\rm{o}}$\citet{2014A&A...566A.145R}.
%\end{tablenotes}
\label{table:atomium-sources}
\end{table*}

\section{Observations and data preparation}
\label{sec:data}
The observations were obtained in three different antenna configurations, 16 sub-bands (i.e. spectral windows, spw), covering 27 GHz. The selected sources cover many (circum)stellar parameters and evolutionary stages for studying the chemistry of dust precursors and the stellar characteristics, e.g. pulsation period, variability type, effective temperature, and most notably mass-loss rate, $\dot{M}$ (\citealt{2020Sci...369.1497D}; \citealt{2022A&A...660A..94G}). The project provides one of the first sets of observations that have a sufficient spectral and spatial resolution to study the circumstellar high-$J$ SiO emission at scales of individual clumps and to unravel how the physico-chemical properties of the stellar winds are reflected in maser emission. The ATOMIUM observations are also valuable for investigating how highly excited SiO lines and mass-loss rates may be related to one another.

\subsection{Source sample}
\label{sec:data:source}
The ATOMIUM source sample was chosen from a broad range of oxygen-rich evolved stars that are representative of various pulsational variability types and mass-loss rates (Table \ref{table:atomium-sources}). Most stars are M-type, except $\pi^1$ Gru and W Aql which are S-type, representing a transitional phase where \textit{s}-process enrichment is evident. We note that S-type characteristics can manifest at C/O ratios comparable to M-type stars, rather than strictly C/O $\sim$ 1 (e.g. \citealt{2021A&A...650A.118S}). We exclude the three red supergiants, namely AH Sco, KW Sgr, and VX Sgr, from the current work despite evidence of rather strong SiO emission and maser-like profiles due to the overall complexity they present e.g. line blending and extremely complicated spatial distributions.

To enable consistent comparisons across the ATOMIUM sample, we used the latest estimated mass-loss rates of ATOMIUM AGB stars derived from CO radiative transfer modelling provided in \citet{2025Danilovich} and references therein, all of which have been scaled to the adopted distances (Table \ref{table:atomium-sources}).

The pulsation phase, $\phi$, at the time of extended-configuration observations (Section \ref{sec:data:obs}) for each of the stars (column 10 of Table \ref{table:atomium-sources}) was determined by fitting a polynomial to the optical light curve data available on the American Association of Variable Star Observers (AAVSO)\footnote{\url{https://app.aavso.org/vsp/}} database using the \textsc{vstar}\footnote{\url{https://www.aavso.org/vstar}} package \citep{2012JAVSO..40..852B}. The pulsation phases of SV Aqr and V PsA were estimated using photometric data from the ASAS-SN\footnote{\url{http://asas-sn.ifa.hawaii.edu/skypatrol/}} database, while for IRC$-$10529 and IRC+10011, the phases were derived by extrapolating the results of \citet{1974ApJS...27..331H} and \citet{2016ApJ...822....3Y}, respectively. All results were determined based on the pulsation periods $P$ listed in Table \ref{table:atomium-sources}.

\subsection{Observations}
\label{sec:data:obs}
Millimetre-wavelength interferometric observations were made using ALMA Band 6 ($\sim$214--270 GHz) as part of the ATOMIUM programme between October 2018 and August 2019. Each of the 14 AGB stars in the ATOMIUM sample was observed with three different array configurations: extended (with maximum recoverable scale, MRS, $\sim$0.4--0.6 arcsec; resolution of a few tens mas), mid (MRS $\sim$1.5--4 arcsec; resolution of a few hundred mas) and compact (MRS 8--10 arcsec, resolution of up to 1 arcsec); see Table E.3 of \citet{2022A&A...660A..94G}. In this work, only extended- and mid-configuration data were used. They were observed in 4 tunings, each of 4 spws, within the frequency span, but not contiguous, with a total bandwidth of 26.8 GHz. The extended-configuration data were all observed in June--July 2019, while the mid-configuration observations were taken over several months but in almost all cases with only one epoch per tuning. Details are given in \citet{2022A&A...660A..94G}. Comparing flux densities on different ALMA baselines suggests that the extended configuration (MRS $> \sim$0.4 arcsec) recovers all SiO maser emission for $v>0$, but sensitivity to arcsec scales is needed to recover all from $\varv=0$.

The \ce{^{n}Si}\ce{^{16}O}, where $n$ is 28, 29 or 30, transitions that were observed in this work are listed in the order of rest frequency with their Si isotopes, vibrational states, upper/lower rotational states, and the upper state energy in K in Table \ref{table:line-coverage}. The reference frequencies are taken from the Cologne Database of Molecular Spectroscopy\footnote{\url{https://cdms.astro.uni-koeln.de/}} (CDMS, \citealt{2005JMoSt.742..215M,2013JPCA..11713843M}), with additional high-$J$, high-$\varv$ \ce{^{28}SiO} data from \citet{1991ApJ...368L..19M}. Lines from both minor O isotopes (\ce{^{n}Si^{17}O} and \ce{^{n}Si^{18}O}) were not detected in the ATOMIUM data set. The central frequency and velocity width of the 16 spws as well as the exact spectral coverage of each window are provided in \citet{2022A&A...660A..94G}. The uncertainty in the flux scale is about 10 per cent because each target underwent multiple observations spread across several months, introducing uncertainties at each instance. 

The source coordinates were taken from the position of the 2D-Gaussian fitted emission peak of the stellar continuum observed at around 241.8 GHz with the ALMA extended configuration (see Table E.2 of \citet{2022A&A...660A..94G} for details).

\subsection{Data processing}
\label{sec:data:processing}
The data were processed using the standard data reduction methodology via the ALMA calibration and imaging pipelines and scripts \citep{2016alma.confE...1H} implemented in the Common Astronomy Software Applications package or \textsc{casa}\footnote{\protect\url{https://casa.nrao.edu/}}. We took the pipeline-calibrated data and used \textsc{lumberjack}\footnote{\url{https://github.com/adam-avison/LumberJack/}} to select the line-free continuum. We combined the selection for all tunings, for each configuration, and self-calibrated and imaged the continuum. We fitted 2D Gaussian components to the stellar emission and used the coordinates measured from the extended configuration as the reference position for each star (see Table E2 of \citet{2022A&A...660A..94G} for details). The absolute astrometric accuracy is approximately 2 and 7 mas for extended- and mid-configuration observations, respectively. After self-calibration, the relative astrometry between SiO transitions and the stellar position is limited only by the S/N ratio (Appendix \ref{appendix:fitting}). Only the original resolutions of the extended- and mid-configuration data were used in the current work, with the focus being primarily on the former, as we were interested in resolving as small a spatial scale as possible when it comes to individual maser clumps. A closer look at the continuum maps of the ATOMIUM sources is also available in \citet{2025Danilovich}.

Spectral cubes were made after continuum subtraction, adjusted to constant velocity in the Local Standard of Rest (LSR) convention. ATOMIUM used a spectral resolution of $\sim$1.1--1.35 \kms\ suitable for thermal lines. Individual maser spectral features can be narrow and hence are not always spectrally resolved in this data. However, the total velocity spans are well-sampled, being typically in the range 5--20 \kms. Typical rms per channel were 1--2 mJy and 2--4 mJy for extended and mid configurations, respectively (see Table E.2 of \citet{2022A&A...660A..94G} for values for each spectral window). The individually attained noise levels were contingent upon the elevation of the source, weather conditions, and the total integration time on target. For masers brighter than about 1 Jy, the noise was also limited by dynamic range effects.

\begin{table}
 \caption{Detected \ce{^{n}SiO} transitions within the frequency coverage of\\ATOMIUM.}\centering
 \begin{tabular}{llccccr}
 \hline \hline
 Line \# & $^n$Si & \textit{J}$_\text{u}$ & \textit{J}$_\text{l}$ & \textit{v} & $f$ & \multicolumn{1}{c}{$E_{\rm{u}}$} \\ [0.5ex]
 & & & & & (GHz) & \multicolumn{1}{c}{(K)} \\
 \hline 
 1 & 28 & 5 & 4 & 2 & 214.08857 & 3551.7 \\
 2 & 29 & 5 & 4 & 0 & 214.38575 & 30.9 \\
 3 & 28 & 5 & 4 & 1 & 215.59602 & 1800.1 \\
 4 & 28 & 5 & 4 & 0 & 217.10492 & 31.3 \\
 5 & 30 & 6 & 5 & 5 & 245.50644 & 8613.0 \\
 6 & 28 & 6 & 5 & 8 & 246.07886 & 13717.2 \\
 7 & 29 & 6 & 5 & 6 & 246.61956 & 10336.9 \\
 8 & 30 & 6 & 5 & 4 & 247.24533 & 6932.1 \\
 9 & 29 & 6 & 5 & 3 & 251.93012 & 5265.6 \\
 10 & 30 & 6 & 5 & 1 & 252.47137 & 1790.1 \\ 
 11 & 28 & 6 & 5 & 4 & 253.28559 & 7016.1 \\
 12 & 30 & 6 & 5 & 0 & 254.21665 & 42.7 \\
 13 & 28 & 6 & 5 & 3 & 255.09116 & 5298.6 \\
 14 & 29 & 6 & 5 & 1 & 255.47849 & 1800.9 \\
 15 & 28 & 6 & 5 & 1 & 258.70732 & 1812.5 \\
 16 & 28 & 6 & 5 & 0 & 260.51801 & 43.8 \\
 \hline 
 \end{tabular}
 \\
 \raggedright
 \textbf{Notes:} The columns give the line number, the isotope of Si (denoted by $n$), the upper and lower rotational levels, the vibrational state, the line frequency in GHz taken from the CDMS catalogue \citep{1991ApJ...368L..19M,2005JMoSt.742..215M,2013JPCA..11713843M}, and the upper state energy in K \citep{2013MNRAS.434.1469B}. The oxygen isotope for all listed transitions is \ce{^{16}O}.
 \label{table:line-coverage}
\end{table}

\section{Identification and characterisation of high-$J$ S\MakeLowercase{i}O lines}
\label{sec:SiO-id}

\begin{figure*}
  \centering
  \begin{subfigure}[b]{0.31\textwidth}
    \includegraphics[width=\textwidth]{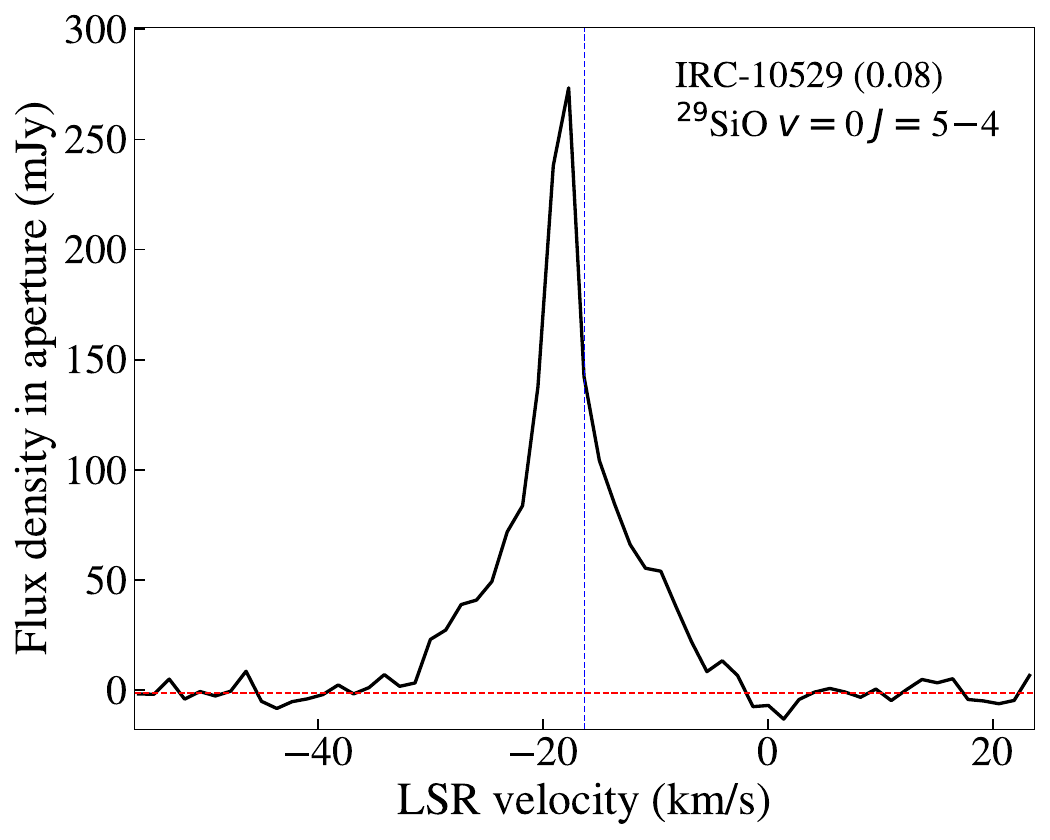}
  \end{subfigure}
  \begin{subfigure}[b]{0.31\textwidth}
    \includegraphics[width=\textwidth]{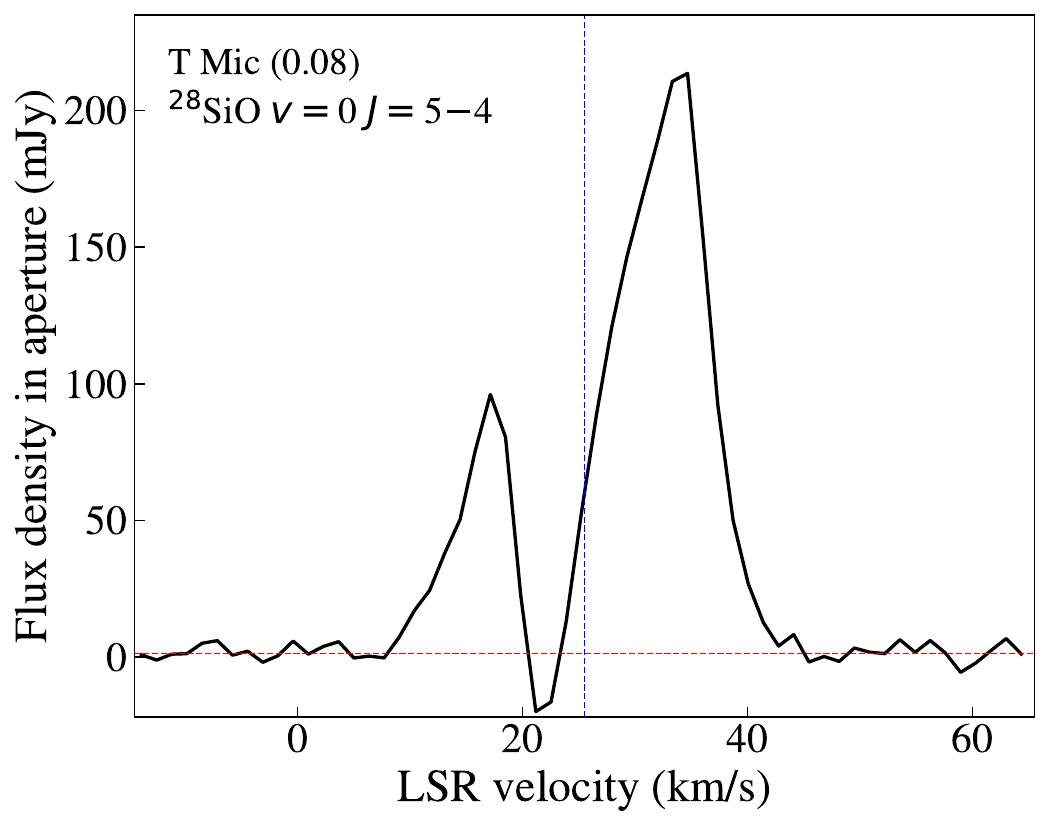}
  \end{subfigure}
  \begin{subfigure}[b]{0.31\textwidth}
    \includegraphics[width=\textwidth]{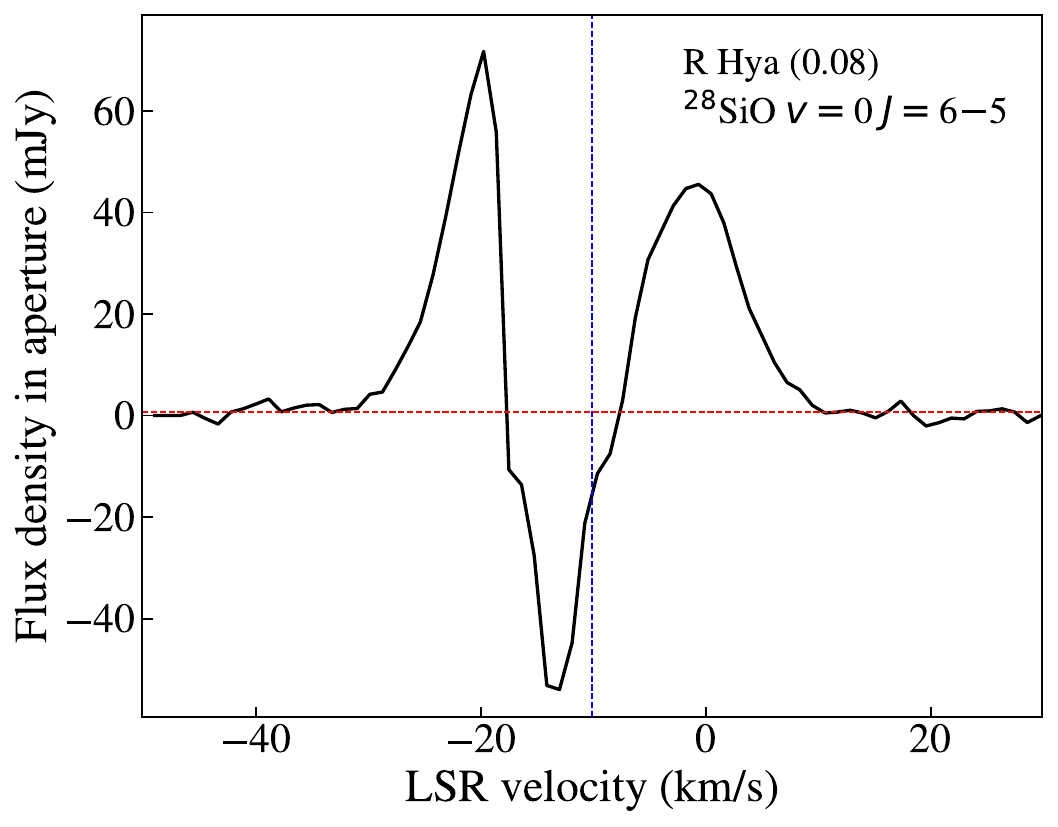}
  \end{subfigure}
  
  \begin{subfigure}[b]{0.31\textwidth}
    \includegraphics[width=\textwidth]{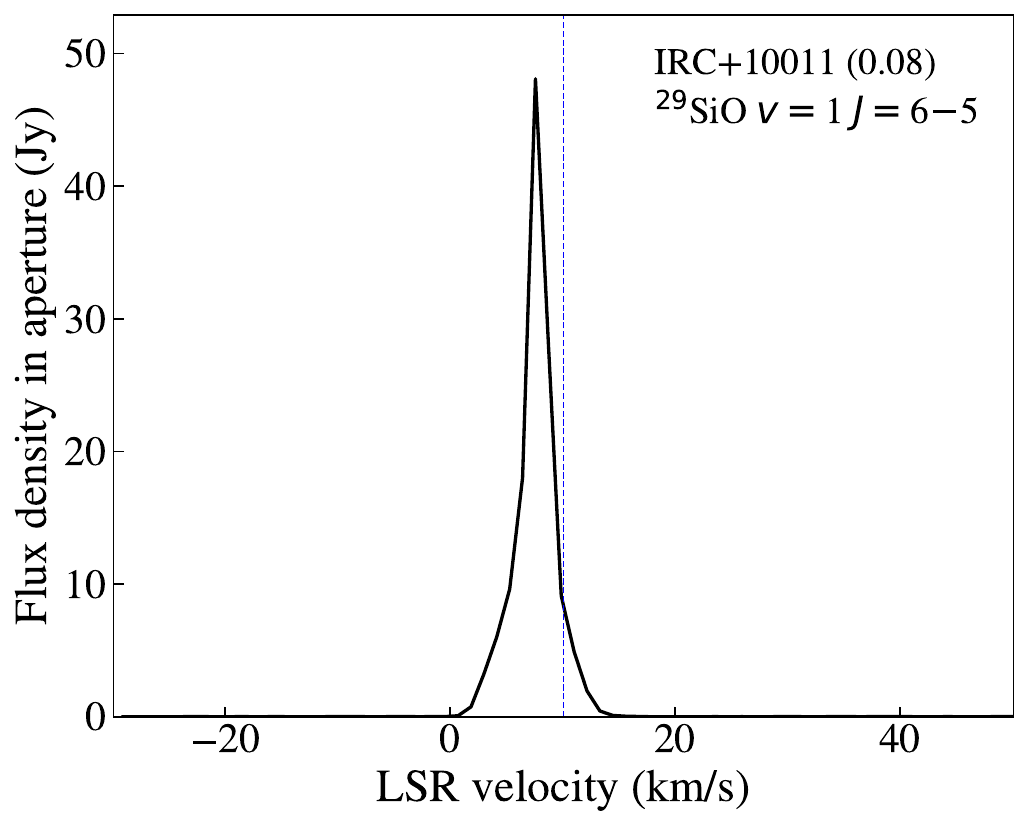}
  \end{subfigure}
  \begin{subfigure}[b]{0.315\textwidth}
    \includegraphics[width=\textwidth]{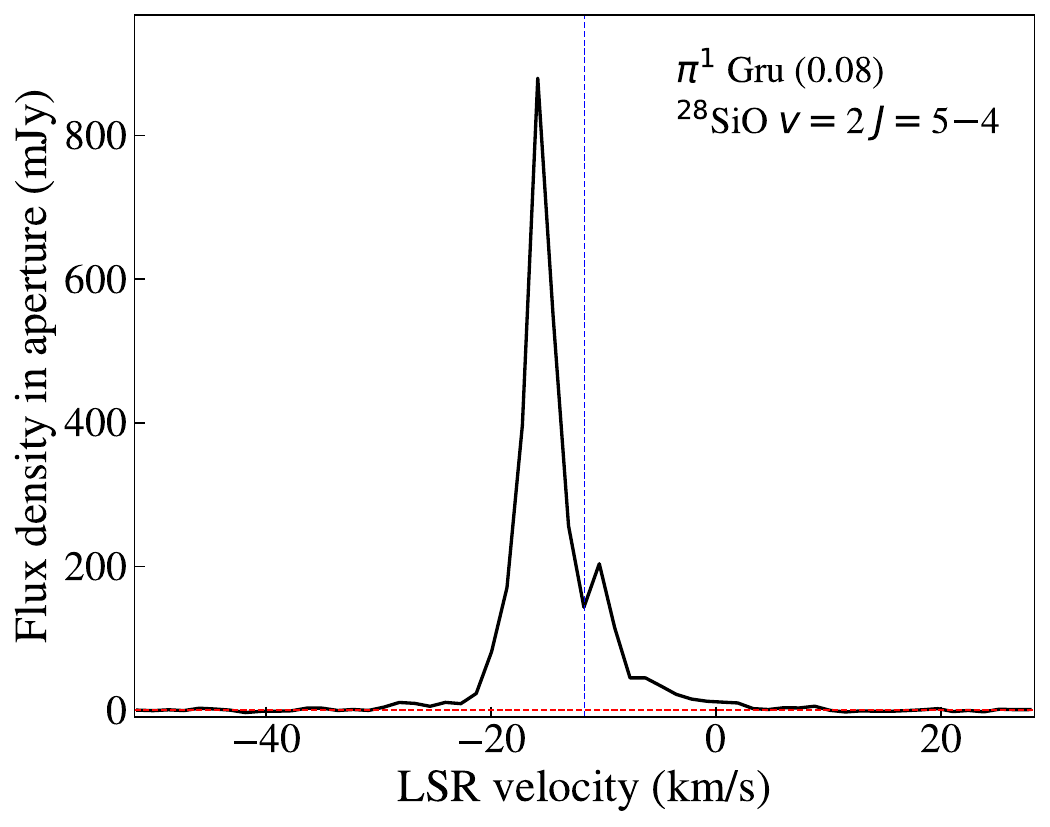}
  \end{subfigure}
  \begin{subfigure}[b]{0.31\textwidth}
    \includegraphics[width=\textwidth]{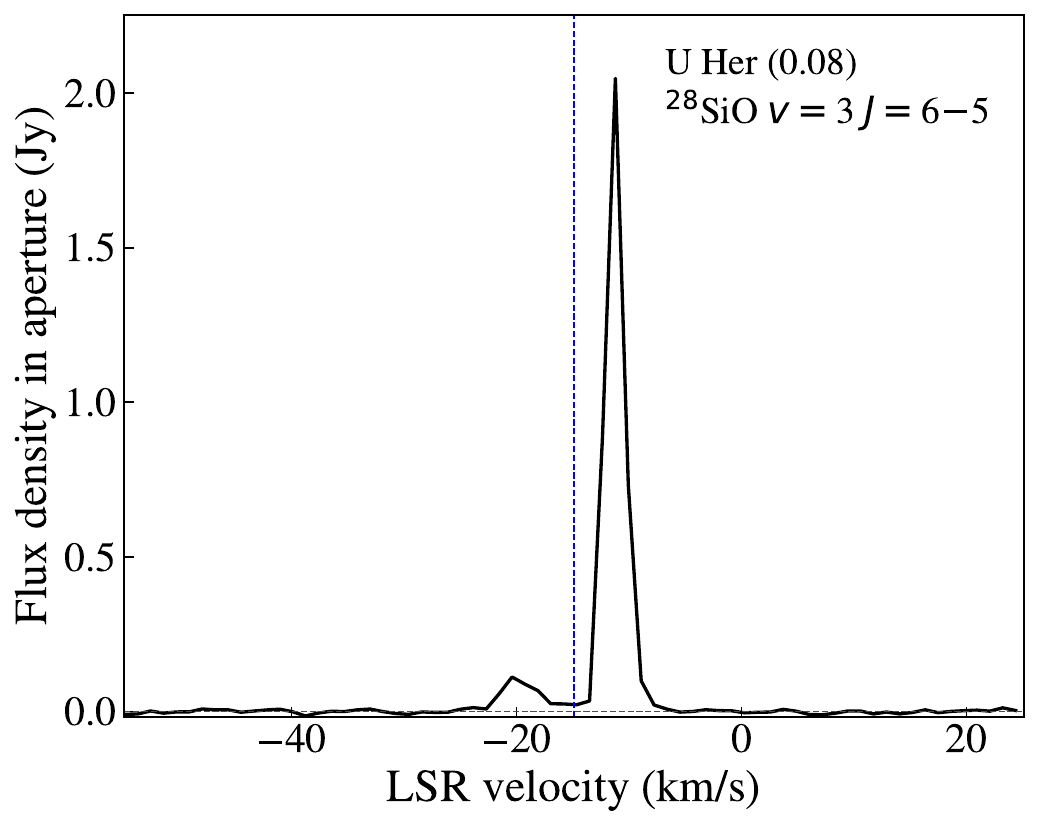}
  \end{subfigure}
  
  \begin{subfigure}[b]{0.31\textwidth}
    \includegraphics[width=\textwidth]{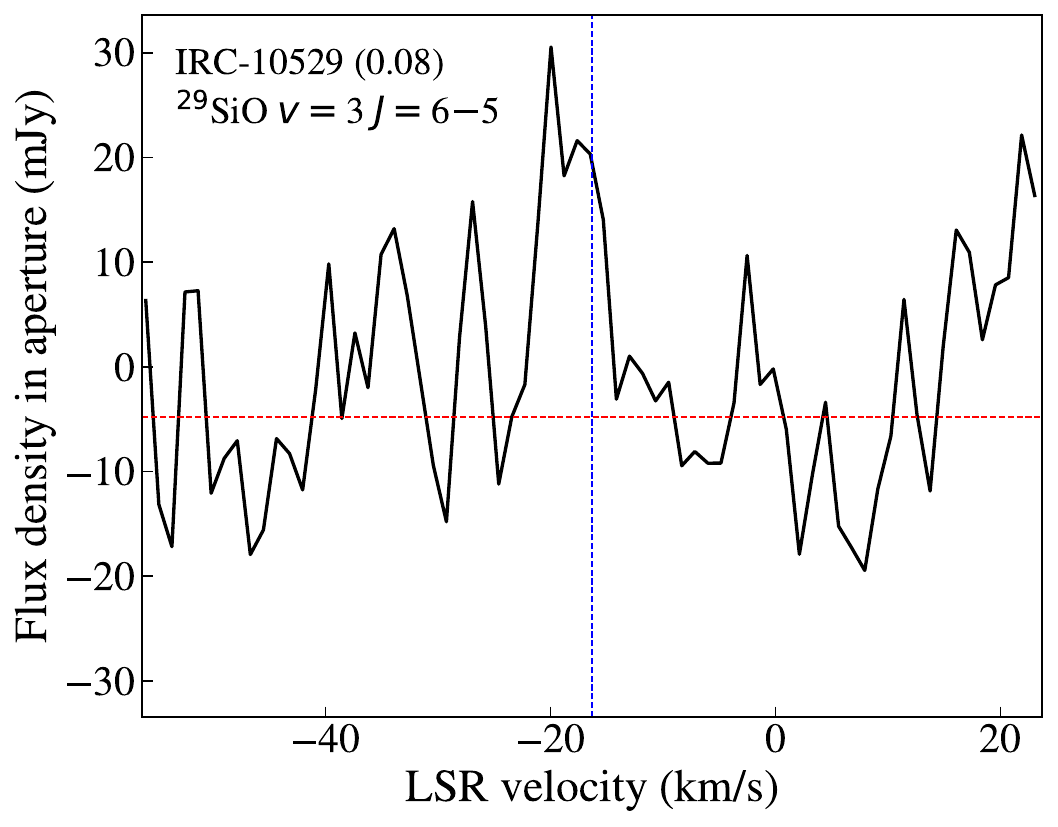}
  \end{subfigure}
  \begin{subfigure}[b]{0.31\textwidth}
    \includegraphics[width=\textwidth]{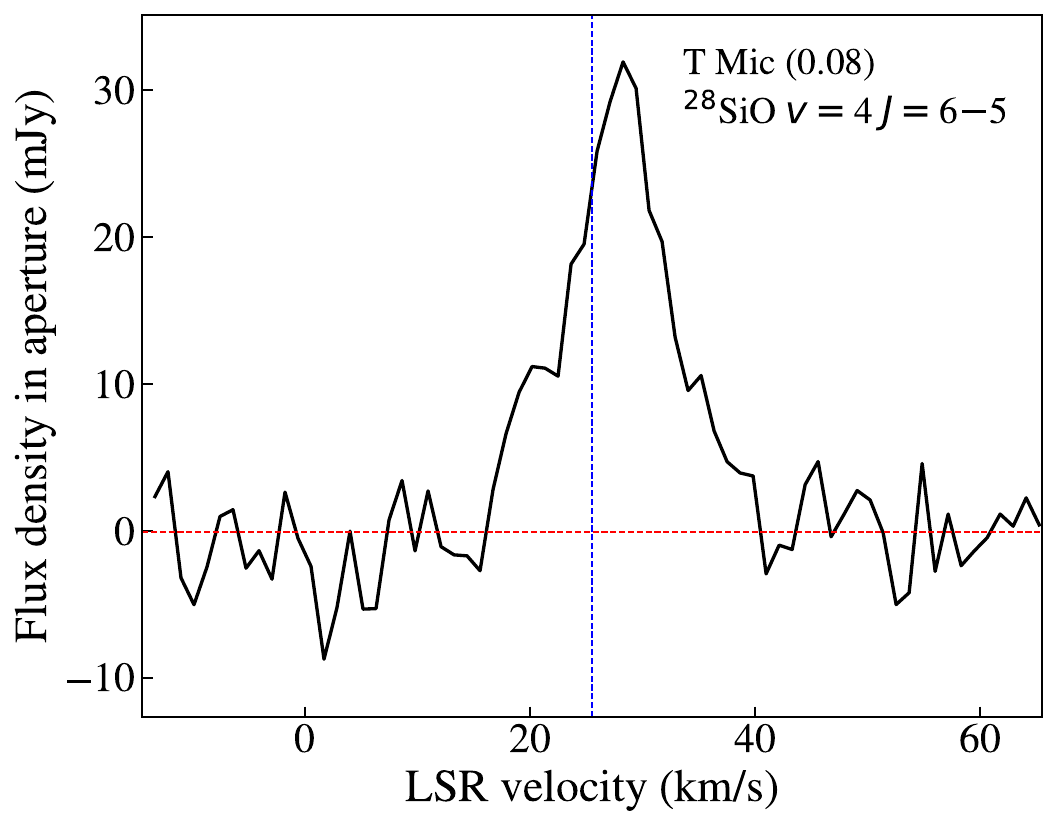}
  \end{subfigure}
  \begin{subfigure}[b]{0.31\textwidth}
    \includegraphics[width=\textwidth]{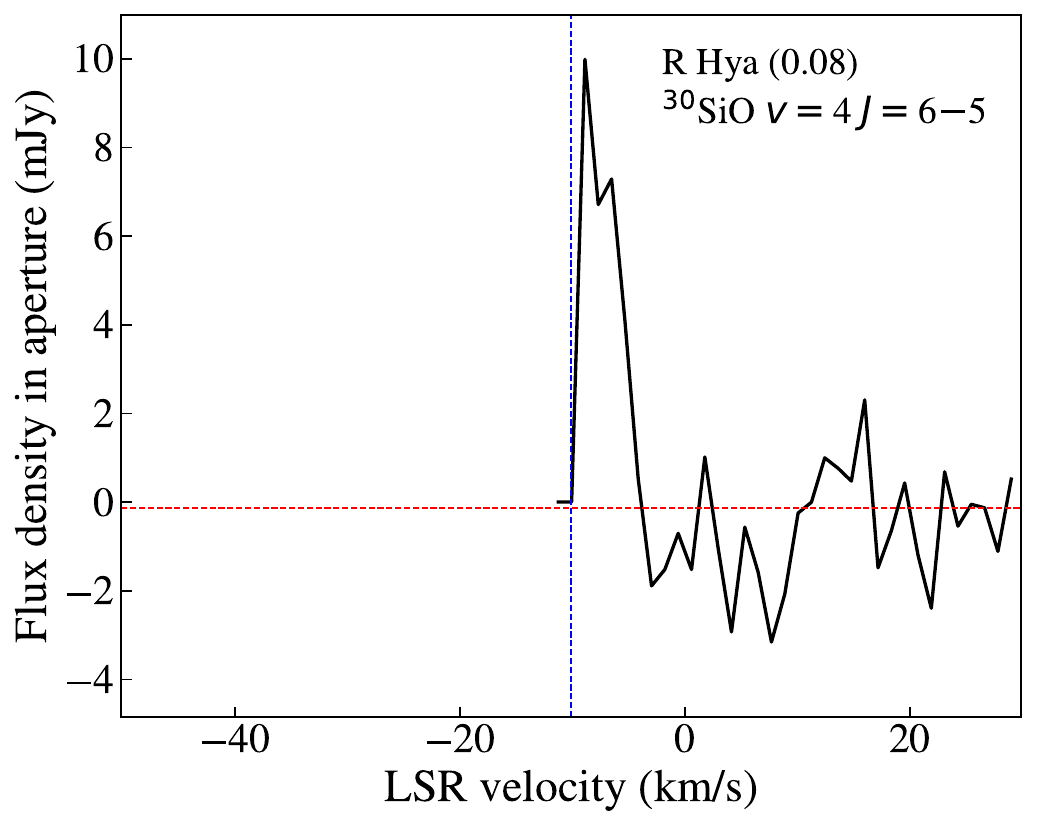}
  \end{subfigure}
  
  \begin{subfigure}[b]{0.32\textwidth}
    \includegraphics[width=\textwidth]{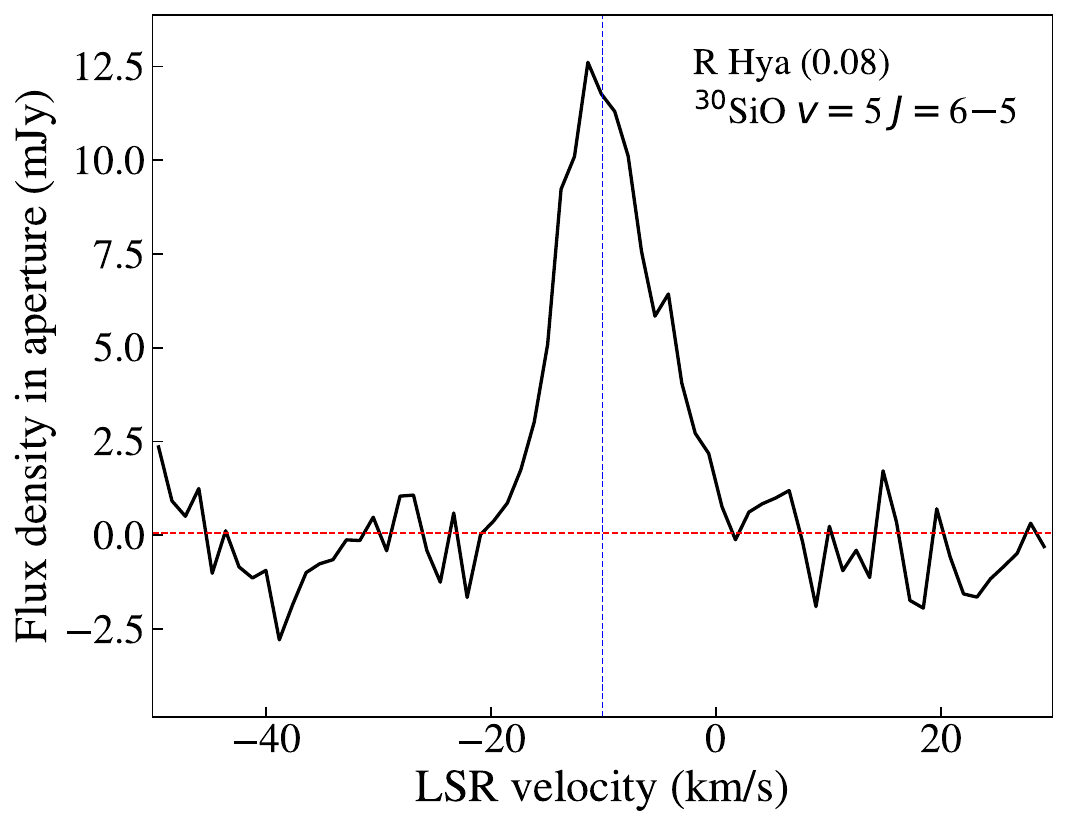}
  \end{subfigure}
  \begin{subfigure}[b]{0.31\textwidth}
    \includegraphics[width=\textwidth]{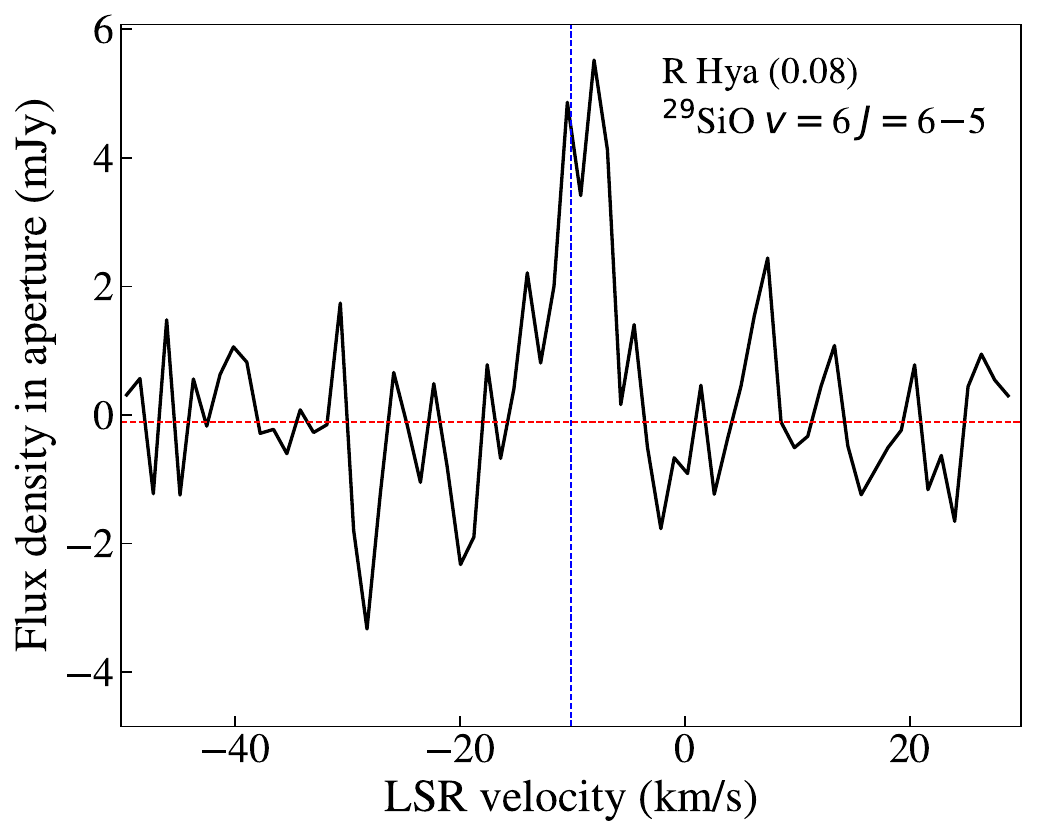}
  \end{subfigure}
  \begin{subfigure}[b]{0.31\textwidth}
    \includegraphics[width=\textwidth]{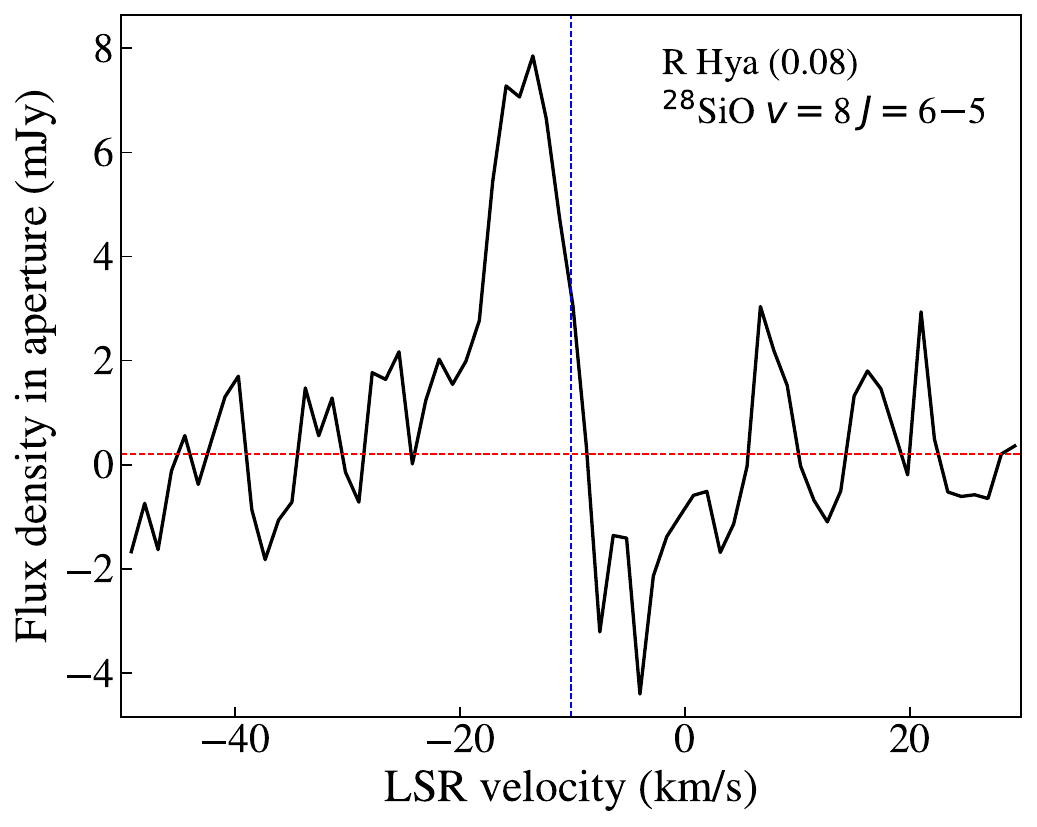}
  \end{subfigure}

  \caption{Continuum-subtracted SiO line profiles from the extended-configuration data for $\varv=$ 0 to $\varv=8$ transitions, extracted with an aperture of 0.08 arcsec in diameter (shown in brackets). The source and transition are noted atop each panel. The spectra are centred on the systemic velocity for each source (blue dashed line) as given in Table \ref{table:atomium-sources}. The red horizontal line shows the fitted spectral baseline used for rms and S/N estimates. Note that the lack of flux below $-10$ \kms\ in the \ce{^{30}SiO} $\varv=4$, $J=6-5$ panel is owing to the line being situated right at the boundary of the spectral window (see Fig. 1 of \citealt{2022A&A...660A..94G}).}
  \label{fig:example-spectra}
\end{figure*}

% % All detections of SiO transitions
\begin{table*}
\centering
\caption{Peak flux density (in mJy) of observed SiO lines extracted with an aperture diameter of 0.08 arcsec from the extended-configuration ATOMIUM data. The velocity extent of the spectral feature in \kms\ is highlighted in italics. The third line in the header shows the vibrational state $\varv$.}
\begin{tabular}{l|llll|llllllllllllll}
\hline \hline
& \textbf{$J=5-4$} & & & & \textbf{$J=6-5$} & & & & & & & & & & & \\
Source & \ce{^{28}SiO} & & & \ce{^{29}SiO} & \ce{^{28}SiO} & & & & &  \ce{^{29}SiO} & & & \ce{^{30}SiO} & & &\\
& 0 & 1 & 2 & 0 & 0 & 1 & 3 & 4 & 8 & 1 & 3 & 6 & 0 & 1 & 4 & 5\\
\hline

S Pav & 120 & \textbf{222} & 45 & 31 & 91 & 135 & * & 57 & * & 162 & * & -- & 99 & 44 & -- & 12?\\
& \textit{24.3} & \textit{23.1} & \textit{19.1} & \textit{19.1} & \textit{28.1} & \textit{22.6} & & \textit{12.7} & & \textit{16.0} & & & \textit{18.4} & \textit{16.2} & & \vspace{2mm} \\

U Del & 24 & \textbf{58} & -- & 10? & 39 & 26 & -- & -- & -- & -- & -- & -- & 12 & -- & -- & --\\
& \textit{14.8} & \textit{19.0} & & & \textit{23.6} & \textit{10.2} & & & & & & & 1.2 & & & \vspace{2mm}\\

V PsA & 150 & \textbf{1471} & 9? & 34 & 98 & 67 & -- & -- & -- & 27 & -- & -- & 48 & 14? & -- & --\\
& \textit{36.4} & \textit{28.5} & & \textit{24.6} & \textit{34.8} & \textit{46.4} & & & & \textit{14.9} & & & \textit{24.2} & & & \vspace{2mm}\\

U Her & 97 & 4007 & 1744 & 67 & 110 & 2104 & 2047 & 19 & -- & \textbf{11827} & -- & -- & 133 & 30 & 8? & --\\
& \textit{21.6} & \textit{24.4} & \textit{13.7} & \textit{17.7} & \textit{22.5} & \textit{45.3} & \textit{13.8} & \textit{10.4} & & \textit{18.3} & & & \textit{35.7} & \textit{8.1} & & \vspace{2mm}\\

SV Aqr & \textbf{88} & 20 & -- & 40 & 80 & 28 & -- & -- & -- & -- & -- & -- & 42 & -- & -- & --\\
& \textit{28.3} & \textit{17.7} & & \textit{19.1} & \textit{28.1} & \textit{14.7} & & & & & & & \textit{20.7} & & & \vspace{2mm}\\

RW Sco & 88 & \textbf{4141} & 134 & 81 & 104 & 3943 & -- & -- & -- & 1587 & -- & -- & 103 & 17? & -- & --\\
& \textit{45.8} & \textit{20.4} & \textit{6.8} & \textit{5.5} & \textit{16.9} & \textit{19.2} & & & & \textit{13.8} & & & \textit{15.0} & & & \vspace{2mm}\\

T Mic & 213 & \textbf{804} & 32 & 77 & 228 & 134 & 17* & 32 & -- & 149 & -- & -- & 85 & 25 & -- & --\\
& \textit{31.0} & \textit{24.4} & \textit{17.8} & \textit{19.1} & \textit{32.6} & \textit{33.9} & \textit{5.7} & \textit{12.7} & & \textit{20.6} & & & \textit{27.6} & \textit{17.4} & & \vspace{2mm}\\

$\pi^1$ Gru & 111 & \textbf{885} & 879 & 47 & 120 & 400 & 365 & -- & -- & 27 & -- & -- & 72 & 28 & -- & --\\
& \textit{39.1} & \textit{31.2} & \textit{30.1} & \textit{24.6} & \textit{52.8} & \textit{53.2} & \textit{3.4} & & & \textit{9.2} & & & \textit{28.8} & \textit{15.1} & & \vspace{2mm}\\

R Hya & 79 & \textbf{887} & 76 & 46 & 72 & 127 & * & 43* & 8? & 69 & 9?* & 6? & 62 & 48 & 10 & 13\\
& \textit{27.0} & \textit{31.2} & \textit{26.0} & \textit{20.5} & \textit{28.1} & \textit{45.3} & & \textit{12.7} & \textit{3.6} & \textit{16.0} & & \textit{3.6} & \textit{23.0} & \textit{19.7} & \textit{2.4} & \textit{9.5} \vspace{2mm}\\

W Aql & \textbf{184} & 69 & 22 & 73 & 85 & 80 & 22 & 10? & -- & 38 & -- & -- & 66 & 24 & -- & --\\
& \textit{29.7} & \textit{14.9} & \textit{13.7} & \textit{20.5} & \textit{29.2} & \textit{17.0} & \textit{5.7} & \textit{2.3} & & \textit{13.7} & & & \textit{27.6} & \textit{10.4} & & \vspace{2mm}\\

R Aql & 67 & \textbf{2213} & 17 & 34 & 39 & 39 & * & 25* & * & 648 & * & -- & 23 & 21 & -- & --\\
& \textit{6.7} & \textit{20.4} & \textit{1.4} & \textit{5.5} & \textit{2.2} & \textit{14.7} & & \textit{8.1} & & \textit{12.6} & & & \textit{11.5} & \textit{10.5} & &  \vspace{2mm}\\

IRC & 141 & 114 & 192 & 273 & 168 & 147 & 52 & 40 & -- & \textbf{1601} & 31? & -- & 137 & 71 & -- & --\\
$-$10529 & \textit{32.3} & \textit{16.3} & \textit{9.6} & \textit{23.2} & \textit{32.6} & \textit{17.0} & \textit{5.7} & \textit{3.5} & & \textit{10.3} & & & \textit{23.0} & \textit{9.3} & & \vspace{2mm}\\

GY Aql & 86 & \textbf{10218} & 279 & 57 & 103 & 2647 & 42? & 41? & -- & 859 & -- & -- & 182 & 29? & -- & --\\
& \textit{16.2} & \textit{19.0} & \textit{10.9} & \textit{13.6} & \textit{13.5} & \textit{13.6} & \textit{2.3} & \textit{4.6} & & \textit{11.5} & & & \textit{27.6} & \textit{7.0} & &\vspace{2mm}\\

IRC & 508 & 927 & 11040 & 1284 & 178 & 192 & 760 & 75 & -- & \textbf{48101} & -- & -- & 142 & 296 & 24? & 17?\\
+10011 & \textit{32.3} & \textit{23.1} & \textit{13.7} & \textit{28.7} & \textit{32.6} & \textit{52.0} & \textit{11.5} & \textit{9.2} & & \textit{14.9} & & & \textit{33.4} & \textit{16.2} & & \vspace{2mm}\\

\hline
\end{tabular}
\raggedright
\textbf{Notes:} Typical peak flux density uncertainties are 1--4 mJy. The strongest SiO line detection in the source is highlighted in \textbf{bold}. The velocity extent is determined by the difference between the red and blue velocities for line emission exceeding 2.5$\sigma$ \citep{2023A&A...674A.125B}. No detection is shown as --, whereas ? marks a possible detection (i.e. $3.0\lesssim \mathrm{S/N} < 5.0$). Asterisks (*) are used here to denote apparent absorption.
\label{table:SiO-flux}
\end{table*}

In this section, we present an overview of the SiO spectral profiles, spatial distributions, and line identifications obtained from the extended-configuration ALMA observations. We describe the criteria used for line detection and summarise trends observed across different isotopologues and vibrational states from $\varv=0$ to $\varv=8$. More discussion on the classifications of (quasi-)thermal\footnote{We use “quasi-thermal emission” throughout the manuscript to better reflect the likely non-LTE conditions of the SiO excitation. In the present work, we do not attempt to distinguish rigorously between purely thermal and weakly non-LTE excitation.} vs maser emission is presented later in Section \ref{sec:dist:analysis}.

\subsection{Overview: spectral profiles}
\label{sec:SiO-id:overview-spec}
Considering that SiO molecules are excited in the inner CSE and that the ALMA Band 6 can provide us with high sensitivity and high spectral resolution, we used the extended-configuration ATOMIUM data to search for the SiO lines in the sources. Spectral extraction from the extended-configuration data cubes was performed  using circular apertures of 0.04, 0.08, 0.2, 1.2, 2 arcsec in diameter. In general, larger extraction diameters did not yield additional flux for vibrationally excited ($\varv>0$) lines. Consequently, an extraction diameter of 0.08 arcsec was systematically adopted for $\varv>0$ transitions, as it ensures we have considered most if not all flux density, and gives the optimal S/N ratio. For the $\varv=0$ lines, where emission is spatially more extended, a 1.2-arcsec diameter aperture generally recovers almost all the flux. However, such a large aperture can make compact maser components less prominent in the extracted spectra by including additional extended emission and noise. As a result, for discussions involving the most compact SiO emission, we keep the extraction aperture of 0.08 arcsec across the sample unless explicitly stated otherwise. 

The sources of uncertainty come from the limited spectral resolution of $\sim$1 MHz (i.e. 1.4--1.1 \kms\ at 214--260 GHz), and the estimation of stellar velocity, $V_*$ \citep{2022A&A...660A..94G}, although they do not cause any ambiguity in identifying the line transitions in the data. 

A summary of the SiO line detections, including both thermal and maser emission, in the ATOMIUM data set is given in Table \ref{table:SiO-flux} and in the text below, categorised in the order of vibrational states. In this work, thermal emission refers to the radiation emitted by SiO lines whose profiles are dominated by Doppler broadening and/or local turbulence at the velocity of the bulk motion of the CSE. Maser emission exhibits narrower line profiles resulting from stimulated amplification of the radiation. Examples of the spectral characteristics discussed in this section are shown in Fig. \ref{fig:example-spectra}, where $\varv$ increases progressively from top to bottom. A complete compilation of the SiO line detections in the ATOMIUM data is available in the online supplementary material, which we encourage the reader to consult.

Choosing the flux cut-off level for line detection is somewhat arbitrary and can be different from one study to another. Here, we decided that 3$\sigma_{\rm{rms}}$ and 5$\sigma_{\rm{rms}}$ (written as $\sigma$, hereafter) in 2 consecutive channels were the appropriate cutoffs for tentative and clear line detection, respectively, in our data based on the inspection of the extracted spectra. We note that the observed flux densities can include both thermal and maser contributions and can therefore be biased in the transitions heavily dominated by thermal emission such as those originating from the $\varv=0$ state. In this work, we therefore classify masing strictly on empirical grounds (brightness temperature and ancillary diagnostics), and treat high-$\varv$ detections as maser candidates only when they meet those criteria (Section \ref{sec:dist:analysis:Tb}) or when discussed otherwise. It should also be noted that maser excitation is phase-dependent, so detections that fail to meet the maser criteria are not necessarily evidence of absence. 

Table \ref{table:SiO-flux} shows the peak flux densities (in mJy) and velocity extents of line emission $>$ 3$\sigma$ in the ATOMIUM extended-configuration observations extracted with an aperture diameter of 0.08 arcsec, where the values in bold highlight the strongest SiO line emission detected towards the ATOMIUM sources. Possible detections (3.0 $\lesssim$ S/N $<$ 5.0) and absorption features are also marked with ? and *, respectively. Velocity extents are measured as the separation between red- and blue-shifted velocities where the line emission exceeds 2.5$\sigma$ \citep{2023A&A...674A.125B}. We acknowledge that the full characterisation of these highly excited SiO lines is inconclusive without polarisation and time variability information and the border between a weak and a strong maser remains open to discussion. The spatial extents and $T_{\rm{b}}$ estimates, which strongly characterise maser emission, are discussed later in Sections \ref{sec:dist} and \ref{sec:dist:analysis:Tb}, respectively.

In the ATOMIUM source sample, we detected high-$J$ SiO emission in the ground, and up to the eighth excited vibrational state (i.e. $\varv=0$ to $\varv=8$). The number of sources for which we have a (possible) spectral identification as defined above, varies from 1 (\ce{^{28}SiO} $\varv=8$, $J=6-5$ and \ce{^{29}SiO} $\varv=6$, $J=6-5$) to all 14 (e.g. \ce{^{28}SiO} $\varv=0, 1$, $J=5-4, 6-5$). The source with the highest number of possible SiO line detections is R Hya, where 15 out of 16 transitions were potentially detected at a level of 3$\sigma$ or better. The only exception is the \ce{^{28}SiO} $\varv=3$, $J=6-5$ line, which appears in absorption. The most SiO line-poor sources are undoubtedly SV Aqr and U Del, both exhibiting only the $\varv \leq 1$ lines and are the only two sources where all detections are below 100 mJy, despite the stars having comparable distances to some of the other sources e.g. GY Aql. The richest highly excited SiO spectra are observed in IRC+10011, with most of the detected peak flux densities being in the top three compared to other sources, even though IRC+10011 is the second most distant source. The \ce{^{29}SiO} $\varv=1$, $J=6-5$ line detected towards this source reaches a striking 48 Jy, the strongest SiO line emission in the ATOMIUM data set. 

It is of note that transitions from higher-lying states were less frequently observed among the sampled sources, as when the energy of the transition increases (Table  \ref{table:line-coverage}), populating the corresponding energy levels becomes progressively more challenging. In Section \ref{sec:compare:common}, we discuss additional, potential factors specific to maser excitation. Line identification did not suffer from spectral confusion problems, but we also compared the list of SiO detections with the full chemical inventory of the ATOMIUM observations \citep{2024A&A...681A..50W} as a double check. Further discussions regarding how our detections may be affected by pulsation period and phase and how the SiO emission, especially masers, from different vibrational states are related are given in Sections \ref{sec:compare:detection-rate}--\ref{sec:compare:common}.

\begin{figure*} 
   \includegraphics[width=0.87\textwidth, height=0.5\textwidth]{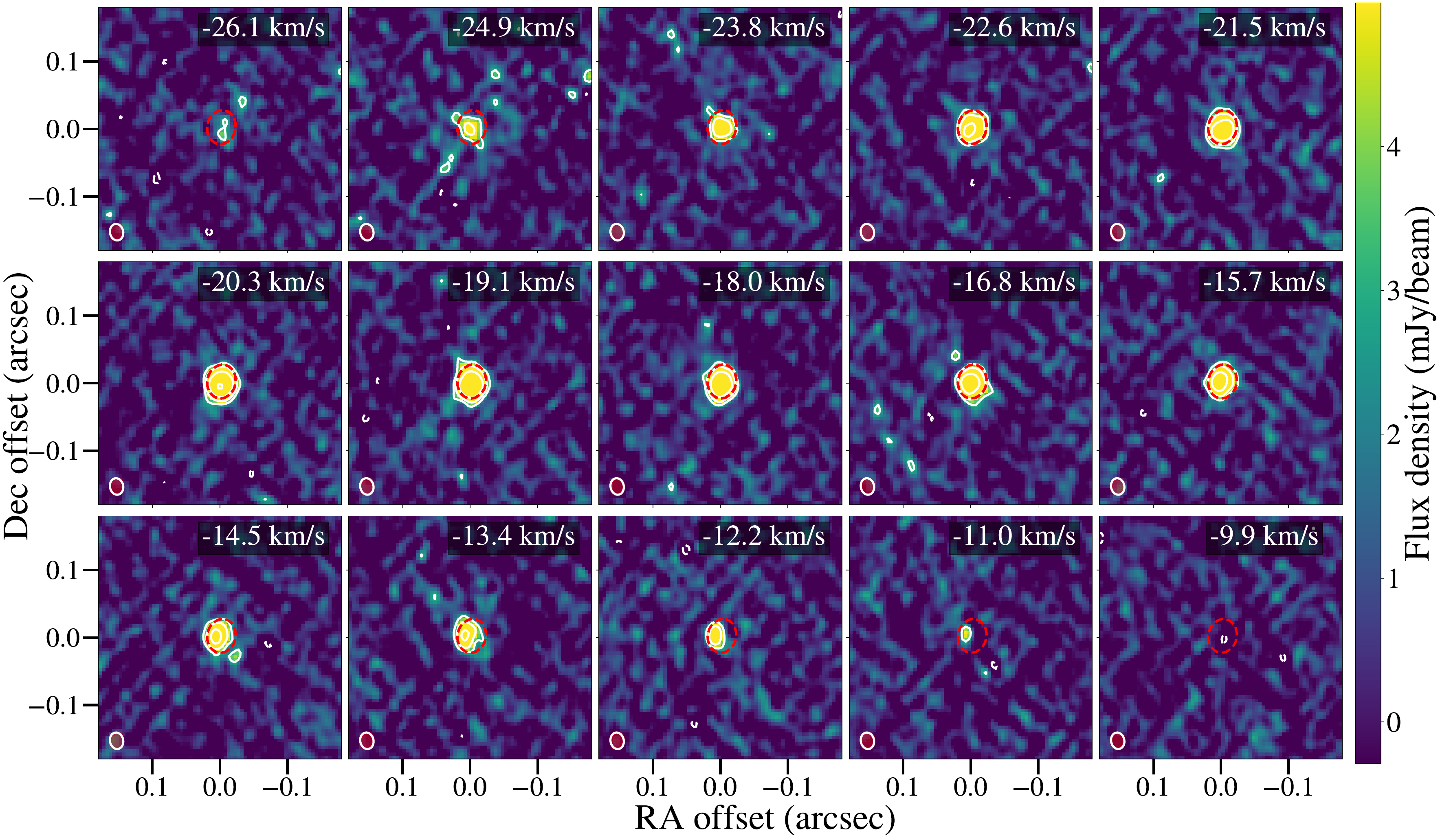}
   \includegraphics[width=0.87\textwidth, height=0.5\textwidth]{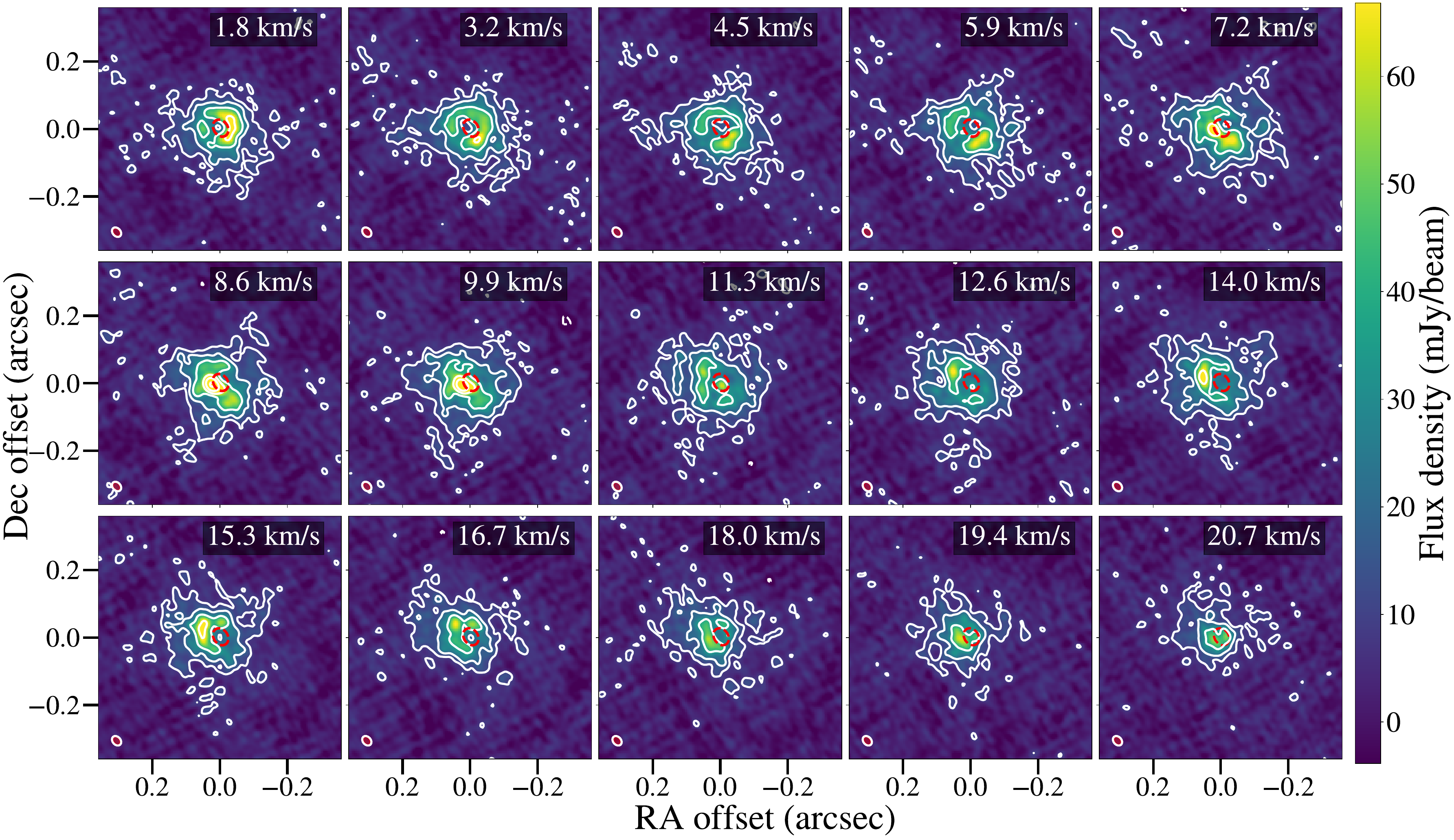} 
   \caption{High-resolution channel maps of the \ce{^{28}SiO} $\varv=4$, $J=6-5$ line at 253.286 GHz (top, S Pav) and the \ce{^{28}SiO} $\varv=0$, $J=5-4$ line at 217.105 GHz (bottom, IRC+10011), which show RA and Dec offsets (arcsec) relative to the continuum peak over an area of 0.36 $\times$ 0.36 arcsec (top) and 0.72 $\times$ 0.72 arcsec (bottom). A red dashed contour at (0,0) indicates the extent of 10 per cent continuum level. Velocities are centred around the systemic velocities in Table \ref{table:atomium-sources}. The colour bar indicates line flux density in mJy beam$^{-1}$. White contours mark $-$3$\sigma$, 3$\sigma$, 6$\sigma$, and 12$\sigma$. Negative contours, when present, are shown with dashed lines. The line peak and typical rms noise are 24 and 1 mJy beam$^{-1}$ (S Pav) and 318 and 3 mJy beam$^{-1}$ (IRC+10011). Synthesised beams for line and continuum are shown in white and red, respectively, in the lower-left corner. For the line observations, these are (25 $\times$ 20) mas at PA 12$^\circ$ for S Pav and (32 $\times$ 24) mas at PA 42$^\circ$ for IRC+10011, respectively. Similarly, the continuum parameters are (25 $\times$ 20) mas at PA 13$^\circ$ (S Pav) and (27 $\times$ 19) mas at PA 31$^\circ$  (IRC+10011).}
   \label{fig:channel-maps}
\end{figure*}

\subsection{Overview: channel maps}
\label{sec:SiO-id:overview-maps}
Channel maps were created for each source in the sample for all the detected transitions listed in Table \ref{table:SiO-flux}. As with the spectra, these are primarily extracted from the extended-configuration data cubes. Here, the focus is on the angular sizes and structures of the SiO emitting regions, while the absorption features of the $\varv>0$ lines from \ce{^{28}SiO} and \ce{^{29}SiO} are discussed in Section \ref{sec:absorption}. The following selected channel maps, used in tandem with the spectra (Fig. \ref{fig:example-spectra} and online supplementary material) during discussions about the detected SiO emission, were chosen to highlight the different characteristics observed in different $\varv$ states. More examples of the channel maps can be found in Appendix \ref{appendix:additional-maps}.

Most of the highly excited SiO lines observed in the ATOMIUM sources appear to be spatially concentrated near or coincident with the continuum peak (e.g. S Pav, top panel of Fig. \ref{fig:channel-maps}). The emission is often slightly displaced in individual channels and can appear ring-like and clumpy in nature (e.g. IRC+10011, bottom panel of Fig. \ref{fig:channel-maps}). In a few cases, there is also a systematic offset relative to the continuum (e.g. $\pi^1$ Gru and T Mic, Fig. \ref{fig:add-maps-1}). For $\pi^1$ Gru, the morphology and direction of the offset is probably due to the close companion \citep{2025A&A...699A..22M}, which introduces complicated wind interaction (see also \citealt{2020A&A...644A..61H,2026NatAs..10..124E}). Note that AGB stars are typically optically thick, so emission (or indeed absorption) towards the star must be from the near side to the observer.

Following the technique employed in \citet{2021A&A...655A..80D} and \citet{2023A&A...674A.125B}, we estimated the angular sizes of the regions exhibiting SiO emission by calculating the azimuthal average of the 3$\sigma$ contour representing the lower limit of line detection in the zeroth moment (mom0) maps. The angular radii determined with this method range from approximately 10 to 40 mas for all stars and detected lines. The variations in angular radii in the units of $R_*$ can be as much as a factor of 2 to 3 for different lines in a given star, and 3 to 4 times for the most widespread $\varv=1$ transitions across different stars. Comparing these to the sizes derived from highly excited H$_2$O lines in \citet{2023A&A...674A.125B}, the upper limit of the angular radii observed with SiO appears to be $\sim$20 per cent smaller. We then considered the maximum angular separation from the central star to the 3$\sigma$ contour to be approximately the SiO excitation size in this work, which is between 15 and 50 mas (i.e. $\sim$2--5 $R_*$ from the centre of the star) for most lines and stars. These radii place the ATOMIUM high-$J$ SiO emission within the chemically active dust-formation zone, consistent with co-spatial distributions of SiO/H$_2$O/AlO emission and dust clouds in W Hya \citep{2025A&A...704A..18O}.    \citet{1993AJ....105..595S} showed that rapid depletion of SiO from the gas phase occurs at radii larger than about 1--2 $\times$ $10^{15}$ cm, initially through adhesion onto dust grains and subsequently through photodissociation by interstellar UV radiation, while modelling of Mira \citep{2016A&A...590A.127W} indicates that major SiO depletion does not begin until beyond $\sim$4 $R_*$. The high-$J$ SiO emission therefore likely traces gas immediately preceding, and partly overlapping with, the onset of condensation.

\subsection{Comments on maser variability}
\label{sec:SiO-id:variability}
Time variability is a well-known characteristic of maser emission and is often used as a way to identify a maser (e.g. \citealt{1977MNRAS.180..415B}; \citealt{1994A&A...286..501P}; \citealt{2002A&A...386..256H}; \citealt{2004A&A...424..145P}; \citealt{2010MNRAS.406..395G}, \citeyear{2013MNRAS.433.3133G}; \citealt{2020A&A...642A.213G}). Changes in the flux density of circumstellar masers suggest variations in maser pumping or amplification due to changing physical conditions in the CSEs, typically caused by pulsation-driven shock waves or variable levels of radiation of the host star. In this work, however, there is only one epoch of observations available per configuration per target, which means we have no information regarding the variability of the detected high-$J$ SiO masers (see Section \ref{sec:dist:analysis:Tb} for discussions concerning the distinction between quasi-thermal and maser emission in the current work). Any conclusion drawn from the list of SiO detections (Table \ref{table:SiO-flux}) and the discussions below should be considered with caution. The differences in line intensity ratios between the sources could arise from the fact that these observations were taken at different $\phi$, which have been found to directly influence how the SiO maser intensity varies (see e.g. \citealt{2009MNRAS.394...51G} and references therein) as the maser clumps react to the changes in number density, temperature and velocity brought about by pulsation shocks and variation in stellar and dust infrared radiation. A dedicated study on the variability of high-$J$ SiO masers is therefore encouraged to verify the findings obtained from these ATOMIUM observations. 

\subsection{Specific transitions}
To facilitate a direct comparison of intrinsic line strengths across sources located at varying distances (from 126 to 930 pc, Table \ref{table:atomium-sources}), we refer throughout this section to the specific line luminosity, $L_{\nu}$. This is defined in the isotropic approximation as $L_{\nu} \sim S_{\nu}D^2$, where $S_{\nu}$ is the observed peak flux density and $D$ is the distance to the star \citep{2016ApJ...817..115C,2019ARA&A..57..417C}. Note that we omit the 4$\pi$ geometric factor corresponding to isotropic emission. In SI units, 1 mJy kpc$^2$ is equivalent to $9.52\times10^9$ W Hz$^{-1}$.

\subsubsection{$\varv=0$ transitions}
\label{sec:SiO-id:v=0}
In general, most of the detected SiO emission from the vibrational ground state can be identified as thermal as the energies required for these transitions are only between 30 and 44 K (Table \ref{table:line-coverage}), well below the typical kinetic gas temperature, $T_{\rm{k}}$, in inner CSEs. The wide bases of $>$ 10 \kms\ in width in the $\varv=0$ spectra extracted from the image cubes are due to the projection effect of radial expansion seen from our line of sight plus some local perturbations. The $\varv=0$ transitions were detected with specific line luminosities ranging from 0.7 mJy kpc$^2$ (\ce{^{29}SiO} $J=5-4$ in R Hya, 46 mJy at 126 pc) to 666 mJy kpc$^2$ (\ce{^{29}SiO} $J=5-4$ in IRC+10011, 1284 mJy at 720 pc). The velocity line widths range from 1.2 \kms\ (\ce{^{30}SiO} $J=6-5$ in U Del) to 52.8 \kms\ (\ce{^{28}SiO} $J=6-5$ in $\pi^1$ Gru). The $\varv=0$ lines are generally not the strongest SiO line detections in a particular source, with two exceptions being the \ce{^{28}SiO} $J=5-4$ line in SV Aqr and W Aql at the time of observations (Table \ref{table:SiO-flux}). While R Hya is often detected due to its proximity (126 pc), it is intrinsically one of the faintest sources in the $\varv=0$ state, peaking at only 1.3 mJy kpc$^2$ in the \ce{^{28}SiO} $J=5-4$ line.

There is evidence of a narrow line component blended with the broader, less bright emission in e.g. IRC$-$10529 (Fig. \ref{fig:example-spectra}, top row, left). In this source, the \ce{^{29}SiO} $J=5-4$ line has a specific line luminosity of 236 mJy kpc$^2$ ($S_{\nu}$ = 273 mJy at 930 pc) at $-$18.2 \kms, 1.9 \kms\ to the blue of $V_*$ ($-$16.3 \kms). It exemplifies a typical shift (within $V_* \pm$ 5 \kms) in the line of sight velocities of the narrow $\varv=0$ line peaks, where present. The wide wings of the spectral feature (of flux density $\sim$50 mJy) are characteristic of thermal emission and may also include contributions from faint masers with velocities within $\pm$10 \kms\ from the systemic velocity. We acknowledge that precise separation of the thermal and masing components through detailed spectral fitting, especially for the $\varv=0$ lines, would provide valuable insights. However, such a decomposition is beyond the scope of this study, as it requires multi-epoch observations and extensive analysis to robustly characterise the stable thermal emission. Possible masing in the $\varv=0$ state of the \ce{^{28}SiO} and \ce{^{29}SiO} isotopes, identified by the narrow line profile atop the wider thermal base, appears to be more common in the observations towards the sampled AGB stars with $\dot{M} \gtrsim $10$^{-6}$ M$_\odot$ yr$^{-1}$. 

Another feature often observed in the $\varv=0$ spectra is the presence of absorption, with values ranging between $-$10 and $-$40 mJy. The velocity of this absorption is typically offset by approximately $-$5 \kms\ from $V_*$, as clearly demonstrated in the $\varv=0$ spectra of T Mic and R Hya (Fig \ref{fig:example-spectra}, top). Both sources also exhibit prominent absorption in the $\varv=3$ transition (see Section \ref{sec:absorption}).

\subsubsection{$\varv=1$ transitions}
\label{sec:SiO-id:v=1}
Emission from the first vibrationally excited ($\varv=1$) state of SiO ($E_{\rm{u}} \sim$1800 K) is, by far, the most commonly detected among the $\varv>0$ lines and all 14 targets likely exhibit masing activity from this vibrational state to varying degrees. Overall, the peak flux densities of the high-$J$ SiO $\varv=1$ lines are greater than other vibrational states within a source. 

We detected all four $\varv=1$ transitions (\ce{^{28}SiO} $J=5-4$, $J=6-5$; \ce{^{29}SiO} $J=6-5$; \ce{^{30}SiO} $J=6-5$) covered in the ATOMIUM observations, although each of the $\varv=1$ lines have different detection rates among the sampled AGB stars (Table \ref{table:SiO-flux}). The $\varv=1$ lines were observed with specific line luminosities ranging from less than 1 mJy kpc$^2$ (in $\pi^1$ Gru and T Mic) to a few tens of Jy kpc$^2$ (in IRC+10011), i.e. four orders of magnitude variation. The velocity extents varied between 7.0 \kms\ (for \ce{^{30}SiO} $J=6-5$ in GY Aql) and 52.0 \kms\ (for \ce{^{28}SiO} $J=6-5$ in IRC+10011).

The \ce{^{28}SiO} $J=5-4$ line contributes the most luminous detected SiO emission in 9 sources. When comparing the two observed \ce{^{28}SiO} $\varv=1$ transitions, the specific line luminosity of $J=5-4$ exceeds that of $J=6-5$ by a factor of a few for some stars (e.g. S Pav, $\pi^1$ Gru) and by over an order of magnitude for others (e.g. V PsA, R Aql). However, SV Aqr, W Aql and IRC$-$10529 are exceptions with specific luminosity ratios between the two transitions ($J=5-4$/$J=6-5$) $<1$ (i.e. 0.71, 0.86 and 0.78, respectively).

We also found the $J=6-5$ transitions from the \ce{^{29}SiO} and \ce{^{30}SiO} isotopologues in all stars except SV Aqr and U Del. Overall, the SiO line emission from these isotopologues can vary significantly with the specific luminosity ratio of \ce{^{29}SiO}:\ce{^{30}SiO} reaching as low as 1.4 (R Hya) and as high as 160 (IRC+10011). The \ce{^{29}SiO} $J=6-5$ is the brightest emission line in three stars, including U Her, IRC$-$10529 and IRC+10011. The highest specific luminosity of any high-$J$ SiO emission in the ATOMIUM observations is the \ce{^{29}SiO} $J=6-5$ transition observed towards IRC+10011 at 24.9 Jy kpc$^2$, $S_{\nu}$ = 48 Jy at 720 pc (Fig. \ref{fig:example-spectra}, 2nd row, left). This is over 50 times its second brightest $\varv=1$ line, \ce{^{28}SiO} $J=5-4$, which peaks at only 0.48 Jy kpc$^2$ ($S_{\nu}$ = 0.9 Jy). Note that the rest frequencies of the $J=5-4$ lines from the SiO isotopologues are outside the frequency range covered by the ATOMIUM tunings.

\subsubsection{$\varv=2$ transitions}
\label{sec:SiO-id:v=2}
The only SiO $\varv=2$ line covered by the ATOMIUM tunings (see Table \ref{table:line-coverage}) is \ce{^{28}SiO} $J=5-4$, which has $E_{\rm{u}} \sim3550$ K. It was observed in every ATOMIUM source except SV Aqr and U Del, though we note that the detection in V PsA was only tentative. The strongest $\varv=2$ emission was seen towards IRC+10011 at 5.72 Jy kpc$^2$ ($S_{\nu}$ $\sim$11 Jy at 720 pc), which is an order of magnitude higher in specific luminosity compared to the second brightest $\varv=2$ detection among the ATOMIUM sample in U Her (128 mJy kpc$^2$, $S_{\nu}$ $\sim$1.7 Jy at 271 pc). An example spectrum of a clear detection of the $\varv=2$ line towards $\pi^1$ Gru is shown in Fig. \ref{fig:example-spectra} (2nd row, middle panel). The angular size of $\varv=2$ emission in the channel maps of $\pi^1$ Gru and IRC$-$10529 is comparable to the extent of the continuum contour of the AGB star (see online supplementary material).

\subsubsection{$\varv=3$ transitions}
\label{sec:SiO-id:v=3}
There are two $\varv=3$ SiO lines detected by ATOMIUM: \ce{^{28}SiO} $J=6-5$ (e.g. Fig. \ref{fig:example-spectra}, 2nd row, right) and \ce{^{29}SiO} $J=6-5$ (e.g. Fig. \ref{fig:example-spectra}, 3rd row, left). Six out of 14 sources, namely U Her, T Mic, $\pi^1$ Gru, W Aql, IRC$-$10529, and IRC+10011, hosted unambiguous \ce{^{28}SiO} $J=6-5$ line emission, while emission in the \ce{^{29}SiO} $J=6-5$ transition was only tentatively detected in R Hya (0.14 mJy kpc$^2$, $S_{\nu}$ = 9 mJy at 126 pc) and IRC$-$10529 (27 mJy kpc$^2$, $S_{\nu}$ = 31 mJy at 930 pc), with the former present alongside an absorption (Section \ref{sec:absorption}). The seven stars exhibiting the \ce{^{28}SiO} $J=6-5$ line cover a wide range of mass-loss rate (1.9$\times$10$^{-7}$ $\lesssim \dot{M} \lesssim$ 4.0$\times$10$^{-5}$ M$_\odot$ yr$^{-1}$), where the observed specific luminosities of this transition vary from 0.5 mJy kpc$^2$ ($S_{\nu}$ = 17 mJy in T Mic, 175 pc) to 394 mJy kpc$^2$ ($S_{\nu}$ = 760 mJy in IRC+10011). Notably, while the detection towards U Her is the brightest in terms of flux density ($S_{\nu}$ $\sim$2 Jy), the detection towards IRC+10011 is intrinsically the most luminous, exceeding that of U Her by a factor of 2.6 (394 vs 150 mJy kpc$^2$).

\subsubsection{$\varv = 4, 5$ transitions}
\label{sec:SiO-id:v>3}
A total of three transitions originating from $\varv = 4, 5$ states were observed: \ce{^{28}SiO} $\varv=4$, $J=6-5$ ($E_{\rm{u}}$ = 7016 K), \ce{^{30}SiO} $\varv=4$, $J=6-5$ ($E_{\rm{u}}$ = 6932 K) and \ce{^{30}SiO} $\varv=5$, $J=6-5$ ($E_{\rm{u}}$ = 8613 K), albeit at low luminosity levels. From Table \ref{table:SiO-flux}, we see that the \ce{^{28}SiO} $\varv=4$, $J=6-5$ line (e.g. Fig. \ref{fig:example-spectra}, 3rd row, middle) was unambiguously detected in 7 ATOMIUM sources (or 9 sources if tentative detections are included). Amongst these, two sources (R Hya and R Aql) also showed signs of absorption. Both the \ce{^{28}SiO} and \ce{^{30}SiO} $\varv=4$ lines were completely absent in U Del, V PsA, SV Aqr, RW Sco, and $\pi^1$ Gru. The specific luminosities of the $\varv=4$ lines were generally low, reaching a maximum of 39 mJy kpc$^2$ ($S_{\nu}$ = 75 mJy) in IRC+10011. Across the 9 targets, these luminosities were generally 1--4 orders of magnitude lower than those of most low-$\varv$ state lines observed towards the same source. As for the \ce{^{30}SiO} $\varv=5$ line, it was only clearly observed in R Hya with a specific line luminosity of 0.2 mJy kpc$^2$ ($S_{\nu}$ = 13 mJy) (Fig. \ref{fig:example-spectra}, bottom row, left). S Pav and IRC+10011 also showed tentative detections of this $\varv=5$ line, with measured specific luminosities of 0.4 mJy kpc$^2$ ($S_{\nu}$ = 12 mJy at 184 pc) and 9 mJy kpc$^2$ ($S_{\nu}$ = 17 mJy), respectively. We note that while the detection in R Hya is the most robust signal-to-noise wise, the tentative feature in IRC+10011 represents an intrinsic luminosity nearly 45 times greater.

\subsubsection{$\varv \geq 6$ transitions}
\label{sec:SiO-id:v>6}
In this work, we also identified two unknown lines in R Hya, listed in \citet{2024A&A...681A..50W} as U246.619 and U246.076, to be the \ce{^{29}SiO} $\varv=6$, $J=6-5$ and \ce{^{28}SiO} $\varv=8$, $J=6-5$ transitions, respectively. The $\varv=6$ line of \ce{^{29}SiO} at 246.619 GHz was weakly detected with a specific luminosity of 0.10 mJy kpc$^2$ ($S_{\nu}$ = 6 mJy) and a velocity extent of 3.6 \kms, while the \ce{^{28}SiO} $\varv=8$ line at 246.078 GHz was found at 0.13 mJy kpc$^2$ ($S_{\nu}$ = 8 mJy) with the same velocity extent, as shown in the bottom row of Fig. \ref{fig:example-spectra} (middle and right panels). We also point out that there is a tentative absorption feature in the \ce{^{28}SiO} $\varv=8$, $J=6-5$ line towards S Pav and R Aql (see Section \ref{sec:absorption}). To the best of our knowledge, the \ce{^{28}SiO} $\varv=8$, $J=6-5$ emission line observed towards R Hya is the highest vibrational state ever observed for SiO. Our $\varv=8$, $J=6-5$ line profile is similar to that of the \ce{^{29}SiO} $\varv=3$, $J=6-5$ transition in this source (Section \ref{sec:absorption:line-profiles}). We discuss the possible excitation mechanism from these high-$\varv$ states further in Section \ref{sec:dist:non-LTE:pop-diagrams}.

\begin{figure*}
   \begin{center}
   \includegraphics[width=0.38\textwidth, height=0.3\textwidth]{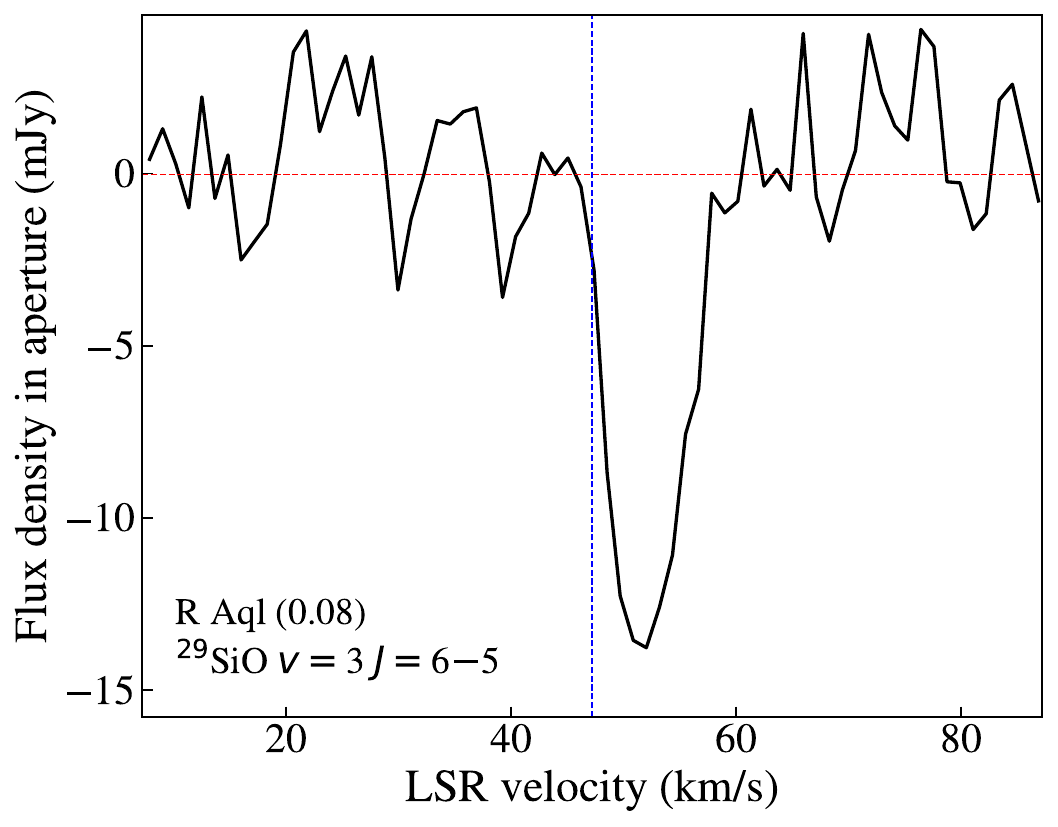} 
   \includegraphics[width=0.38\textwidth, height=0.3\textwidth]{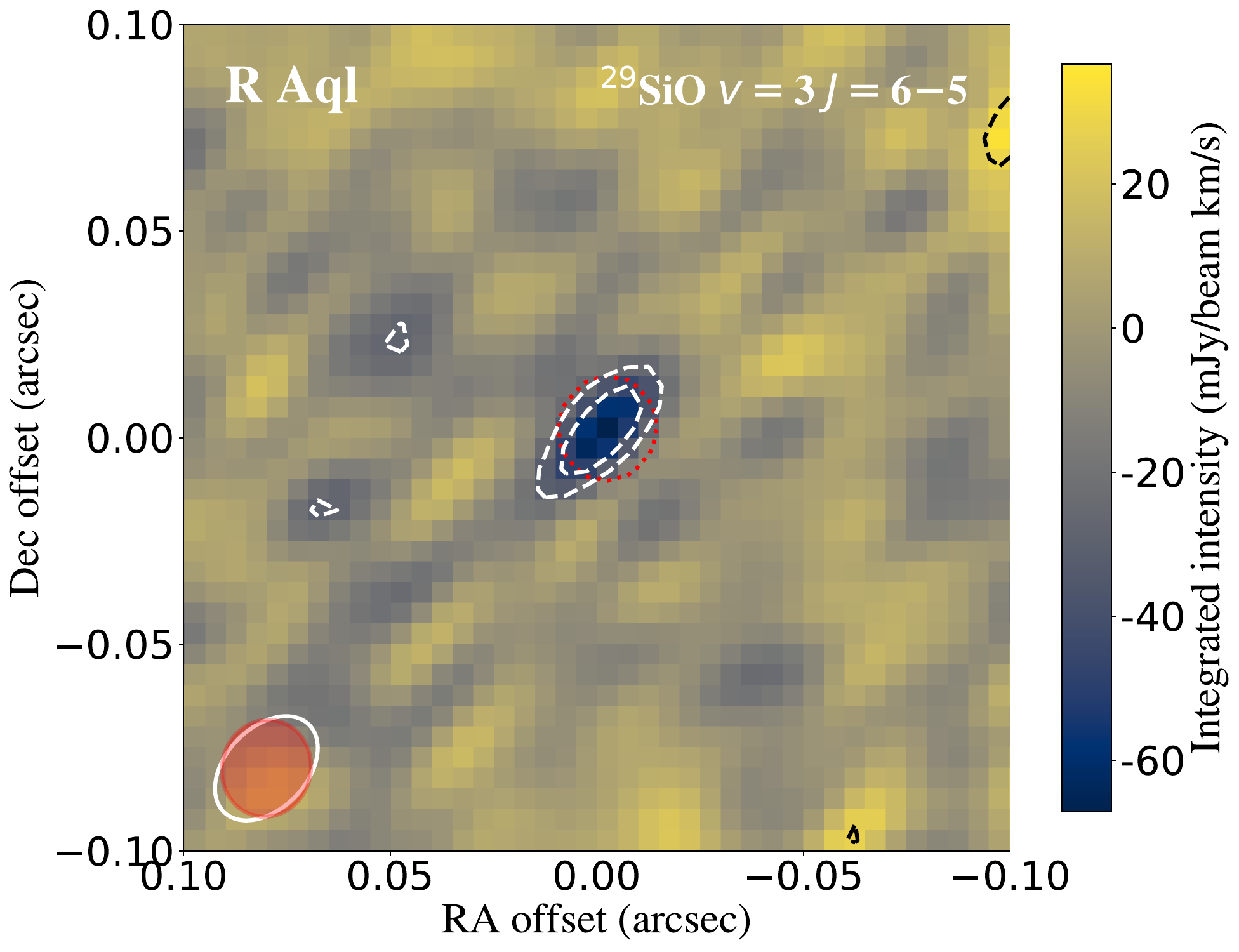}
   \includegraphics[width=0.38\textwidth, height=0.3\textwidth]{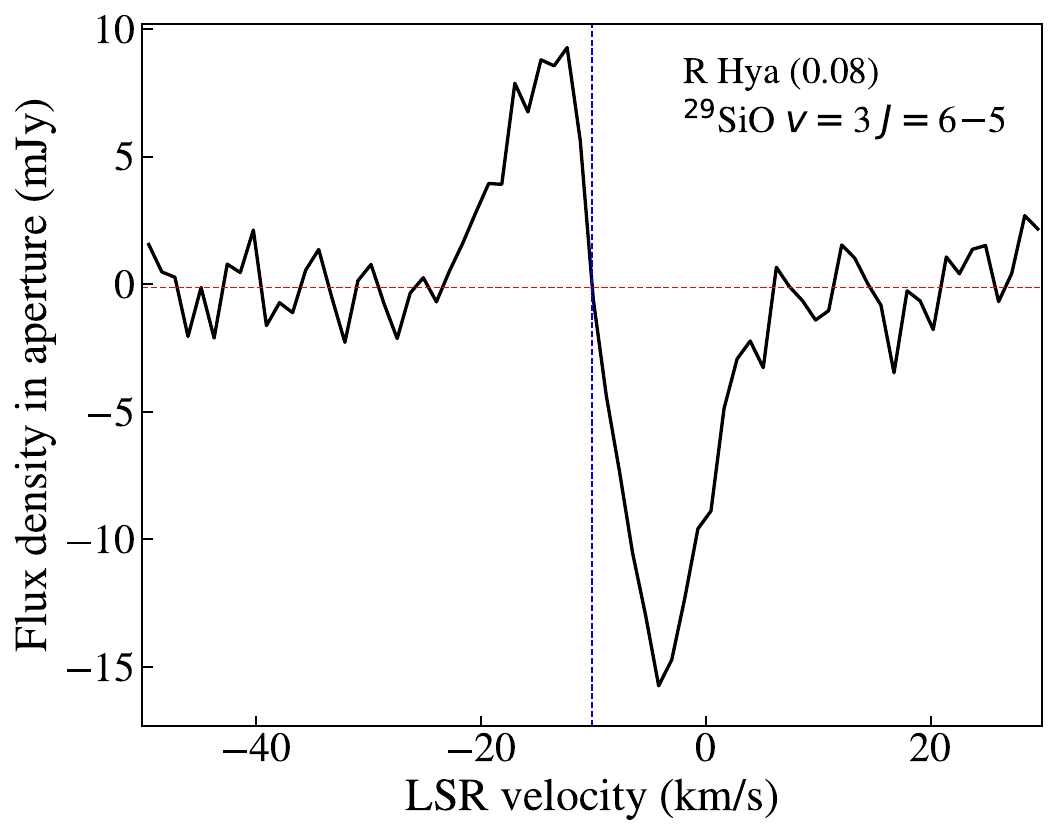} 
   \includegraphics[width=0.38\textwidth, height=0.3\textwidth]{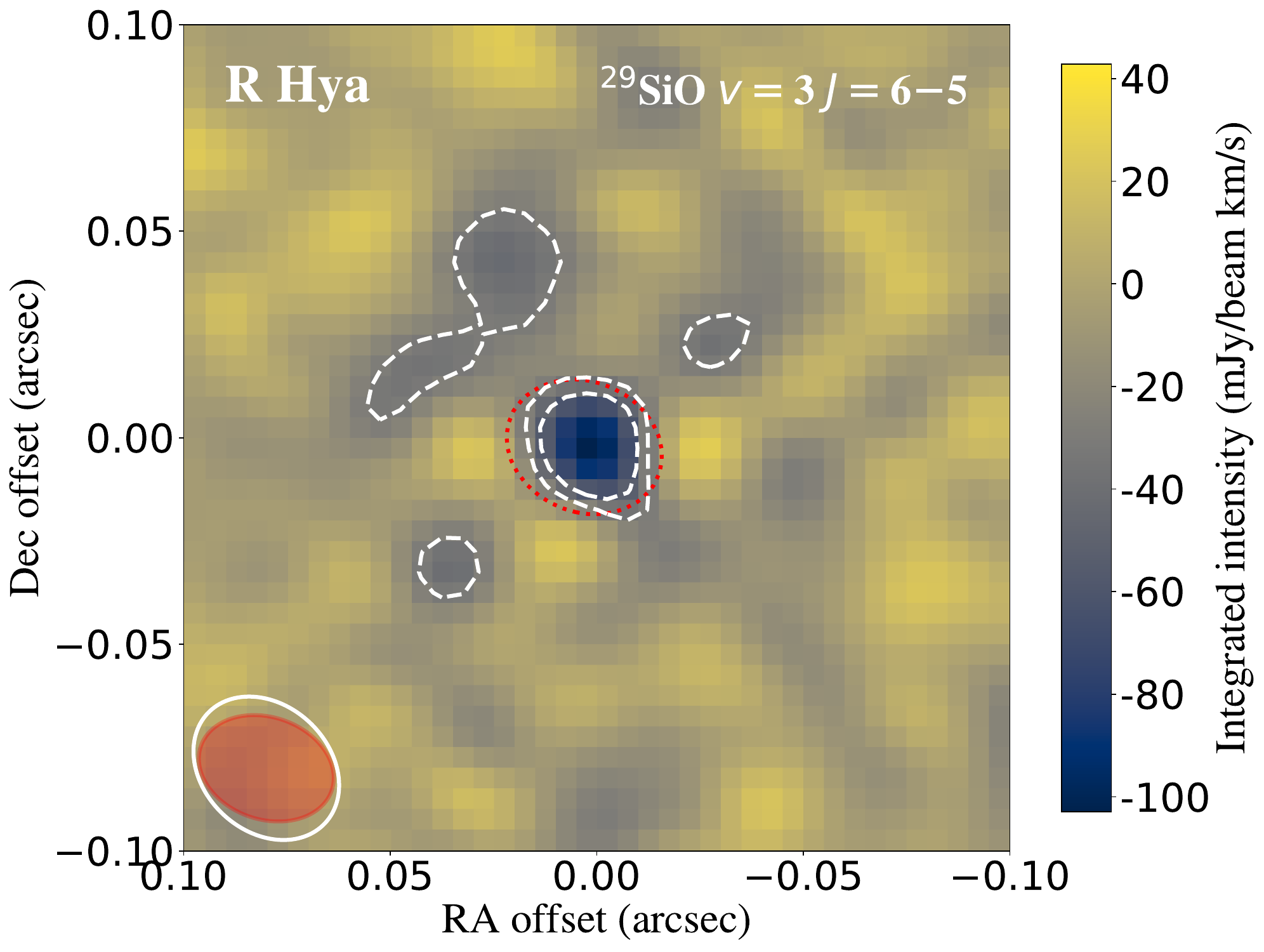}
   \includegraphics[width=0.38\textwidth, height=0.3\textwidth]{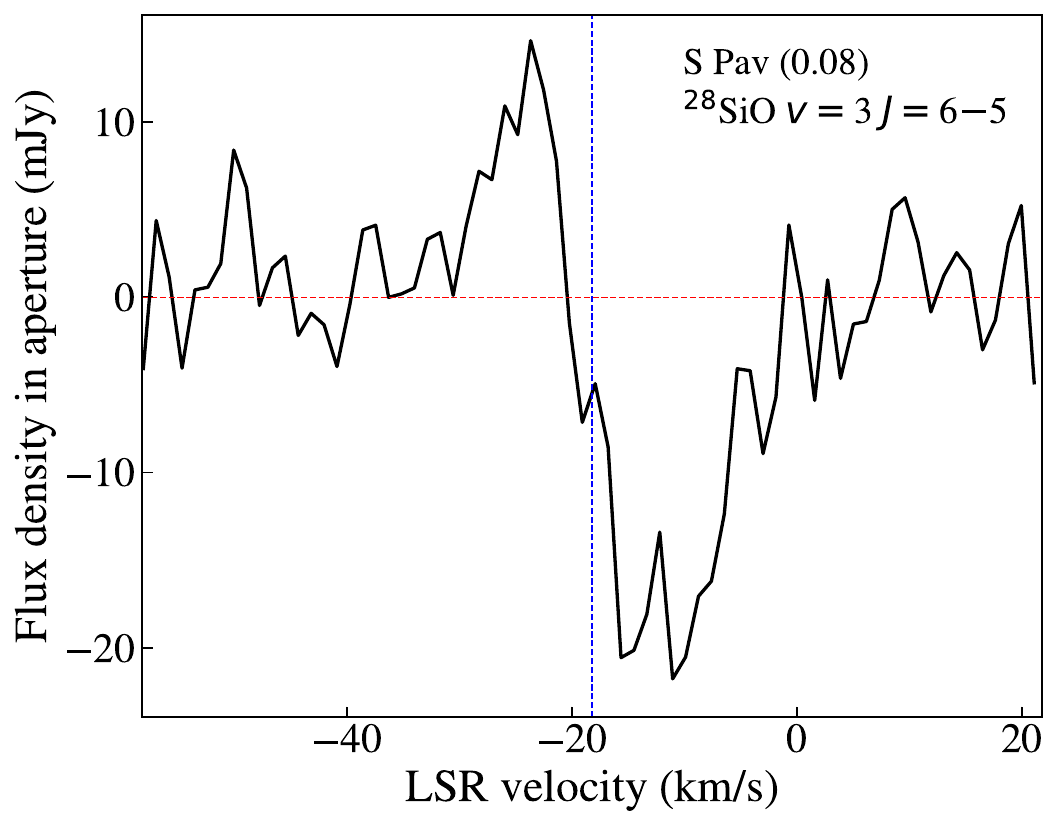} 
   \includegraphics[width=0.38\textwidth, height=0.3\textwidth]{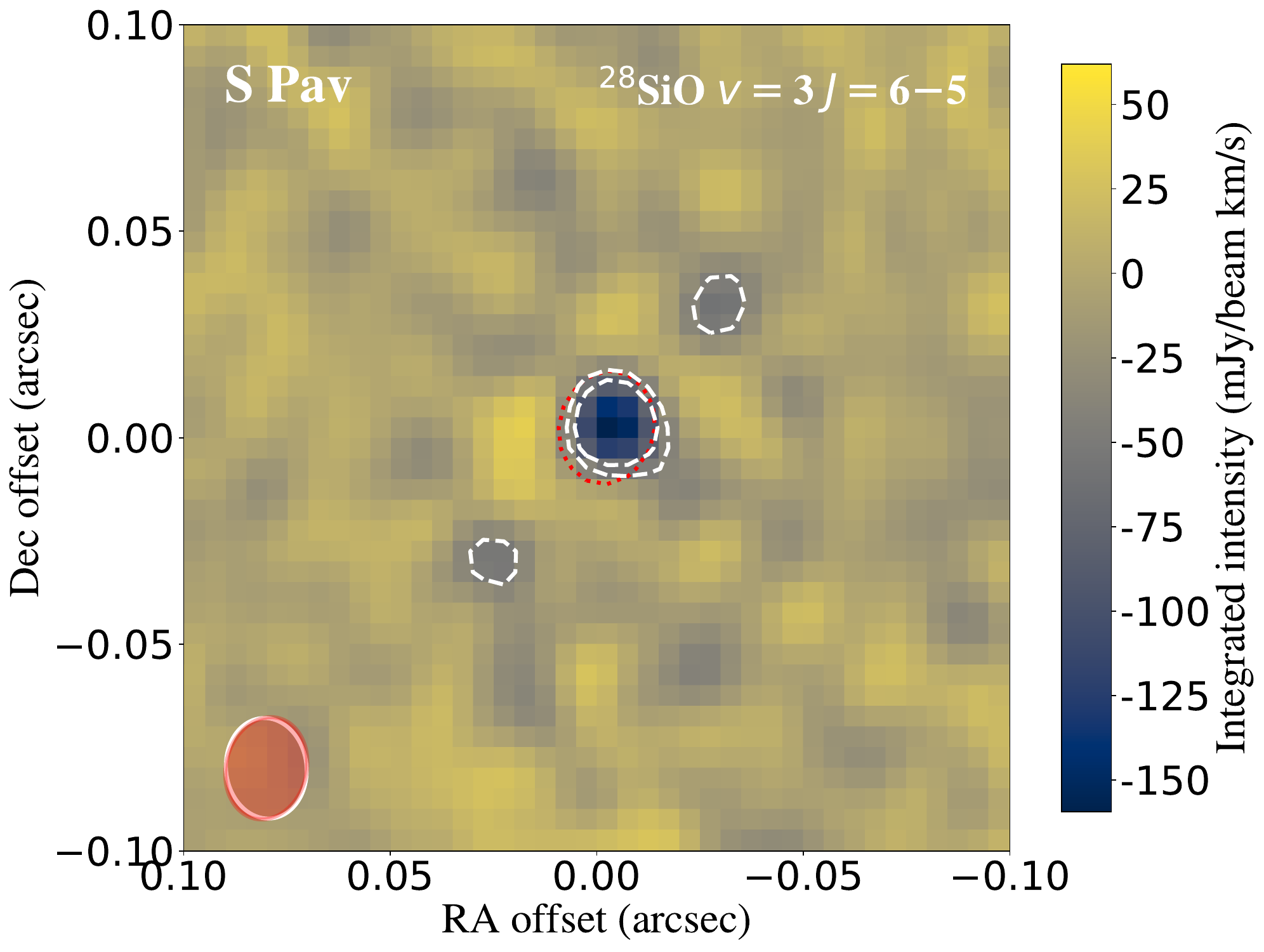}
   \includegraphics[width=0.38\textwidth, height=0.3\textwidth]{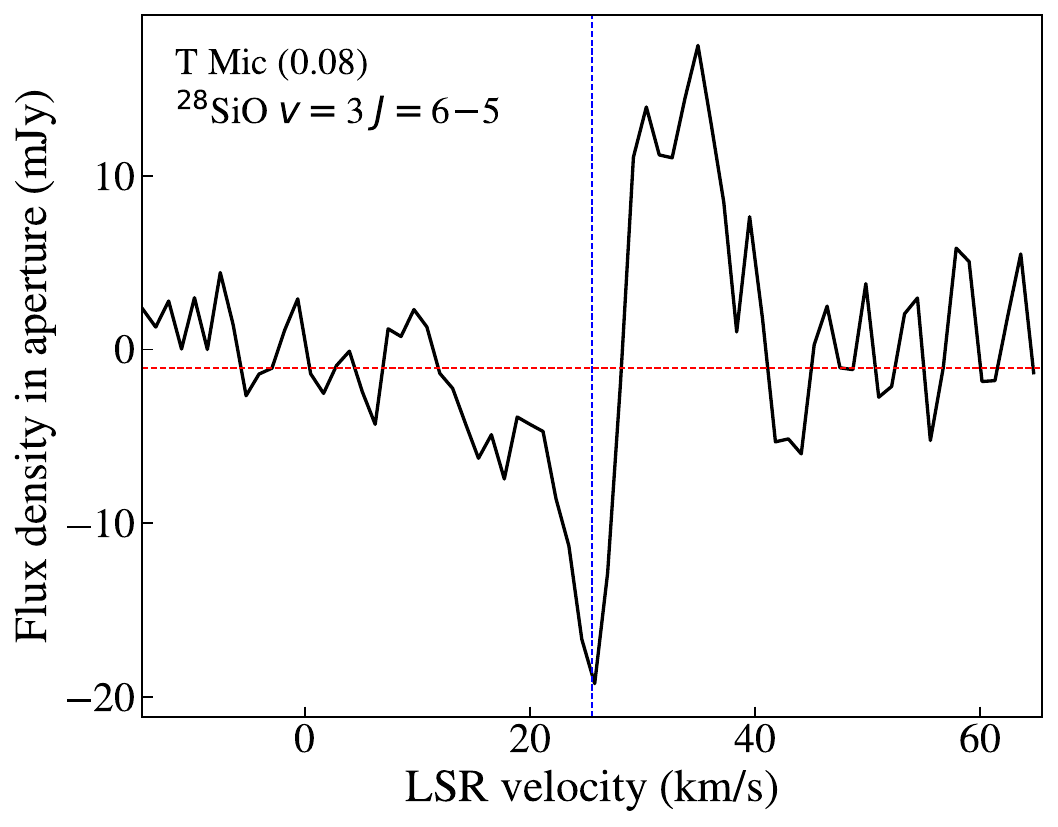} 
   \includegraphics[width=0.38\textwidth, height=0.3\textwidth]{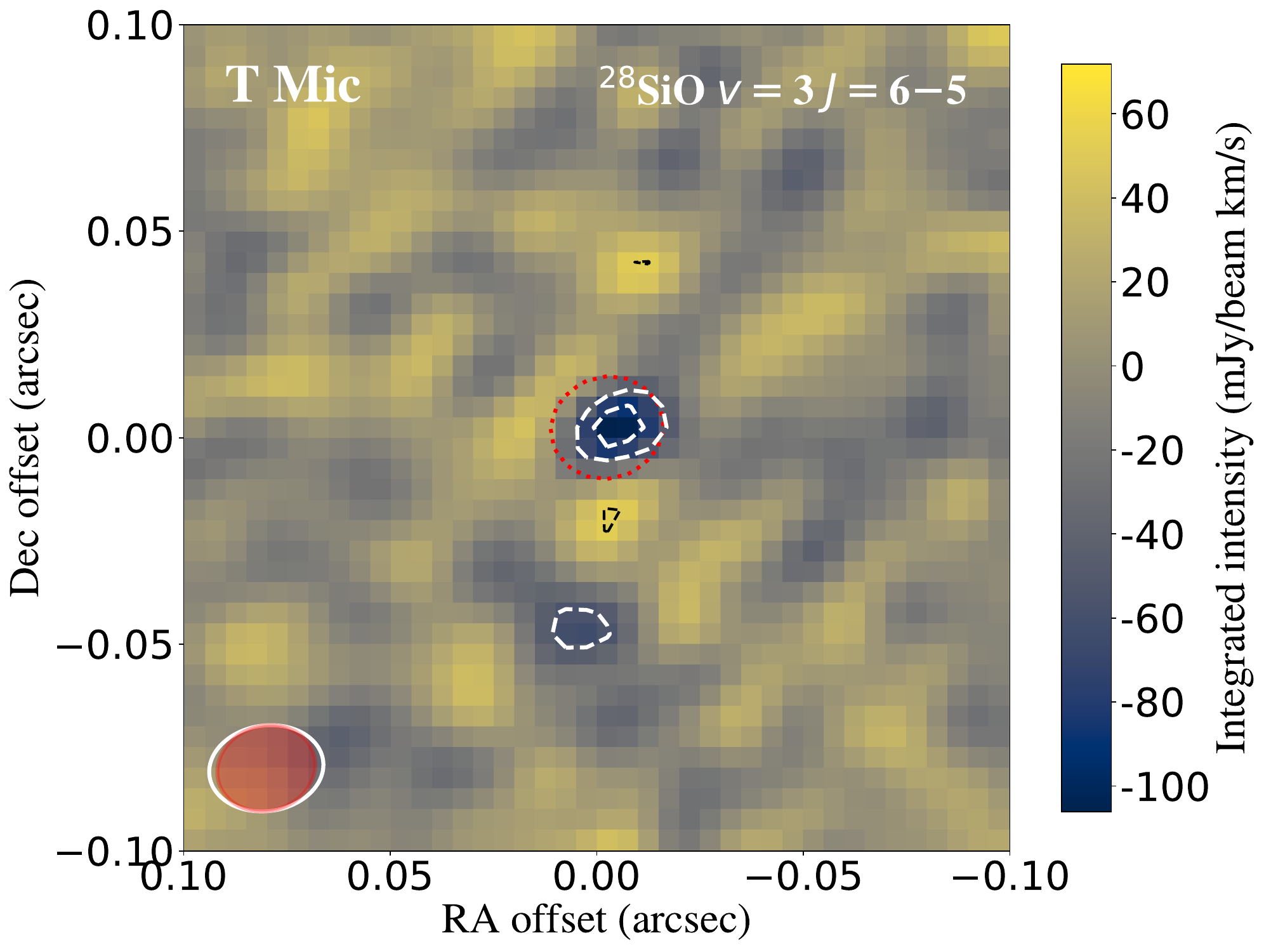}
   \caption{Left: Absorption spectra of the \ce{^{29}SiO} $\varv=3$, $J=6-5$ line at 251.930 GHz towards R Aql and R Hya as well as the \ce{^{28}SiO} $\varv=3$, $J=6-5$ line at 255.091 GHz towards S Pav and T Mic. The observed frequencies are converted to velocity using the rest frequencies listed in Table \ref{table:line-coverage}. The spectrum was extracted from the extended-configuration data cubes using an aperture diameter of 0.08 arcsec. The vertical blue dashed line indicates $V_*$ (see Table \ref{table:atomium-sources}) while the horizontal red dashed lines show the fitted spectral baselines. Right: Zeroth moment maps of the same transitions towards the same source. The map field of view is 200$\times$200 mas. The white dashed lines show the $-$3\rm{$\sigma$} and $-$5\rm{$\sigma$} absorption contours. The red dotted line illustrates the half-intensity extent of the continuum emission. The white and red ellipses in the bottom left corner are the reference sizes of the half-power beam width (HPBW) beam and the continuum beam, respectively.}
   \label{fig:absorption}
   \end{center}
\end{figure*}

\section{S\MakeLowercase{i}O line absorption}
\label{sec:absorption}
Since the advent of ALMA, the absorption of a millimetre-wavelength SiO line from the vibrational ground state towards AGB stars has been observed towards e.g. $o$ Ceti \citep{2016A&A...590A.127W} and R Dor \citep{2021MNRAS.504.2687N}. \citet{2025A&A...704A..18O} recently reported inhomogeneous absorption of SiO $\varv>0$ isotopologue lines in W Hya, a nearby (93 pc) semi-regular variable star of type a (SRa), caused by the cooler layers of gas along the line of sight to the stellar disc. In this work, we identified absorption in $\varv=0$ transitions as well as four $\varv>0$ SiO lines, including \ce{^{28}SiO} $\varv=3$, $J=6-5$ at 255.091 GHz, \ce{^{28}SiO} $\varv=4$, $J=6-5$ at 253.286 GHz, \ce{^{29}SiO} $\varv=3$, $J=6-5$ at 251.930 GHz and potentially \ce{^{28}SiO} $\varv=8$, $J=6-5$ at 246.078 GHz, in our extended-configuration data extracted with a 0.08 arcsec diameter. The absorption is evaluated at the $-$3\rm{$\sigma$} and $-$5\rm{$\sigma$} levels in front of the central star and can be considered unresolved even with the high-resolution ALMA data. No absorption of other $\varv < 3$ lines was detected at the time of observations. In the following sections, we discuss the characteristics of the line profiles and mom0 maps of $\varv>0$ SiO, which are summarised in Table \ref{table:SiO-flux}, and show a comparison between these and the H$_2$O and OH absorption lines detected in the same data set reported by \citet{2023A&A...674A.125B}. 

\subsection{Line profiles}
\label{sec:absorption:line-profiles}
Absorption in the SiO $\varv=0$ state is a well-documented phenomenon often attributed to cooler, overlying gas layers in the circumstellar environment absorbing emission from the hotter, inner regions (see e.g. \citealt{2017AJ....153..176L} and references therein, \citealt{2021MNRAS.504.2687N}). It is likely to be blue-shifted at velocities typical of the wind at many tens of $R_*$ as per e.g. \citet{2022A&A...660A..94G} and \citet{2016A&A...590A.127W}. On the other hand, red-shifted absorption or very narrow absorption, at velocities close to $V_*$, are likely to be from inner layers where masers can arise.

Among the ATOMIUM sources, S Pav, T Mic, R Hya and R Aql exhibit clear $\varv>0$ absorption features, often accompanied by line emission. Fig. \ref{fig:absorption} shows a few examples of the absorption spectra (left panels) and mom0 maps (right panels) of the four sources. Additional plots can be found in Appendix \ref{appendix:absorption}. Broadly speaking, these spectral profiles fall into three categories: red-shifted absorption alone, red-shifted absorption paired with blue-shifted emission and the reverse of this latter profile. In each star, the absorption feature appears at a roughly consistent velocity across multiple detected lines, although the levels of absorption can be up to $\sim$25 per cent different. The SiO absorption profiles are often spectrally narrow with the widths at half minimum of $\sim$5 \kms. Overall, S Pav, R Hya and R Aql are tied for exhibiting the highest number of absorption lines (3) in the ATOMIUM sample, see Table \ref{table:SiO-flux}.

In R Aql, we detected absorption in the \ce{^{29}SiO} $\varv=3$ line at 251.930 GHz (Fig. \ref{fig:absorption}, top) and the \ce{^{28}SiO} $\varv=3$ line at 255.091~GHz (Fig. \ref{fig:add-absorption-RAql}), with minimum flux densities of $-13$ and $-12$~mJy occurring at LSR velocities of $+5$ and $+7$ \kms\ relative to $V_*$ of 47.2 \kms, respectively. The \ce{^{29}SiO} $\varv=3$ absorption line was detected in S Pav (Fig. \ref{fig:add-absorption-SPav}), with the minimum flux density reaching $-$27 mJy, red-shifted by 8 \kms\ with respect to $V_*$. We also report tentative detections of \ce{^{28}SiO} $\varv=8$, $J=6-5$ absorption towards these two stars (Figs.~\ref{fig:add-absorption-RAql} and \ref{fig:add-absorption-SPav}, bottom). The minimum flux densities observed were $-5$ mJy at $\sim$$V_*$ (R Aql) and $-10$ mJy at $+4$~\kms\ relative to $V_*$ (S Pav). However, no corresponding absorption feature can be produced in the mom0 maps, which may indicate that the feature is either not real or remains unresolved within the primary beam. Only these two stars exhibit such a single-feature, absorption-only profile.

Blue-shifted line emission with red-shifted absorption was observed in the \ce{^{29}SiO} $\varv=3$ line towards R Hya (Fig. \ref{fig:absorption}, second row). The \ce{^{28}SiO} $\varv=3$ line profile (Fig. \ref{fig:add-absorption-RHya}) exhibits a double red-shifted absorption feature of similar levels, where both minima are within 10 per cent of one another (24--26 mJy) and separated by $\sim$7 \kms. A similar absorption profile is also present in the same line observed towards S Pav (Fig. \ref{fig:absorption}, third row), with the absorption feature reaching $-$20 and $-$21 mJy at 2 \kms\ and 8 \kms\ to the red of $V_*$, respectively. These detections are characteristic of an inverse P Cygni profile. In addition, in R Aql and R Hya, there is a sign of absorption ($-$12 mJy) seen $\sim$10 \kms\ to the red of $V_*$ alongside a line detection at $V_*$ (+25 and +43 mJy, respectively) in the \ce{^{28}SiO} $\varv=4$ transition at 253.286 GHz (Figs. \ref{fig:add-absorption-RAql} and \ref{fig:add-absorption-RHya}). 

In contrast, T Mic shows the reverse profile where line emission, likely quasi-thermal or weakly masing, of $\sim$14--17 mJy is seen between +5 and +15 \kms\ of $V_*$ and an absorption feature sits right at $V_*$ in the \ce{^{28}SiO} $\varv=3$ line (Fig. \ref{fig:absorption}, bottom panels). This is the only evidence of absorption seen in this source. Such characteristics are likely an indication of a P Cygni profile, which signifies outward motions of the line-emitting gas, although the fact that the absorption is at $V_*$ may suggest the possible existence of a stationary layer of absorbing gas in the CSE, as previously observed in high-resolution observations of H$_2$O rovibrational lines by e.g. \citet{1979ApJ...227..923H}. We emphasise, however, that our current data do not allow us to draw firm conclusions, and this interpretation remains speculative. The presence of both moderate optical-depth P Cygni profiles and inverse P Cygni profiles for the $\varv=3$ $J=6-5$ transition provides direct evidence of SiO-emitting regions tracing outflows in some stars and infalls in other stars, similar to what was observed towards W Hya \citep{2025A&A...704A..18O}. This is consistent with the emission coming from the pulsation zone of these stars' extended atmospheres.

\subsection{Absorption maps}
\label{sec:absorption:maps}
The mom0 maps of the same high-$J$ $\varv>0$ lines discussed above, which are 200 $\times$ 200 mas in size, are shown in the right panels of Fig. \ref{fig:absorption}. The intensity is integrated over the whole velocity range shown in each spectrum. In all cases, the main absorption feature is mostly featureless, spatially compact, and strictly coincident with the continuum peak position \citep{2025Danilovich}, exhibiting angular diameters in the range of $\sim$20--30 mas i.e. similar to the half-intensity extent of the continuum emission illustrated by red dotted ellipses (Fig. \ref{fig:absorption}).

In several maps, apparent extended negative features are visible at larger radial offsets. For instance, the \ce{^{29}SiO} $\varv=3$ mom0 map of R Hya shows an apparent absorption feature 15 mas NE of the continuum peak, while the \ce{^{28}SiO} $\varv=3$ map exhibits extended features resembling projected "spirals" and a separate $-$5$\sigma$ island 50 mas to the SE (Fig. \ref{fig:add-absorption-RHya}). These are likely interferometric imaging artifacts as the extended-configuration continuum map for R Hya also exhibits symmetric negative regions around the central peak resulting from amplitude errors. In the case of the \ce{^{29}SiO} $\varv=3$ map of S Pav, there is an apparent extension to the $-$3$\sigma$ contour towards the SW, roughly 25 mas from the continuum peak (Fig. \ref{fig:add-absorption-SPav}). However, this is still well within the extent of the continuum emission of angular diameter $\sim$50 mas (Fig. 11 of \citealt{2025Danilovich}).

\subsection{Comparison with \texorpdfstring{H$_2$O}{H2O} and OH line absorption}
\label{sec:absorption:H2O_OH}
It is interesting then to compare these highly excited SiO absorption features to the H$_2$O and OH absorption lines detected in the extended-configuration ATOMIUM data, all of which were taken within a few weeks of each other \citep{2023A&A...674A.125B}. Among the four sources showing SiO absorption, S Pav, R Hya, and R Aql show clear absorption from para-H$_2$O ($\nu_1,\nu_2,\nu_3$ = (1,1,0)--(0,1,1), $J_{K_a,K_c}$ = $4_{2,2} \rightarrow 3_{2,1}$) at 244.330 GHz (their Figs. 8 and B.2). All four sources also show signs of absorption in one or both H$_2$O lines at 254.040 and 254.053 GHz (their Figs. B.4 and D.14), based on their extracted spectra and mom0 maps retrieved using the same approach and convention as those presented in this work.

Absorption in the $\varv=3$ line from \ce{^{28}SiO} is detected towards the same four sources listed above while that from \ce{^{29}SiO} is present in all but T Mic (Table \ref{table:SiO-flux}). The amplitudes of the absorption in H$_2$O are usually between 5--10 mJy, roughly 1.5--2.5 times smaller than the SiO absorption features detected in the same stars. The redshifted absorption components in the SiO $\varv=3$ lines across R Hya, S Pav, and R Aql show velocity offsets that are highly consistent with those observed in the 244.330 GHz H$_2$O line, all falling within the 5--9 \kms\ range. This agreement reinforces the interpretation of systematic inward motions in the inner CSEs of these stars at this epoch based on the model predictions by synthetic CO and CN line profiles observed in the dust-free regions of the CSE \citep{2010A&A...514A..35N}. In addition, although H$_2$O absorption (also at 244.330 GHz) has been identified in IRC+10011, no such line profiles are detected in the ATOMIUM SiO data of this source.

In the case of OH, \citet{2023A&A...674A.125B} reported red-shifted absorption features as well as line emission in the $J=27/2$ (221.335 and 221.353 GHz) and $J=29/2$ (252.127 and 252.145 GHz) levels detected towards S Pav, R Hya, and R Aql (their Fig. 20), i.e. the same three sources showing H$_2$O absorption. They suggested that the OH gas falls towards the central star with velocities between 5--10 \kms, similar in range to the velocities of the absorption features in highly excited SiO and H$_2$O lines discussed above. Based on their mom0 maps, the angular extent of the OH absorption is approximately 30--40~mas on average. Although the size estimates are approximate due to the limited S/N, they likely trace only the hot component of the OH envelope. Within this context, the inferred OH emission region appears roughly twice as extended as that derived for SiO in the same sources.

\section{Inner circumstellar envelopes as probed by highly excited S\MakeLowercase{i}O}
\label{sec:dist}

In this section, we use the high-$J$ SiO emission observed in the ATOMIUM sources to investigate the spatial and kinematic structures of these inner CSEs, focusing on the brightness temperatures, morphology, velocity distribution of the maser and quasi-thermal components that reveal the immediate environments around the central stars.

\begin{figure*}
  \centering
  \includegraphics[width=0.99\textwidth, height=0.75\textwidth]{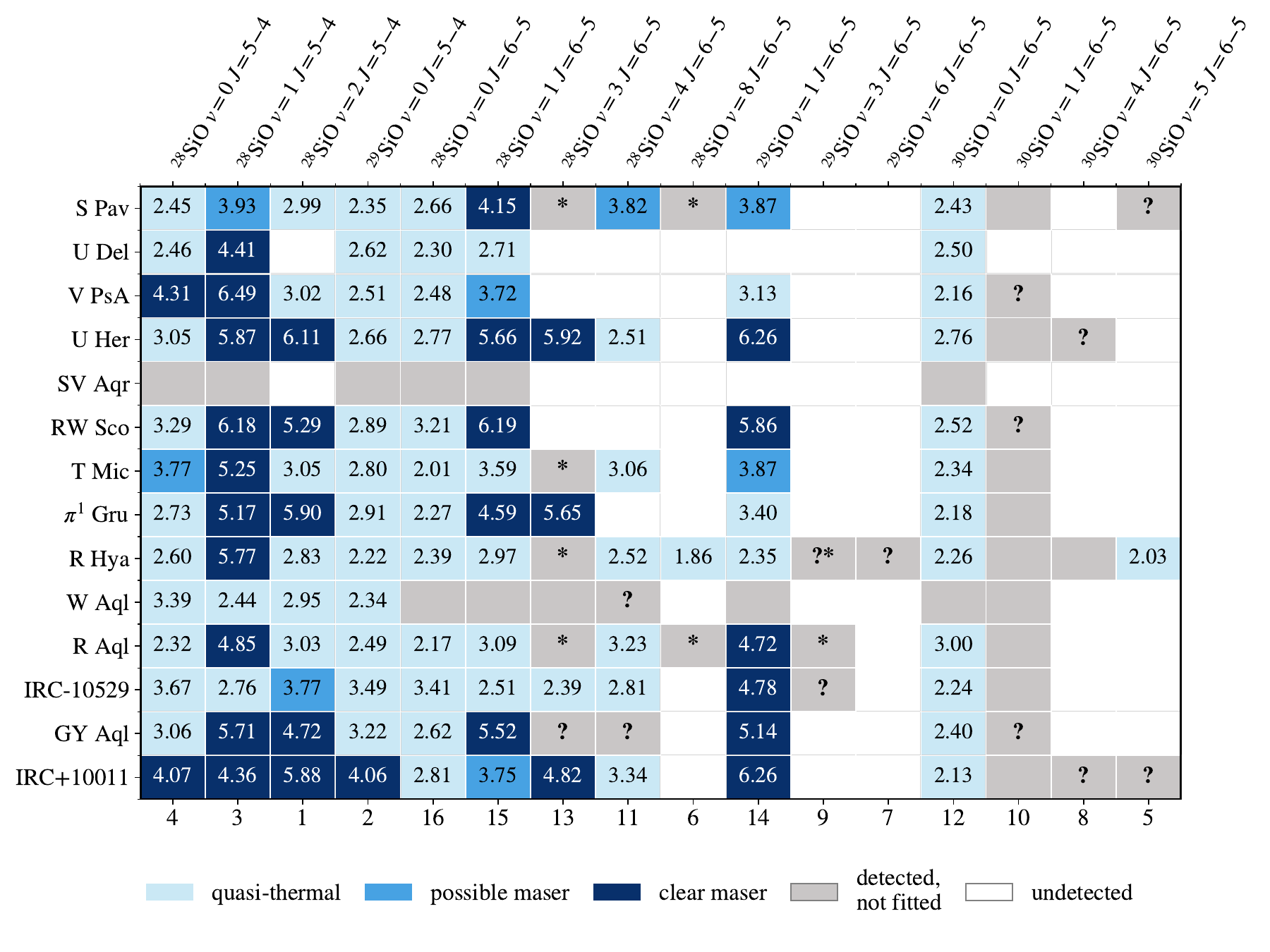}
  \caption{Lower limits to the brightness temperatures of the detected SiO transitions, derived from 2D Gaussian component fits as described in Section~\ref{sec:dist:analysis:SAD}. Each cell corresponds to the maximum $T_{\mathrm{b}}$ measured for a given transition in each ATOMIUM source. Colours indicate $\log_{10}(T_{\mathrm{b}}/\mathrm{K})$, with the emission classified as quasi-thermal for $\log_{10}(T_{\mathrm{b}}) \lesssim 3.7$ (light blue), as a possible maser for $3.7\lesssim\log_{10}(T_{\mathrm{b}})<4.0$ (blue), and as a clear maser for $\log_{10}(T_{\mathrm{b}})\geq4.0$ (navy). Grey cells represent transitions which are detected but 2D Gaussian fitting does not result in any components due to their weak intensities, while blank cells correspond to undetected transitions. Question marks indicate tentative detections ($3.0\lesssim \mathrm{S/N} < 5.0$), and asterisks denote the lines where absorption is detected. The line numbers from Table \ref{table:line-coverage} are given in the bottom. Note that higher resolution observations might allow higher $T_{\rm b}$ to be measured and promote some lines to maser classification. We recommend viewing the digital version of this article for optimal legibility.}
  \label{fig:Tb}
\end{figure*}

\begin{figure*}
   \begin{center} 
   \includegraphics[width=0.396\textwidth, height=0.306\textwidth]{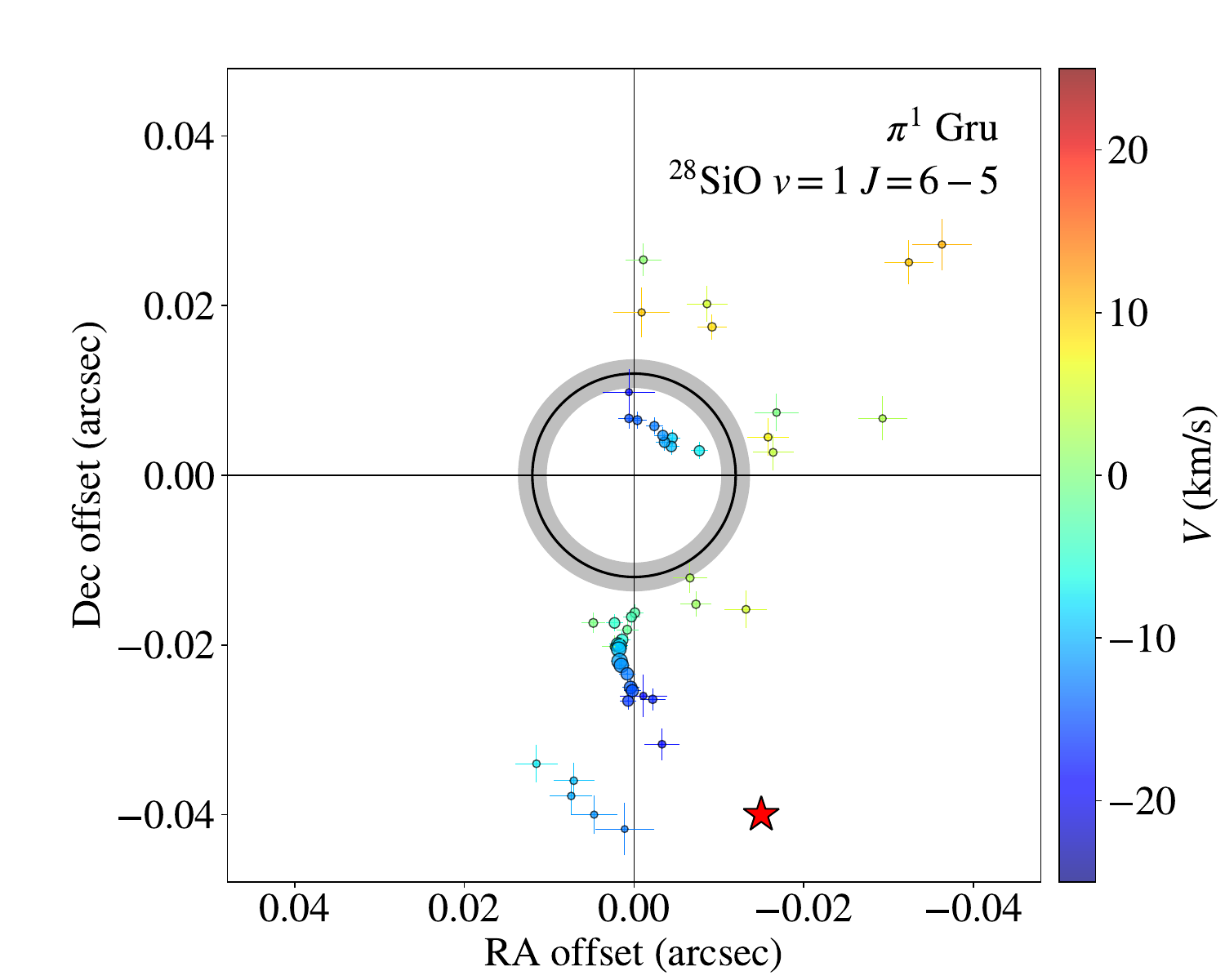}
   \includegraphics[width=0.396\textwidth, height=0.306\textwidth]{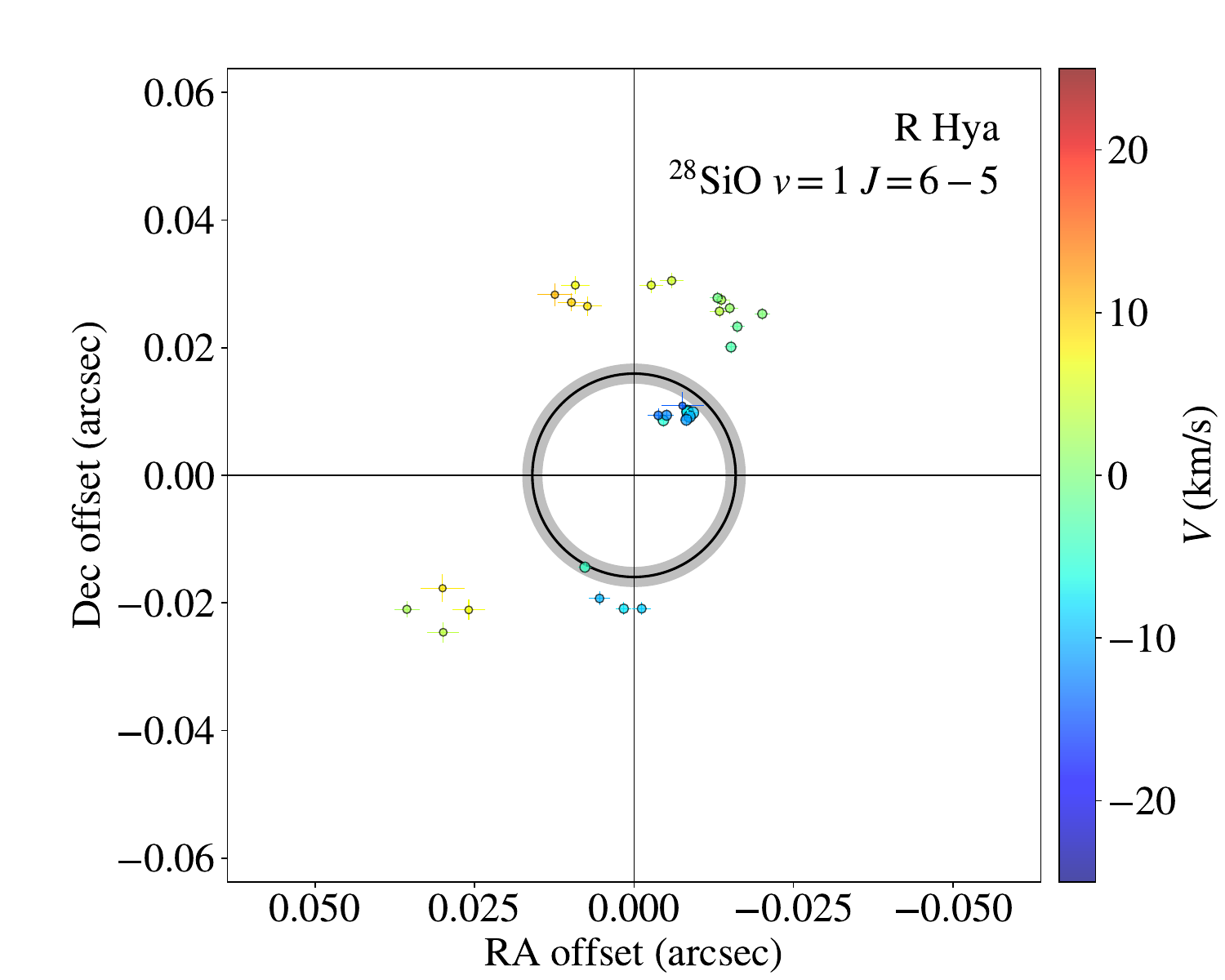}
   \includegraphics[width=0.396\textwidth, height=0.306\textwidth]{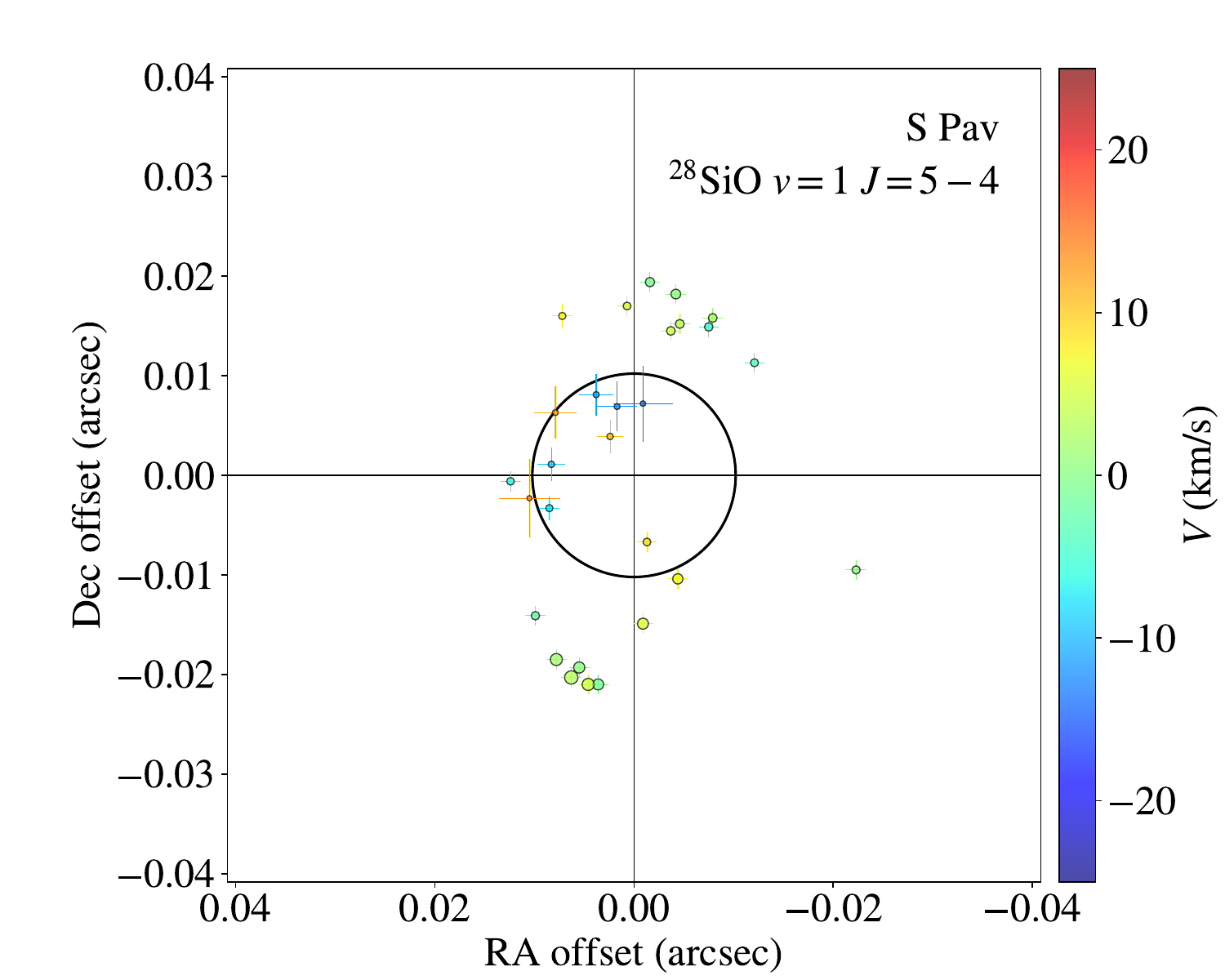}
   \includegraphics[width=0.396\textwidth, height=0.306\textwidth]{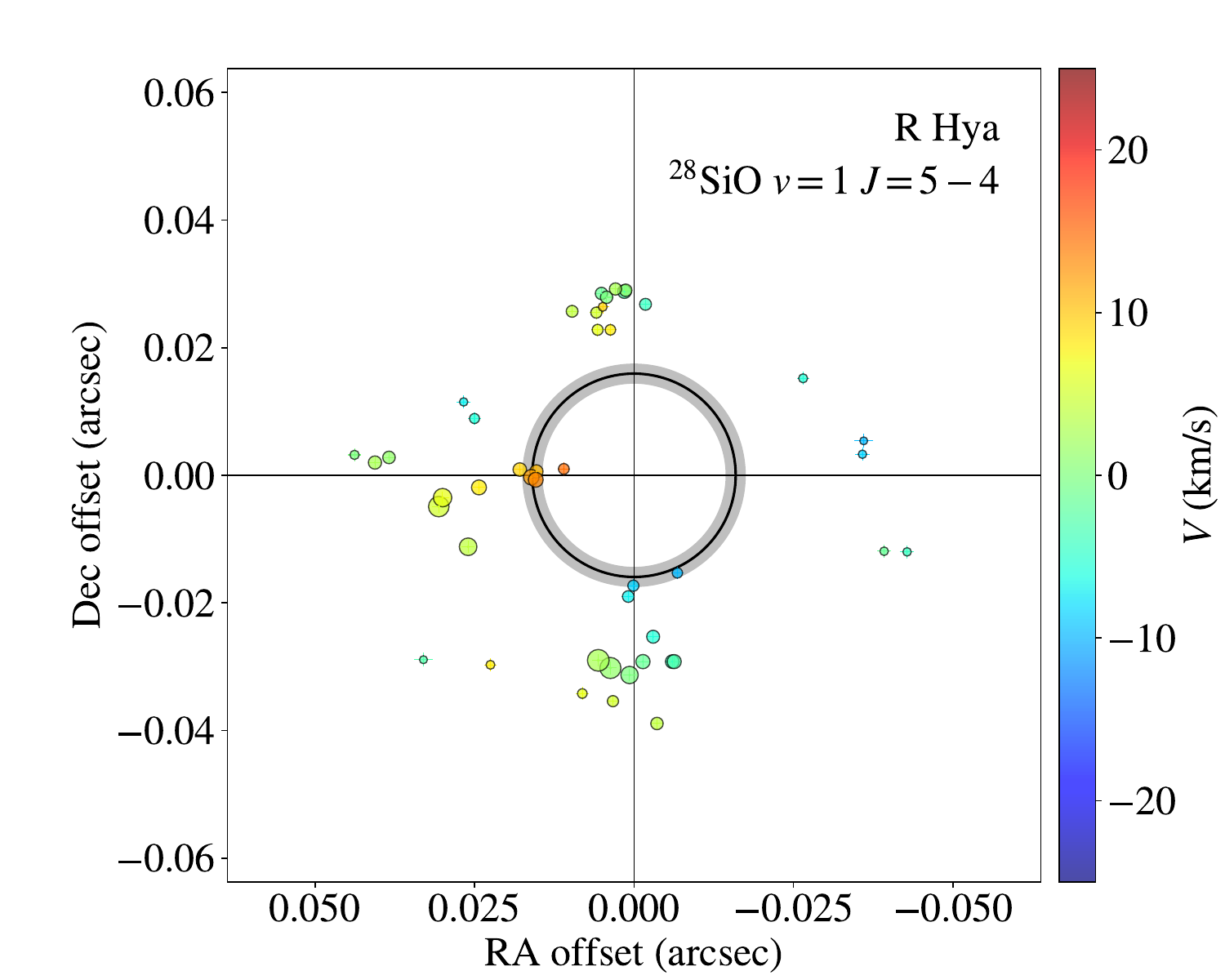}
   \caption{Position and velocity (with respect to $V_*$) of the extended-configuration SiO components around $\pi^1$ Gru (top left), R Hya (top right), S Pav (bottom left) and R Hya (bottom right). Note that the red star in the $\pi^1$ Gru panel marks the position of the close companion revealed by the ATOMIUM data set (see e.g. \citealt{2020A&A...644A..61H}). The SiO line of interest is listed in the top right corner of each panel. The size of the symbols represents log(integrated flux) and crosses indicate the position uncertainties. The black circle and grey shaded area represent the size of the uniform disc (UD) fit of the source and its corresponding uncertainty (Table \ref{table:atomium-sources}), respectively. The symbols are velocity colour-coded as in the right-hand side colour bar. The x- and y-axis ranges are set dynamically to be equivalent to 4 times the diameter of the UD fit.}
   \label{fig:maps-paper}
   \end{center}
\end{figure*}

\begin{figure*}
   \begin{center}
   \includegraphics[width=0.396\textwidth, height=0.306\textwidth]{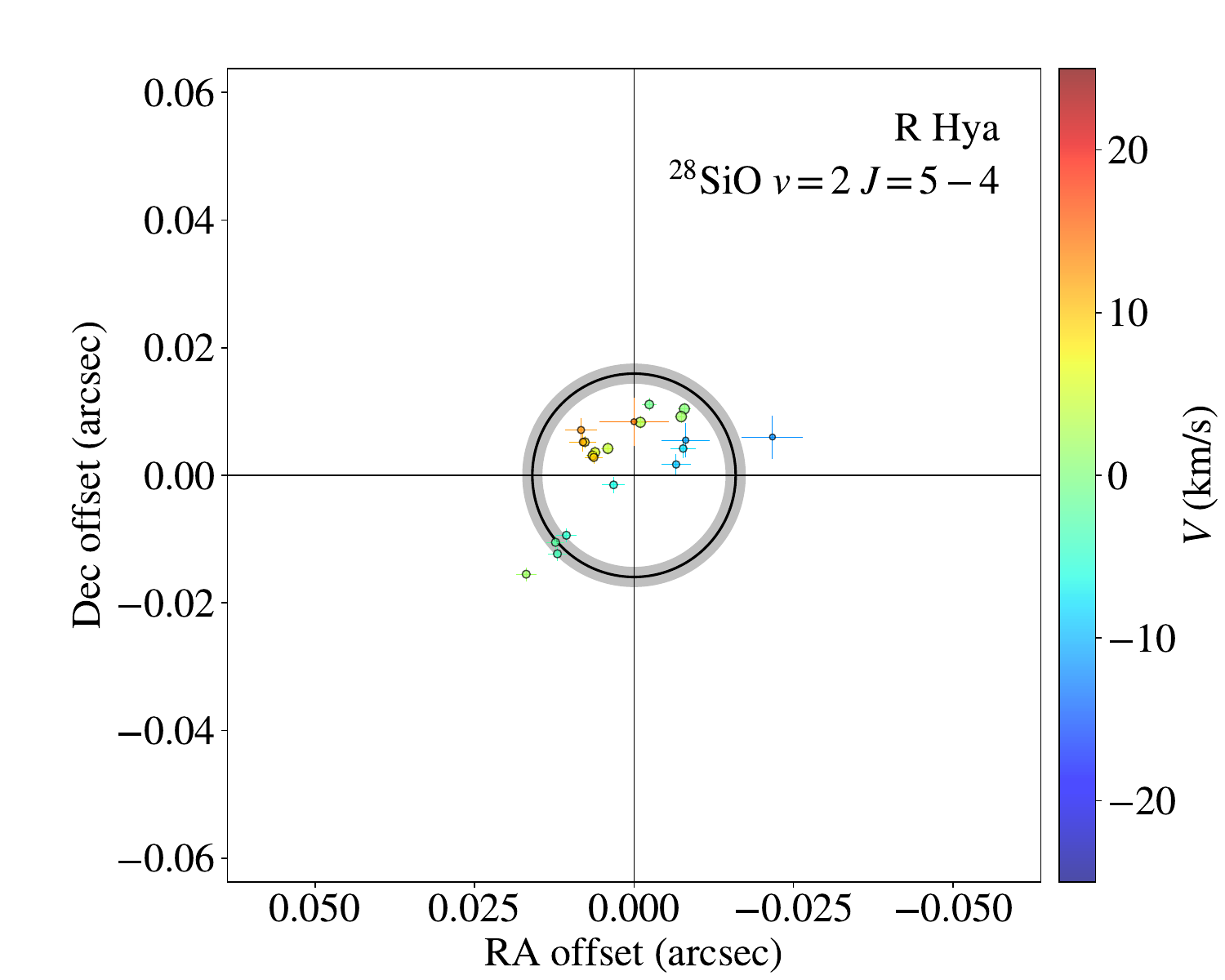}
   \includegraphics[width=0.396\textwidth, height=0.306\textwidth]{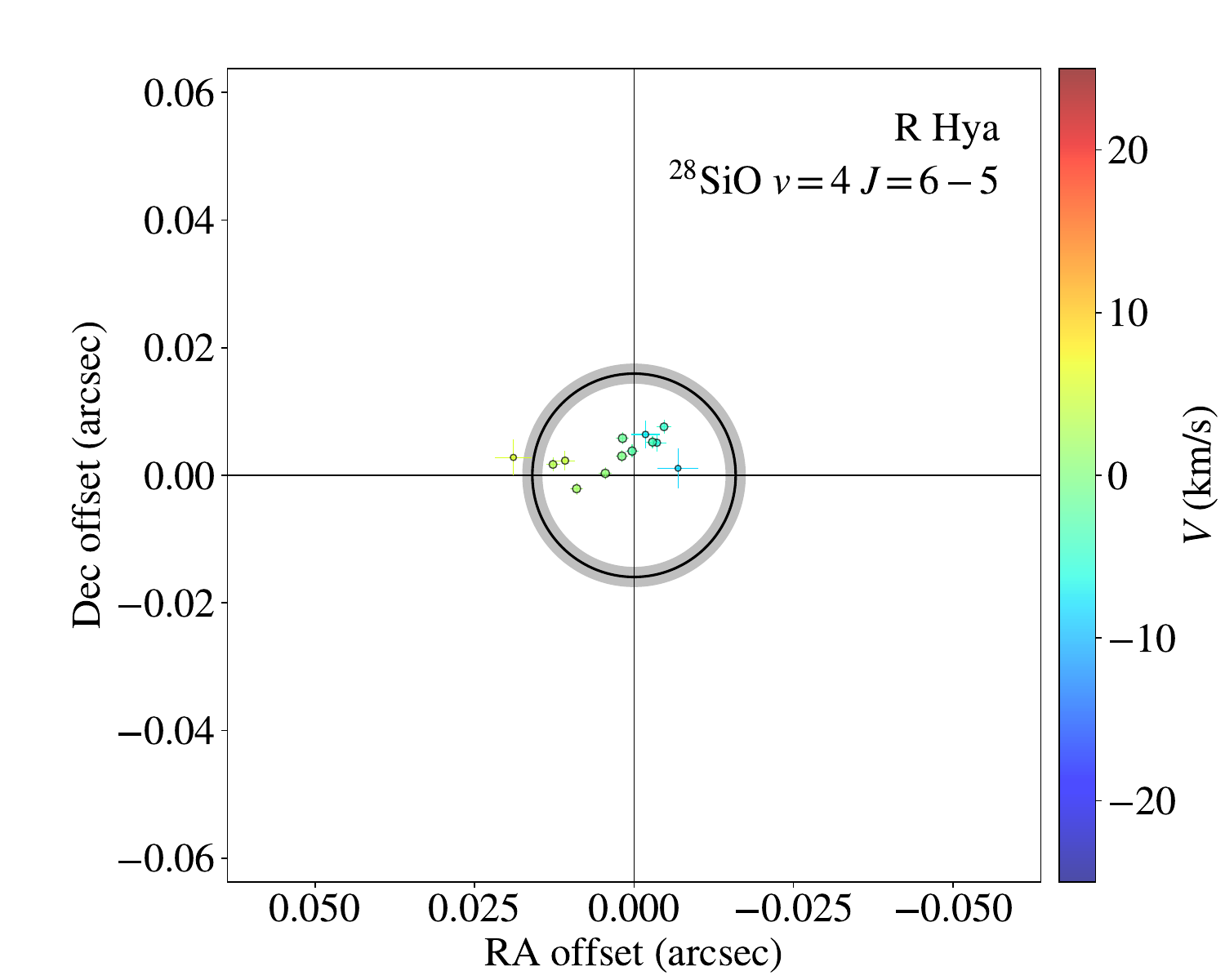}
   \includegraphics[width=0.396\textwidth, height=0.306\textwidth]{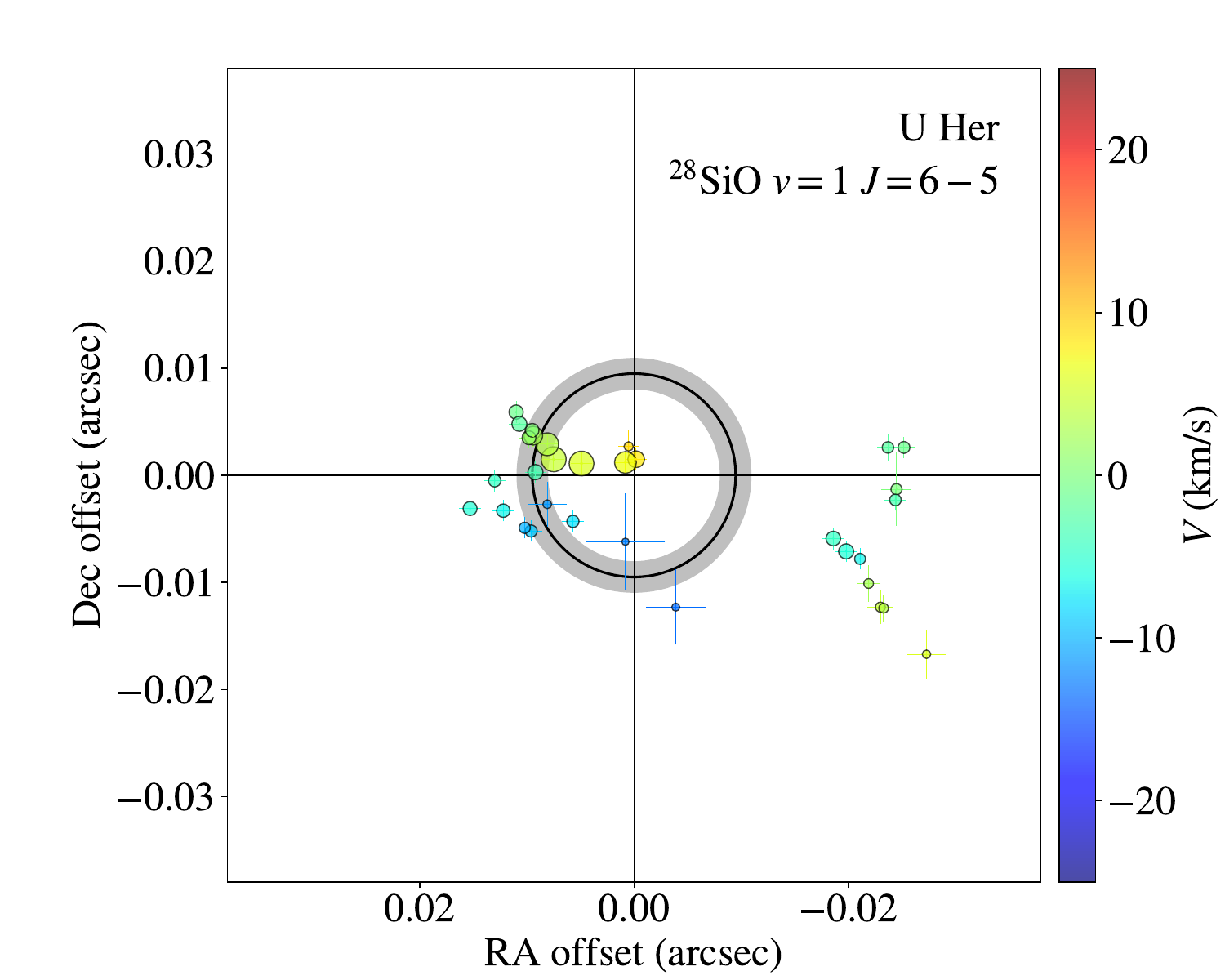}
   \includegraphics[width=0.396\textwidth, height=0.306\textwidth]{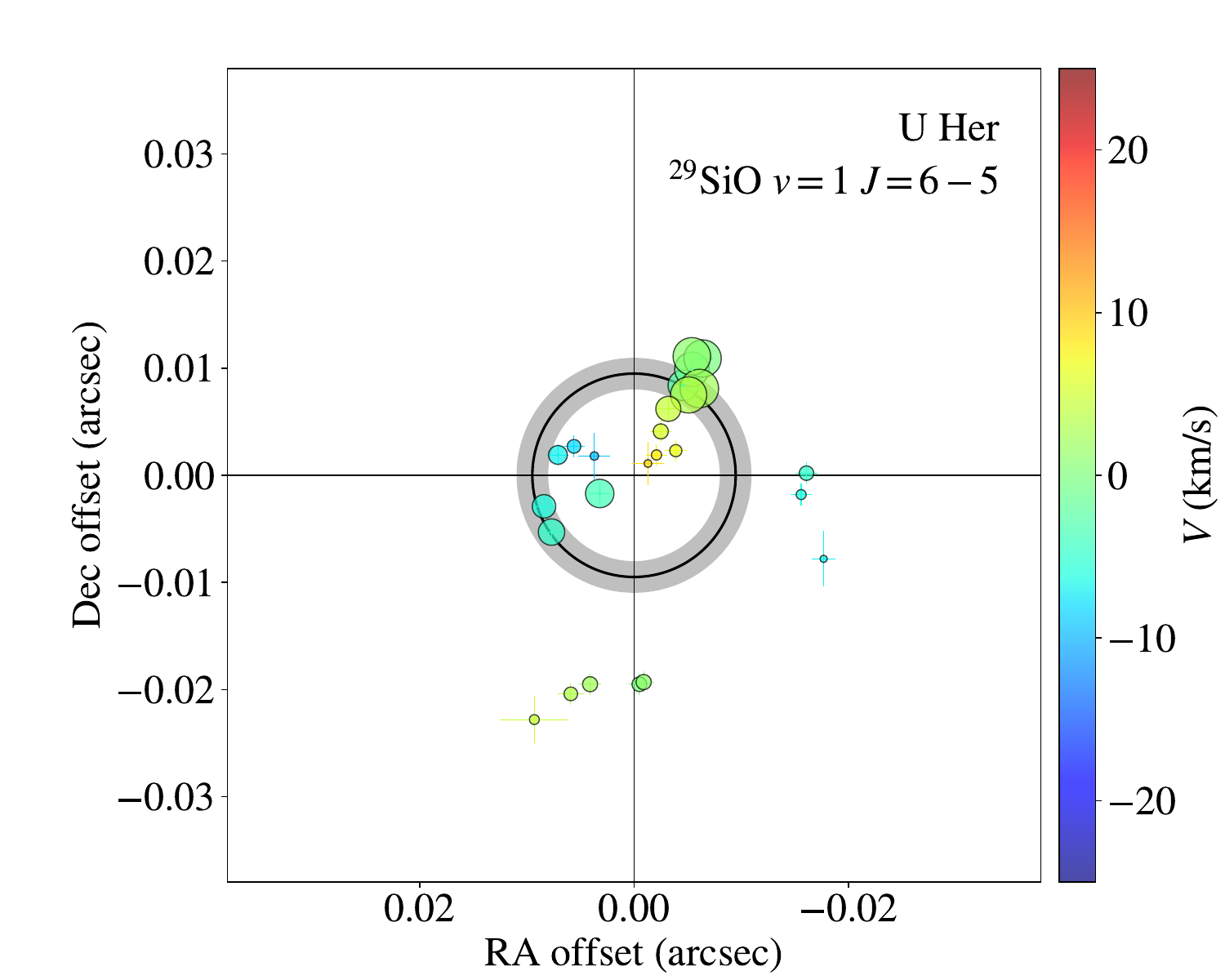}
   \includegraphics[width=0.396\textwidth, height=0.306\textwidth]{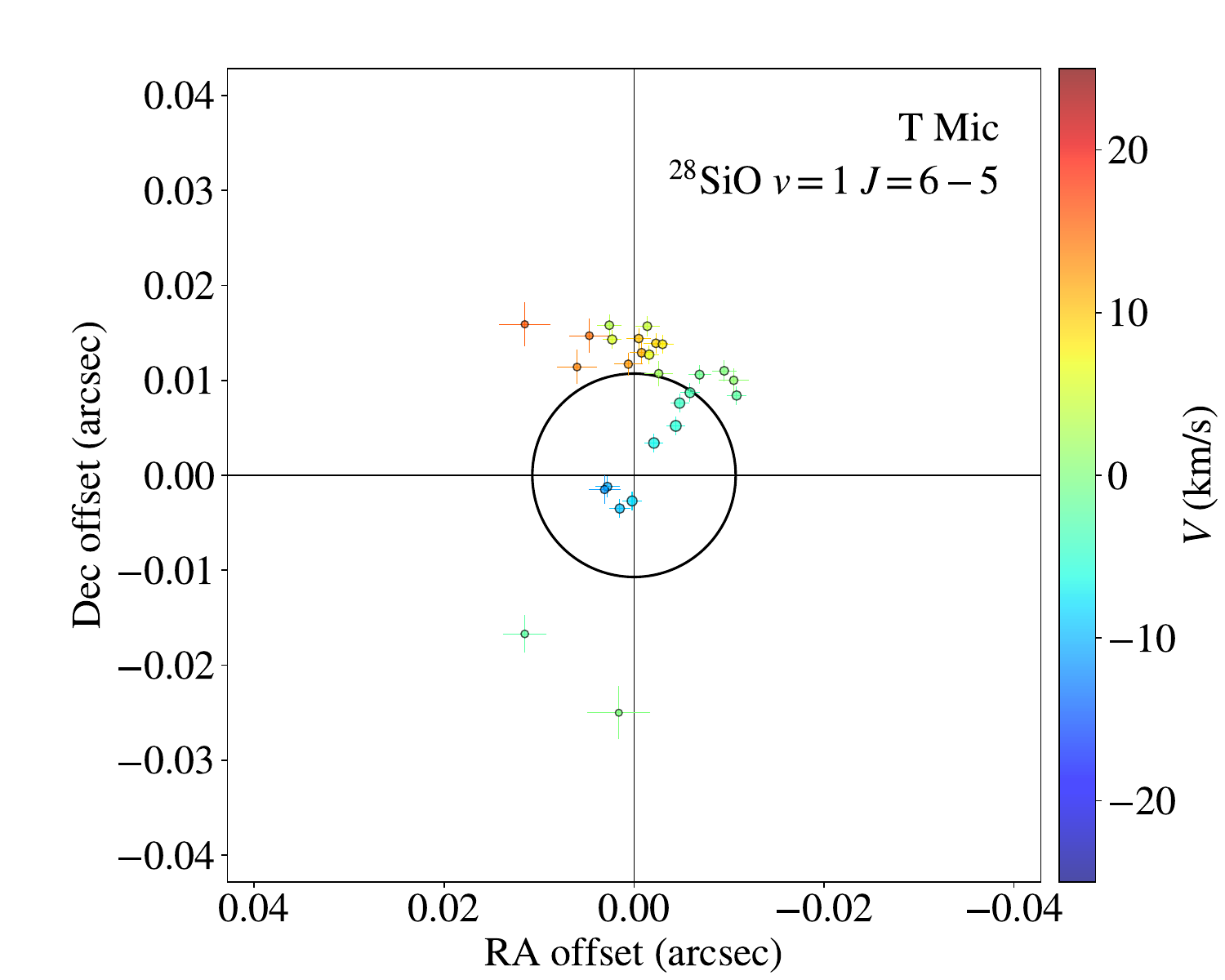}
   \includegraphics[width=0.396\textwidth, height=0.306\textwidth]{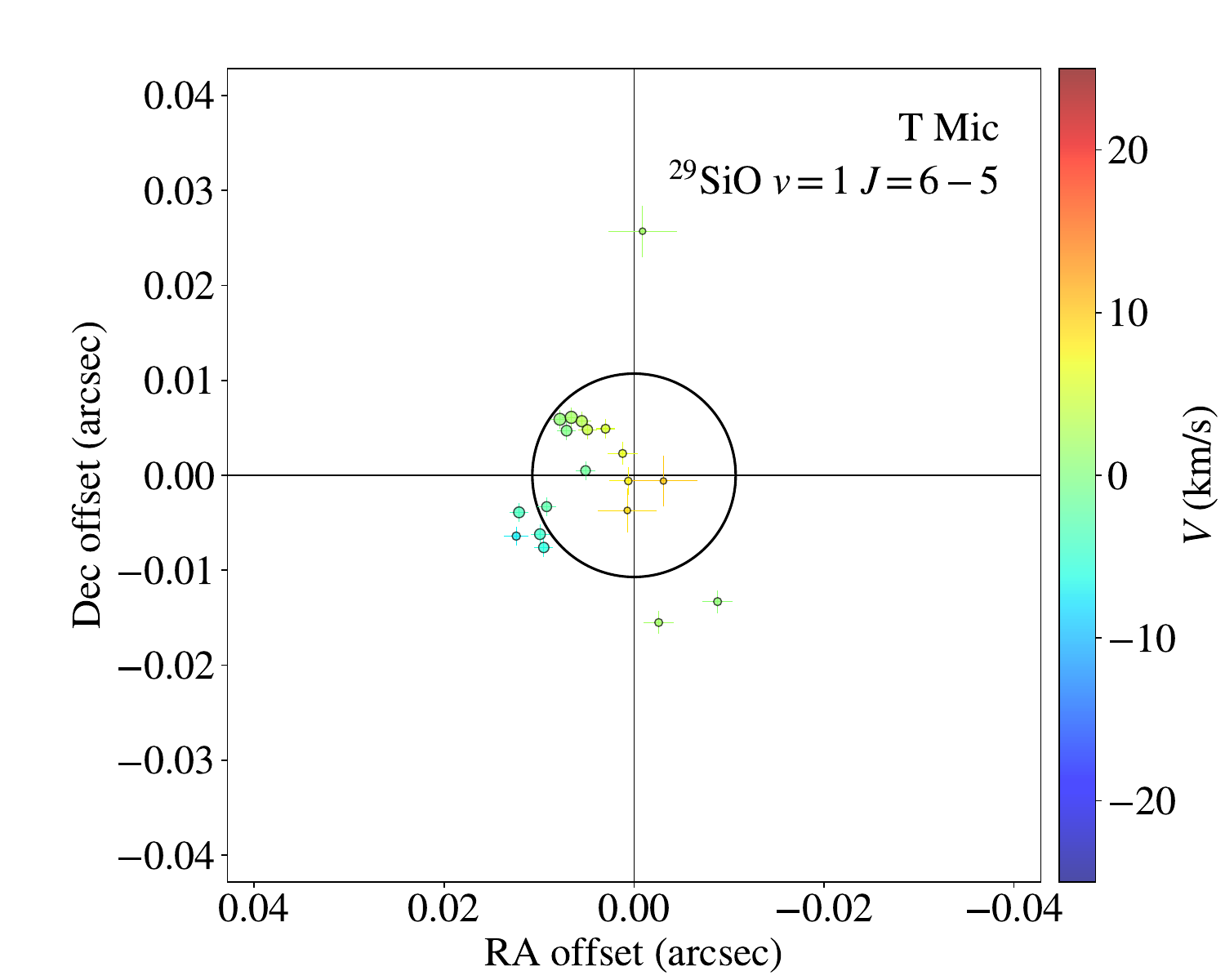}
   \caption{Position and velocity (with respect to $V_*$) of the SiO components around R Hya (top), U Her (middle), and T Mic (bottom). The rest of the caption is the same as in Fig. \ref{fig:maps-paper}.}
   \label{fig:maps-compare-H2O}
   \end{center}
\end{figure*}

\begin{figure*}
   \begin{center}
   \includegraphics[width=0.45\textwidth, height=0.32\textwidth]{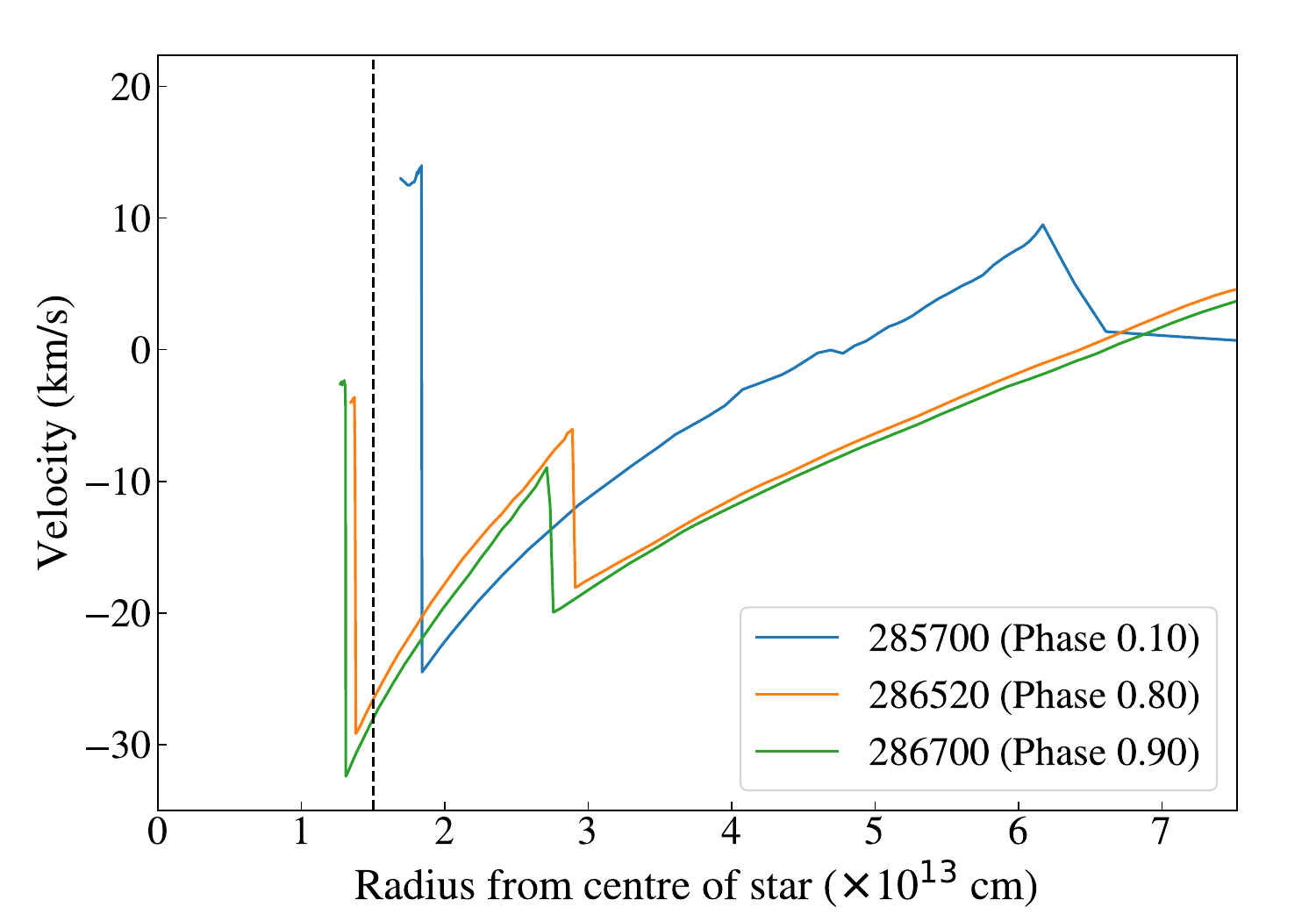}
   \includegraphics[width=0.43\textwidth, height=0.31\textwidth]{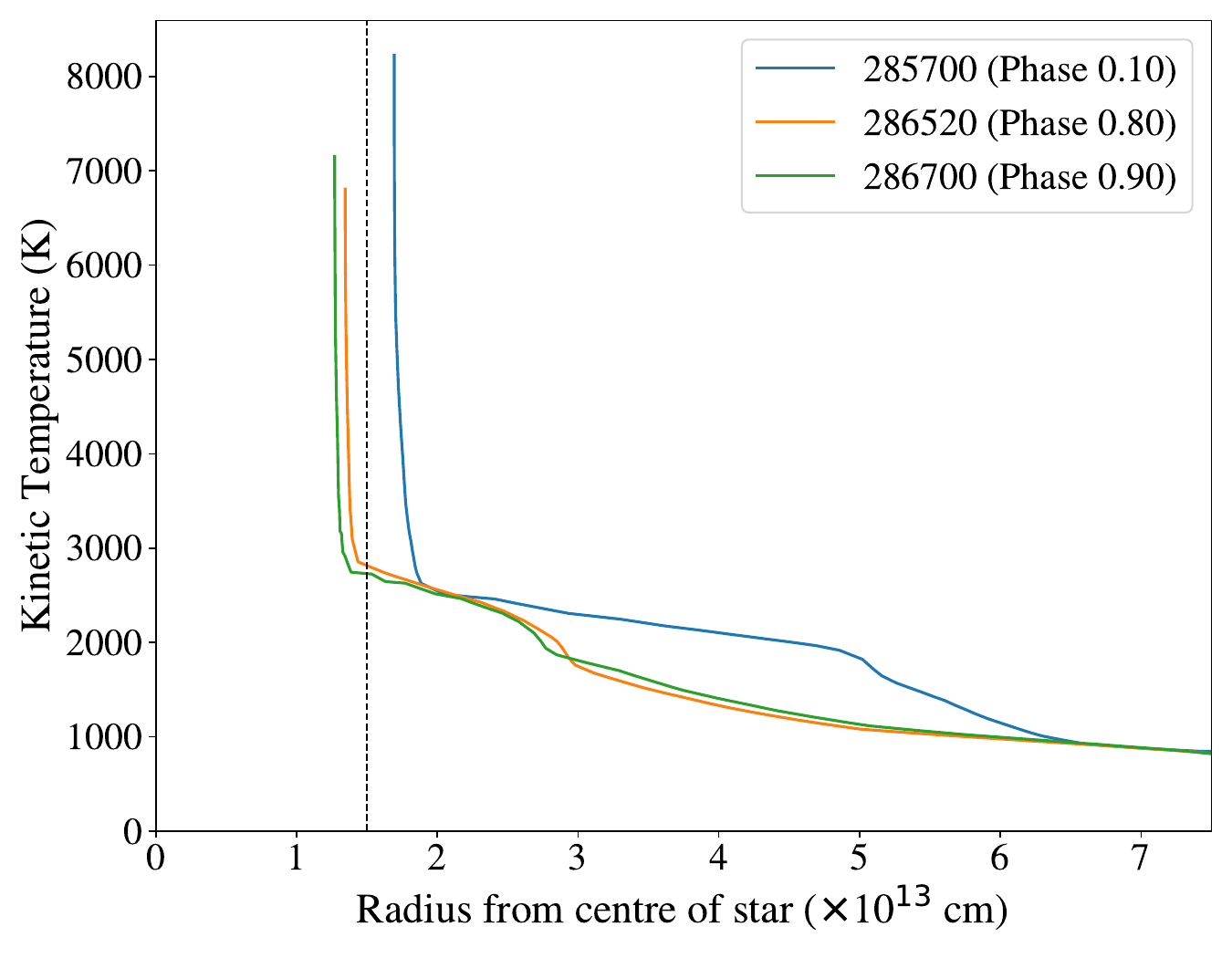}
   \caption{Plots of radial velocity (left) and kinetic temperature (right) vs radius from the centre of star for the CODEX models \citep{2011MNRAS.418..114I} numbered 285700 (blue, $\phi$ = 0.1), 286520 (orange, $\phi$ = 0.8) and 286700 (green, $\phi$ = 0.9), which are the closest in phase to R Hya, U Her and T Mic, respectively. A positive velocity means outward with respect to the star. Note that 1 modelled radius here ($R_{\rm{s}}$ = 215 R$_\odot$ or $\sim$1.5$\times$10$^{13}$ cm) is marked by a vertical black dashed line and corresponds to 8.0 mas at 126 pc (R Hya), 3.7 mas at 271 pc (U Her), and 5.7 mas at 175 pc (T Mic).}
   \label{fig:infall-codex}
   \end{center}
\end{figure*}

\subsection{Analysis of interferometric images}
\label{sec:dist:analysis}

\subsubsection{2D Gaussian component maps}
\label{sec:dist:analysis:SAD}
In order to investigate the kinematics and morphologies of the inner CSEs on au scales, our study has employed two-dimensional (2D) Gaussian fitting (see e.g. \citealt{2004MNRAS.348...34E}, \citealt{2011MNRAS.415.1083A}, \citealt{2011A&A...525A..56R}) to analyse the highly concentrated, non-diffuse zones of high-$J$ SiO emission. We utilised the {\small{\MakeUppercase{SAD}}} (Search And Destroy) task within the \textsc{aips}\footnote{\url{http://www.aips.nrao.edu/index.shtml}} package \citep{1985daa..conf..195W} and produced plots of the fitted component positions with the colours representing the radial velocities relative to the systemic velocity (hereafter, component maps). In this work, we pay particular attention to the smallest-scale SiO emission by following the 'maser feature' approach described in \citet{2011A&A...525A..56R}, Section 2.2. The maser amplification process produces emission in each velocity channel which usually has a Gaussian angular distribution (\citealt{1989ApJ...346..983E}; \citealt{2011A&A...525A..56R}), and 2D Gaussian components can be fitted with a precision that is contingent upon (beam size)/(S/N ratio) \citep{1997PASP..109..166C}. We defined our initial typical threshold to extract components as 5 times the rms noise level, 5$\sigma$ (or higher at 10$\sigma$ if spurious components need to be excluded further, see Appendix \ref{appendix:fitting}). A component is deemed real if there are 2 other components within a beam width of it in at least 3 successive channels.

\subsubsection{Brightness temperature}
\label{sec:dist:analysis:Tb}
Brightness temperature ($T_{\rm{b}}$) is a useful quantity for contextualising the relatively bright nature of masers in comparison with quasi-thermal emission and for determining the conditions required for the line excitations. In this work, we estimated the brightness temperatures (Eq. 1 of \citealt{2011A&A...525A..56R}) by determining the sizes of the deconvolved spots and their corresponding peak flux densities from 2D Gaussian component fitting obtained with {\small{\MakeUppercase{SAD}}} (see Section \ref{sec:dist:analysis:SAD} and Appendix \ref{appendix:fitting}). The uncertainties in $T_{\rm{b}}$ are the error propagated from the fitting errors in peak flux density and the semi-major and semi-minor axes of the components and are typically in the range of 20--35 per cent across the transitions and sources. In cases where the emission is detected at lower dynamic range, the fitted component size represents an upper limit to the true emitting region, and consequently, the derived \(T_{\rm{b}}\) values should be regarded as lower limits to the actual brightness temperature. 

In general, SiO maser emission is expected when the brightness temperature significantly exceeds the kinetic gas temperature, $T_{\rm{k}}$, in the excitation region. Although the exact gas temperature is difficult to determine observationally, it is likely comparable to, or slightly higher than, the continuum brightness temperature. \citet{2025Danilovich} derived continuum brightness temperature values for the ATOMIUM sources based on the stellar uniform disc fits measured at 241.25 GHz and found them to be less than $\sim$2300 K. We therefore adopt 2500 K, which is also the average T$_{\rm{eff}}$ from Table \ref{table:atomium-sources}, as a representative upper limit to $T_{\rm{k}}$ in the SiO-emitting layers. Using this reference, we arbitrarily classify emission as a probable maser when $5000 \lesssim T_{\rm{b}} < 10^4$ K, and as a clear maser when \( T_{\rm{b}} \geq 10^4\) K. Since the derived $T_{\rm{b}}$ are lower limits, emission with \( T_{\rm{b}} < 5000\) K cannot be definitively categorised; while it falls below our maser threshold, the true brightness temperature could still be high enough to represent maser emission. We present in Fig.~\ref{fig:Tb} the lower limits to the peak brightness temperatures derived for all detected SiO transitions in the ATOMIUM sample. Each value represents $\log_{10}$($T_{\rm{b}}$) for a particular transition in each ATOMIUM AGB star.

In this sample, the logarithmic scale of $T_{\rm b}$ spans more than four orders of magnitude, from approximately $10^{2.0}$ to $10^{6.5}$ K. Of the 16 detected SiO transitions, seven reach $T_{\rm b}\ge 10^{4}$ K in at least one ATOMIUM source and are therefore classified as clear masers. These are primarily associated with the \ce{^{28}SiO} $\varv=1,2,3$ and \ce{^{29}SiO} $\varv=1$ states. Clear masing in $\varv=0$ states is found only in V~PsA (\ce{^{28}SiO} $J=5-4$) and IRC+10011 (\ce{^{28}SiO} $J=5-4$ and \ce{^{29}SiO} $J=5-4$), both of which show $T_{\rm b}\sim10^{4}$~K. Possible \ce{^{28}SiO} $\varv=0$, $J=5-4$ maser emission is also seen in T~Mic ($T_{\rm b}\sim5900$~K). The \ce{^{28}SiO} $\varv=1$, $J=5-4$ transition shows the strongest and most widespread masing activity, being classified as a clear maser in 10 sources; S~Pav, W~Aql and IRC$-$10529 show no components above the clear masing threshold, and in SV~Aqr no Gaussian components could be fitted above the 5$\sigma$ level. The highest brightness temperature in the sample, $T_{\rm b} = 3.1\times10^6$~K, is measured in V~PsA for this same transition. Similarly, high brightness temperatures exceeding $10^6$~K are also measured in U~Her, RW~Sco, and IRC+10011, all of which display intense emission in one or more of the \ce{^{28}SiO} and \ce{^{29}SiO} lines. IRC+10011 shows the highest number of possible and clear masing transitions (7). The \ce{^{29}SiO} $\varv=6$, $J=6-5$ line in R Hya is weak and results in no fitted Gaussian components, but its narrow velocity range is good supporting evidence for maser action (see Section \ref{sec:SiO-id:v>6}).

Only one higher-vibrational transition, \ce{^{28}SiO} $\varv=4$, $J=6-5$ in S~Pav, attains an intermediate brightness temperature ($T_{\rm b} = 6600$ K), making it the sole $\varv>3$ line in the sample with evidence for probable masing. Among the detected \ce{^{30}SiO} transitions, only the $\varv=0$, $J=6-5$ line provides a statistically meaningful $T_{\rm b}$ estimate; the remaining lines are too weak for reliable component fitting. The derived $T_{\rm b}$ values for this line remain below 1000 K, consistent with predominantly quasi-thermal emission.

\subsection{Small-scale SiO emission}
\label{sec:dist:overview}
Fig. \ref{fig:maps-paper}--\ref{fig:maps-compare-H2O} present examples of fitted component maps in five ATOMIUM sources: $\pi^1$ Gru, R Hya, S Pav, U Her and T Mic. The SiO transition is labelled in each of the panels alongside the source name and the colour bar shows radial velocities with respect to $V_*$. The error bars on each component represent the fitted position uncertainty from 2D Gaussian fitting. We measured the diameter of the stellar continuum (central frequency 241.75 GHz) of each star (2$R_*^{\rm{mm}}$, Table \ref{table:atomium-sources}) by fitting a uniform disc (UD) to the visibilities as described in \citet{2025Danilovich}. This is shown as a black circle in the component maps with its uncertainty represented by a grey shaded area. Velocities are displayed as ($V_{\rm{LSR}}-V_*$) where $V_*$ is given in Table \ref{table:atomium-sources}. In the case of RW Sco, since the UD diameter has large uncertainty, we have opted to use the calculated diameter (6.0 mas) in the plots instead. The uncertainty in component positions obtained from the fitting is shown as error bars. A compilation of SiO component maps can be found as online supplementary material. 

We focus here on identifying specific details of the morphological features of the inner CSEs in the ATOMIUM sources as traced by the high-$J$ SiO, especially when compared to H$_2$O, OH \citep{2023A&A...674A.125B} and CO (\citealt{2020Sci...369.1497D}, \citealt{2022A&A...660A..94G}) at larger spatial scales. We also emphasise that the results and discussions that follow only focus on what can be deduced from single-epoch component maps, e.g. gas motion at the time of observation. We do not aim to describe nor explain in detail the physico-chemical processes that lead to such detections as single-epoch observations of SiO-emitting sources often display varying morphologies, such as in IK Tau (\citealt{2005ApJ...625..978B}, \citealt{2008PASJ...60.1039M}) and R Cas (\citealt{2001AJ....122.2679P}, \citealt{2004ApJ...608..480B}, \citealt{2011MNRAS.415.1083A}). In the following discussion, we examine the observed morphological and kinematic structures without trying to draw a strict distinction between quasi-thermal and maser emission except in the case of strong maser action.

By far, the most common feature found in the millimetre-wavelength SiO component distribution towards the ATOMIUM sources is velocity-position gradients. They are identifiable by their characteristic gradual change in line-of-sight velocity, from red to blue, usually seen as an arc-like feature of size (roughly defined as the angular size covered from one end to the other) of $\sim$5 mas to a few tens of mas. Such a pattern may not accurately represent the overall structure of the maser-emitting cloud, as beaming can lead to linear or curved alignments of components even when originating from a spherical clump \citep{2011A&A...525A..56R}, but it informs us of gas kinematics near the stellar surface, sometimes leading to a more complete interpretation of the wind morphology. 

Some of the fitted component data have already been published as part of the ATOMIUM series. In $\pi^1$ Gru, we found evidence of rotation and a stream of gas accelerating from the surface of the AGB star to a close companion in the $\pi^1$ Gru system (Fig. \ref{fig:maps-paper}, top left; Figs. 11--12 of \citealt{2020A&A...644A..61H}). The mid- and extended-configuration component maps of $\varv=0$ lines from the \ce{^{28}SiO}, \ce{^{29}SiO} and \ce{^{30}SiO} isotopologues towards R Hya also helped identify an inclined, differentially rotating equatorial density enhancement (EDE) which implies the presence of a second nearby companion (Figs. 12--13 of \citealt{2021A&A...651A..82H}). The top right panel of Fig. \ref{fig:maps-paper} shows the \ce{^{28}SiO} $\varv=1$, $J=6-5$ components which trace parts of this inclined, sub-arcsecond scale EDE.

While complex velocity structures are common in the CSE, SiO components near $V_*$ can still appear symmetrically distributed around the star within a few $R_*^{\rm{mm}}$, as seen in the \ce{^{28}SiO} $\varv=1$, $J=5-4$ maps of S Pav and R Hya. In S Pav (Fig. \ref{fig:maps-paper}, bottom left), two prominent arcs ($\sim$7 mas) located NW and SE of the star span $-5$ to $+5$ \kms, lying at 13.7 and 11.1 mas from the centre, consistent with the size of the UD radius \citep{2025Danilovich}. Additional blue- and red-shifted components ($\pm13$ \kms) appear scattered in the NE and SW. R Hya (Fig. \ref{fig:maps-paper}, bottom right) shows similar clumpy, arc-like structures at comparable distances, especially along the N-S axis, with velocities between $-7$ and $+10$~\kms, and weak blue-shifted features ($-11$ to $-2$~\kms) seen westward at offsets of $-30$ to $-40$~mas. There is also a position-velocity structure extending between RA offsets +30 and +12 mas going from $\sim$7 to $\sim$14 \kms\ at the east. These show that compact high-$J$ SiO emission can appear in a sparsely filled ring, similar to low-\( J \) masers that trace a spherical shell where radial motions favour tangential amplification \citep{1994ApJ...430L..61D,2004A&A...414..275C}. However, with one epoch, we cannot say that this is a general feature expected of observation of high-$J$ SiO masers, as these position-velocity structures may be indicative of a more complicated gross morphology.

In R Hya (Fig. \ref{fig:maps-compare-H2O}, top), it is clear that the components are distributed in the NW-SE (\ce{^{28}SiO} $\varv=2$, $J=5-4$) and E-W (\ce{^{28}SiO} $\varv=4$, $J=6-5$) directions across the UD diameter, where the velocity shifts from $\sim$3--5 \kms\ in the SE to roughly $-$9 \kms\ in the NW and from 6 \kms\ in the E to $-$8 \kms\ in the W, respectively. This range is similar to what has been reported in \citet{2023A&A...674A.125B} for the 262.898 and 268.149 GHz H$_2$O lines. Based on the spectra, absorption maps and these component maps, we point out that our SiO data not only support the matter infall first deduced in H$_2$O and CO towards this source and other AGB stars such as R Aqr and R Scl, but the velocities also fall in the same range of 6--9 \kms\ (\citealt{2023A&A...674A.125B}, \citealt{2016A&A...591A..70K}). They are in favour of the line formation models by \citet{2010A&A...514A..35N}, which suggest that velocity variations observed in the dust-free region of 1--2 $R_*^{\rm{mm}}$, i.e. where SiO masers are usually seen, in AGB CSEs require infall/outflow velocities of 5--10 \kms. Resolved case studies of Mira \citep{2016A&A...590A.127W}, R Dor \citep{2024A&A...685A..11K}, and W Hya \citep{2025A&A...704A..18O} suggest that these velocities may trace a shock-levitated, largely gravitationally bound inner atmosphere in which infall and outflow coexist within a few $R_*$, rather than a simple monotonic acceleration into the terminal wind.

In U Her (middle panels of Fig. \ref{fig:maps-compare-H2O}), both transitions show a velocity gradient: the velocity is near $V_*$ at $\sim$0.8$R_*^{\rm{mm}}$ and increases to 9.5 \kms\ toward the source centre. However, the gradients differ in orientation, NE–W for \ce{^{28}SiO} $\varv=1$, $J=6-5$ and NW–SE for \ce{^{29}SiO} $\varv=1$, $J=6-5$. The presence of such velocity gradients on the near side of the star might point towards a possible gas infall. A cluster of blue-shifted components (between $-$10 and $-$3 \kms) in the east of the AGB star, $\sim$0.5$R_*^{\rm{mm}}$ in the sky plane (Table \ref{table:atomium-sources}), is visible in both transitions. These components likely arise due to the localised complex, irregular dynamics in the inner CSEs where SiO is hosted \citep{2020Sci...369.1497D}, or, to a lesser extent, a streamer in the direction towards the observer.

As for T Mic, it is unclear from the component maps of the \ce{^{28}SiO} $\varv=1$, $J=6-5$ and \ce{^{29}SiO} $\varv=1$, $J=6-5$ lines (the bottom panels of Fig.~\ref{fig:maps-compare-H2O}) whether the SiO masers arise in an expanding CSE or a streamer as suggested by the line profile and mom0 map of the \ce{^{28}SiO} $\varv=3$, $J=6-5$ line (Fig. \ref{fig:absorption}, bottom row). However, this does not immediately rule out the scenario, as absorption may still occur in the warmer inner parts of the CSE, given the moderate excitation energies involved, $E_{\rm{u}} \sim$1800 K, and we may simply be observing SiO emission from different layers or radii along the line of sight. Interestingly, we also observe opposing velocity gradients in the two transitions: red to blue (13 to $-$12 \kms) and blue to red ($-$7 to 10 \kms), despite both gradients appearing to originate from $\sim$1.2$R_*^{\rm{mm}}$ and end near the stellar position.

\subsection{Interpreting SiO emission through shocks and stellar phases}
\label{sec:dist:shocks}

At the time of their respective observations, the stellar phases of R Hya ($P$ = 359 d), U Her ($P$ = 406 d) and T Mic ($P$ = 352 d) were 0.07, 0.76 and 0.84 respectively (Table \ref{table:atomium-sources}). These stars have comparable luminosities of order a few $10^{3}$--$10^{4}$ $L_\odot$ \citep{2010A&A...523A..18D,2024A&A...686A.251W}, placing them in a similar parameter space to typical Mira variables considered in the CODEX models by \citet{2011MNRAS.418..114I}. Based on the four-cycle continuous \textsc{o54} CODEX model series for \textit{o} Cet (5400 L$_\odot$, $P$ = 332 d; their Table 4), we can describe a possible scenario responsible for the SiO emission based on the relative positions between the modelled shock fronts and the fitted components in the observed phases. The pulsation period of the model is comparable to those of the three ATOMIUM targets. Given the complex dynamical structure of Mira atmospheres, different layers of material may be interacting with multiple shock fronts simultaneously, with inner regions being accelerated by one shock while outer layers respond to another. To account for this, we focus on the velocities and kinetic temperatures of the three closest (in phase) CODEX models, numbered 285700, 286520 and 286700, which are plotted in Fig. \ref{fig:infall-codex}. A positive velocity in this figure is in the outward direction from the central star. One modelled stellar radius, $R_{\rm{s}}$, in the CODEX models is 215 R$_\odot$ i.e. $\sim$1.5$\times$10$^{13}$ cm ($\sim$1 au), which will be used as a reference in the discussion throughout this section. Note that $R_{\rm{s}}$ here is the constant 'parent' radius, and that the photospheric radius is itself a function of phase (see e.g. Fig. 1 of \citealt{2011MNRAS.418..114I}). Based on the modelled kinetic temperatures, any quasi-thermal SiO emission with $E_{\rm{u}} \gtrsim 2500$ K, i.e. $\varv > 1$ (Table \ref{table:line-coverage}), may be emitted from regions $<$ 2 $R_{\rm{s}}$.

For R Hya (CODEX model 285700), it is apparent that while a new shock front was being launched, previously shocked material at radii up to $\sim$3 $R_{\rm{s}}$ was already falling back toward the star. This provides strong support for the inferred gas infall seen in the absorption of the \ce{^{29}SiO} $\varv=3$, $J=6-5$ line (Fig. \ref{fig:absorption}, second row). The line-of-sight velocity range in the CODEX 285700 model appears consistent with that of the \ce{^{28}SiO} $\varv=2$, $J=5-4$ and \ce{^{28}SiO} $\varv=4$, $J=6-5$ components shown in the top panels of Fig. \ref{fig:maps-compare-H2O}. The red-shifted components observed toward R Hya likely originated from regions around $\sim$4 $R_{\rm{s}}$ from the stellar surface. The blue-shifted components likely trace regions where the shock, situated at $\sim$1.1 $R_{\rm{s}}$, is actively accelerating material outwards. Given that the shock front had not yet reached the more extended layers of the CSE, we expect the strongest blue-shifted emission to originate from these inner regions, where the post-shock velocity field was still well-defined. The predicted kinetic temperatures at $\sim$a few $R_{\rm{s}}$ are in the range 2000--2800 K. This appears consistent with the excitation temperature derived from the population diagram, although it should be interpreted cautiously because the excitation is likely affected by radiative pumping, optical-depth effects, line overlaps, and other non-LTE processes (see Section \ref{sec:dist:non-LTE:pop-diagrams}). The similar observed absorption depths in the $\varv=3$ transitions of \ce{^{28}SiO} and \ce{^{29}SiO} are also likely to reflect the combined effects of these processes rather than a simple abundance-scaled opacity relation. 

Based on the CODEX models 286520 and 286700, i.e. $\phi$ at which U Her and T Mic were observed, the stellar phases of 0.8--0.9 suggest a contraction-dominated phase of the CSE. They indicate that the envelope was undergoing infall at all radii smaller than $\lesssim$4.5 $R_{\rm{s}}$, meaning that no large-scale outward-moving shock had yet disrupted the collapse. This is in agreement with the observed red-shifted components in U Her, which likely trace material still falling inwards (Fig. \ref{fig:maps-compare-H2O}, middle panels). The presence of blue-shifted SiO emission, however, suggests that some emission arises from larger stellocentric distances, i.e. $\gtrsim$ 4.5 $R_{\rm{s}}$, beyond the infall-dominated region. Material may be undergoing slower infall or has already been influenced by earlier shock activity, where the inward velocity field diminishes or reverses, as $E_{\rm{u}}$ of the $\varv=1$ transitions $\sim$1800 K. We did not detect any SiO absorption in the ATOMIUM data set towards U Her, unlike some of the stars discussed in Section \ref{sec:absorption}. In those cases, the absorbing transitions are $\varv=3$ or higher, found situated close to or in front of the star, whereas the detected compact $\varv=1$ emission in U Her likely traces more extended regions (see Fig. \ref{fig:maps-compare-H2O}, middle panels). This suggests the infalling layer, if present, may lie farther out compared to the other sources where absorption was observed. 

The origin of the arc-like distributions of the \ce{^{28}SiO} $\varv=1$, $J=6-5$ and \ce{^{29}SiO} $\varv=1$, $J=6-5$ components around T Mic at phase $\phi=0.84$ (Fig. \ref{fig:maps-compare-H2O}, bottom) is unclear from the CODEX models. Despite the P Cygni profile in \ce{^{28}SiO} $\varv=3$, $J=6-5$, neither modelled radial velocities nor fitted components in these $\varv=1$ lines seem to strengthen the argument that the star exhibits radially outward streaming gas. One should also consider, however, the limitations of the 1D models, which do not capture the complex 3D morphology and simultaneous infall and outflow in this region as observed and modelled previously (\citealt{2023A&A...669A.155F} and references therein). The large difference in $E_{\rm{u}}$ (1800 vs 5300 K) means they may not originate from similar radii from the star. The existence of the $\varv=3$ emission and absorption could arise from local density or temperature inhomogeneities \citep{2009MNRAS.394...51G} or from transient shocks interacting with a pre-existing, non-uniform CSE \citep{2002A&A...386..256H}. The structured arcs support a scenario in which localised shock-driven compression or flow divergence leads to favourable conditions for maser amplification along curved paths. 
From these results, further temporal monitoring at multiple epochs and high angular resolutions would be needed to distinguish between persistent asymmetries and phase-dependent features.

\begin{figure*}
   \begin{center} 
   \includegraphics[width=0.396\textwidth, height=0.306\textwidth]{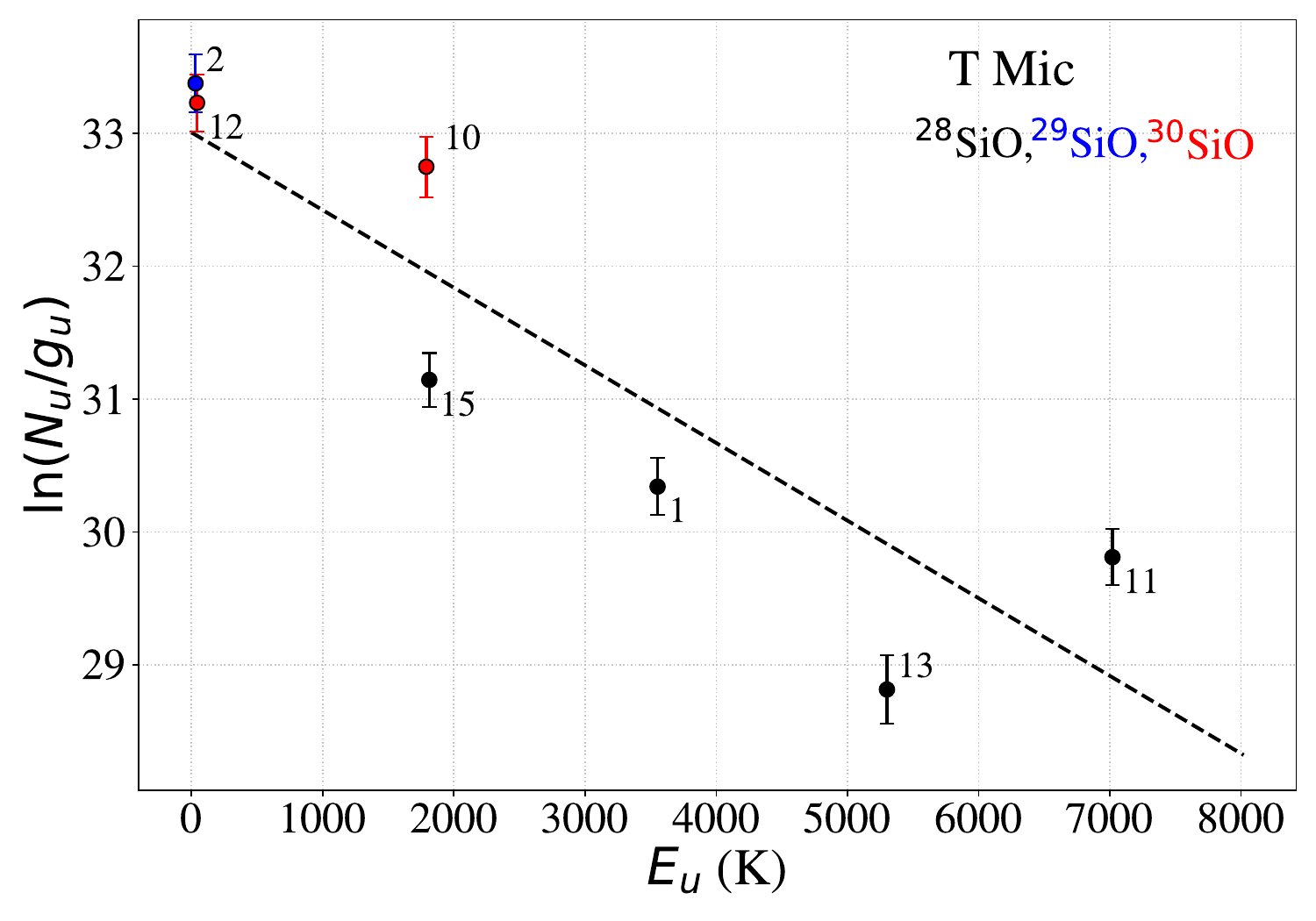}
   \includegraphics[width=0.396\textwidth, height=0.306\textwidth]{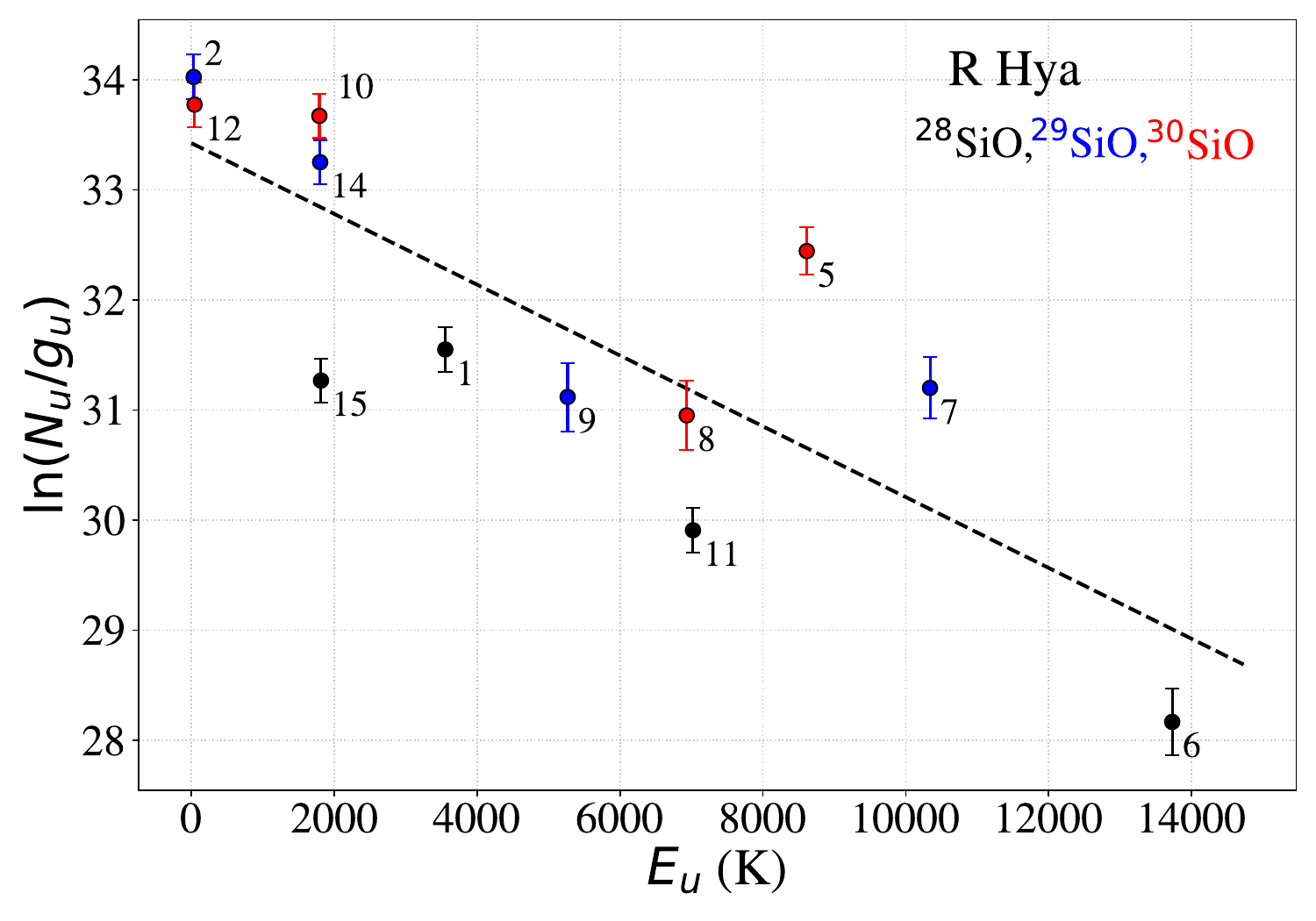}
   \includegraphics[width=0.396\textwidth, height=0.306\textwidth]{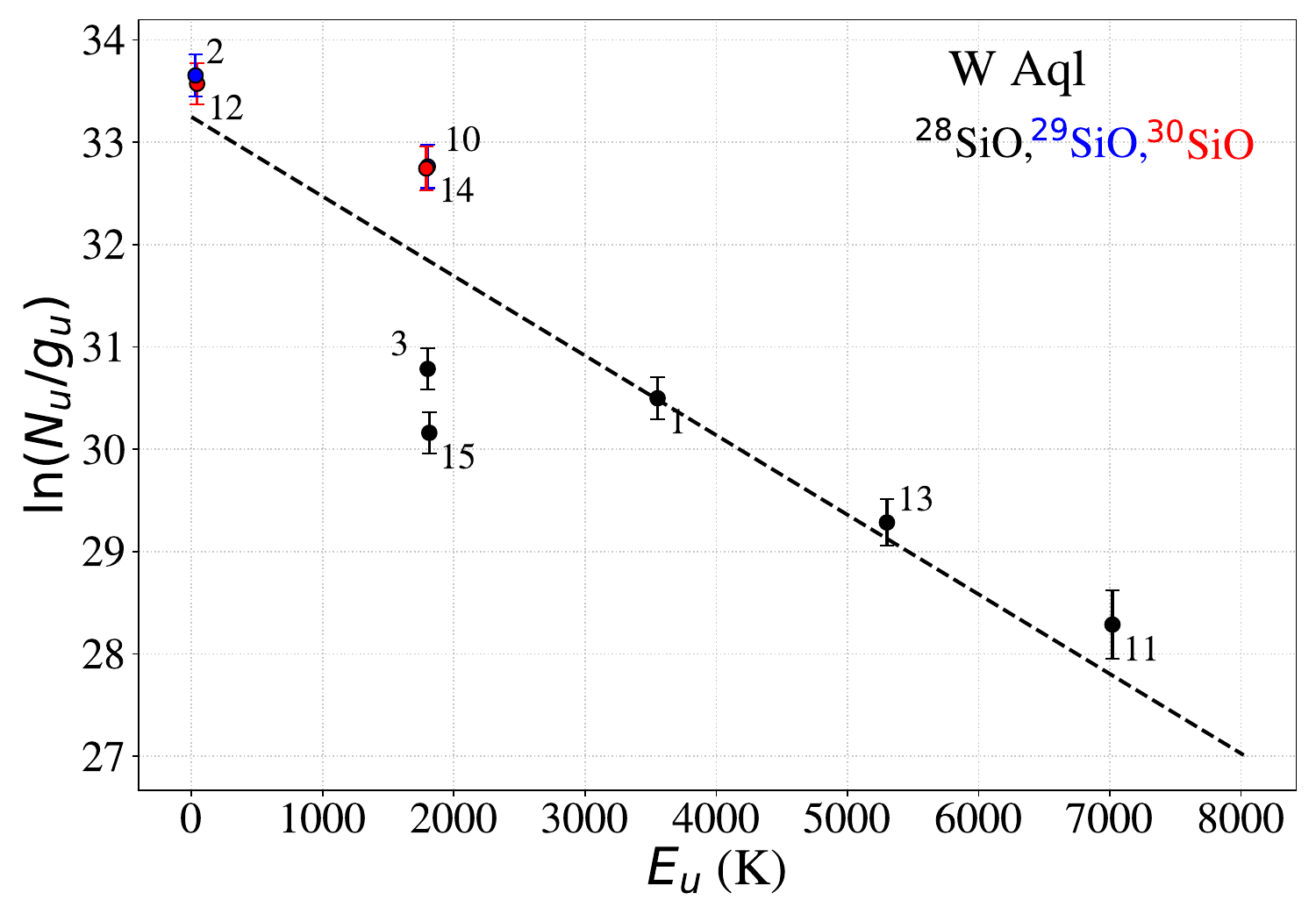}
   \includegraphics[width=0.396\textwidth, height=0.306\textwidth]{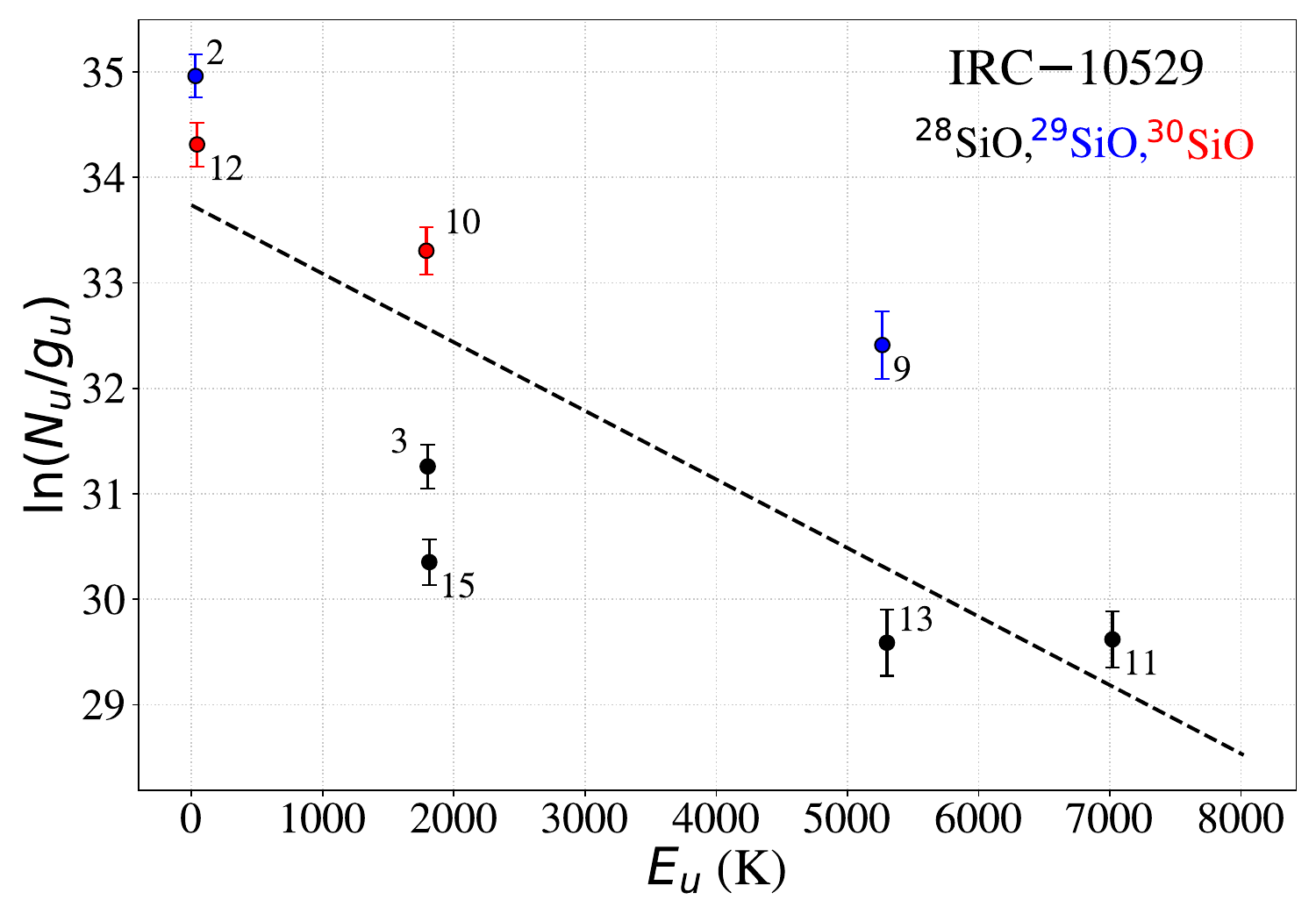}
   \caption{Population diagrams of SiO in T~Mic, R~Hya, W~Aql and IRC$-$10529). The natural logarithm of the upper state column density in cm$^{-2}$ divided by its statistical weight, $\ln(N_{\rm u}/g_{\rm u})$, is plotted against the upper state energy, $E_{\rm u}$ in K. Data points are colour-coded by isotopologue: $^{28}$SiO (black), $^{29}$SiO (blue), and $^{30}$SiO (red). The column densities of the optically thin minor isotopologues ($^{29}$SiO and $^{30}$SiO) have been scaled by assumed solar elemental abundance ratios ($^{28}$Si/$^{29}$Si = 19.7 and $^{28}$Si/$^{30}$Si = 30.0). Numbers adjacent to the data points correspond to the line numbers listed in Table \ref{table:line-coverage}. See the fitted results in the text.}
   \label{fig:pop-diagrams}
   \end{center}
\end{figure*}

\section{Properties of highly excited S\MakeLowercase{i}O emission in oxygen-rich AGB stars}
\label{sec:compare}
Based on the extracted spectra, channel maps and Gaussian component maps already introduced and discussed in Sections \ref{sec:SiO-id}--\ref{sec:dist}, we present here a discussion about line intensity ratios and population diagram analysis, a summary of how different transitions from various $\varv$-states compare, and the effects of the stellar pulsation and phase on SiO detections.

\subsection{Line intensity ratios}
\label{sec:dist:line-ratios}

The two most ubiquitous $\varv>0$, high-$J$ SiO transitions in the ATOMIUM observations, \ce{^{28}SiO} $\varv=1$, $J=5-4$ (215.596~GHz) and $J=6-5$ (258.707~GHz), provide a useful diagnostic for deviations from LTE. Because both lines originate from the same vibrational level, their expected LTE intensity ratio is set primarily by the Einstein A-coefficients and the line frequencies $\nu$:

\begin{equation}
\left( \frac{S_{5-4}}{S_{6-5}} \right)_{\rm LTE}
\simeq
\left( \frac{A_{5-4}}{A_{6-5}} \right)^{1/2}
\left( \frac{\nu_{6-5}}{\nu_{5-4}} \right)^{3/2}
\approx 0.992.
\end{equation}

\noindent{Therefore, for excitation described by a Boltzmann distribution}, the flux ratio $S_{5-4}/S_{6-5}$ is expected to be on the order of unity.

The observations, however, show a much wider range of ratios. In most stars, the $\varv=1$, $J=5-4$ line is significantly brighter than LTE predicts, with $S_{5-4}/S_{6-5}$ ranging from just over 1 (RW Sco) to 56.7 (R Aql). Conversely, the $J=6-5$ transition dominates in SV Aqr, W Aql and IRC$-$10529, again by factors larger than compatible with LTE opacity scaling ($S_{5-4}/S_{6-5}$ = 0.71, 0.86 and 0.78, respectively). Because these two transitions differ only in rotational quantum number, such large departures from the LTE opacity ratio imply strong non-LTE population redistribution within the $\varv=1$ ladder, i.e. likely evidence of maser action. These deviations also correlate with the brightness-temperature analysis (Fig.~\ref{fig:Tb}). In the sources where $S_{5-4}/S_{6-5}$ is anomalously high, the $J=5-4$ transition typically reaches \(\log_{10} T_{\rm b} \gtrsim 4\). However, for the sources that show $S_{5-4}/S_{6-5} < 1$, the detected $J=6-5$ emission results in either zero fitted components or quasi-thermal classification. The asymmetry between the two transitions thus reflects rotation-dependent pumping or selective radiative coupling, both of which require non-LTE excitation.

\subsection{Population diagram analysis}
\label{sec:dist:non-LTE:pop-diagrams}

To estimate the physical conditions of the SiO-emitting gas and to investigate the excitation of the detected lines from highly excited vibrational states, we performed a population diagram analysis \citep{1999ApJ...517..209G}. Under the assumption of LTE and optically thin emission, the column density of the upper state in cm$^{-2}$, $N_{\rm u}$, is related to the excitation temperature, $T_{\rm ex}$, and the total column density, $N_{\rm tot}$, by the standard relation:

\begin{equation}
\ln \left( \frac{N_{\rm u}}{g_{\rm u}} \right) = \ln \left( \frac{N_{\rm tot}}{Q(T_{\rm ex})} \right) - \frac{E_{\rm u}}{k_{\rm B} T_{\rm ex}},
\label{eq:pop_diagram}
\end{equation}

\noindent where $g_{\rm u}$ is the statistical weight of the upper level, $E_{\rm u}$ is the upper state energy, $k_{\rm B}$ is the Boltzmann constant, and $Q(T_{\rm ex})$ is the partition function. The upper state column density, $N_{\rm u}$, is derived directly from the velocity-integrated flux density of each transition. Here, we assume that the detected emission uniformly fills a circular aperture of 0.08 arcsec in diameter, and note that different transitions can have different opacities and excitation temperatures in the CSE. We do not attempt such in-depth considerations in the current work and assume that the optically thin case is dominant as shown later in this Section.

In Fig. \ref{fig:pop-diagrams}, we plot population diagrams using the extended-configuration data for four of the sources with a large number of lines not classified as possible or clear masers in Fig. \ref{fig:Tb}: T Mic, R Hya, W Aql and IRC$-$10529. The dashed line is the regression line across the data points from which the excitation temperature and the column density are derived. We have excluded the $\varv=0$ lines from \ce{$^{28}$SiO}, as they are likely optically thick, and only considered emission lines in this analysis. We incorporated the detected transitions from the less abundant \ce{$^{29}$SiO} and $^{30}$SiO isotopologues into our regression by scaling the measured $N_{\rm u}$ values by solar abundance ratios of $^{28}$Si/$^{29}$Si = 19.7 and $^{28}$Si/$^{30}$Si = 30.0, i.e. $^{29}$Si/$^{30}$Si $\sim$1.52 \citep{2021A&A...653A.141A}; these ratios are strongly supported by several studies in oxygen-rich and S-type AGB stars, where $^{29}$Si/$^{30}$Si generally falls within the range of 1.0 to 2.0 \citep{2013A&A...559L...8P,2018A&A...615A...8D,2019MNRAS.484..494D}. In T~Mic, R~Hya, W~Aql and IRC$-$10529, we found $T_{\rm ex}$ $\sim 1700 \pm 400 $~K, $3100 \pm 800$~K, $1300 \pm 300$~K, and $1500 \pm 500$~K and SiO column densities $N_{\rm tot} \sim$ 0.6, 2.5, 0.5 and 1.0 $\times$ $10^{18}$ cm$^{-2}$, respectively, from the regression line fits. Our conservative uncertainty in the column densities is $\sim$25--30 per cent due to the varying numbers of lines considered, the range of $E_{\rm u}$, and the possible systematic overcorrection of $N_{\rm u}$ for the minor SiO isotopologues. It is worth noting that excluding the merely detected lines (grey in Fig. \ref{fig:Tb}) or the highly excited $\varv=6$ and $\varv=8$ lines (in the case of R Hya) from the population diagrams does not affect the overall derived results.

From the derived $N_{\rm tot}$, we determined the average SiO number density, where the extent of the emission is estimated as the maximum angular separation from the central star to the 3$\sigma$ contour in the channel maps discussed in Section \ref{sec:SiO-id:overview-maps}. In R Hya, the SiO-emitting region is $\sim$50 mas in radius, i.e. 6.3 au at 126 pc. Assuming a uniform shell surrounding the central star of radius $\sim$1.6 au, we obtained an average number density of roughly 3.5 $\times$ $10^4$ cm$^{-3}$. For an adopted pre-depletion fractional SiO abundance of 3.3 $\times$ $10^{-6}$ \citep{2020A&A...641A..57M}, which is broadly consistent with earlier estimates of (1--few) $\times$ $10^{-5}$ from an interferometric study of SiO $\varv=0$, $J=2-1$ \citep{1993AJ....105..595S}, the H$_2$
density is therefore 1.1 $\times$ $10^{10}$ cm$^{-3}$. This estimate is comparable to the ATOMIUM \ce{H_2O} and OH study of R~Hya \citep{2023A&A...674A.125B}, which derived a H$_2$ density of $\sim$(4--5) $\times$ $10^{10}$ cm$^{-3}$. Applying the same methodology to the rest of this sub-sample, we derived the H$_2$ densities to be 1.4, 0.7, and 0.2 $\times$ $10^{10}$ cm$^{-3}$ in T~Mic, W~Aql and IRC$-$10529, respectively. These values are within an order of magnitude of implied H$_2$ densities from ALMA observations of the inner regions of AGB stars \citep{2016A&A...590A.127W,2017NatAs...1..848V,2024A&A...685A..11K,2025A&A...704A..18O}. The exact H$_2$ densities deduced from this SiO fractional abundance remain debatable, however, due to the large uncertainty. 

We also estimated the line-centre opacity ($\tau_{\rm c}$) for each transition following the expression used by \citet{2023A&A...674A.125B}. The opacity can be approximated by:

\begin{equation}
\tau_{\rm c} \sim 5.145 \times 10^{-4} \frac{ A_{\rm ul} \int n_{\rm u} dl}{\nu^2 \Delta V T_{\rm ex}},
\label{eq:opacity}
\end{equation}

\noindent where $\nu$ is the frequency in GHz of the transition, $A_{\rm ul}$ is the spontaneous emission rate in s$^{-1}$, $T_{\rm {ex}}$ is the line excitation temperature, $\Delta V$ is the line width at half intensity and $\int n_{\rm u} dl$ is the integrated density in cm$^{-2}$. We estimated the latter quantity from the integrated Boltzmann population of the upper energy level using $N_{\rm tot}$/$Q$ and $T_{\rm ex}$ from the population diagrams above. In general, $\tau_{\rm c}$ were derived to be $<$ 0.6 across all four sources, with the two $\varv=1$ lines from \ce{^{28}SiO} being responsible for this upper limit. In R Hya, for example, all of the remaining detected lines give $\tau_{\rm c}$ between 0.002 (\ce{^{30}SiO} $\varv=5$, $J=6-5$) and 0.2 (\ce{^{28}SiO} $\varv=2$, $J=5-4$).

The detection of high vibrational state SiO lines in our data does not by itself require extreme thermal conditions, nor does it by itself demonstrate maser amplification. Collisions and radiative excitation can together populate these high vibrational states to the required degree even if the $T_{\rm ex}$ is relatively low. We also note that these estimates only represent macroscopic averages relying on the aforementioned assumptions, without any consideration of chemical networks present in the inner CSE. In reality, molecular excitation in these regions may be significantly influenced by local heating driven by shocks or other mechanisms as seen in e.g. compact hot spots \citep{2015ApJ...808...36M, 2017NatAs...1..848V} or chromospheric pockets of $T_{\rm b}$ $>$3000 to 10000 K \citep{2019A&A...626A..81V}. In any case, the true physical conditions almost certainly deviate from this idealised approximation and require multi-level, non-LTE, radiative transfer modelling to constrain.

\begin{figure*}
   \begin{center}
   \includegraphics[width=0.8\textwidth, height=0.38\textwidth]{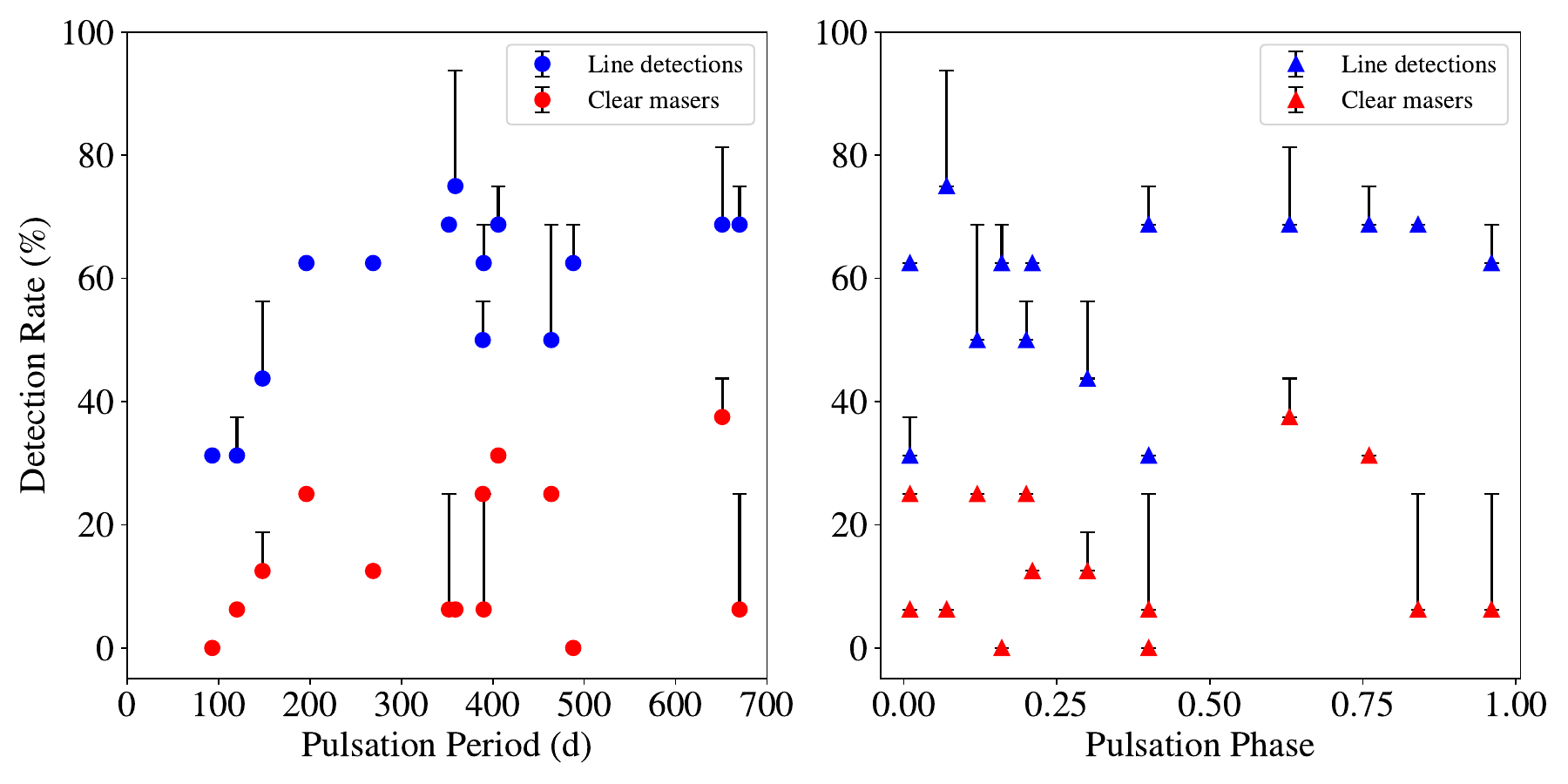}
   \caption{Left: Detection rates of the SiO lines covered by ATOMIUM as a function of pulsation period (circles). Right: Same as the left panel, but as a function of pulsation phase (triangles). The detection rates of total SiO emission lines are shown in blue, while the maser ($\log_{10}(T_{\rm{b}}) \geq 4.0$, Fig.~\ref{fig:Tb}) detection rates are in red. Upper error bars on blue and red data points indicate unclear line detections and probable masers with $3.7 < \log_{10}(T_{\rm{b}}) < 4.0$, respectively.}
   \label{fig:detection-rate}
   \end{center}
\end{figure*}

\begin{figure*}
   \begin{center}
   \includegraphics[width=0.396\textwidth, height=0.332\textwidth]{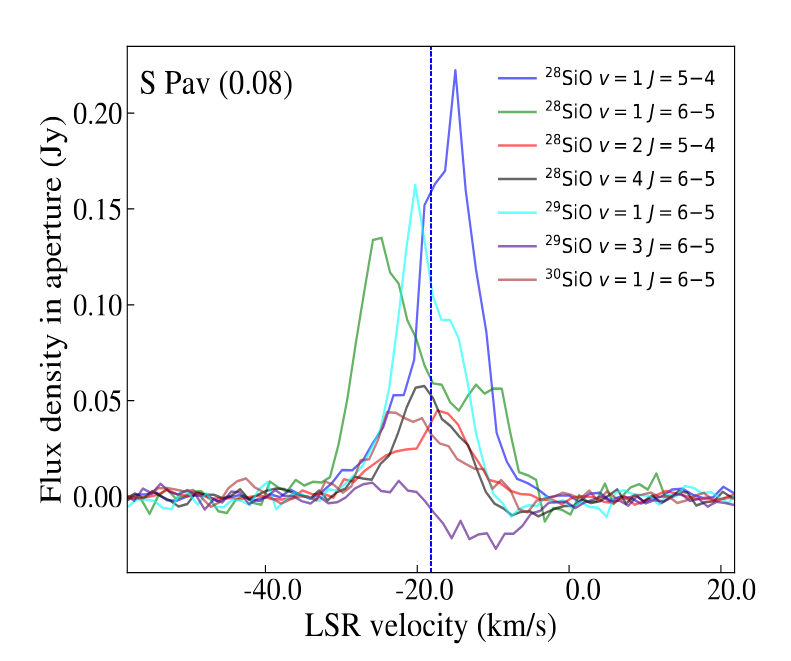}
   \includegraphics[width=0.384\textwidth, height=0.32\textwidth]{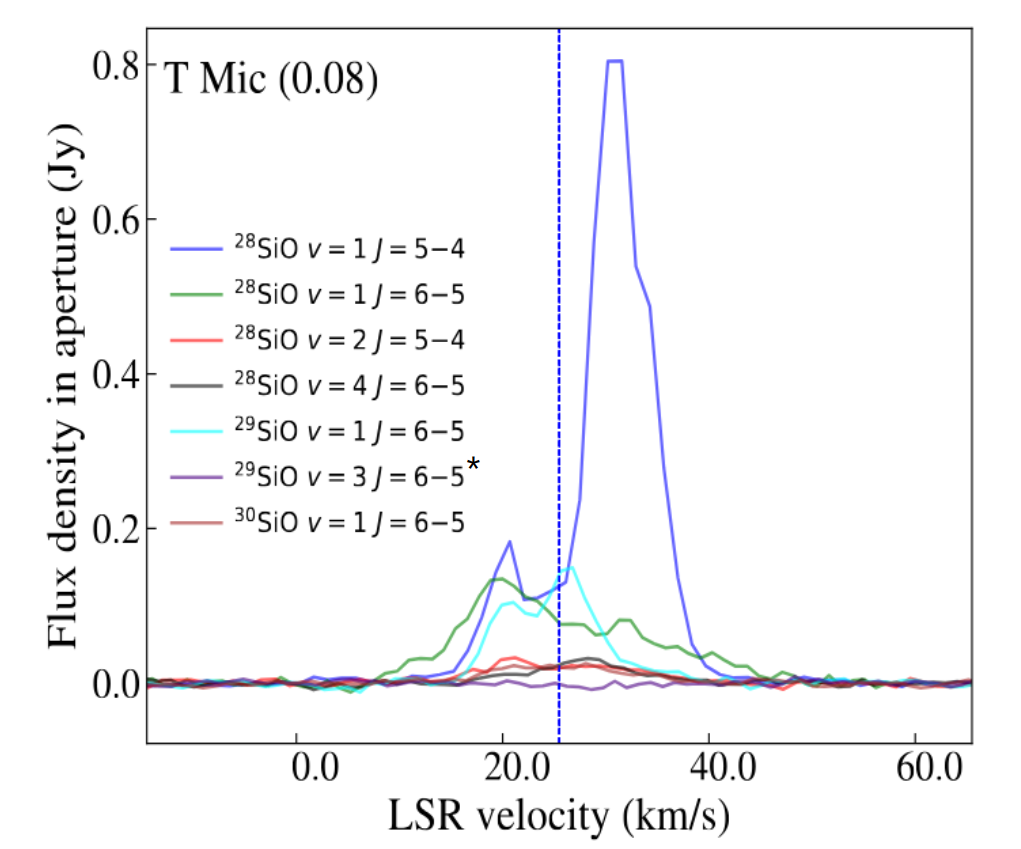}
   \includegraphics[width=0.396\textwidth, height=0.328\textwidth]{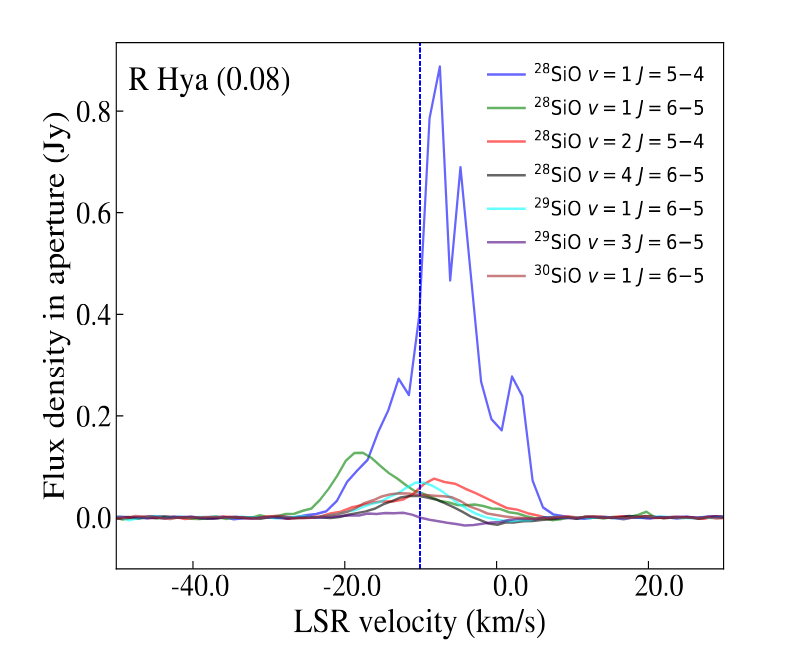}
   \includegraphics[width=0.384\textwidth, height=0.32\textwidth]{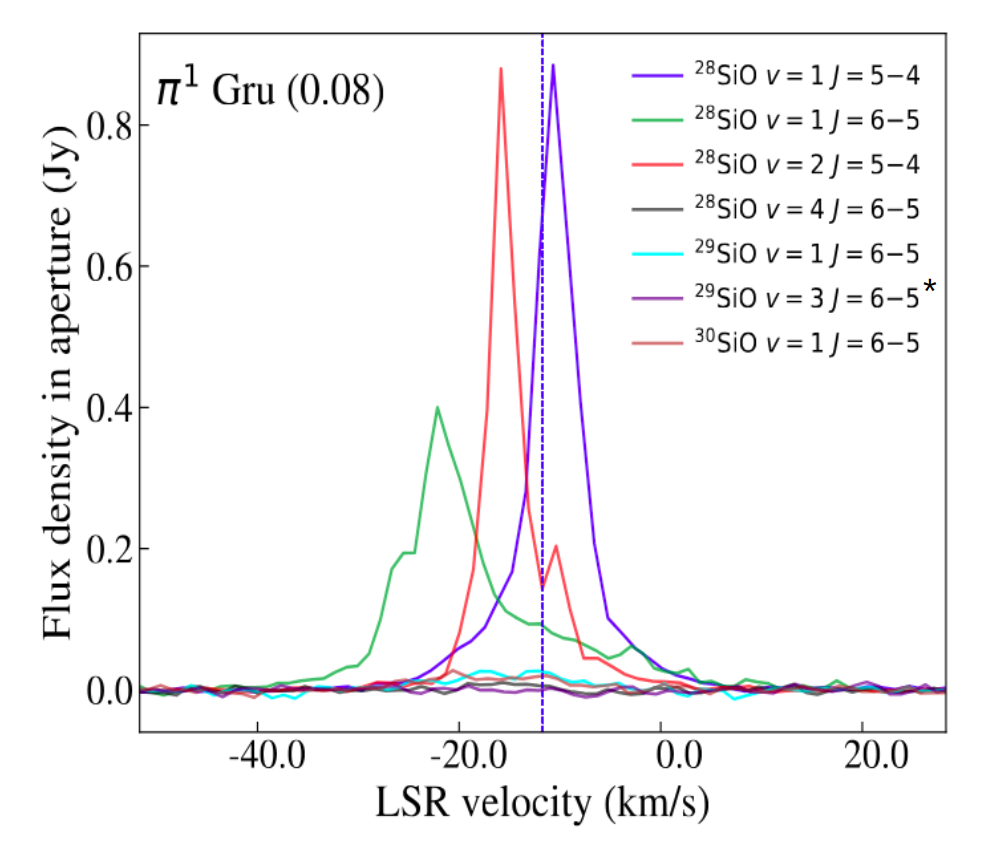}
   \includegraphics[width=0.384\textwidth, height=0.32\textwidth]{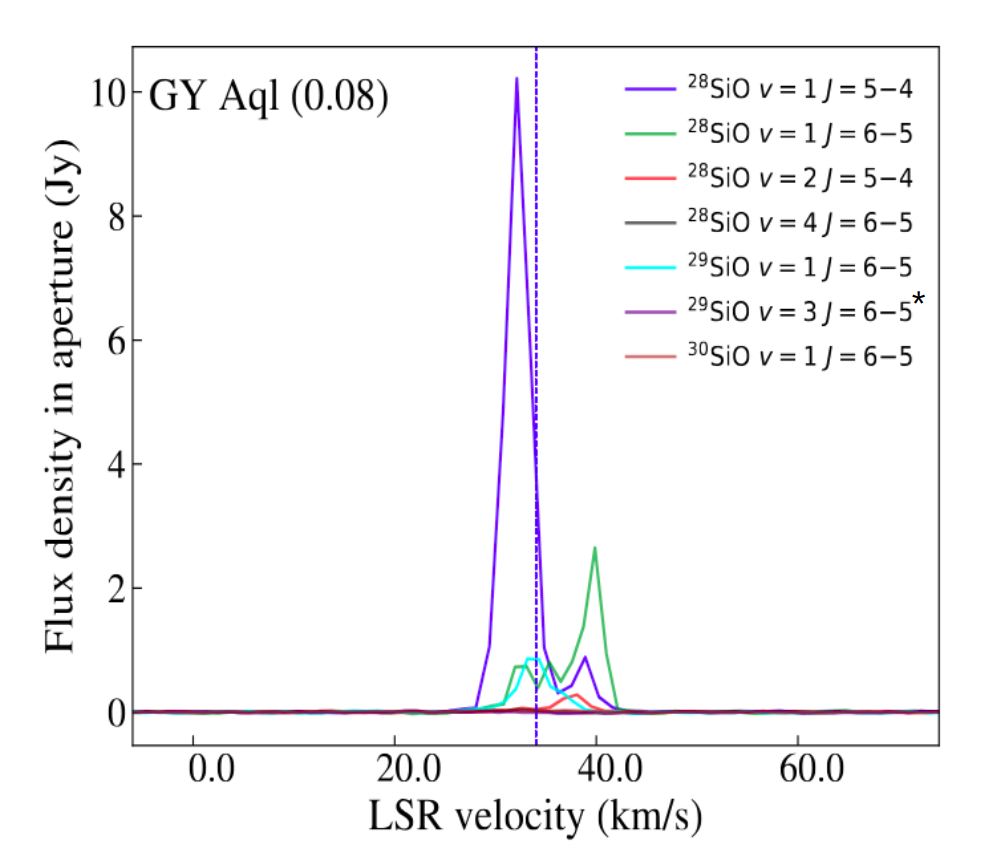} 
   \includegraphics[width=0.396\textwidth, height=0.328\textwidth]{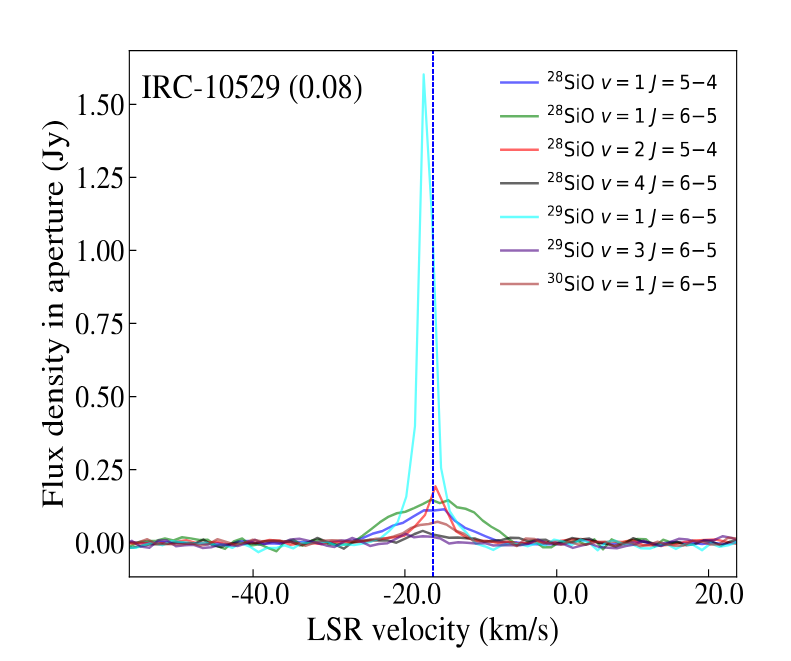} 
   \caption{Spectra of seven SiO lines in the ATOMIUM data towards S Pav, T Mic, R Hya, $\pi^1$ Gru, GY Aql and IRC$-$10529. The transitions along with their corresponding colours are given in the legend in the top right corner. The dashed blue line once again indicates $V_*$. Note that non-detection, marked by *, has been included in the plot for completeness and ease of direct comparison. The same plots for the rest of the ATOMIUM sample can be found in Appendix \ref{appendix:seven-spec}.}
   \label{fig:spectra-compare}
   \end{center}
\end{figure*}
%For R Hya, the only star with $\varv=8$ emission, the population diagram does not indicate that the $\varv=8$ line strongly deviates from LTE conditions; therefore, it cannot be a strong maser. However, based on this analysis alone, we cannot confidently rule out weak maser activity.

\subsection{Possible effects of pulsation period and phase}
\label{sec:compare:detection-rate}

Recently, \citet{2024IAUS..380..314L} found that the detection rates of seven $\sim$43 GHz SiO lines in AGB stars covered by the Bulge Asymmetries and Dynamical Evolution (BAaDE) survey \citep{2019ApJS..244...25S} strongly depend on the length of the stellar period, where the \ce{^{28}SiO} \( \varv=1, 2 \), \( J=1-0 \) lines show a rise in the number of detections up to a period of 400 d, while that of the less common isotopologue lines can continue increasing until a period of about 600 d. In addition, they found that the \( \varv=3 \) line likely has the strongest dependence on phase with the rate being highest near an infrared maximum and lowest near a minimum. We have performed a similar analysis on the high-resolution ATOMIUM data to see if highly excited SiO emission in the sampled sources depends on pulsation period and phase. We acknowledge that the sample size in this study is limited, and extending our analysis to a larger number of sources, such as those included in broader surveys like BAaDE, would help further validate and generalise our findings across the diverse oxygen-rich AGB star population in the Galaxy.

The ATOMIUM sample, excluding SV Aqr, covers (primary) pulsation periods from 120 (U Del) to 670 d (IRC$-$10529). However, the distribution is not smooth, with a notable absence of sources in the period 490--650 days (see Table \ref{table:atomium-sources}). The detection rates are defined as the percentage of clear line and maser detections per transition out of 16 observed high-$J$ SiO lines based on Table \ref{table:SiO-flux} and $T_{\rm{b}}$ estimated from the 2D Gaussian-fitted components (Fig. \ref{fig:Tb}), respectively.

Fig. \ref{fig:detection-rate} shows the detection rates of the SiO emission and maser lines as a function of pulsation period (left panel) and pulsation phase (right panel). Note that we have assigned the data points to represent the number of clear detections and the upper error bars the tentative detections marked with ? in Table \ref{table:SiO-flux} and those with 3.7 $\leq$ $\log_{10}$($T_{\rm{b}}$) $<$ 4.0 for total emission and maser lines, respectively. We find that, for the 16 SiO transitions of interest, the detection rates in the ATOMIUM AGB stars tend to rise with the pulsation periods from $\sim$31 per cent (U Del, primary $P$ = 120 d) to $\sim$69 per cent (IRC+10011, $P$ = 651 d), which is consistent with the trend reported by \citet{2024IAUS..380..314L} and could be a preliminary sign that this relationship may extend even further in high-$J$ SiO emission. However, the main contributor to the increase in detection rate in the sources with larger pulsation periods ($>$ 400 d) is, in this case, the \ce{^{28}SiO} $\varv=0$ lines (most notably in the two OH/IR stars) and not necessarily the isotopologue lines. Within the intermediate period range of 350--500 d,  we observe a wide spread in high-$J$ SiO detection rates, ranging from 50 per cent in RW Sco and GY Aql to 75 per cent in R Hya, with a median value of 62.5 per cent across the intermediate period sub-sample. For clear masers, it is ambiguous whether the same tentative trend is present. Although there are 6 clear SiO masers in IRC+10011, we can see a wide spread of maser detection rates across the sampled pulsation periods; e.g. SV Aqr (primary $P$ = 93 d) and W Aql ($P$ = 488 d) show no confirmed masers, while $\sim$31 per cent of SiO lines detected towards U Her ($P$ = 406 d) exhibit clear masing action.

From inspection of the right-hand panel of Fig. \ref{fig:detection-rate}, there is no clear monotonic or periodic trend between pulsation phase and the detection rates of the high-$J$ SiO transitions. Line detections are observed across the entire pulsation cycle, with detection fractions ranging from $\sim$30 per cent to close to 80 per cent at all phases, suggesting that reduced detectability may not be confined to a particular segment of the pulsation cycle. A similar absence of phase-dependent behaviour is also seen for the clear masers. However, due to the small sample size and lack of multi-epoch observations, any underlying phase dependence, if present, is not recoverable with just the current set of ATOMIUM data. Further surveys are needed to understand the effects of the pulsation period and phase on highly excited SiO emission.

%For IRC+10011 ($P$ = 651 d, $\phi$ = 0.63)}, despite having just undergone an optical minimum at the time of observation, the presence of strong SiO masers (Table \ref{table:SiO-flux} and Fig. \ref{fig:Tb})} suggests that population inversion may have been sustained or even enhanced within localised overdense clumps. In such clumps, collisional excitation can maintain inversion even when the large-scale infrared radiation field is weak, allowing amplification of background photons from spontaneous line emission or dust emission from nearby clumps in the line of sight (if the dust temperature is a few hundred K or more). Such conditions are especially plausible in IRC+10011, where episodic mass-loss and CSE asymmetries have been previously reported (e.g. \citealt{2007PASJ...59..799I}; \citealt{2016ApJ...822....3Y}). There is no clear trend or phase-dependence of the detection rate in Fig. \ref{fig:detection-rate} (right). Due to the small sample size and lack of multi-epoch observations, we cannot confirm nor deny if high-$J$ maser detection rates follow the same trend as lower-\( J \) masers. Further surveys are needed to understand the effects of the pulsation period and phase on highly excited SiO} emission.

\subsection{Comparison between highly excited SiO transitions}
\label{sec:compare:common}

\subsubsection{Statistical check of line detections and mass-loss rate}
\label{sec:compare:common:U-test}

\begin{table}
 \caption{Results of Mann-Whitney U test at 95 per cent confidence level for SiO line detections in low-$\dot{M}$ and high-$\dot{M}$ stars.}
 \label{table:U-test}
 \centering
 \begin{tabular}{ccc}
 \hline \hline
$\varv$ state & P-value & Reject $H_0?$ \\ 
\hline 
0 & 0.391 & No \\
1 & 0.099 & No \\
2 & 0.073 & No \\
3 & 0.141 & No \\
4 & 0.520 & No \\
 \hline
 \end{tabular}
 \\ \raggedright \textbf{Note.} $H_0$: The distributions of SiO line detection rates are the same for stars with low and high mass-loss rates.
\end{table}

We first examined the detection rates of SiO lines with $\varv \leq 4$ to determine whether there is any apparent correlation with mass-loss rate (see Table \ref{table:atomium-sources}), by dividing the sample into two halves with mass-loss rates more than and less than the median $\dot{M} = 6.0 \times 10^{-7}$ M$_{\sun}$ yr$^{-1}$, which is excluded from either group. From Table \ref{table:SiO-flux}, S Pav, the star with the smallest $\dot{M}$, appears to have the same number of clear line detections as R Aql, which has the 11th largest $\dot{M}$ in the sample, in line with the lack of trends in the ATOMIUM H$_2$O lines as reported by \citet{2023A&A...674A.125B}. We assess the statistical significance of this result by performing a Mann-Whitney U test, a non-parametric statistical test used to evaluate whether there is a significant difference between two independent groups. Specifically, we test for differences in clear SiO line detections across different $\varv$ states between stars with low and high mass-loss rates (i.e., under the null hypothesis $H_0$ that no difference exists). To interpret the results, we consider a standard significance threshold of P-value, i.e. the probability of obtaining a result at least as extreme as the one observed given that the null hypothesis is true, less than 0.05. This would allow us to reject $H_0$ at a 95 per cent confidence level. Table \ref{table:U-test} summarises the P-values and their corresponding statistical interpretations. It is clear that, from the ATOMIUM data, there is no evidence of a significant difference in line detections in low mass-loss rate and high mass-loss rate stars at a 95 per cent confidence level, although $\varv=2$ detections (P-value = 0.073) come closest to satisfying the condition to reject the hypothesis (P-value $<$ 0.05). When considering all transitions from $\varv=0$ to $\varv=4$ together and applying the same test, the resulting p-value is 0.150. This indicates that there is still no statistically significant difference between the distributions of high-$J$ SiO line detections in stars with low and high mass-loss rates. We acknowledge that applying statistical tests to a small sample size carries inherent limitations and caution must be exercised in interpreting results from this small data set.

\subsubsection{Line shapes}
\label{sec:compare:common:line-shapes}

Figure \ref{fig:spectra-compare} presents a comparison of seven high-$J$ SiO transitions observed toward S Pav, T Mic, R Hya, $\pi^1$ Gru, GY Aql, and IRC$-$10529. The selection includes the \ce{^{28}SiO} $\varv=1$ lines, which are the most ubiquitous and typically the brightest across the sample, as well as the \ce{^{29}SiO} and \ce{^{30}SiO} $\varv=1$ transitions to enable direct comparison with the main isotope. Higher vibrational transitions ($\varv=3$ and $\varv=4$) are included to investigate how the line profiles in these more highly excited states compare with the $\varv=1$ lines, and how these trends vary between sources. Their overall velocity extents are generally similar, typically within 20 per cent of one another, although the peak velocities of individual transitions often do not align. Double-peaked features are present, particularly in the \ce{^{28}SiO} $\varv=1$ and $\varv=2$, $J=5-4$ transitions, despite the two peaks sometimes being blended. These double peaks generally consist of a blue- and a red-shifted component, with one often dominating in intensity (line intensity ratios between $\sim$1 and 20). Which side is brighter varies by source, with no clear systematic trend. Notably, in $\pi^1$ Gru, GY Aql and RW Sco (Fig. \ref{fig:add-spectra-compare-1}), the $\varv=1$ and $\varv=2$ $J=5-4$ lines appear to exhibit reversed profiles, where one is blue-dominated, the other is red-dominated, whereas in S Pav and U Her, they share similar shapes. The \ce{^{28}SiO} $\varv=1$, $J=6-5$ line appears to be emitted from regions where the peak radial velocities are at least $\pm$ 10 \kms\ relative to $V_*$ in most sources except IRC$-$10529, SV Aqr and W Aql, as well as peaking further from $V_*$ than the other transitions.

\subsubsection{Spectral line overlaps and beyond}
\label{sec:compare:common:line-overlap}

From the peak flux densities listed in Table \ref{table:SiO-flux}, the high-$J$ SiO emission from the ATOMIUM sources exhibits significant differences in intensity ratio from one rotational line to the next within the same vibrational state. This is most apparent in the \ce{^{28}SiO} $\varv=1$, $J=5-4$ and $J=6-5$ lines, where the ratio between the two lines can be $\sim$1 (e.g., RW Sco, IRC$-$10529) up to a few tens (e.g., V PsA, R Aql). When combined with the brightness-temperature classifications (Fig.~\ref{fig:Tb}), these ratios reveal that large variations occur in both quasi-thermal and masing regimes. Several sources with ratios near unity nevertheless show clear maser action in both transitions, indicating that matched fluxes do not imply thermalised excitation. Conversely, the most extreme ratios often occur where only one of the two transitions displays masing levels of $T_{\rm b}$. Standard radiative and collisional pumping models (e.g. \citealt{1996A&A...314..883B}) do not explain differences in intensity ratios of nearby transitions for which we expect the conditions required for pumping and amplifying the masers to be similar between these lines.

Such intensity variations have sometimes been attributed to overlaps between the ro-vibrational lines of isotopes \ce{^{28}SiO}, \ce{^{29}SiO}, and \ce{^{30}SiO} (\citealt{1992ApJ...401L.109C}; \citealt{1993ApJ...407L..33C}). Following the modelling results derived by \citet{2000A&A...359.1117H}, the influence of a line overlap between two transitions is significantly determined by their respective intensities. Typically, during an overlap between \ce{^{28}SiO} and its isotopologues \ce{^{29}SiO} or \ce{^{30}SiO}, the rarer isotopic transitions, which exhibit lower optical thickness, are more susceptible to the effects than the \ce{^{28}SiO} transitions. This impact is more pronounced on the transitions with lower vibrational states. There are a couple of good examples from the ATOMIUM data that agree with these findings. The \ce{^{29}SiO} $\varv=0$, $J=5-4$ line was detected as possible and clear masers in IRC$-$10529 and IRC+10011, respectively, as predicted by the model. This result suggests that the line may overlap the rovibrational transitions \ce{^{29}SiO} $\varv=1-0$, $J=1-0$ and \ce{^{28}SiO} $\varv=2-1$, $J=4-3$. For the \ce{^{29}SiO} $\varv=1$, $J=6-5$ line, its immense intensity in U~Her and IRC+10011 compared to other sources may be explained by the number of overlaps (4) of infrared lines (see Table 1 of \citealt{2000A&A...359.1117H}), where three out of the four scenarios involve transitions that require similar or lower excitation temperatures.

%For masers where line overlap is not involved, the underlying reason for the decreasing prevalence and weaker intensity of successively higher vibrational levels (in cases without line overlap) likely lies in the thermal distribution, as the progressively lower population of higher vibrational states within a given rotational transition results in a smaller population for inversion (Section \ref{sec:dist:non-LTE})}. In $\pi^1$ Gru, the \ce{^{28}SiO} $J=5-4$, $\varv=1$ and $\varv=2$ transitions have a peak flux density of 0.9 Jy and $T_{\rm b} > 10^5$~K}, both of which are the brightest among the detected lines; it is the only source in the ATOMIUM sample where this is true. These detections turn out to be strong evidence of a spiral and mass transfer between the primary and the newly detected close companion (see \citealt{2020A&A...644A..61H} for detailed discussions).

For transitions where line overlap does not play a dominant role, the declining occurrence and typically weaker intensities of successively higher vibrational states can be understood qualitatively in terms of the underlying LTE population distribution: higher-$\varv$ levels contain fewer molecules available for inversion, reducing the likelihood or efficiency of masing. In most ATOMIUM sources this trend is reflected in the relative line strengths, but $\pi^1$ Gru represents a clear exception. In this star, the \ce{^{28}SiO} $\varv=1$ and $\varv=2$ $J=5-4$ transitions reach peak flux densities of 0.9 Jy and exhibit brightness temperatures exceeding $10^{5}$ K, making $\pi^1$ Gru a good example where maser flux densities between the lines in the $\varv=1$ and $\varv=2$ states are comparable. These detections turn out to be strong evidence of a spiral and mass transfer between the primary and the newly detected close companion (see \citealt{2020A&A...644A..61H}, \citealt{2025A&A...699A..22M} and \citealt{2026NatAs..10..124E} for detailed discussions).

In some cases, line overlap is involved. For the $\varv=2$ line, \citet{1981ApJ...247L..81O} suggested that, for the low-$J$ counterpart (\ce{^{28}SiO} $\varv=2$, $J=2-1$), the inversion of a rovibrational pumping line, the \ce{^{28}SiO} $\varv=1-2$, $J=0-1$ transition may be suppressed by overlap with the $(\varv_1,\varv_2,\varv_3)=(0,0,0)$, $J_{K_a,K_c}=12_{7,5} \rightarrow (0,1,0)$ $11_{6,6}$ water line. If this H$_2$O transition is in emission with respect to the local radiation field as seen by SiO, the population transfer would lead to the suppression of the \ce{^{28}SiO} $\varv=2$, $J=2-1$ line. \citet{1996A&A...314..883B}
used the IRAM 30m telescope to observe 47 M-type and S-type stars and reported the line detection rate of the $\varv=2$, $J=2-1$ line
to be $\sim$17 per cent, compared with 86 per cent for the $\varv=2$, $J=5-4$ line in the ATOMIUM sample. The \ce{^{28}SiO} $J=5-4$, $\varv=2$/$\varv=1$ line intensity ratios for the detections are $\leq1$ in 12 sources, excluding the two OH/IR stars IRC$-$10529 (1.58) and IRC+10011 (12.2). The
$\varv=2$/$\varv=1$ line ratios in M-type stars do not show any particular trend compared to those in S-type stars, unlike the results for the \ce{^{28}SiO} $J=2-1$ transitions, where these ratios are about ten times larger in S-type stars \citep{1996A&A...314..883B}. It is of note, however,
that the lowest rms noise achieved in \citet{1996A&A...314..883B} is 0.2 Jy, roughly two orders of magnitude larger than the current work
despite not resolving out any flux. \citet{1998A&A...329..219P} also observed the $J=5-4$ lines from the $\varv=1$ and $\varv=2$ states of \ce{^{28}SiO} among others towards 12 evolved stars and suggested three possible infrared overlaps based on the model presented in \citet{1997A&A...322..938G}. They considered the $(0,1,0)$ $11_{7,5} \rightarrow (0,0,0)$ $12_{8,4}$ H$_2$O line overlap affecting the population of the upper $\varv=2$, $J=5$ level, but noted the velocity difference ($\sim$16 \kms) is likely too large to be effective in most of their observed objects. They also identified overlaps involving the \ce{^{29}SiO} $\varv=1-0$, $J=1-0$ and $J=2-1$ lines that could affect the $\varv=2$, $J=4$ population level. However, they concluded the influence of these mechanisms remains doubtful, as their effectiveness is severely limited if the involved \ce{^{29}SiO} transitions are optically thin.

Beyond line overlap effects, additional mechanisms may contribute to maser flaring in certain sources. Localised shocks can temporarily enhance population inversion, leading to transient brightening, while the orientation and distribution of clumps within the CSE can affect the observed maser intensity by altering the line-of-sight amplification conditions. Recent 3D radiation-hydrodynamical models of AGB stars, which naturally produce complex velocity fields and clumpy winds \citep{2023A&A...669A.155F}, may offer valuable insight into the distribution and geometrical properties of the clumpy structures. A detailed parameterisation of the clump morphology may help interpreting the observed maser intensities and detection statistics with radiative transfer models (\citealt{Ahmad2025}, their Section 3.3).

\begin{figure*}
   \begin{center}
   \includegraphics[width=0.75\textwidth, height=0.3\textwidth]{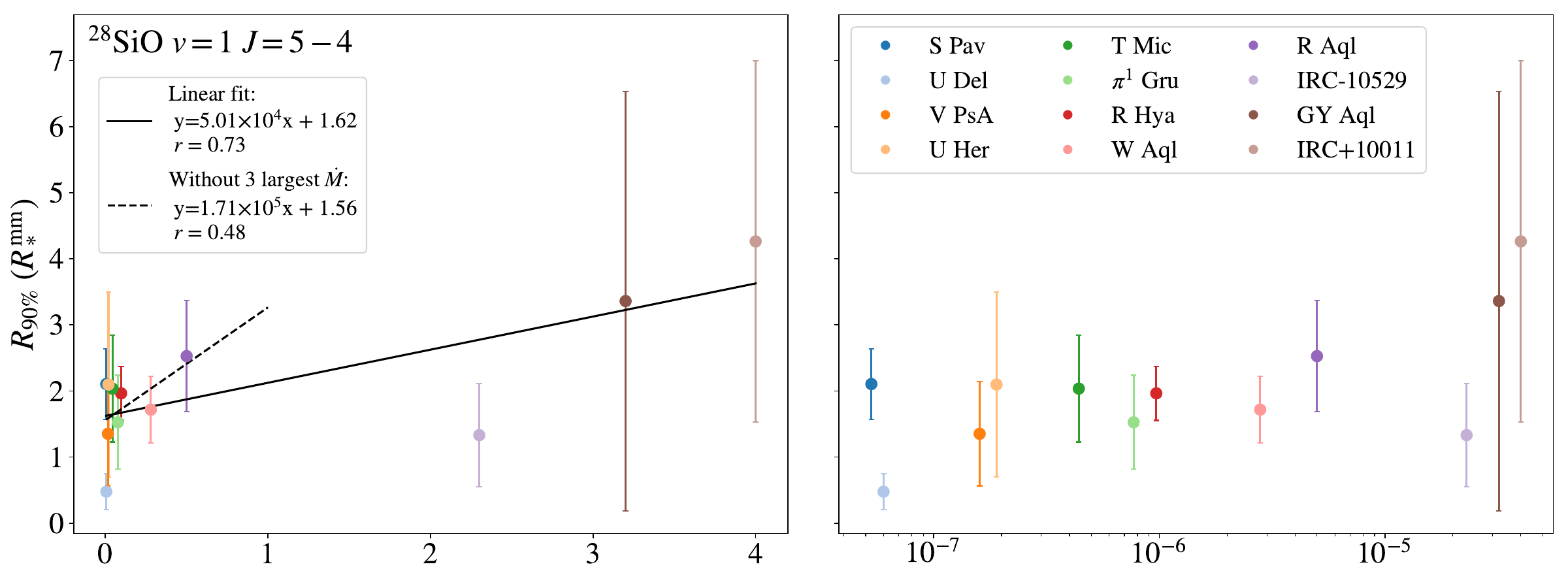}
   \includegraphics[width=0.75\textwidth, height=0.3\textwidth]{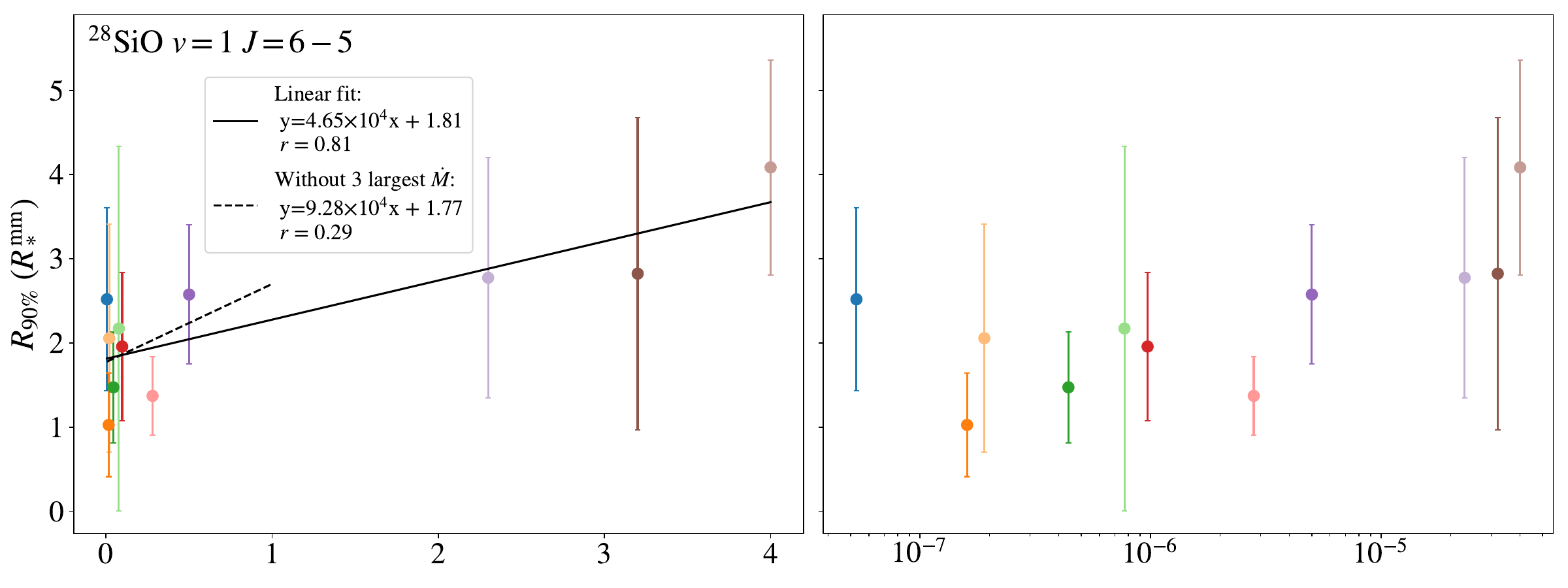}
   \includegraphics[width=0.75\textwidth, height=0.3\textwidth]{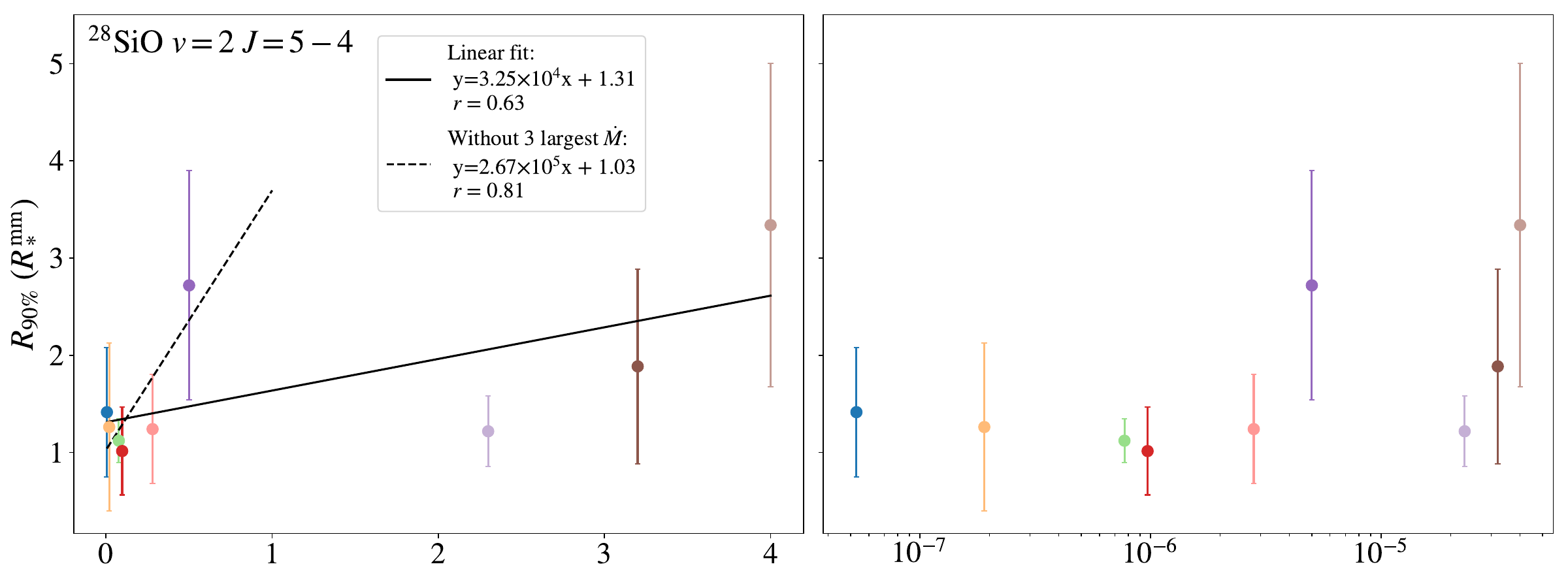}
   \begin{overpic}[width=0.76\textwidth, height=0.3\textwidth]{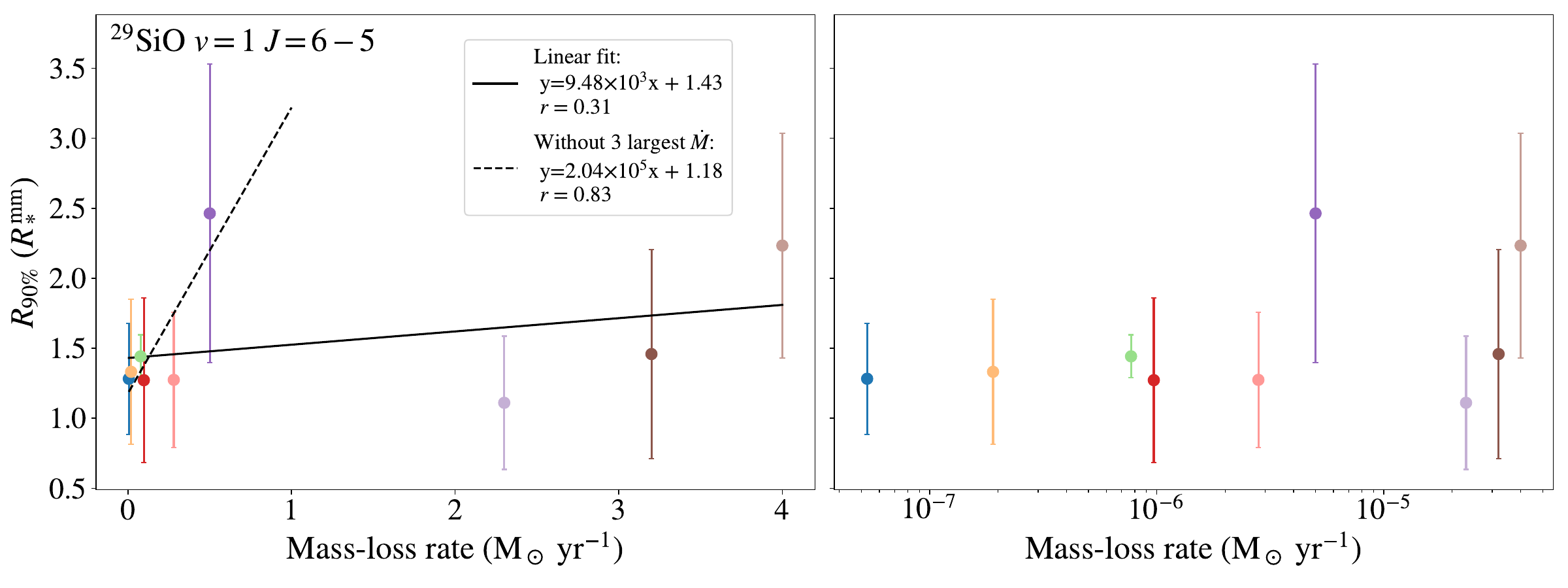}
       \put(48, 2){$\times 10^{-5}$} 
   \end{overpic}
   \caption{Plots of radius which encloses 90 per cent of the total emission ($R_{90\%}$) in multiples of $R_*^{\rm{mm}}$ vs mass-loss rate ($\dot{M}$) for the \ce{^{28}SiO} $\varv=1$, $J=5-4$, \ce{^{28}SiO} $\varv=1$, $J=6-5$, \ce{^{28}SiO} $\varv=2$, $J=5-4$ and \ce{^{29}SiO} $\varv=1$, $J=6-5$ transitions, shown in linear (left, in $10^{-5}$ M$_\odot$ yr$^{-1}$) and log (right) scales of mass-loss rate. The correlation coefficient $r$ and the best-fitted linear regression model are presented and plotted (black solid line) for each of the transitions. The same set of output is also given in the case where the three highest $\dot{M}$ stars are excluded (dashed line). Note that the regression line for this case has been extrapolated for better readability. Only the AGB sources with $>$10 fitted components were selected as the sample.}
   \label{fig:radMdot-1}
   \end{center}
\end{figure*}

\section{Mass-loss rate and angular extent of S\MakeLowercase{i}O emission}
\label{sec:correlation}

\subsection{Motivation}
\label{sec:correlation:motivation}
Although we have demonstrated that the difference in line detection between low and high mass-loss rate AGB stars may not be statistically significant (Table \ref{table:U-test}), this does not preclude meaningful differences in other characteristics, particularly their spatial distribution. The large span of $\dot{M}$ covered by the ATOMIUM sample ($10^{-8}$--$10^{-5}$ M$_\odot$ yr$^{-1}$, Table \ref{table:atomium-sources}) offers a unique opportunity to investigate whether the location of compact SiO emission, quantified by the angular extent from the central star, correlates with mass-loss rate, even in the absence of a detection-rate difference. While detection alone may depend on many factors including variability, pulsation phase, and observational sensitivity, the spatial distribution of SiO emission probes the physical scale of emitting zones. Identifying a trend between $\dot{M}$ and the angular distance of Gaussian components observed could thus provide new constraints on the excitation conditions and radiative transfer mechanisms at work. Observationally, such a correlation would help estimate the desired spatial resolution required for a particular target and thus allow the observer to roughly calculate the required integration time, as well as set up other relevant observational parameters accordingly.

\subsection{Angular extent of Gaussian components from the star}
\label{sec:correlation:Mdot}
The distributions of SiO emission at $> 5 \sigma$,  were measured using 2D Gaussian fitting as described in Section \ref{sec:dist}. We adapted the method of determining angular extents from the AGB star described by \citet{2011MNRAS.415.1083A} and used it on the single-epoch ATOMIUM extended-configuration data set. In each source, the angular extent from the central star enclosing 90 per cent of the total emission (in $R_*^{\rm{mm}}$), $R_{90\%}$, was calculated, where the flux cutoff was chosen as such to avoid being biased by very weak components with large position errors. One can interpret this maximum radial extent at a single epoch, which typically ranges between 2--4 $R_*^{\rm{mm}}$, as a lower limit to the outer radius of the SiO-emitting region. Using a slightly different threshold such as 75 per cent  would not substantially alter the interpretation, as the bulk of the bright components is typically confined to a compact region ($\lesssim R_*^{\rm{mm}}$ in the sky plane from the centre of the star). This differs from lower-\( J\) SiO masers, which often exhibit clearer ring-like distributions even in single-epoch observations (e.g.,\citealt{2004A&A...414..275C}; \citealt{2011MNRAS.415.1083A}, \citealt{2013MNRAS.433.3133G}). Any emission seen on top of the star can be inferred to emanate from the near side. The uncertainty in $R_{90\%}$ was estimated to be the flux-weighted standard deviation of individual fitted components from $R_{90\%}$, which is generally 1--2 orders of magnitude larger than the observational errors.

\subsection{Results and discussion}
\label{sec:correlation:results}
The plots of $R_{90\%}$ (in multiples of $R_*^{\rm{mm}}$) against $\dot{M}$ for four of the most detected highly excited SiO transitions, i.e. \ce{^{28}SiO} $\varv=1, 2$, $J=5-4$, \ce{^{28}SiO} $\varv=1$, $J=6-5$ and \ce{^{29}SiO} $\varv=1$, $J=6-5$, are shown in Fig.~\ref{fig:radMdot-1}. RW Sco is excluded from this analysis because its UD diameter is unreliable due to a low S/N ratio and a small angular size (see Table \ref{table:atomium-sources} and \citealt{2025Danilovich}), while SV Aqr is omitted due to the absence of fitted components in our data. The Pearson correlation coefficient, $r$, is given in each panel to indicate how strong the relationship between $R_{90\%}$ and $\dot{M}$ is for that transition. In the current work, we selected only the sources with more than 10 fitted components (for that particular transition) to be analysed such that the sample size is large enough to minimise any bias that may arise due to the apparent sporadicity of high-$J$ SiO emission at the time of observations.

From Fig.~\ref{fig:radMdot-1}, all four transitions show varying levels of correlation between $R_{90\%}$ and $\dot{M}$ from 0.31 in \ce{^{29}SiO} $\varv=1$, $J=6-5$ to 0.81 in \ce{^{28}SiO} $\varv=1$, $J=6-5$. It is of note that the coefficients for \ce{^{28}SiO} $\varv=2$, $J=5-4$ and \ce{^{29}SiO} $\varv=1$, $J=6-5$ are fitted based on the data from 9 sources as opposed to 12 and 11 sources for \ce{^{28}SiO} $\varv=1$, $J=5-4$ and $J=6-5$, respectively. We estimate the underlying linear relationship for each of the transitions using linear regression. The regression lines are obtained using the method of least squares and are shown as solid black lines, illustrating the best-fitted linear models between $R_{90\%}$ and $\dot{M}$. We note that, across all selected transitions, the three largest $\dot{M}$ stars have a significant impact on the apparent correlation between $\dot{M}$ and $R_{90\%}$. Without these, the updated correlation coefficients provide a more nuanced view of the relationship between the two quantities across the three transitions (dashed lines in Fig.~\ref{fig:radMdot-1}). For the \ce{^{28}SiO} $\varv=2$, $J=5-4$ and  \ce{^{29}SiO} $\varv=1$, $J=6-5$ lines, the correlation coefficient strengthens to 0.81 and 0.83, respectively. The \ce{^{28}SiO} $\varv=1$, $J=6-5$ line shows a weakened positive correlation of 0.29 (vs 0.81 when taking into account all the stars), indicating that the stars with high $\dot{M}$ are the major contributors to the apparent correlation in this transition. In the case of the \ce{^{28}SiO} $\varv=1$, $J=5-4$ line, the supposed correlation would also weaken, albeit less sharply, from 0.73 to 0.48. Overall, while tentative trends between mass-loss rate and the spatial distribution of small-scale SiO emission are observed, the limited number of available sources, especially in the higher mass-loss rate regime ($5.0\times10^{-6} < \dot{M} < 2.0\times10^{-5}$ M$_{\odot}~\mathrm{yr}^{-1}$), restricts the robustness of our conclusions. More observations targeting AGB stars across this critical mass-loss range are therefore essential to better constrain the relationship between $\dot{M}$ and the spatial extent of high-$J$ SiO emission.

In addition, for most of the high-$J$ SiO lines, the maser distribution is highly lopsided or sporadic to the point that a typical ring-like structure cannot be defined. Multi-epoch observations of low-$J$ SiO masers around well-studied AGB stars (e.g. R Cas, TX Cam) suggest that such asymmetries are transient and, over several stellar cycles, the average SiO maser distribution is ring-like without a preferred direction \citep{2011MNRAS.415.1083A,2013MNRAS.433.3133G}. As many bright fitted components are situated on top of the star in the line-of-sight direction, this leads to (severe) underestimation of $R_{90\%}$ from the AGB star owing to the weighting bias towards smaller radii. The main source of uncertainty is the absence of the line-of-sight direction’s contribution to $R_{90\%}$, which may be addressed in future work through monitoring and inferring the 3D motion of emission features. The usual radial expansion approximation is not applicable at such small radii due to the inner wind complexity, as exemplified by e.g. Fig. \ref{fig:maps-paper}.

Despite our best efforts to minimise the uncertainties, $\dot{M}$ measurements in the literature represent long-term average rates derived from extended species, commonly CO, tracing centuries of mass loss. On the other hand, the innermost few $R_*^{\rm{mm}}$ contain just a few stellar cycles worth of mass loss, ejected variably during the pulsation period and possibly anisotropically. This means our best estimates of mass-loss rates in turn may not represent the current level of mass loss in each of the sources, which could be continually changing as a delayed response to a 'recent' thermal pulse event ($<$ a few hundred yr), e.g. in R Hya (\citealt{2002MNRAS.334..498Z}; \citealt{2008A&A...484..401D}) and R Aql \citep{1981ApJ...247..247W,2000JBAA..110..131G}. With these limitations as well as the lack of variability information, the ATOMIUM observations are unfortunately not enough for us to confidently say whether the correlation between $R_{90\%}$ and $\dot{M}$ is intrinsic to the transitions or simply a coincidence. Additional epochs of high-resolution interferometric observations, both of the ATOMIUM sources and a broader sample of AGB stars, are needed to robustly test and support our tentative findings.

\section{Conclusions}
\label{sec:conclusions}
We have reported ALMA Band 6 observations (214--270 GHz) of highly excited SiO lines towards 14 selected AGB stars taken as part of the ATOMIUM Large Programme. Sixteen $\varv=0$ to $\varv=8$ transitions were detected in total in at least one of the sources. Among the observed lines, the $\varv=1$, $J=5-4$ transition from \ce{^{28}SiO} was by far the most common amongst the $\varv>0$ lines in the sample, followed by three other $\varv=1$ lines including from the \ce{^{29}SiO} and \ce{^{30}SiO} isotopologues. We detected clear $\varv=0$ maser emission from \ce{^{28}SiO} (in V PsA and IRC+10011) and \ce{^{29}SiO} (in IRC+10011), alongside tentative \ce{^{28}SiO} $\varv=0$ masing in T Mic. We also report the first high-resolution detections of the \ce{^{28}SiO} $\varv=3, 4, 8$, $J=6-5$, \ce{^{29}SiO} $\varv=6$, $J=6-5$ and \ce{^{30}SiO} $\varv=4, 5$, $J=6-5$ lines in oxygen-rich AGB stars. The $\varv=8$ transition represents the highest vibrationally excited SiO line detected in AGB stars to date. Overall, the higher the $\varv$ state, the less common the emission in that transition becomes.

We detected absorption in the $\varv=3$ and $\varv=4$ SiO lines towards R~Aql, R~Hya, S~Pav and T~Mic, and tentative absorption in the $\varv=8$ line towards R~Aql and S~Pav. The deepest absorption is observed in R~Hya, reaching a minimum flux density of $-25$ mJy. These features are spatially compact ($\sim$10--25 mas) and kinematically narrow ($\sim$a few \kms), with R Hya and S Pav exhibiting spectral profiles indicative of infall, while T Mic shows evidence of outflow dynamics through inverse P Cygni and P Cygni profiles, respectively. The SiO absorption is detected in the same sources as H$_2$O absorption, with amplitudes 1.5--2.5 times stronger in SiO. These results also indicate that highly excited SiO absorption likely traces regions of mass-loss variability near AGB stars ($\sim$a few $R_*^{\rm{mm}}$).

Analysis of 2D Gaussian-fitted component maps of the regions hosting the most compact SiO emission has revealed that the position-velocity plots of the high-$J$ masers show a range of features, including velocity-position gradients and signs of a close companion in $\pi^1$ Gru, influencing the inner wind morphologies. Red-shifted components situated on top of the star suggest that the SiO-emitting regions are on the near side relative to Earth and are falling towards the star. At the time of observations, R Hya and U Her are the best evidence supporting this hypothesis as they are consistent with the implied shock front positions within the velocity profiles from the CODEX models. However, with no spatial information in the line-of-sight direction, it is impossible to tell whether the emission was produced near the stellar surface or farther out. Additional interferometric observations would help trace these SiO clumps over time and inform us of the changes to the physical conditions as the clumps move closer to the star during an infall. Many of the brightest components appear directly above the central star and skew the measured radial extents to within 2--3 $R_*^{\rm{mm}}$.

The relationships between SiO line detection rates and physical parameters, such as pulsation period, pulsation phase and mass-loss rate, do not appear straightforward. We found that the SiO line detection rates in the ATOMIUM AGB stars tend to rise with the pulsation periods if $P<$ 400 d. The same trend was not seen in clear maser detections and no discernible correlation with pulsation phase could be inferred. The non-parametric Mann-Whitney U test, comparing SiO line detections in different vibrational states between low and high mass-loss rate stars, indicates no significant difference in line detections between the two groups at a 95 per cent confidence level. In terms of relative intensities, the high-$J$ SiO lines vary significantly within the same vibrational state. Line overlaps between ro-vibrational lines of isotopes \ce{^{28}SiO}, \ce{^{29}SiO}, and \ce{^{30}SiO} may explain the observed intensity variations, with rarer isotopic transitions being more affected, but further investigation involving overlaps from other molecules should also be considered. The main limitation of these statistical results is the small sample size.

The ATOMIUM data set reveals a tentative correlation between the radius that encloses 90 per cent of the total emission and mass-loss rate for the most detected, most compact high-$J$ SiO emission but at varying degrees. The correlation either strengthens or weakens depending on the transition if the three highest mass-loss rate stars are excluded from the analysis, hinting at this finding potentially being due to chance. Biases due to the lopsidedness and lack of information in the line-of-sight direction are the main sources of uncertainty in the current work. 

While our results point to maser emission as the origin of many observed spectral features, particularly those with narrow velocity extents and high brightness temperatures, a more robust separation of thermal and masing components, especially for the $\varv=0$ lines, requires multi-epoch data and detailed spectral modelling. This analysis, along with an investigation into the variability and spatial structure of high-$J$, high-$\varv$ SiO masers, will be pursued in future work. Furthermore, establishing any robust correlation between SiO emission behaviour and AGB (circum)stellar parameters will require a much larger statistical sample, ideally on the order of hundreds of stars. The link between SiO emission and dust formation will also be explored further in a forthcoming publication (Gottlieb et al., in prep.).

\section*{Acknowledgements}
The authors would like to thank the reviewer for helpful comments and suggestions. This paper makes use of the ALMA Main Array data: ADS/JAO.ALMA\#2018.1.00659.L (ATOMIUM: ALMA tracing the origins of molecules in dust forming oxygen-rich M-type stars). ALMA is a partnership of ESO (representing its member states), NSF (USA), and NINS (Japan), together with NRC (Canada), NSC and ASIAA (Taiwan), and KASI (Republic of Korea), in cooperation with the Republic of Chile. The Joint ALMA Observatory is operated by ESO, AUI/NR.A.O, and NAOJ. This work is supported by the Fundamental Fund of Thailand Science Research and Innovation (TSRI) through the National Astronomical Research Institute of Thailand (Public Organization) (FFB680072/0269 and FFB690078/0269). BP acknowledges the support from the Royal Thai Government through the DPST scholarship during his PhD studies, which form part of the current work. TD is supported in part by the Australian Research Council through a Discovery Early Career Researcher Award (DE230100183). LD acknowledges support from the KU Leuven C1 excellence grant BRAVE C16/23/009, KU Leuven Methusalem grant SOUL METH/24/012, and the FWO research grants
G099720N and G0B3823N. IEM acknowledges support from ANID/FONDECYT (grant 11240206) and funding via the BASAL Centro de Excelencia en Astrofisica y Tecnologias Afines (CATA) grant PFB06/2007. HSPM thanks the Deutsche Forschungsgemeinschaft (DFG) for support through the collaborative research center SFB 1601 (project ID 500700252) subprojects A4 and Inf. RS’s contribution to the research described in this publication was carried out at the Jet Propulsion Laboratory, California Institute of Technology, under a contract with NASA (80NM0018D0004). RS thanks NASA for financial support via GALEX GO and ADAP awards. KTW acknowledges support from the European Research Council (ERC) under the European Union’s Horizon 2020 Research and Innovation programme (grant agreement number 883867, project EXWINGS). This work is funded by the French National Research Agency (ANR) project PEPPER (ANR-20-CE31-0002).  We thank IRIS for provision of high-performance computing facilities. STFC IRIS is investing in the UK’s Radio and mm/sub-mm Interferometry Services in order to improve the data quality and allow much more data
to be processed. This paper makes use of the Cologne Database for Molecular Spectroscopy (CDMS; https://cdms.astro.uni-koeln.de/classic/)
and the spectral line catalogues of the Jet Propulsion Laboratory (JPL, https://spec.jpl.nasa.gov). This research has made use of the SIMBAD database and the VizieR catalogue access tool, CDS, Strasbourg Astronomical Observatory, France (\citealt{2000A&AS..143....9W}; \citealt{2000A&AS..143...23O}). This research has made use of the International Variable Star Index (VSX) database, operated at AAVSO, Cambridge, Massachusetts,
USA. This research made use of Astropy (\citealt{2013A&A...558A..33A}, \citeyear{2018AJ....156..123A}), SciPy \citep{2020NatMe..17..261V}, Matplotlib \citep{2007CSE.....9...90H} and NumPy \citep{2020Natur.585..357H}.

%%%%%%%%%%%%%%%%%%%%%%%%%%%%%%%%%%%%%%%%%%%%%%%%%%

\section*{Data Availability}

The observed data, ALMA pipeline products and ATOMIUM enhanced data products (including spectra and image cubes after self-calibration) are available via the ALMA archive. The analysed data underlying this article are available on a case-by-case basis based on reasonable requests to the corresponding author (BP). Additional spectra, channel maps and component maps of the highly excited SiO emission can be found in the online supplementary material. 

%%%%%%%%%%%%%%%%%%%% REFERENCES %%%%%%%%%%%%%%%%%%

% The best way to enter references is to use BibTeX:

\bibliographystyle{mnras}
\bibliography{Atomium_SiO_overview.bib} % if your bibtex file is called example.bib

@ARTICLE{2020Sci...369.1497D,
       author = {{Decin}, L. and {Montarg{\`e}s}, M. and {Richards}, A.~M.~S. and {Gottlieb}, C.~A. and {Homan}, W. and {McDonald}, I. and {El Mellah}, I. and {Danilovich}, T. and {Wallstr{\"o}m}, S.~H.~J. and {Zijlstra}, A. and {Baudry}, A. and {Bolte}, J. and {Cannon}, E. and {De Beck}, E. and {De Ceuster}, F. and {de Koter}, A. and {De Ridder}, J. and {Etoka}, S. and {Gobrecht}, D. and {Gray}, M. and {Herpin}, F. and {Jeste}, M. and {Lagadec}, E. and {Kervella}, P. and {Khouri}, T. and {Menten}, K. and {Millar}, T.~J. and {M{\"u}ller}, H.~S.~P. and {Plane}, J.~M.~C. and {Sahai}, R. and {Sana}, H. and {Van de Sande}, M. and {Waters}, L.~B.~F.~M. and {Wong}, K.~T. and {Yates}, J.},
        title = "{(Sub)stellar companions shape the winds of evolved stars}",
      journal = {Science},
         year = 2020,
        month = sep,
       volume = {369},
       number = {6510},
        pages = {1497-1500},
          doi = {10.1126/science.abb1229},
archivePrefix = {arXiv},
       eprint = {2009.11694},
 primaryClass = {astro-ph.SR},
       adsurl = {https://ui.adsabs.harvard.edu/abs/2020Sci...369.1497D}
}

@ARTICLE{2022A&A...660A..94G,
       author = {{Gottlieb}, C.~A. and {Decin}, L. and {Richards}, A.~M.~S. and {De Ceuster}, F. and {Homan}, W. and {Wallstr{\"o}m}, S.~H.~J. and {Danilovich}, T. and {Millar}, T.~J. and {Montarg{\`e}s}, M. and {Wong}, K.~T. and {McDonald}, I. and {Baudry}, A. and {Bolte}, J. and {Cannon}, E. and {De Beck}, E. and {de Koter}, A. and {El Mellah}, I. and {Etoka}, S. and {Gobrecht}, D. and {Gray}, M. and {Herpin}, F. and {Jeste}, M. and {Kervella}, P. and {Khouri}, T. and {Lagadec}, E. and {Maes}, S. and {Malfait}, J. and {Menten}, K.~M. and {M{\"u}ller}, H.~S.~P. and {Pimpanuwat}, B. and {Plane}, J.~M.~C. and {Sahai}, R. and {Van de Sande}, M. and {Waters}, L.~B.~F.~M. and {Yates}, J. and {Zijlstra}, A.},
        title = "{ATOMIUM: ALMA tracing the origins of molecules in dust forming oxygen rich M-type stars. Motivation, sample, calibration, and initial results}",
      journal = {\aap},
         year = {2022},
       volume = {660},
          eid = {A94},
        pages = {A94},
          doi = {10.1051/0004-6361/202140431},
archivePrefix = {arXiv},
       eprint = {2112.04399},
 primaryClass = {astro-ph.SR},
       adsurl = {https://ui.adsabs.harvard.edu/abs/2022A&A...660A..94G}
}

@ARTICLE{1962MNRAS.124..417H,
       author = {Hoyle, F. and Wickramasinghe, N. C.},
        title = "{On graphite particles as interstellar grains}",
      journal = {\mnras},
         year = {1962},
       volume = {124},
        pages = {417},
          doi = {10.1093/mnras/124.5.417},
       adsurl = {https://ui.adsabs.harvard.edu/abs/1962MNRAS.124..417H}
}

@ARTICLE{2018A&ARv..26....1H,
       author = {H{\"o}fner, Susanne and Olofsson, Hans},
        title = "{Mass loss of stars on the asymptotic giant branch. Mechanisms, models and measurements}",
      journal = {\aapr},
         year = {2018},
       volume = {26},
       number = {1},
          eid = {1},
        pages = {1},
          doi = {10.1007/s00159-017-0106-5},
       adsurl = {https://ui.adsabs.harvard.edu/abs/2018A&ARv..26....1H}
}

@ARTICLE{2020A&A...642A.213G,
       author = {{G{\'o}mez-Garrido}, M. and {Bujarrabal}, V. and {Alcolea}, J. and {Soria-Ruiz}, R. and {de Vicente}, P. and {Desmurs}, J. -F.},
        title = "{Very fast variations of SiO maser emission in evolved stars}",
      journal = {\aap},
         year = {2020},
       volume = {642},
          eid = {A213},
        pages = {A213},
          doi = {10.1051/0004-6361/202037499},
archivePrefix = {arXiv},
       eprint = {2009.09771},
 primaryClass = {astro-ph.SR},
       adsurl = {https://ui.adsabs.harvard.edu/abs/2020A&A...642A.213G}
}

@ARTICLE{2009MNRAS.394...51G,
       author = {{Gray}, M.~D. and {Wittkowski}, M. and {Scholz}, M. and {Humphreys}, E.~M.~L. and {Ohnaka}, K. and {Boboltz}, D.},
        title = "{SiO maser emission in Miras}",
      journal = {\mnras},
         year = 2009,
        month = mar,
       volume = {394},
       number = {1},
        pages = {51-66},
          doi = {10.1111/j.1365-2966.2008.14237.x},
archivePrefix = {arXiv},
       eprint = {0811.2770},
 primaryClass = {astro-ph},
       adsurl = {https://ui.adsabs.harvard.edu/abs/2009MNRAS.394...51G}
}

@ARTICLE{2019MNRAS.484.4678M,
       author = {{McDonald}, I. and {Trabucchi}, M.},
        title = "{The onset of the AGB wind tied to a transition between sequences in the period-luminosity diagram}",
      journal = {\mnras},
         year = 2019,
        month = apr,
       volume = {484},
       number = {4},
        pages = {4678-4682},
          doi = {10.1093/mnras/stz324},
archivePrefix = {arXiv},
       eprint = {1901.06325},
 primaryClass = {astro-ph.SR},
       adsurl = {https://ui.adsabs.harvard.edu/abs/2019MNRAS.484.4678M}
}

@BOOK{2004agbs.book.....H,
       author = {{Habing}, Harm J. and {Olofsson}, Hans},
        title = "{Asymptotic Giant Branch Stars}",
         year = 2004,
    publisher = "Springer New York",
          doi = {10.1007/978-1-4757-3876-6},
       adsurl = {https://ui.adsabs.harvard.edu/abs/2004agbs.book.....H}
}

@ARTICLE{1997A&A...322..938G,
       author = {{Gonzalez-Alfonso}, E. and {Cernicharo}, J.},
        title = "{Explanation of \^29\^SiO, \^30\^SiO and high-v \^28\^SiO maser emission.}",
      journal = {\aap},
         year = 1997,
        month = jun,
       volume = {322},
        pages = {938-942},
       adsurl = {https://ui.adsabs.harvard.edu/abs/1997A&A...322..938G}
}

@ARTICLE{2021ApJS..253...44R,
       author = {{Rizzo}, J.~R. and {Cernicharo}, J. and {Garc{\'\i}a-Mir{\'o}}, C.},
        title = "{SiO, $^{29}$SiO, and $^{30}$SiO Emission from 67 Oxygen-rich Stars: A Survey of 61 Maser Lines from 7 to 1 mm}",
      journal = {\apjs},
         year = 2021,
        month = apr,
       volume = {253},
       number = {2},
          eid = {44},
        pages = {44},
          doi = {10.3847/1538-4365/abe469},
archivePrefix = {arXiv},
       eprint = {2102.03548},
 primaryClass = {astro-ph.SR},
       adsurl = {https://ui.adsabs.harvard.edu/abs/2021ApJS..253...44R}
}

@ARTICLE{1992ApJ...401L.109C,
       author = {{Cernicharo}, J. and {Bujarrabal}, V.},
        title = "{High-Excitation 29SiO and 30SiO Maser Emission}",
      journal = {\apjl},
         year = 1992,
        month = dec,
       volume = {401},
        pages = {L109},
          doi = {10.1086/186683},
       adsurl = {https://ui.adsabs.harvard.edu/abs/1992ApJ...401L.109C}
}

@ARTICLE{1982ApJ...256L..55S,
       author = {{Schwartz}, P.~R. and {Zuckerman}, B. and {Bologna}, J.~M.},
        title = "{Nearly simultaneous observations of vibrationally excited J=1-O, J=2-1,J=3-2, and J=4-3 SiO masers.}",
      journal = {\apjl},
         year = 1982,
        month = may,
       volume = {256},
        pages = {L55-L59},
          doi = {10.1086/183795},
       adsurl = {https://ui.adsabs.harvard.edu/abs/1982ApJ...256L..55S}
}

@ARTICLE{2002A&A...393..115M,
       author = {{Messineo}, M. and {Habing}, H.~J. and {Sjouwerman}, L.~O. and {Omont}, A. and {Menten}, K.~M.},
        title = "{86 GHz SiO maser survey of late-type stars in the Inner Galaxy. I. Observational data}",
      journal = {\aap},
         year = 2002,
        month = oct,
       volume = {393},
        pages = {115-128},
          doi = {10.1051/0004-6361:20021017},
archivePrefix = {arXiv},
       eprint = {astro-ph/0207284},
 primaryClass = {astro-ph},
       adsurl = {https://ui.adsabs.harvard.edu/abs/2002A&A...393..115M}
}

@ARTICLE{1996A&AS..115..117C,
       author = {{Cho}, S. -H. and {Kaifu}, N. and {Ukita}, N.},
        title = "{SiO maser survey of late-type stars. I. Simultaneous observations of six transitions of \^28\^SiO and \^29\^SiO.}",
      journal = {\aaps},
         year = 1996,
        month = jan,
       volume = {115},
        pages = {117},
       adsurl = {https://ui.adsabs.harvard.edu/abs/1996A&AS..115..117C}
}

@ARTICLE{2019ApJS..244...25S,
       author = {{Stroh}, Michael C. and {Pihlstr{\"o}m}, Ylva M. and {Sjouwerman}, Lor{\'a}nt O. and {Lewis}, Megan O. and {Claussen}, Mark J. and {Morris}, Mark R. and {Rich}, R. Michael},
        title = "{The Bulge Asymmetries and Dynamical Evolution (BAaDE) SiO Maser Survey at 86 GHz with ALMA}",
      journal = {\apjs},
         year = 2019,
        month = oct,
       volume = {244},
       number = {2},
          eid = {25},
        pages = {25},
          doi = {10.3847/1538-4365/ab3c35},
archivePrefix = {arXiv},
       eprint = {1909.02090},
 primaryClass = {astro-ph.GA},
       adsurl = {https://ui.adsabs.harvard.edu/abs/2019ApJS..244...25S}
}

@INPROCEEDINGS{2016alma.confE...1H,
       author = {{Humphreys}, Elizabeth and {Miura}, Rie and {Brogan}, Crystal L. and {Hibbard}, John and {Hunter}, Todd R. and {Indebetouw}, Remy},
        title = "{The ALMA Science Pipeline: Current Status}",
    booktitle = {Proceedings of the 2016 ALMA Conference},
         year = 2016,
        month = sep,
          eid = {1},
        pages = {1},
       adsurl = {https://ui.adsabs.harvard.edu/abs/2016alma.confE...1H}
}

@ARTICLE{1996A&ARv...7...97H,
       author = {{Habing}, H.~J.},
        title = "{Circumstellar envelopes and Asymptotic Giant Branch stars}",
      journal = {\aapr},
         year = 1996,
        month = jan,
       volume = {7},
       number = {2},
        pages = {97-207},
          doi = {10.1007/PL00013287},
       adsurl = {https://ui.adsabs.harvard.edu/abs/1996A&ARv...7...97H}
}

@ARTICLE{2005ARA&A..43..435H,
       author = {{Herwig}, Falk},
        title = "{Evolution of Asymptotic Giant Branch Stars}",
      journal = {\araa},
         year = 2005,
        month = sep,
       volume = {43},
       number = {1},
        pages = {435-479},
          doi = {10.1146/annurev.astro.43.072103.150600},
       adsurl = {https://ui.adsabs.harvard.edu/abs/2005ARA&A..43..435H}
}

@ARTICLE{2011MNRAS.415.1083A,
       author = {{Assaf}, K.~A. and {Diamond}, P.~J. and {Richards}, A.~M.~S. and {Gray}, M.~D.},
        title = "{The 43-GHz SiO maser in the circumstellar envelope of the asymptotic giant branch star R Cassiopeiae}",
      journal = {\mnras},
         year = 2011,
        month = aug,
       volume = {415},
       number = {2},
        pages = {1083-1092},
          doi = {10.1111/j.1365-2966.2011.18629.x},
archivePrefix = {arXiv},
       eprint = {1105.1089},
 primaryClass = {astro-ph.GA},
       adsurl = {https://ui.adsabs.harvard.edu/abs/2011MNRAS.415.1083A}
}

@ARTICLE{2020A&A...644A..61H,
       author = {{Homan}, Ward and {Montarg{\`e}s}, Miguel and {Pimpanuwat}, Bannawit and {Richards}, Anita M.~S. and {Wallstr{\"o}m}, Sofia H.~J. and {Kervella}, Pierre and {Decin}, Leen and {Zijlstra}, Albert and {Danilovich}, Taissa and {de Koter}, Alex and {Menten}, Karl and {Sahai}, Raghvendra and {Plane}, John and {Lee}, Kelvin and {Waters}, Rens and {Baudry}, Alain and {Wong}, Ka Tat and {Millar}, Tom J. and {Van de Sande}, Marie and {Lagadec}, Eric and {Gobrecht}, David and {Yates}, Jeremy and {Price}, Daniel and {Cannon}, Emily and {Bolte}, Jan and {De Ceuster}, Frederik and {Herpin}, Fabrice and {Nuth}, Joe and {Philip Sindel}, Jan and {Kee}, Dylan and {Grey}, Malcolm D. and {Etoka}, Sandra and {Jeste}, Manali and {Gottlieb}, Carl A. and {Gottlieb}, Elaine and {McDonald}, Iain and {El Mellah}, Ileyk and {M{\"u}ller}, Holger S.~P.},
        title = "{ATOMIUM: A high-resolution view on the highly asymmetric wind of the AGB star {\ensuremath{\pi}}$^{1}$Gruis. I. First detection of a new companion and its effect on the inner wind}",
      journal = {\aap},
         year = 2020,
        month = dec,
       volume = {644},
          eid = {A61},
        pages = {A61},
          doi = {10.1051/0004-6361/202039185},
archivePrefix = {arXiv},
       eprint = {2010.05509},
 primaryClass = {astro-ph.SR},
       adsurl = {https://ui.adsabs.harvard.edu/abs/2020A&A...644A..61H}
}

@ARTICLE{2021A&A...651A..82H,
       author = {{Homan}, Ward and {Pimpanuwat}, Bannawit and {Herpin}, Fabrice and {Danilovich}, Taissa and {McDonald}, Iain and {Wallstr{\"o}m}, Sofia H.~J. and {Richards}, Anita M.~S. and {Baudry}, Alain and {Sahai}, Raghvendra and {Millar}, Tom J. and {de Koter}, Alex and {Gottlieb}, C.~A. and {Kervella}, Pierre and {Montarg{\`e}s}, Miguel and {Van de Sande}, Marie and {Decin}, Leen and {Zijlstra}, Albert and {Etoka}, Sandra and {Jeste}, Manali and {M{\"u}ller}, Holger S.~P. and {Maes}, Silke and {Malfait}, Jolien and {Menten}, Karl and {Plane}, John and {Lee}, Kelvin and {Waters}, Rens and {Wong}, Ka Tat and {Lagadec}, Eric and {Gobrecht}, David and {Yates}, Jeremy and {Price}, Daniel and {Cannon}, Emily and {Bolte}, Jan and {De Ceuster}, Frederik and {Nuth}, Joe and {Philip Sindel}, Jan and {Kee}, Dylan and {Gray}, Malcolm D. and {El Mellah}, Ileyk},
        title = "{ATOMIUM: The astounding complexity of the near circumstellar environment of the M-type AGB star R Hydrae. I. Morpho-kinematical interpretation of CO and SiO emission}",
      journal = {\aap},
         year = 2021,
        month = jul,
       volume = {651},
          eid = {A82},
        pages = {A82},
          doi = {10.1051/0004-6361/202140512},
archivePrefix = {arXiv},
       eprint = {2104.07297},
 primaryClass = {astro-ph.SR},
       adsurl = {https://ui.adsabs.harvard.edu/abs/2021A&A...651A..82H}
}

@ARTICLE{1993ApJ...407L..33C,
       author = {{Cernicharo}, J. and {Bujarrabal}, V. and {Santaren}, J.~L.},
        title = "{High-Excitation SiO Maser Emission in VY Canis Majoris: Detection of the V = 4 J = 5--4 Transition}",
      journal = {\apjl},
         year = 1993,
        month = apr,
       volume = {407},
        pages = {L33},
          doi = {10.1086/186799},
       adsurl = {https://ui.adsabs.harvard.edu/abs/1993ApJ...407L..33C}
}

@ARTICLE{1983ApJ...264L..65D,
       author = {{Deguchi}, S. and {Good}, J. and {Fan}, Y. and {Mao}, X. and {Wang}, D. and {Ukita}, N.},
        title = "{SiO isotopic maser emission from VY Canis Majoris.}",
      journal = {\apjl},
         year = 1983,
        month = jan,
       volume = {264},
        pages = {L65-L68},
          doi = {10.1086/183945},
       adsurl = {https://ui.adsabs.harvard.edu/abs/1983ApJ...264L..65D}
}

@ARTICLE{2007A&A...470..191W,
       author = {{Wittkowski}, M. and {Boboltz}, D.~A. and {Ohnaka}, K. and {Driebe}, T. and {Scholz}, M.},
        title = "{The Mira variable S Orionis: relationships between the photosphere, molecular layer, dust shell, and SiO maser shell at 4 epochs}",
      journal = {\aap},
         year = 2007,
        month = jul,
       volume = {470},
       number = {1},
        pages = {191-210},
          doi = {10.1051/0004-6361:20077168},
archivePrefix = {arXiv},
       eprint = {0705.4614},
 primaryClass = {astro-ph},
       adsurl = {https://ui.adsabs.harvard.edu/abs/2007A&A...470..191W}
}

@ARTICLE{2022A&A...667A..74A,
       author = {{Andriantsaralaza}, M. and {Ramstedt}, S. and {Vlemmings}, W.~H.~T. and {De Beck}, E.},
        title = "{Distance estimates for AGB stars from parallax measurements}",
      journal = {\aap},
         year = 2022,
        month = nov,
       volume = {667},
          eid = {A74},
        pages = {A74},
          doi = {10.1051/0004-6361/202243670},
archivePrefix = {arXiv},
       eprint = {2209.03906},
 primaryClass = {astro-ph.SR},
       adsurl = {https://ui.adsabs.harvard.edu/abs/2022A&A...667A..74A}
}

@BOOK{2005pcim.book.....T,
       author = {{Tielens}, A.~G.~G.~M.},
        title = "{The Physics and Chemistry of the Interstellar Medium}",
         year = 2005,
    publisher = "Cambridge University Press",
       adsurl = {https://ui.adsabs.harvard.edu/abs/2005pcim.book.....T}
}

@ARTICLE{2002A&A...391.1053O,
       author = {{Olofsson}, H. and {Gonz{\'a}lez Delgado}, D. and {Kerschbaum}, F. and {Sch{\"o}ier}, F.~L.},
        title = "{Mass loss rates of a sample of irregular and semiregular M-type AGB-variables}",
      journal = {\aap},
         year = 2002,
        month = sep,
       volume = {391},
        pages = {1053-1067},
          doi = {10.1051/0004-6361:20020841},
archivePrefix = {arXiv},
       eprint = {astro-ph/0206172},
 primaryClass = {astro-ph},
       adsurl = {https://ui.adsabs.harvard.edu/abs/2002A&A...391.1053O}
}

@ARTICLE{1999A&AS..140..197G,
       author = {{Groenewegen}, M.~A.~T. and {Baas}, F. and {Blommaert}, J.~A.~D.~L. and {Stehle}, R. and {Josselin}, E. and {Tilanus}, R.~P.~J.},
        title = "{Millimeter and some near infra-red observations of short-period Miras and other AGB stars}",
      journal = {\aaps},
         year = 1999,
        month = dec,
       volume = {140},
        pages = {197-224},
          doi = {10.1051/aas:1999418},
       adsurl = {https://ui.adsabs.harvard.edu/abs/1999A&AS..140..197G}
}

@ARTICLE{2010A&A...523A..18D,
       author = {{De Beck}, E. and {Decin}, L. and {de Koter}, A. and {Justtanont}, K. and {Verhoelst}, T. and {Kemper}, F. and {Menten}, K.~M.},
        title = "{Probing the mass-loss history of AGB and red supergiant stars from CO rotational line profiles. II. CO line survey of evolved stars: derivation of mass-loss rate formulae}",
      journal = {\aap},
         year = 2010,
        month = nov,
       volume = {523},
          eid = {A18},
        pages = {A18},
          doi = {10.1051/0004-6361/200913771},
archivePrefix = {arXiv},
       eprint = {1008.1083},
 primaryClass = {astro-ph.SR},
       adsurl = {https://ui.adsabs.harvard.edu/abs/2010A&A...523A..18D}
}

@ARTICLE{2020A&A...638A..19B,
       author = {{Bergman}, P. and {Humphreys}, E.~M.~L.},
        title = "{Submillimetre water masers at 437, 439, 471, and 474 GHz towards evolved stars. APEX observations and radiative transfer modelling}",
      journal = {\aap},
         year = 2020,
        month = jun,
       volume = {638},
          eid = {A19},
        pages = {A19},
          doi = {10.1051/0004-6361/202037774},
archivePrefix = {arXiv},
       eprint = {2005.12624},
 primaryClass = {astro-ph.SR},
       adsurl = {https://ui.adsabs.harvard.edu/abs/2020A&A...638A..19B}
}

@ARTICLE{2017A&A...605A..28D,
       author = {{Doan}, L. and {Ramstedt}, S. and {Vlemmings}, W.~H.~T. and {H{\"o}fner}, S. and {De Beck}, E. and {Kerschbaum}, F. and {Lindqvist}, M. and {Maercker}, M. and {Mohamed}, S. and {Paladini}, C. and {Wittkowski}, M.},
        title = "{The extended molecular envelope of the asymptotic giant branch star {\ensuremath{\pi}}$^{1}$ Gruis as seen by ALMA. I. Large-scale kinematic structure and CO excitation properties}",
      journal = {\aap},
         year = 2017,
        month = sep,
       volume = {605},
          eid = {A28},
        pages = {A28},
          doi = {10.1051/0004-6361/201730703},
archivePrefix = {arXiv},
       eprint = {1709.09435},
 primaryClass = {astro-ph.SR},
       adsurl = {https://ui.adsabs.harvard.edu/abs/2017A&A...605A..28D}
}

@ARTICLE{2017A&A...605A.126R,
       author = {{Ramstedt}, S. and {Mohamed}, S. and {Vlemmings}, W.~H.~T. and {Danilovich}, T. and {Brunner}, M. and {De Beck}, E. and {Humphreys}, E.~M.~L. and {Lindqvist}, M. and {Maercker}, M. and {Olofsson}, H. and {Kerschbaum}, F. and {Quintana-Lacaci}, G.},
        title = "{The circumstellar envelope around the S-type AGB star W Aql. Effects of an eccentric binary orbit}",
      journal = {\aap},
         year = 2017,
        month = sep,
       volume = {605},
          eid = {A126},
        pages = {A126},
          doi = {10.1051/0004-6361/201730934},
archivePrefix = {arXiv},
       eprint = {1709.07327},
 primaryClass = {astro-ph.SR},
       adsurl = {https://ui.adsabs.harvard.edu/abs/2017A&A...605A.126R}
}

@ARTICLE{2015A&A...581A..60D,
       author = {{Danilovich}, T. and {Teyssier}, D. and {Justtanont}, K. and {Olofsson}, H. and {Cerrigone}, L. and {Bujarrabal}, V. and {Alcolea}, J. and {Cernicharo}, J. and {Castro-Carrizo}, A. and {Garc{\'\i}a-Lario}, P. and {Marston}, A.},
        title = "{New observations and models of circumstellar CO line emission of AGB stars in the Herschel SUCCESS programme}",
      journal = {\aap},
         year = 2015,
        month = sep,
       volume = {581},
          eid = {A60},
        pages = {A60},
          doi = {10.1051/0004-6361/201526705},
archivePrefix = {arXiv},
       eprint = {1506.09065},
 primaryClass = {astro-ph.SR},
       adsurl = {https://ui.adsabs.harvard.edu/abs/2015A&A...581A..60D}
}

@ARTICLE{2003A&A...411..123G,
       author = {{Gonz{\'a}lez Delgado}, D. and {Olofsson}, H. and {Kerschbaum}, F. and {Sch{\"o}ier}, F.~L. and {Lindqvist}, M. and {Groenewegen}, M.~A.~T.},
        title = "{``Thermal'' SiO radio line emission towards M-type AGB stars: A probe of circumstellar dust formation and dynamics}",
      journal = {\aap},
         year = 2003,
        month = nov,
       volume = {411},
        pages = {123-147},
          doi = {10.1051/0004-6361:20031068},
archivePrefix = {arXiv},
       eprint = {astro-ph/0302179},
 primaryClass = {astro-ph},
       adsurl = {https://ui.adsabs.harvard.edu/abs/2003A&A...411..123G}
}

@ARTICLE{2019A&A...623A.158H,
       author = {{H{\"o}fner}, Susanne and {Freytag}, Bernd},
        title = "{Exploring the origin of clumpy dust clouds around cool giants. A global 3D RHD model of a dust-forming M-type AGB star}",
      journal = {\aap},
         year = 2019,
        month = mar,
       volume = {623},
          eid = {A158},
        pages = {A158},
          doi = {10.1051/0004-6361/201834799},
archivePrefix = {arXiv},
       eprint = {1902.04074},
 primaryClass = {astro-ph.SR},
       adsurl = {https://ui.adsabs.harvard.edu/abs/2019A&A...623A.158H}
}

@ARTICLE{1997ApJ...487L.147B,
       author = {{Boboltz}, D.~A. and {Diamond}, P.~J. and {Kemball}, A.~J.},
        title = "{R Aquarii: First Detection of Circumstellar SiO Maser Proper Motions}",
      journal = {\apjl},
         year = 1997,
        month = oct,
       volume = {487},
       number = {2},
        pages = {L147-L150},
          doi = {10.1086/310896},
       adsurl = {https://ui.adsabs.harvard.edu/abs/1997ApJ...487L.147B}
}

@ARTICLE{1996MNRAS.282.1359H,
       author = {{Humphreys}, E.~M.~L. and {Gray}, M.~D. and {Yates}, J.~A. and {Field}, D. and {Bowen}, G. and {Diamond}, P.~J.},
        title = "{SiO masers in Mira variables at a single stellar phase}",
      journal = {\mnras},
         year = 1996,
        month = oct,
       volume = {282},
       number = {4},
        pages = {1359-1371},
          doi = {10.1093/mnras/282.4.1359},
       adsurl = {https://ui.adsabs.harvard.edu/abs/1996MNRAS.282.1359H}
}

@ARTICLE{2004ApJ...608..480B,
       author = {{Boboltz}, D.~A. and {Claussen}, M.~J.},
        title = "{Ground-State SiO Maser Emission toward Evolved Stars}",
      journal = {\apj},
         year = 2004,
        month = jun,
       volume = {608},
       number = {1},
        pages = {480-488},
          doi = {10.1086/386541},
archivePrefix = {arXiv},
       eprint = {astro-ph/0403217},
 primaryClass = {astro-ph},
       adsurl = {https://ui.adsabs.harvard.edu/abs/2004ApJ...608..480B}
}

@ARTICLE{2016A&A...589A..74D,
       author = {{de Vicente}, P. and {Bujarrabal}, V. and {D{\'\i}az-Pulido}, A. and {Albo}, C. and {Alcolea}, J. and {Barcia}, A. and {Barbas}, L. and {Bola{\~n}o}, R. and {Colomer}, F. and {Diez}, M.~C. and {Gallego}, J.~D. and {G{\'o}mez-Gonz{\'a}lez}, J. and {L{\'o}pez-Fern{\'a}ndez}, I. and {L{\'o}pez-Fern{\'a}ndez}, J.~A. and {L{\'o}pez-P{\'e}rez}, J.~A. and {Malo}, I. and {Moreno}, A. and {Patino}, M. and {Serna}, J.~M. and {Tercero}, F. and {Vaquero}, B.},
        title = "{$^{28}$SiO v = 0 J = 1-0 emission from evolved stars}",
      journal = {\aap},
         year = 2016,
        month = may,
       volume = {589},
          eid = {A74},
        pages = {A74},
          doi = {10.1051/0004-6361/201527174},
archivePrefix = {arXiv},
       eprint = {1603.01163},
 primaryClass = {astro-ph.GA},
       adsurl = {https://ui.adsabs.harvard.edu/abs/2016A&A...589A..74D}
}

@ARTICLE{1994ApJ...430L..61D,
       author = {{Diamond}, P.~J. and {Kemball}, A.~J. and {Junor}, W. and {Zensus}, A. and {Benson}, J. and {Dhawan}, V.},
        title = "{Observation of a Ring Structure in SiO Maser Emission from Late-Type Stars}",
      journal = {\apjl},
         year = 1994,
        month = jul,
       volume = {430},
        pages = {L61},
          doi = {10.1086/187438},
       adsurl = {https://ui.adsabs.harvard.edu/abs/1994ApJ...430L..61D}
}

@ARTICLE{2004A&A...426..131S,
       author = {{Soria-Ruiz}, R. and {Alcolea}, J. and {Colomer}, F. and {Bujarrabal}, V. and {Desmurs}, J. -F. and {Marvel}, K.~B. and {Diamond}, P.~J.},
        title = "{High resolution observations of SiO masers: Comparing the spatial distribution at 43 and 86 GHz}",
      journal = {\aap},
         year = 2004,
        month = oct,
       volume = {426},
        pages = {131-144},
          doi = {10.1051/0004-6361:20041139},
archivePrefix = {arXiv},
       eprint = {astro-ph/0409467},
 primaryClass = {astro-ph},
       adsurl = {https://ui.adsabs.harvard.edu/abs/2004A&A...426..131S}
}

@ARTICLE{2020MNRAS.495.1284Y,
       author = {{Yang}, Haneul and {Cho}, Se-Hyung and {Yun}, Youngjoo and {Yoon}, Dong-Hwan and {Kim}, Dong-Jin and {Kim}, Hyosun and {Yoon}, Sung-Chul and {Dodson}, Richard and {Rioja}, Mar{\'\i}a J. and {Imai}, Hiroshi},
        title = "{Asymmetric distributions of H$_{2}$O and SiO masers towards V627 Cas}",
      journal = {\mnras},
         year = 2020,
        month = jun,
       volume = {495},
       number = {1},
        pages = {1284-1290},
          doi = {10.1093/mnras/staa1206},
archivePrefix = {arXiv},
       eprint = {2006.01344},
 primaryClass = {astro-ph.SR},
       adsurl = {https://ui.adsabs.harvard.edu/abs/2020MNRAS.495.1284Y}
}

@ARTICLE{2016ApJ...817..115C,
       author = {{Chibueze}, James O. and {Miyahara}, Takeshi and {Omodaka}, Toshihiro and {Ohta}, Takashi and {Fujii}, Takahiro and {Tanaka}, Masuo and {Motohara}, Kentaro and {Makoto}, Miyoshi},
        title = "{Near-infrared Observations of SiO Maser-emitting Asymptotic Giant Branch (AGB) Stars}",
      journal = {\apj},
         year = 2016,
        month = feb,
       volume = {817},
       number = {2},
          eid = {115},
        pages = {115},
          doi = {10.3847/0004-637X/817/2/115},
       adsurl = {https://ui.adsabs.harvard.edu/abs/2016ApJ...817..115C}
}

@ARTICLE{2011MNRAS.418..114I,
       author = {{Ireland}, M.~J. and {Scholz}, M. and {Wood}, P.~R.},
        title = "{Dynamical opacity-sampling models of Mira variables - II. Time-dependent atmospheric structure and observable properties of four M-type model series}",
      journal = {\mnras},
         year = 2011,
        month = nov,
       volume = {418},
       number = {1},
        pages = {114-128},
          doi = {10.1111/j.1365-2966.2011.19469.x},
archivePrefix = {arXiv},
       eprint = {1107.3619},
 primaryClass = {astro-ph.SR},
       adsurl = {https://ui.adsabs.harvard.edu/abs/2011MNRAS.418..114I}
}

@ARTICLE{2013MNRAS.433.3133G,
       author = {{Gonidakis}, I. and {Diamond}, P.~J. and {Kemball}, A.~J.},
        title = "{A long-term VLBA monitoring campaign of the v = 1, J = 1 {\textrightarrow}0 SiO masers towards TX Cam - I. Morphology and shock waves}",
      journal = {\mnras},
         year = 2013,
        month = aug,
       volume = {433},
       number = {4},
        pages = {3133-3151},
          doi = {10.1093/mnras/stt954},
archivePrefix = {arXiv},
       eprint = {1306.0274},
 primaryClass = {astro-ph.SR},
       adsurl = {https://ui.adsabs.harvard.edu/abs/2013MNRAS.433.3133G}
}

@ARTICLE{2010MNRAS.406..395G,
       author = {{Gonidakis}, I. and {Diamond}, P.~J. and {Kemball}, A.~J.},
        title = "{Kinematics of the v=1, J=1->0 SiO masers at 43 GHz towards TX Cam - a new 73-frame movie}",
      journal = {\mnras},
         year = 2010,
        month = jul,
       volume = {406},
       number = {1},
        pages = {395-408},
          doi = {10.1111/j.1365-2966.2010.16716.x},
       adsurl = {https://ui.adsabs.harvard.edu/abs/2010MNRAS.406..395G}
}

@ARTICLE{2002A&A...386..256H,
       author = {{Humphreys}, E.~M.~L. and {Gray}, M.~D. and {Yates}, J.~A. and {Field}, D. and {Bowen}, G.~H. and {Diamond}, P.~J.},
        title = "{Numerical simulations of stellar SiO maser variability. Investigation of the effect of shocks}",
      journal = {\aap},
         year = 2002,
        month = apr,
       volume = {386},
        pages = {256-270},
          doi = {10.1051/0004-6361:20020202},
archivePrefix = {arXiv},
       eprint = {astro-ph/0202426},
 primaryClass = {astro-ph},
       adsurl = {https://ui.adsabs.harvard.edu/abs/2002A&A...386..256H}
}

@ARTICLE{1994A&A...286..501P,
       author = {{Pijpers}, F.~P. and {Pardo}, J.~R. and {Bujarrabal}, V.},
        title = "{Short time scale monitoring of SiO sources.}",
      journal = {\aap},
         year = 1994,
        month = jun,
       volume = {286},
        pages = {501-507},
          doi = {10.48550/arXiv.astro-ph/9402061},
archivePrefix = {arXiv},
       eprint = {astro-ph/9402061},
 primaryClass = {astro-ph},
       adsurl = {https://ui.adsabs.harvard.edu/abs/1994A&A...286..501P}
}

@ARTICLE{2004A&A...424..145P,
       author = {{Pardo}, J.~R. and {Alcolea}, J. and {Bujarrabal}, V. and {Colomer}, F. and {del Romero}, A. and {de Vicente}, P.},
        title = "{$^{28}$SiO v = 1 and v = 2, J = 1-0 maser variability in evolved stars. Eleven years of short spaced monitoring}",
      journal = {\aap},
         year = 2004,
        month = sep,
       volume = {424},
        pages = {145-156},
          doi = {10.1051/0004-6361:20040309},
       adsurl = {https://ui.adsabs.harvard.edu/abs/2004A&A...424..145P}
}

@ARTICLE{1977MNRAS.180..415B,
       author = {{Balister}, M. and {Batchelor}, R.~A. and {Haynes}, R.~F. and {Knowles}, S.~H. and {McCulloch}, M.~G. and {Robinson}, B.~J. and {Wellington}, K.~J. and {Yabsley}, D.~E.},
        title = "{Observations of SiO masers at 43 GHz with the Parkes radio telescope.}",
      journal = {\mnras},
         year = 1977,
        month = aug,
       volume = {180},
        pages = {415-427},
          doi = {10.1093/mnras/180.3.415},
       adsurl = {https://ui.adsabs.harvard.edu/abs/1977MNRAS.180..415B}
}

@ARTICLE{1981ApJ...247L..81O,
       author = {{Olofsson}, H. and {Rydbeck}, O.~E.~H. and {Lane}, A.~P. and {Predmore}, C.~R.},
        title = "{Detection of the SIO (V=2,J=2-1) maser.}",
      journal = {\apjl},
         year = 1981,
        month = jul,
       volume = {247},
        pages = {L81-L84},
          doi = {10.1086/183594},
       adsurl = {https://ui.adsabs.harvard.edu/abs/1981ApJ...247L..81O}
}

@ARTICLE{1996A&A...314..883B,
       author = {{Bujarrabal}, V. and {Alcolea}, J. and {Sanchez Contreras}, C. and {Colomer}, F.},
        title = "{The anomalous SiO maser transition v=2 J=2-1.}",
      journal = {\aap},
         year = 1996,
        month = oct,
       volume = {314},
        pages = {883-895},
       adsurl = {https://ui.adsabs.harvard.edu/abs/1996A&A...314..883B}
}

@ARTICLE{2021A&A...655A..80D,
       author = {{Danilovich}, T. and {Van de Sande}, M. and {Plane}, J.~M.~C. and {Millar}, T.~J. and {Royer}, P. and {Amor}, M.~A. and {Hammami}, K. and {Decock}, L. and {Gottlieb}, C.~A. and {Decin}, L. and {Richards}, A.~M.~S. and {De Beck}, E. and {Baudry}, A. and {Bolte}, J. and {Cannon}, E. and {De Ceuster}, F. and {de Koter}, A. and {Etoka}, S. and {Gobrecht}, D. and {Gray}, M. and {Herpin}, F. and {Homan}, W. and {Jeste}, M. and {Kervella}, P. and {Khouri}, T. and {Lagadec}, E. and {Maes}, S. and {Malfait}, J. and {McDonald}, I. and {Menten}, K.~M. and {Montarg{\`e}s}, M. and {M{\"u}ller}, H.~S.~P. and {Pimpanuwat}, B. and {Sahai}, R. and {Wallstr{\"o}m}, S.~H.~J. and {Waters}, L.~B.~F.~M. and {Wong}, K.~T. and {Yates}, J. and {Zijlstra}, A.},
        title = "{ATOMIUM: halide molecules around the S-type AGB star W Aquilae}",
      journal = {\aap},
         year = 2021,
        month = nov,
       volume = {655},
          eid = {A80},
        pages = {A80},
          doi = {10.1051/0004-6361/202141757},
archivePrefix = {arXiv},
       eprint = {2109.04747},
 primaryClass = {astro-ph.SR},
       adsurl = {https://ui.adsabs.harvard.edu/abs/2021A&A...655A..80D}
}

@ARTICLE{2006A&A...452..257M,
       author = {{Mauron}, N. and {Huggins}, P.~J.},
        title = "{Imaging the circumstellar envelopes of AGB stars}",
      journal = {\aap},
         year = 2006,
        month = jun,
       volume = {452},
       number = {1},
        pages = {257-268},
          doi = {10.1051/0004-6361:20054739},
archivePrefix = {arXiv},
       eprint = {astro-ph/0602623},
 primaryClass = {astro-ph},
       adsurl = {https://ui.adsabs.harvard.edu/abs/2006A&A...452..257M}
}

@INPROCEEDINGS{2004ASPC..313..141S,
       author = {{Sahai}, R.},
        title = "{Sowing the Seeds of Asymmetry: Jet-like Outflows in Proto-Planetary Nebulae and AGB Stars}",
    booktitle = {Asymmetrical Planetary Nebulae III: Winds, Structure and the Thunderbird},
         year = 2004,
       editor = {{Meixner}, Margaret and {Kastner}, Joel H. and {Balick}, Bruce and {Soker}, Noam},
       series = {Astronomical Society of the Pacific Conference Series},
       volume = {313},
        month = jul,
        pages = {141},
       adsurl = {https://ui.adsabs.harvard.edu/abs/2004ASPC..313..141S}
}

@ARTICLE{2016A&A...596A..92K,
       author = {{Kervella}, P. and {Homan}, W. and {Richards}, A.~M.~S. and {Decin}, L. and {McDonald}, I. and {Montarg{\`e}s}, M. and {Ohnaka}, K.},
        title = "{ALMA observations of the nearby AGB star L$_{2}$ Puppis. I. Mass of the central star and detection of a candidate planet}",
      journal = {\aap},
         year = 2016,
        month = dec,
       volume = {596},
          eid = {A92},
        pages = {A92},
          doi = {10.1051/0004-6361/201629877},
archivePrefix = {arXiv},
       eprint = {1611.06231},
 primaryClass = {astro-ph.SR},
       adsurl = {https://ui.adsabs.harvard.edu/abs/2016A&A...596A..92K}
}

@ARTICLE{1996ApJS..103...81C,
       author = {{Condon}, J.~J. and {Helou}, G. and {Sanders}, D.~B. and {Soifer}, B.~T.},
        title = "{A 1.425 GHz Atlas of the IRAS Bright Galaxy Sample, Part II}",
      journal = {\apjs},
         year = 1996,
        month = mar,
       volume = {103},
        pages = {81-108},
          doi = {10.1086/192270},
       adsurl = {https://ui.adsabs.harvard.edu/abs/1996ApJS..103...81C}
}

@ARTICLE{2011A&A...525A..56R,
       author = {{Richards}, A.~M.~S. and {Elitzur}, M. and {Yates}, J.~A.},
        title = "{Observational evidence for the shrinking of bright maser spots}",
      journal = {\aap},
         year = 2011,
        month = jan,
       volume = {525},
          eid = {A56},
        pages = {A56},
          doi = {10.1051/0004-6361/201015397},
archivePrefix = {arXiv},
       eprint = {1010.4419},
 primaryClass = {astro-ph.GA},
       adsurl = {https://ui.adsabs.harvard.edu/abs/2011A&A...525A..56R}
}

@ARTICLE{2023A&A...674A.125B,
       author = {{Baudry}, A. and {Wong}, K.~T. and {Etoka}, S. and {Richards}, A.~M.~S. and {M{\"u}ller}, H.~S.~P. and {Herpin}, F. and {Danilovich}, T. and {Gray}, M.~D. and {Wallstr{\"o}m}, S. and {Gobrecht}, D. and {Khouri}, T. and {Decin}, L. and {Gottlieb}, C.~A. and {Menten}, K.~M. and {Homan}, W. and {Millar}, T.~J. and {Montarg{\`e}s}, M. and {Pimpanuwat}, B. and {Plane}, J.~M.~C. and {Kervella}, P.},
        title = "{ATOMIUM: Probing the inner wind of evolved O-rich stars with new, highly excited H$_{2}$O and OH lines}",
      journal = {\aap},
         year = 2023,
        month = jun,
       volume = {674},
          eid = {A125},
        pages = {A125},
          doi = {10.1051/0004-6361/202245193},
archivePrefix = {arXiv},
       eprint = {2305.03171},
 primaryClass = {astro-ph.SR},
       adsurl = {https://ui.adsabs.harvard.edu/abs/2023A&A...674A.125B}
}

@ARTICLE{2024A&A...681A..50W,
       author = {{Wallstr{\"o}m}, S.~H.~J. and {Danilovich}, T. and {M{\"u}ller}, H.~S.~P. and {Gottlieb}, C.~A. and {Maes}, S. and {Van de Sande}, M. and {Decin}, L. and {Richards}, A.~M.~S. and {Baudry}, A. and {Bolte}, J. and {Ceulemans}, T. and {De Ceuster}, F. and {de Koter}, A. and {El Mellah}, I. and {Esseldeurs}, M. and {Etoka}, S. and {Gobrecht}, D. and {Gottlieb}, E. and {Gray}, M. and {Herpin}, F. and {Jeste}, M. and {Kee}, D. and {Kervella}, P. and {Khouri}, T. and {Lagadec}, E. and {Malfait}, J. and {Marinho}, L. and {McDonald}, I. and {Menten}, K.~M. and {Millar}, T.~J. and {Montarg{\`e}s}, M. and {Nuth}, J.~A. and {Plane}, J.~M.~C. and {Sahai}, R. and {Waters}, L.~B.~F.~M. and {Wong}, K.~T. and {Yates}, J. and {Zijlstra}, A.},
        title = "{ATOMIUM: Molecular inventory of 17 oxygen-rich evolved stars observed with ALMA}",
      journal = {\aap},
         year = 2024,
        month = jan,
       volume = {681},
          eid = {A50},
        pages = {A50},
          doi = {10.1051/0004-6361/202347632},
archivePrefix = {arXiv},
       eprint = {2312.03467},
 primaryClass = {astro-ph.SR},
       adsurl = {https://ui.adsabs.harvard.edu/abs/2024A&A...681A..50W}
}

@ARTICLE{2012JAVSO..40..852B,
       author = {{Benn}, D.},
        title = "{Algorithms + Observations = VStar}",
      journal = {\javso},
         year = 2012,
        month = may,
       volume = {40},
       number = {2},
        pages = {852},
       adsurl = {https://ui.adsabs.harvard.edu/abs/2012JAVSO..40..852B}
}

@INPROCEEDINGS{2024IAUS..380..314L,
       author = {{Lewis}, Megan O. and {Pihlstr{\"o}m}, Ylva M. and {Sjouwerman}, Lor{\'a}nt O.},
        title = "{SiO maser line ratios in the BAaDE Survey}",
    booktitle = {Cosmic Masers: Proper Motion Toward the Next-Generation Large Projects},
       series = {IAU Symposium},
         year = 2024,
       editor = {{Hirota}, Tomoya and {Imai}, Hiroshi and {Menten}, Karl and {Pihlstr{\"o}m}, Ylva},
       volume = {380},
        month = jan,
        pages = {314-318},
          doi = {10.1017/S1743921323002375},
       adsurl = {https://ui.adsabs.harvard.edu/abs/2024IAUS..380..314L}
}

@ARTICLE{2020ApJ...892...52L,
       author = {{Lewis}, Megan O. and {Pihlstr{\"o}m}, Ylva M. and {Sjouwerman}, Lor{\'a}nt O. and {Stroh}, Michael C. and {Morris}, Mark R. and {BAaDE Collaboration}},
        title = "{Carbon- and Oxygen-rich Asymptotic Giant Branch (AGB) Stars in the Bulge Asymmetries and Dynamical Evolution (BAaDE) Survey}",
      journal = {\apj},
         year = 2020,
        month = mar,
       volume = {892},
       number = {1},
          eid = {52},
        pages = {52},
          doi = {10.3847/1538-4357/ab7920},
archivePrefix = {arXiv},
       eprint = {2002.11095},
 primaryClass = {astro-ph.SR},
       adsurl = {https://ui.adsabs.harvard.edu/abs/2020ApJ...892...52L}
}

@ARTICLE{2000A&A...359.1117H,
       author = {{Herpin}, F. and {Baudry}, A.},
        title = "{$^{28}$SiO, $^{29}$SiO and $^{30}$SiO excitation: effects of infrared line overlaps}",
      journal = {\aap},
         year = 2000,
        month = jul,
       volume = {359},
        pages = {1117-1123},
          doi = {10.48550/arXiv.astro-ph/0005456},
archivePrefix = {arXiv},
       eprint = {astro-ph/0005456},
 primaryClass = {astro-ph},
       adsurl = {https://ui.adsabs.harvard.edu/abs/2000A&A...359.1117H}
}

@ARTICLE{2021MNRAS.504.2687N,
       author = {{Nhung}, P.~T. and {Hoai}, D.~T. and {Tuan-Anh}, P. and {Darriulat}, P. and {Diep}, P.~N. and {Ngoc}, N.~B. and {Phuong}, N.~T. and {Thai}, T.~T.},
        title = "{Morpho-kinematics of the circumstellar envelope of the AGB star R Dor: a global view}",
      journal = {\mnras},
         year = 2021,
        month = jun,
       volume = {504},
       number = {2},
        pages = {2687-2706},
          doi = {10.1093/mnras/stab954},
archivePrefix = {arXiv},
       eprint = {2101.05455},
 primaryClass = {astro-ph.SR},
       adsurl = {https://ui.adsabs.harvard.edu/abs/2021MNRAS.504.2687N}
}

@ARTICLE{2016A&A...590A.127W,
       author = {{Wong}, K.~T. and {Kami{\'n}ski}, T. and {Menten}, K.~M. and {Wyrowski}, F.},
        title = "{Resolving the extended atmosphere and the inner wind of Mira (o Ceti) with long ALMA baselines}",
      journal = {\aap},
         year = 2016,
        month = may,
       volume = {590},
          eid = {A127},
        pages = {A127},
          doi = {10.1051/0004-6361/201527867},
archivePrefix = {arXiv},
       eprint = {1603.03371},
 primaryClass = {astro-ph.SR},
       adsurl = {https://ui.adsabs.harvard.edu/abs/2016A&A...590A.127W}
}

@ARTICLE{2022MNRAS.513.4405A,
       author = {{Aydi}, Elias and {Mohamed}, Shazrene},
        title = "{3D models of the circumstellar environments of evolved stars: Formation of multiple spiral structures}",
      journal = {\mnras},
         year = 2022,
        month = jul,
       volume = {513},
       number = {3},
        pages = {4405-4430},
          doi = {10.1093/mnras/stac749},
archivePrefix = {arXiv},
       eprint = {2203.08318},
 primaryClass = {astro-ph.SR},
       adsurl = {https://ui.adsabs.harvard.edu/abs/2022MNRAS.513.4405A}
}

@ARTICLE{2017MNRAS.468.4465C,
       author = {{Chen}, Zhuo and {Frank}, Adam and {Blackman}, Eric G. and {Nordhaus}, Jason and {Carroll-Nellenback}, Jonathan},
        title = "{Mass transfer and disc formation in AGB binary systems}",
      journal = {\mnras},
         year = 2017,
        month = jul,
       volume = {468},
       number = {4},
        pages = {4465-4477},
          doi = {10.1093/mnras/stx680},
archivePrefix = {arXiv},
       eprint = {1702.06160},
 primaryClass = {astro-ph.SR},
       adsurl = {https://ui.adsabs.harvard.edu/abs/2017MNRAS.468.4465C}
}

@ARTICLE{1999ApJ...523..357M,
       author = {{Mastrodemos}, Nikos and {Morris}, Mark},
        title = "{Bipolar Pre-Planetary Nebulae: Hydrodynamics of Dusty Winds in Binary Systems. II. Morphology of the Circumstellar Envelopes}",
      journal = {\apj},
         year = 1999,
        month = sep,
       volume = {523},
       number = {1},
        pages = {357-380},
          doi = {10.1086/307717},
       adsurl = {https://ui.adsabs.harvard.edu/abs/1999ApJ...523..357M}
}

@ARTICLE{2020A&A...642A..93H,
       author = {{Homan}, Ward and {Cannon}, Emily and {Montarg{\`e}s}, Miguel and {Richards}, Anita M.~S. and {Millar}, Tom J. and {Decin}, Leen},
        title = "{A detailed view on the circumstellar environment of the M-type AGB star EP Aquarii. I. High-resolution ALMA and SPHERE observations}",
      journal = {\aap},
         year = 2020,
        month = oct,
       volume = {642},
          eid = {A93},
        pages = {A93},
          doi = {10.1051/0004-6361/202038255},
archivePrefix = {arXiv},
       eprint = {2008.08394},
 primaryClass = {astro-ph.SR},
       adsurl = {https://ui.adsabs.harvard.edu/abs/2020A&A...642A..93H}
}

@ARTICLE{2004ApJ...600..992G,
       author = {{Garc{\'\i}a-Arredondo}, F. and {Frank}, Adam},
        title = "{Collimated Outflow Formation via Binary Stars: Three-Dimensional Simulations of Asymptotic Giant Branch Wind and Disk Wind Interactions}",
      journal = {\apj},
         year = 2004,
        month = jan,
       volume = {600},
       number = {2},
        pages = {992-1003},
          doi = {10.1086/379821},
archivePrefix = {arXiv},
       eprint = {astro-ph/0307454},
 primaryClass = {astro-ph},
       adsurl = {https://ui.adsabs.harvard.edu/abs/2004ApJ...600..992G}
}

@ARTICLE{2021AJ....161..147B,
       author = {{Bailer-Jones}, C.~A.~L. and {Rybizki}, J. and {Fouesneau}, M. and {Demleitner}, M. and {Andrae}, R.},
        title = "{Estimating Distances from Parallaxes. V. Geometric and Photogeometric Distances to 1.47 Billion Stars in Gaia Early Data Release 3}",
      journal = {\aj},
         year = 2021,
        month = mar,
       volume = {161},
       number = {3},
          eid = {147},
        pages = {147},
          doi = {10.3847/1538-3881/abd806},
archivePrefix = {arXiv},
       eprint = {2012.05220},
 primaryClass = {astro-ph.SR},
       adsurl = {https://ui.adsabs.harvard.edu/abs/2021AJ....161..147B}
}

@ARTICLE{2020ApJS..247...23A,
       author = {{Asaki}, Yoshiharu and {Maud}, Luke T. and {Fomalont}, Edward B. and {Phillips}, Neil M. and {Hirota}, Akihiko and {Sawada}, Tsuyoshi and {Barcos-Mu{\~n}oz}, Loreto and {Richards}, Anita M.~S. and {Dent}, William R.~F. and {Takahashi}, Satoko and {Corder}, Stuartt and {Carpenter}, John M. and {Villard}, Eric and {Humphreys}, Elizabeth M.},
        title = "{ALMA High-frequency Long Baseline Campaign in 2017: Band-to-band Phase Referencing in Submillimeter Waves}",
      journal = {\apjs},
         year = 2020,
        month = mar,
       volume = {247},
       number = {1},
          eid = {23},
        pages = {23},
          doi = {10.3847/1538-4365/ab6b20},
archivePrefix = {arXiv},
       eprint = {2003.07472},
 primaryClass = {astro-ph.IM},
       adsurl = {https://ui.adsabs.harvard.edu/abs/2020ApJS..247...23A}
}

@INPROCEEDINGS{2024IAUS..380..324Y,
       author = {{Yun}, Youngjoo and {Cho}, Se-Hyung and {Yoon}, Dong-Hwan and {Yang}, Haneul and {Dodson}, Richard and {Rioja}, Mar{\'\i}a J. and {Imai}, Hiroshi},
        title = "{Results of KVN Key Science Program for evolved stars}",
    booktitle = {Cosmic Masers: Proper Motion Toward the Next-Generation Large Projects},
       series = {IAU Symposium},
         year = 2024,
       editor = {{Hirota}, Tomoya and {Imai}, Hiroshi and {Menten}, Karl and {Pihlstr{\"o}m}, Ylva},
       volume = {380},
        month = jan,
        pages = {324-327},
          doi = {10.1017/S174392132300217X},
       adsurl = {https://ui.adsabs.harvard.edu/abs/2024IAUS..380..324Y}
}

@INPROCEEDINGS{1985daa..conf..195W,
       author = {{Wells}, D.~C.},
        title = "{NRAO's Astronomical Image Processing System (AIPS)}",
    booktitle = {Data Analysis in Astronomy},
         year = 1985,
       editor = {{di Gesu}, V. and {Scarsi}, L. and {Crane}, P. and {Friedman}, J.~H. and {Levialdi}, S.},
        month = jan,
        pages = {195},
       adsurl = {https://ui.adsabs.harvard.edu/abs/1985daa..conf..195W}
}

@ARTICLE{1997PASP..109..166C,
       author = {{Condon}, J.~J.},
        title = "{Errors in Elliptical Gaussian Fits}",
      journal = {\pasp},
         year = 1997,
        month = feb,
       volume = {109},
        pages = {166-172},
          doi = {10.1086/133871},
       adsurl = {https://ui.adsabs.harvard.edu/abs/1997PASP..109..166C}
}

@ARTICLE{2002MNRAS.334..498Z,
       author = {{Zijlstra}, Albert A. and {Bedding}, T.~R. and {Mattei}, J.~A.},
        title = "{The evolution of the Mira variable R Hydrae}",
      journal = {\mnras},
         year = 2002,
        month = aug,
       volume = {334},
       number = {3},
        pages = {498-510},
          doi = {10.1046/j.1365-8711.2002.05467.x},
archivePrefix = {arXiv},
       eprint = {astro-ph/0203328},
 primaryClass = {astro-ph},
       adsurl = {https://ui.adsabs.harvard.edu/abs/2002MNRAS.334..498Z}
}

@ARTICLE{2008A&A...484..401D,
       author = {{Decin}, L. and {Blomme}, L. and {Reyniers}, M. and {Ryde}, N. and {Hinkle}, K.~H. and {de Koter}, A.},
        title = "{Probing the mass-loss history of the unusual Mira variable <ASTROBJ>R Hydrae</ASTROBJ> through its infrared CO wind}",
      journal = {\aap},
         year = 2008,
        month = jun,
       volume = {484},
       number = {2},
        pages = {401-412},
          doi = {10.1051/0004-6361:20079312},
       adsurl = {https://ui.adsabs.harvard.edu/abs/2008A&A...484..401D}
}

@ARTICLE{2000JBAA..110..131G,
       author = {{Greaves}, J. and {Howarth}, J.~J.},
        title = "{Further investigations of R Aquilae}",
      journal = {Journal of the British Astronomical Association},
         year = 2000,
        month = jun,
       volume = {110},
        pages = {131-142},
       adsurl = {https://ui.adsabs.harvard.edu/abs/2000JBAA..110..131G}
}

@ARTICLE{2013MNRAS.434.1469B,
       author = {{Barton}, Emma J. and {Yurchenko}, Sergei N. and {Tennyson}, Jonathan},
        title = "{ExoMol line lists - II. The ro-vibrational spectrum of SiO}",
      journal = {\mnras},
         year = 2013,
        month = sep,
       volume = {434},
       number = {2},
        pages = {1469-1475},
          doi = {10.1093/mnras/stt1105},
archivePrefix = {arXiv},
       eprint = {1307.2300},
 primaryClass = {astro-ph.SR},
       adsurl = {https://ui.adsabs.harvard.edu/abs/2013MNRAS.434.1469B}
}

@ARTICLE{2005JMoSt.742..215M,
       author = {{M{\"u}ller}, Holger S.~P. and {Schl{\"o}der}, Frank and {Stutzki}, J{\"u}rgen and {Winnewisser}, Gisbert},
        title = "{The Cologne Database for Molecular Spectroscopy, CDMS: a useful tool for astronomers and spectroscopists}",
      journal = {Journal of Molecular Structure},
         year = 2005,
        month = may,
       volume = {742},
       number = {1-3},
        pages = {215-227},
          doi = {10.1016/j.molstruc.2005.01.027},
       adsurl = {https://ui.adsabs.harvard.edu/abs/2005JMoSt.742..215M}
}

@ARTICLE{2016ApJ...822....3Y,
       author = {{Yun}, Youngjoo and {Cho}, Se-Hyung and {Imai}, Hiroshi and {Kim}, Jaeheon and {Asaki}, Yoshiharu and {Chibueze}, James O. and {Choi}, Yoon Kyung and {Dodson}, Richard and {Kim}, Dong-Jin and {Kusuno}, Kozue and {Matsumoto}, Naoko and {Min}, Cheulhong and {Oyadomari}, Miyako and {Rioja}, Mar{\'\i}a J. and {Yoon}, Dong-Hwan and {Byun}, Do-Young and {Chung}, Hyunsoo and {Chung}, Moon-Hee and {Hagiwara}, Yoshiaki and {Han}, Myoung-Hee and {Han}, Seog-Tae and {Hirota}, Tomoya and {Honma}, Mareki and {Hwang}, Jung-Wook and {Je}, Do-Heung and {Jike}, Takaaki and {Jung}, Dong-Kyu and {Jung}, Taehyun and {Kang}, Ji-Hyun and {Kang}, Jiman and {Kang}, Yong-Woo and {Kan-ya}, Yukitoshi and {Kanaguchi}, Masahiro and {Kawaguchi}, Noriyuki and {Kim}, Bong Gyu and {Ryoung Kim}, Hyo and {Kim}, Hyun-Goo and {Kim}, Jongsoo and {Kim}, Kee-Tae and {Kim}, Mikyoung and {Kobayashi}, Hideyuki and {Kono}, Yusuke and {Kurayama}, Tomoharu and {Lee}, Changhoon and {Lee}, Jeewon and {Lee}, Jeong Ae and {Lee}, Jung-Won and {Lee}, Sang Hyun and {Lee}, Sang-Sung and {Lyo}, A. -Ran and {Minh}, Young Chol and {Oh}, Chungsik and {Oh}, Se-Jin and {Oyama}, Tomoaki and {Roh}, Duk-Gyoo and {Sawada-Satoh}, Satoko and {Shibata}, Katsunori M. and {Sohn}, Bong Won and {Song}, Min-Gyu and {Tamura}, Yoshiaki and {Wi}, Seog-Oh and {Yeom}, Jae-Hwan},
        title = "{SiO Masers around WX Psc Mapped with the KVN and VERA Array (KaVA)}",
      journal = {\apj},
         year = 2016,
        month = may,
       volume = {822},
       number = {1},
          eid = {3},
        pages = {3},
          doi = {10.3847/0004-637X/822/1/3},
       adsurl = {https://ui.adsabs.harvard.edu/abs/2016ApJ...822....3Y}
}

@ARTICLE{1974ApJS...27..331H,
       author = {{Harvey}, Paul M. and {Bechis}, Kenneth P. and {Wilson}, William J. and {Ball}, John A.},
        title = "{Time Variations in the OH Microwave and Infrared Emission from Late-Type Stars}",
      journal = {\apjs},
         year = 1974,
        month = apr,
       volume = {27},
        pages = {331-357},
          doi = {10.1086/190300},
       adsurl = {https://ui.adsabs.harvard.edu/abs/1974ApJS...27..331H}
}

@ARTICLE{2004A&A...414..275C,
       author = {{Cotton}, W.~D. and {Mennesson}, B. and {Diamond}, P.~J. and {Perrin}, G. and {Coud{\'e} du Foresto}, V. and {Chagnon}, G. and {van Langevelde}, H.~J. and {Ridgway}, S. and {Waters}, R. and {Vlemmings}, W. and {Morel}, S. and {Traub}, W. and {Carleton}, N. and {Lacasse}, M.},
        title = "{VLBA observations of SiO masers towards Mira variable stars}",
      journal = {\aap},
         year = 2004,
        month = jan,
       volume = {414},
        pages = {275-288},
          doi = {10.1051/0004-6361:20031597},
       adsurl = {https://ui.adsabs.harvard.edu/abs/2004A&A...414..275C}
}

@ARTICLE{2017AJ....153..176L,
       author = {{Liu}, Jiaming and {Jiang}, Biwei},
        title = "{On the Relation of Silicates and SiO Maser in Evolved Stars}",
      journal = {\aj},
         year = 2017,
        month = apr,
       volume = {153},
       number = {4},
          eid = {176},
        pages = {176},
          doi = {10.3847/1538-3881/aa6334},
archivePrefix = {arXiv},
       eprint = {1702.07820},
 primaryClass = {astro-ph.SR},
       adsurl = {https://ui.adsabs.harvard.edu/abs/2017AJ....153..176L}
}

@ARTICLE{2004MNRAS.348...34E,
       author = {{Etoka}, S. and {Diamond}, P.},
        title = "{First polarimetric images of NML Cyg at 1612 and 1665 MHz}",
      journal = {\mnras},
         year = 2004,
        month = feb,
       volume = {348},
       number = {1},
        pages = {34-45},
          doi = {10.1111/j.1365-2966.2004.07370.x},
       adsurl = {https://ui.adsabs.harvard.edu/abs/2004MNRAS.348...34E}
}

@ARTICLE{2005ApJ...625..978B,
       author = {{Boboltz}, D.~A. and {Diamond}, P.~J.},
        title = "{Axial Symmetry and Rotation in the SiO Maser Shell of IK Tauri}",
      journal = {\apj},
         year = 2005,
        month = jun,
       volume = {625},
       number = {2},
        pages = {978-984},
          doi = {10.1086/429656},
archivePrefix = {arXiv},
       eprint = {astro-ph/0503182},
 primaryClass = {astro-ph},
       adsurl = {https://ui.adsabs.harvard.edu/abs/2005ApJ...625..978B}
}

@ARTICLE{2008PASJ...60.1039M,
       author = {{Matsumoto}, Naoko and {Omodaka}, Toshihiro and {Imai}, Hiroshi and {Shimizu}, Rie and {Bushimata}, Takeshi and {Choi}, Yoon Kyung and {Hirota}, Tomoya and {Honma}, Mareki and {Inomata}, Noritomo and {Iwadate}, Kenzaburo and {Jike}, Takaaki and {Kameno}, Seiji and {Kameya}, Osamu and {Kamohara}, Ryuichi and {Kan-Ya}, Yukitoshi and {Kawaguchi}, Noriyuki and {Kobayashi}, Hideyuki and {Kuji}, Seisuke and {Kurayama}, Tomoharu and {Maeda}, Toshihisa and {Manabe}, Seiji and {Miyaji}, Takeshi and {Nakagawa}, Akiharu and {Nagayama}, Takumi and {Nakashima}, Koichiro and {Oh}, Chung Sik and {Oyama}, Tomoaki and {Sakai}, Satoshi and {Sakakibara}, Seiichiro and {Sasao}, Tetsuo and {Sato}, Katsuhisa and {Shibata}, Katsunori M. and {Shintani}, Motonobu and {Sofue}, Yoshiaki and {Sora}, Kasumi and {Suda}, Hiroshi and {Tamura}, Yoshiaki and {Tsushima}, Miyuki and {Yamashita}, Kazuyoshi},
        title = "{Variable Asymmetry of the Circumstellar Envelope in IK Tauri Traced by SiO Maser Emission}",
      journal = {\pasj},
         year = 2008,
        month = oct,
       volume = {60},
        pages = {1039},
          doi = {10.1093/pasj/60.5.1039},
       adsurl = {https://ui.adsabs.harvard.edu/abs/2008PASJ...60.1039M}
}

@ARTICLE{2016A&A...585A...6G,
       author = {{Gobrecht}, D. and {Cherchneff}, I. and {Sarangi}, A. and {Plane}, J.~M.~C. and {Bromley}, S.~T.},
        title = "{Dust formation in the oxygen-rich AGB star IK Tauri}",
      journal = {\aap},
         year = 2016,
        month = jan,
       volume = {585},
          eid = {A6},
        pages = {A6},
          doi = {10.1051/0004-6361/201425363},
archivePrefix = {arXiv},
       eprint = {1509.07613},
 primaryClass = {astro-ph.SR},
       adsurl = {https://ui.adsabs.harvard.edu/abs/2016A&A...585A...6G}
}

@ARTICLE{2016A&A...591A..70K,
       author = {{Khouri}, T. and {Maercker}, M. and {Waters}, L.~B.~F.~M. and {Vlemmings}, W.~H.~T. and {Kervella}, P. and {de Koter}, A. and {Ginski}, C. and {De Beck}, E. and {Decin}, L. and {Min}, M. and {Dominik}, C. and {O'Gorman}, E. and {Schmid}, H. -M. and {Lombaert}, R. and {Lagadec}, E.},
        title = "{Study of the inner dust envelope and stellar photosphere of the AGB star R Doradus using SPHERE/ZIMPOL}",
      journal = {\aap},
         year = 2016,
        month = jun,
       volume = {591},
          eid = {A70},
        pages = {A70},
          doi = {10.1051/0004-6361/201628435},
archivePrefix = {arXiv},
       eprint = {1605.05504},
 primaryClass = {astro-ph.SR},
       adsurl = {https://ui.adsabs.harvard.edu/abs/2016A&A...591A..70K}
}

@ARTICLE{2001AJ....122.2679P,
       author = {{Phillips}, R.~B. and {Sivakoff}, G.~R. and {Lonsdale}, C.~J. and {Doeleman}, S.~S.},
        title = "{Coordinated Millimeter VLBI Array Observations of R Cassiopeiae: 86 GHz SiO Masers and Envelope Dynamics}",
      journal = {\aj},
         year = 2001,
        month = nov,
       volume = {122},
       number = {5},
        pages = {2679-2685},
          doi = {10.1086/323449},
       adsurl = {https://ui.adsabs.harvard.edu/abs/2001AJ....122.2679P}
}

@ARTICLE{2010A&A...514A..35N,
       author = {{Nowotny}, W. and {H{\"o}fner}, S. and {Aringer}, B.},
        title = "{Line formation in AGB atmospheres including velocity effects. Molecular line profile variations of long period variables}",
      journal = {\aap},
         year = 2010,
        month = may,
       volume = {514},
          eid = {A35},
        pages = {A35},
          doi = {10.1051/0004-6361/200911899},
archivePrefix = {arXiv},
       eprint = {1002.1849},
 primaryClass = {astro-ph.SR},
       adsurl = {https://ui.adsabs.harvard.edu/abs/2010A&A...514A..35N}
}

@article{2025Danilovich,
       author = {{Danilovich}, T. and {Samaratunge}, N. and {Mori}, Y.~L. and {Richards}, A.~M.~S. and {Baudry}, A. and {Etoka}, S. and {Montarg{\`e}s}, M. and {Kervella}, P. and {McDonald}, I. and {Gottlieb}, C.~A. and {Wallace}, A. and {Price}, D.~J. and {Decin}, L. and {Bolte}, J. and {Ceulemans}, T. and {De Ceuster}, F. and {de Koter}, A. and {Dionese}, D. and {El Mellah}, I. and {Esseldeurs}, M. and {Gray}, M. and {Herpin}, F. and {Khouri}, T. and {Lagadec}, E. and {Landri}, C. and {Marinho}, L. and {Menten}, K.~M. and {Millar}, T.~J. and {M{\"u}ller}, H.~S.~P. and {Pimpanuwat}, B. and {Plane}, J.~M.~C. and {Sahai}, R. and {Siess}, L. and {Van de Sande}, M. and {Vermeulen}, O. and {Wong}, K.~T. and {Yates}, J. and {Zijlstra}, A.},
        title = "{ATOMIUM: Continuum emission and evidence of dust enhancement from binary motion}",
      journal = {\aap},
         year = 2025,
        month = dec,
       volume = {704},
          eid = {A341},
        pages = {A341},
          doi = {10.1051/0004-6361/202554878},
archivePrefix = {arXiv},
       eprint = {2504.00517},
 primaryClass = {astro-ph.SR},
       adsurl = {https://ui.adsabs.harvard.edu/abs/2025A&A...704A.341D}
}

@ARTICLE{2026NatAs..10..124E,
       author = {{Esseldeurs}, Mats and {Decin}, Leen and {De Ridder}, Joris and {Mori}, Yoshiya and {Karakas}, Amanda I. and {Malfait}, Jolien and {Danilovich}, Ta{\'\i}ssa and {Mathis}, St{\'e}phane and {Richards}, Anita M.~S. and {Sahai}, Raghvendra and {Yates}, Jeremy and {Van de Sande}, Marie and {Baes}, Maarten and {Baudry}, Alain and {Bolte}, Jan and {Ceulemans}, Thomas and {De Ceuster}, Frederik and {El Mellah}, Ileyk and {Etoka}, Sandra and {Gottlieb}, Carl and {Herpin}, Fabrice and {Kervella}, Pierre and {Landri}, Camille and {Marinho}, Louise and {McDonald}, Iain and {Menten}, Karl and {Millar}, Tom and {Osborn}, Zara and {Pimpanuwat}, Bannawit and {Plane}, John and {Price}, Daniel J. and {Siess}, Lionel and {Vermeulen}, Owen and {Wong}, Ka Tat},
        title = "{Evidence for the Keplerian orbit of a close companion around a giant star}",
      journal = {Nature Astronomy},
         year = 2026,
        month = jan,
       volume = {10},
        pages = {124-143},
          doi = {10.1038/s41550-025-02697-2},
archivePrefix = {arXiv},
       eprint = {2511.11247},
 primaryClass = {astro-ph.SR},
       adsurl = {https://ui.adsabs.harvard.edu/abs/2026NatAs..10..124E}
}

@ARTICLE{2018A&A...617A..23B,
       author = {{Brunner}, M. and {Danilovich}, T. and {Ramstedt}, S. and {Marti-Vidal}, I. and {De Beck}, E. and {Vlemmings}, W.~H.~T. and {Lindqvist}, M. and {Kerschbaum}, F.},
        title = "{Molecular line study of the S-type AGB star W Aquilae. ALMA observations of CS, SiS, SiO and HCN}",
      journal = {\aap},
         year = 2018,
        month = sep,
       volume = {617},
          eid = {A23},
        pages = {A23},
          doi = {10.1051/0004-6361/201832724},
archivePrefix = {arXiv},
       eprint = {1806.01622},
 primaryClass = {astro-ph.SR},
       adsurl = {https://ui.adsabs.harvard.edu/abs/2018A&A...617A..23B}
}

@ARTICLE{2018A&A...615A..28D,
       author = {{Decin}, L. and {Richards}, A.~M.~S. and {Danilovich}, T. and {Homan}, W. and {Nuth}, J.~A.},
        title = "{ALMA spectral line and imaging survey of a low and a high mass-loss rate AGB star between 335 and 362 GHz}",
      journal = {\aap},
         year = 2018,
        month = jul,
       volume = {615},
          eid = {A28},
        pages = {A28},
          doi = {10.1051/0004-6361/201732216},
archivePrefix = {arXiv},
       eprint = {1801.09291},
 primaryClass = {astro-ph.SR},
       adsurl = {https://ui.adsabs.harvard.edu/abs/2018A&A...615A..28D}
}

@ARTICLE{2024A&A...688A..16M,
       author = {{Massalkhi}, S. and {Ag{\'u}ndez}, M. and {Fonfr{\'\i}a}, J.~P. and {Pardo}, J.~R. and {Velilla-Prieto}, L. and {Cernicharo}, J.},
        title = "{Multiline study of the radial extent of SiO, CS, and SiS in asymptotic giant branch envelopes}",
      journal = {\aap},
         year = 2024,
        month = aug,
       volume = {688},
          eid = {A16},
        pages = {A16},
          doi = {10.1051/0004-6361/202450188},
archivePrefix = {arXiv},
       eprint = {2405.19922},
 primaryClass = {astro-ph.SR},
       adsurl = {https://ui.adsabs.harvard.edu/abs/2024A&A...688A..16M}
}

@ARTICLE{2013A&A...559L...8P,
       author = {{Peng}, T. -C. and {Humphreys}, E.~M.~L. and {Testi}, L. and {Baudry}, A. and {Wittkowski}, M. and {Rawlings}, M.~G. and {de Gregorio-Monsalvo}, I. and {Vlemmings}, W. and {Nyman}, L. -A. and {Gray}, M.~D. and {de Breuck}, C.},
        title = "{Silicon isotopic abundance toward evolved stars and its application for presolar grains}",
      journal = {\aap},
         year = 2013,
        month = nov,
       volume = {559},
          eid = {L8},
        pages = {L8},
          doi = {10.1051/0004-6361/201322466},
archivePrefix = {arXiv},
       eprint = {1311.3444},
 primaryClass = {astro-ph.GA},
       adsurl = {https://ui.adsabs.harvard.edu/abs/2013A&A...559L...8P}
}

@ARTICLE{2000A&AS..143....9W,
       author = {{Wenger}, M. and {Ochsenbein}, F. and {Egret}, D. and {Dubois}, P. and {Bonnarel}, F. and {Borde}, S. and {Genova}, F. and {Jasniewicz}, G. and {Lalo{\"e}}, S. and {Lesteven}, S. and {Monier}, R.},
        title = "{The SIMBAD astronomical database. The CDS reference database for astronomical objects}",
      journal = {\aaps},
         year = 2000,
        month = apr,
       volume = {143},
        pages = {9-22},
          doi = {10.1051/aas:2000332},
archivePrefix = {arXiv},
       eprint = {astro-ph/0002110},
 primaryClass = {astro-ph},
       adsurl = {https://ui.adsabs.harvard.edu/abs/2000A&AS..143....9W}
}

@ARTICLE{2000A&AS..143...23O,
       author = {{Ochsenbein}, F. and {Bauer}, P. and {Marcout}, J.},
        title = "{The VizieR database of astronomical catalogues}",
      journal = {\aaps},
         year = 2000,
        month = apr,
       volume = {143},
        pages = {23-32},
          doi = {10.1051/aas:2000169},
archivePrefix = {arXiv},
       eprint = {astro-ph/0002122},
 primaryClass = {astro-ph},
       adsurl = {https://ui.adsabs.harvard.edu/abs/2000A&AS..143...23O}
}

@ARTICLE{2013A&A...558A..33A,
       author = {{Astropy Collaboration} and {Robitaille}, Thomas P. and {Tollerud}, Erik J. and {Greenfield}, Perry and {Droettboom}, Michael and {Bray}, Erik and {Aldcroft}, Tom and {Davis}, Matt and {Ginsburg}, Adam and {Price-Whelan}, Adrian M. and {Kerzendorf}, Wolfgang E. and {Conley}, Alexander and {Crighton}, Neil and {Barbary}, Kyle and {Muna}, Demitri and {Ferguson}, Henry and {Grollier}, Fr{\'e}d{\'e}ric and {Parikh}, Madhura M. and {Nair}, Prasanth H. and {Unther}, Hans M. and {Deil}, Christoph and {Woillez}, Julien and {Conseil}, Simon and {Kramer}, Roban and {Turner}, James E.~H. and {Singer}, Leo and {Fox}, Ryan and {Weaver}, Benjamin A. and {Zabalza}, Victor and {Edwards}, Zachary I. and {Azalee Bostroem}, K. and {Burke}, D.~J. and {Casey}, Andrew R. and {Crawford}, Steven M. and {Dencheva}, Nadia and {Ely}, Justin and {Jenness}, Tim and {Labrie}, Kathleen and {Lim}, Pey Lian and {Pierfederici}, Francesco and {Pontzen}, Andrew and {Ptak}, Andy and {Refsdal}, Brian and {Servillat}, Mathieu and {Streicher}, Ole},
        title = "{Astropy: A community Python package for astronomy}",
      journal = {\aap},
         year = 2013,
        month = oct,
       volume = {558},
          eid = {A33},
        pages = {A33},
          doi = {10.1051/0004-6361/201322068},
archivePrefix = {arXiv},
       eprint = {1307.6212},
 primaryClass = {astro-ph.IM},
       adsurl = {https://ui.adsabs.harvard.edu/abs/2013A&A...558A..33A}
}

@ARTICLE{2018AJ....156..123A,
       author = {{Astropy Collaboration} and {Price-Whelan}, A.~M. and {Sip{\H{o}}cz}, B.~M. and {G{\"u}nther}, H.~M. and {Lim}, P.~L. and {Crawford}, S.~M. and {Conseil}, S. and {Shupe}, D.~L. and {Craig}, M.~W. and {Dencheva}, N. and {Ginsburg}, A. and {VanderPlas}, J.~T. and {Bradley}, L.~D. and {P{\'e}rez-Su{\'a}rez}, D. and {de Val-Borro}, M. and {Aldcroft}, T.~L. and {Cruz}, K.~L. and {Robitaille}, T.~P. and {Tollerud}, E.~J. and {Ardelean}, C. and {Babej}, T. and {Bach}, Y.~P. and {Bachetti}, M. and {Bakanov}, A.~V. and {Bamford}, S.~P. and {Barentsen}, G. and {Barmby}, P. and {Baumbach}, A. and {Berry}, K.~L. and {Biscani}, F. and {Boquien}, M. and {Bostroem}, K.~A. and {Bouma}, L.~G. and {Brammer}, G.~B. and {Bray}, E.~M. and {Breytenbach}, H. and {Buddelmeijer}, H. and {Burke}, D.~J. and {Calderone}, G. and {Cano Rodr{\'\i}guez}, J.~L. and {Cara}, M. and {Cardoso}, J.~V.~M. and {Cheedella}, S. and {Copin}, Y. and {Corrales}, L. and {Crichton}, D. and {D'Avella}, D. and {Deil}, C. and {Depagne}, {\'E}. and {Dietrich}, J.~P. and {Donath}, A. and {Droettboom}, M. and {Earl}, N. and {Erben}, T. and {Fabbro}, S. and {Ferreira}, L.~A. and {Finethy}, T. and {Fox}, R.~T. and {Garrison}, L.~H. and {Gibbons}, S.~L.~J. and {Goldstein}, D.~A. and {Gommers}, R. and {Greco}, J.~P. and {Greenfield}, P. and {Groener}, A.~M. and {Grollier}, F. and {Hagen}, A. and {Hirst}, P. and {Homeier}, D. and {Horton}, A.~J. and {Hosseinzadeh}, G. and {Hu}, L. and {Hunkeler}, J.~S. and {Ivezi{\'c}}, {\v{Z}}. and {Jain}, A. and {Jenness}, T. and {Kanarek}, G. and {Kendrew}, S. and {Kern}, N.~S. and {Kerzendorf}, W.~E. and {Khvalko}, A. and {King}, J. and {Kirkby}, D. and {Kulkarni}, A.~M. and {Kumar}, A. and {Lee}, A. and {Lenz}, D. and {Littlefair}, S.~P. and {Ma}, Z. and {Macleod}, D.~M. and {Mastropietro}, M. and {McCully}, C. and {Montagnac}, S. and {Morris}, B.~M. and {Mueller}, M. and {Mumford}, S.~J. and {Muna}, D. and {Murphy}, N.~A. and {Nelson}, S. and {Nguyen}, G.~H. and {Ninan}, J.~P. and {N{\"o}the}, M. and {Ogaz}, S. and {Oh}, S. and {Parejko}, J.~K. and {Parley}, N. and {Pascual}, S. and {Patil}, R. and {Patil}, A.~A. and {Plunkett}, A.~L. and {Prochaska}, J.~X. and {Rastogi}, T. and {Reddy Janga}, V. and {Sabater}, J. and {Sakurikar}, P. and {Seifert}, M. and {Sherbert}, L.~E. and {Sherwood-Taylor}, H. and {Shih}, A.~Y. and {Sick}, J. and {Silbiger}, M.~T. and {Singanamalla}, S. and {Singer}, L.~P. and {Sladen}, P.~H. and {Sooley}, K.~A. and {Sornarajah}, S. and {Streicher}, O. and {Teuben}, P. and {Thomas}, S.~W. and {Tremblay}, G.~R. and {Turner}, J.~E.~H. and {Terr{\'o}n}, V. and {van Kerkwijk}, M.~H. and {de la Vega}, A. and {Watkins}, L.~L. and {Weaver}, B.~A. and {Whitmore}, J.~B. and {Woillez}, J. and {Zabalza}, V. and {Astropy Contributors}},
        title = "{The Astropy Project: Building an Open-science Project and Status of the v2.0 Core Package}",
      journal = {\aj},
         year = 2018,
        month = sep,
       volume = {156},
       number = {3},
          eid = {123},
        pages = {123},
          doi = {10.3847/1538-3881/aabc4f},
archivePrefix = {arXiv},
       eprint = {1801.02634},
 primaryClass = {astro-ph.IM},
       adsurl = {https://ui.adsabs.harvard.edu/abs/2018AJ....156..123A}
}

@ARTICLE{2020NatMe..17..261V,
       author = {{Virtanen}, Pauli and {Gommers}, Ralf and {Oliphant}, Travis E. and {Haberland}, Matt and {Reddy}, Tyler and {Cournapeau}, David and {Burovski}, Evgeni and {Peterson}, Pearu and {Weckesser}, Warren and {Bright}, Jonathan and {van der Walt}, St{\'e}fan J. and {Brett}, Matthew and {Wilson}, Joshua and {Millman}, K. Jarrod and {Mayorov}, Nikolay and {Nelson}, Andrew R.~J. and {Jones}, Eric and {Kern}, Robert and {Larson}, Eric and {Carey}, C.~J. and {Polat}, {\.I}lhan and {Feng}, Yu and {Moore}, Eric W. and {VanderPlas}, Jake and {Laxalde}, Denis and {Perktold}, Josef and {Cimrman}, Robert and {Henriksen}, Ian and {Quintero}, E.~A. and {Harris}, Charles R. and {Archibald}, Anne M. and {Ribeiro}, Ant{\^o}nio H. and {Pedregosa}, Fabian and {van Mulbregt}, Paul and {SciPy 1. 0 Contributors}},
        title = "{SciPy 1.0: fundamental algorithms for scientific computing in Python}",
      journal = {Nature Methods},
         year = 2020,
        month = feb,
       volume = {17},
        pages = {261-272},
          doi = {10.1038/s41592-019-0686-2},
archivePrefix = {arXiv},
       eprint = {1907.10121},
 primaryClass = {cs.MS},
       adsurl = {https://ui.adsabs.harvard.edu/abs/2020NatMe..17..261V}
}

@ARTICLE{2007CSE.....9...90H,
       author = {{Hunter}, John D.},
        title = "{Matplotlib: A 2D Graphics Environment}",
      journal = {Computing in Science and Engineering},
         year = 2007,
        month = may,
       volume = {9},
       number = {3},
        pages = {90-95},
          doi = {10.1109/MCSE.2007.55},
       adsurl = {https://ui.adsabs.harvard.edu/abs/2007CSE.....9...90H}
}

@ARTICLE{2020Natur.585..357H,
       author = {{Harris}, Charles R. and {Millman}, K. Jarrod and {van der Walt}, St{\'e}fan J. and {Gommers}, Ralf and {Virtanen}, Pauli and {Cournapeau}, David and {Wieser}, Eric and {Taylor}, Julian and {Berg}, Sebastian and {Smith}, Nathaniel J. and {Kern}, Robert and {Picus}, Matti and {Hoyer}, Stephan and {van Kerkwijk}, Marten H. and {Brett}, Matthew and {Haldane}, Allan and {del R{\'\i}o}, Jaime Fern{\'a}ndez and {Wiebe}, Mark and {Peterson}, Pearu and {G{\'e}rard-Marchant}, Pierre and {Sheppard}, Kevin and {Reddy}, Tyler and {Weckesser}, Warren and {Abbasi}, Hameer and {Gohlke}, Christoph and {Oliphant}, Travis E.},
        title = "{Array programming with NumPy}",
      journal = {\nat},
         year = 2020,
        month = sep,
       volume = {585},
       number = {7825},
        pages = {357-362},
          doi = {10.1038/s41586-020-2649-2},
archivePrefix = {arXiv},
       eprint = {2006.10256},
 primaryClass = {cs.MS},
       adsurl = {https://ui.adsabs.harvard.edu/abs/2020Natur.585..357H}
}

@ARTICLE{1985MNRAS.212..375C,
       author = {{Chapman}, J.~M. and {Cohen}, R.~J.},
        title = "{The unusual OH envelope of U Orionis.}",
      journal = {\mnras},
         year = 1985,
        month = jan,
       volume = {212},
        pages = {375-384},
          doi = {10.1093/mnras/212.2.375},
       adsurl = {https://ui.adsabs.harvard.edu/abs/1985MNRAS.212..375C}
}

@ARTICLE{2025A&A...699A..22M,
       author = {{Montarg{\`e}s}, M. and {Malfait}, J. and {Esseldeurs}, M. and {de Koter}, A. and {Baron}, F. and {Kervella}, P. and {Danilovich}, T. and {Richards}, A.~M.~S. and {Sahai}, R. and {McDonald}, I. and {Khouri}, T. and {Shetye}, S. and {Zijlstra}, A. and {Van de Sande}, M. and {El Mellah}, I. and {Herpin}, F. and {Siess}, L. and {Etoka}, S. and {Gobrecht}, D. and {Marinho}, L. and {Wallstr{\"o}m}, S.~H.~J. and {Wong}, K.~T. and {Yates}, J.},
        title = "{An accreting dwarf star orbiting the S-type giant star {\ensuremath{\pi}}$^{1}$ Gru}",
      journal = {\aap},
         year = 2025,
        month = jul,
       volume = {699},
          eid = {A22},
        pages = {A22},
          doi = {10.1051/0004-6361/202452587},
archivePrefix = {arXiv},
       eprint = {2504.16845},
 primaryClass = {astro-ph.SR},
       adsurl = {https://ui.adsabs.harvard.edu/abs/2025A&A...699A..22M}
}

@ARTICLE{1989ApJ...346..983E,
       author = {{Elitzur}, Moshe and {Hollenbach}, David J. and {McKee}, Christopher F.},
        title = "{H 2O Masers in Star-forming Regions}",
      journal = {\apj},
         year = 1989,
        month = nov,
       volume = {346},
        pages = {983},
          doi = {10.1086/168080},
       adsurl = {https://ui.adsabs.harvard.edu/abs/1989ApJ...346..983E}
}

@ARTICLE{2013JPCA..11713843M,
       author = {{M{\"u}ller}, Holger S.~P. and {Spezzano}, Silvia and {Bizzocchi}, Luca and {Gottlieb}, Carl A. and {Degli Esposti}, Claudio and {McCarthy}, Michael C.},
        title = "{Rotational Spectroscopy of Isotopologues of Silicon Monoxide, SiO, and Spectroscopic Parameters from a Combined Fit of Rotational and Rovibrational Data}",
      journal = {Journal of Physical Chemistry A},
         year = 2013,
        month = sep,
       volume = {117},
        pages = {13843-13854},
          doi = {10.1021/jp408391f},
       adsurl = {https://ui.adsabs.harvard.edu/abs/2013JPCA..11713843M}
}

@ARTICLE{1991ApJ...368L..19M,
       author = {{Mollaaghababa}, R. and {Gottlieb}, C.~A. and {Vrtilek}, J.~M. and {Thaddeus}, P.},
        title = "{The Millimeter-Wave Spectrum of Highly Vibrationally Excited SiO}",
      journal = {\apjl},
         year = 1991,
        month = feb,
       volume = {368},
        pages = {L19},
          doi = {10.1086/185938},
       adsurl = {https://ui.adsabs.harvard.edu/abs/1991ApJ...368L..19M}
}

@ARTICLE{2025AJ....170...84B,
       author = {{Baek}, Hyeon and {Cho}, Se-Hyung and {Kim}, Jaeheon and {Son}, Seung-Min and {Yoon}, Dong-Hwan and {Suh}, Kyung-Won},
        title = "{Survey of Oxygen-rich AGB Stars Using the KVN 4 Receiving Bands{\textemdash}SMASTES. I.}",
      journal = {\aj},
         year = 2025,
        month = aug,
       volume = {170},
       number = {2},
          eid = {84},
        pages = {84},
          doi = {10.3847/1538-3881/ade1d1},
       adsurl = {https://ui.adsabs.harvard.edu/abs/2025AJ....170...84B}
}

@ARTICLE{2024ApJS..275...20L,
       author = {{Lim}, Jang-Ho and {Kim}, Jaeheon and {Cho}, Se-Hyung and {Kim}, Hyosun and {Yoon}, Dong-Hwan and {Son}, Seong-Min and {Suh}, Kyung-Won},
        title = "{Long-term Simultaneous Monitoring Observations of SiO and H$_{2}$O Masers toward the Mira Variable WX Serpentis}",
      journal = {\apjs},
         year = 2024,
        month = dec,
       volume = {275},
       number = {2},
          eid = {20},
        pages = {20},
          doi = {10.3847/1538-4365/ad7d80},
archivePrefix = {arXiv},
       eprint = {2501.09369},
 primaryClass = {astro-ph.SR},
       adsurl = {https://ui.adsabs.harvard.edu/abs/2024ApJS..275...20L}
}

@ARTICLE{2023A&A...669A.155F,
       author = {{Freytag}, Bernd and {H{\"o}fner}, Susanne},
        title = "{Global 3D radiation-hydrodynamical models of AGB stars with dust-driven winds}",
      journal = {\aap},
         year = 2023,
        month = jan,
       volume = {669},
          eid = {A155},
        pages = {A155},
          doi = {10.1051/0004-6361/202244992},
archivePrefix = {arXiv},
       eprint = {2301.11836},
 primaryClass = {astro-ph.SR},
       adsurl = {https://ui.adsabs.harvard.edu/abs/2023A&A...669A.155F}
}

@ARTICLE{1981ApJ...247..247W,
       author = {{Wood}, P.~R. and {Zarro}, D.~M.},
        title = "{Helium-shell flashing in low-mass stars and period changes in Mira variables.}",
      journal = {\apj},
         year = 1981,
        month = jul,
       volume = {247},
        pages = {247-256},
          doi = {10.1086/159032},
       adsurl = {https://ui.adsabs.harvard.edu/abs/1981ApJ...247..247W}
}

@phdthesis{Ahmad2025,
  author       = {Ahmad, A.},
  title        = {Dynamical Processes in Red Giants: Pulsations, Convection, and Mass Loss of Cool, Luminous, Evolved Stars in 3D Models},
  school       = {Uppsala University},
  year         = {2025},
  type         = {PhD thesis},
  isbn         = {978-91-513-2457-9},
  url          = {https://urn.kb.se/resolve?urn=urn:nbn:se:uu:diva-553728},
  note         = {Available at: \url{https://urn.kb.se/resolve?urn=urn:nbn:se:uu:diva-553728}}
}

@ARTICLE{1979ApJ...227..923H,
       author = {{Hinkle}, K.~H. and {Barnes}, T.~G.},
        title = "{Infrared spectroscopy of Mira variables. II. R Leonis, the H$_{2}$O vibration-rotation bands.}",
      journal = {\apj},
         year = 1979,
        month = feb,
       volume = {227},
        pages = {923-934},
          doi = {10.1086/156801},
       adsurl = {https://ui.adsabs.harvard.edu/abs/1979ApJ...227..923H}
}

@ARTICLE{2020A&A...641A..57M,
       author = {{Massalkhi}, S. and {Ag{\'u}ndez}, M. and {Cernicharo}, J. and {Velilla-Prieto}, L.},
        title = "{The abundance of S- and Si-bearing molecules in O-rich circumstellar envelopes of AGB stars}",
      journal = {\aap},
         year = 2020,
        month = sep,
       volume = {641},
          eid = {A57},
        pages = {A57},
          doi = {10.1051/0004-6361/202037900},
archivePrefix = {arXiv},
       eprint = {2007.00572},
 primaryClass = {astro-ph.SR},
       adsurl = {https://ui.adsabs.harvard.edu/abs/2020A&A...641A..57M}
}

@ARTICLE{2024A&A...686A.251W,
       author = {{Winnberg}, A. and {Brand}, J. and {Engels}, D.},
        title = "{Water vapour masers in long-period variable stars. III. Mira variables U Her and RR Aql}",
      journal = {\aap},
         year = 2024,
        month = jun,
       volume = {686},
          eid = {A251},
        pages = {A251},
          doi = {10.1051/0004-6361/202348567},
archivePrefix = {arXiv},
       eprint = {2403.00535},
 primaryClass = {astro-ph.GA},
       adsurl = {https://ui.adsabs.harvard.edu/abs/2024A&A...686A.251W}
}

@ARTICLE{2019ARA&A..57..417C,
       author = {{Cordes}, James M. and {Chatterjee}, Shami},
        title = "{Fast Radio Bursts: An Extragalactic Enigma}",
      journal = {\araa},
         year = 2019,
        month = aug,
       volume = {57},
        pages = {417-465},
          doi = {10.1146/annurev-astro-091918-104501},
archivePrefix = {arXiv},
       eprint = {1906.05878},
 primaryClass = {astro-ph.HE},
       adsurl = {https://ui.adsabs.harvard.edu/abs/2019ARA&A..57..417C}
}

@ARTICLE{2014A&A...569A..76D,
       author = {{Danilovich}, T. and {Bergman}, P. and {Justtanont}, K. and {Lombaert}, R. and {Maercker}, M. and {Olofsson}, H. and {Ramstedt}, S. and {Royer}, P.},
        title = "{Detailed modelling of the circumstellar molecular line emission of the S-type AGB star W Aquilae}",
      journal = {\aap},
         year = 2014,
        month = sep,
       volume = {569},
          eid = {A76},
        pages = {A76},
          doi = {10.1051/0004-6361/201322807},
archivePrefix = {arXiv},
       eprint = {1408.1825},
 primaryClass = {astro-ph.SR},
       adsurl = {https://ui.adsabs.harvard.edu/abs/2014A&A...569A..76D}
}

@ARTICLE{2018A&A...615A...8D,
       author = {{De Beck}, E. and {Olofsson}, H.},
        title = "{Circumstellar environment of the M-type AGB star R Doradus. APEX spectral scan at 159.0-368.5 GHz}",
      journal = {\aap},
         year = 2018,
        month = jul,
       volume = {615},
          eid = {A8},
        pages = {A8},
          doi = {10.1051/0004-6361/201732470},
archivePrefix = {arXiv},
       eprint = {1801.07984},
 primaryClass = {astro-ph.SR},
       adsurl = {https://ui.adsabs.harvard.edu/abs/2018A&A...615A...8D}
}

@ARTICLE{2025A&A...704A..18O,
       author = {{Ohnaka}, K. and {Wong}, K.~T. and {Weigelt}, G. and {Hofmann}, K.-H.},
        title = "{High-angular-resolution ALMA imaging of the inhomogeneous dynamical atmosphere of the asymptotic giant branch star W Hya: SiO, H$_{2}$O, SO$_{2}$, SO, HCN, AlO, AlOH, TiO, TiO$_{2}$, and OH lines}",
      journal = {\aap},
         year = 2025,
        month = dec,
       volume = {704},
          eid = {A18},
        pages = {A18},
          doi = {10.1051/0004-6361/202554900},
archivePrefix = {arXiv},
       eprint = {2512.03156},
 primaryClass = {astro-ph.SR},
       adsurl = {https://ui.adsabs.harvard.edu/abs/2025A&A...704A..18O}
}

@ARTICLE{2025AJ....170..266S,
       author = {{Son}, Seong-Min and {Cho}, Se-Hyung and {Kim}, Jaeheon and {Baek}, Hyeon and {Yoon}, Dong-Hwan and {Suh}, Kyung-Won},
        title = "{Survey of S-type AGB Stars Using the KVN Four-band Receiving System: SMASTES II}",
      journal = {\aj},
         year = 2025,
        month = nov,
       volume = {170},
       number = {5},
          eid = {266},
        pages = {266},
          doi = {10.3847/1538-3881/ae0586},
       adsurl = {https://ui.adsabs.harvard.edu/abs/2025AJ....170..266S}
}

@ARTICLE{2021A&A...650A.118S,
       author = {{Shetye}, Shreeya and {Van Eck}, Sophie and {Jorissen}, Alain and {Goriely}, Stephane and {Siess}, Lionel and {Van Winckel}, Hans and {Plez}, Bertrand and {Godefroid}, Michel and {Wallerstein}, George},
        title = "{S stars and s-process in the Gaia era. II. Constraining the luminosity of the third dredge-up with Tc-rich S stars}",
      journal = {\aap},
         year = 2021,
        month = jun,
       volume = {650},
          eid = {A118},
        pages = {A118},
          doi = {10.1051/0004-6361/202040207},
archivePrefix = {arXiv},
       eprint = {2104.03161},
 primaryClass = {astro-ph.SR},
       adsurl = {https://ui.adsabs.harvard.edu/abs/2021A&A...650A.118S}
}

@ARTICLE{2012MNRAS.427..343M,
       author = {{McDonald}, I. and {Zijlstra}, A.~A. and {Boyer}, M.~L.},
        title = "{Fundamental parameters and infrared excesses of Hipparcos stars}",
      journal = {\mnras},
         year = 2012,
        month = nov,
       volume = {427},
       number = {1},
        pages = {343-357},
          doi = {10.1111/j.1365-2966.2012.21873.x},
archivePrefix = {arXiv},
       eprint = {1208.2037},
 primaryClass = {astro-ph.SR},
       adsurl = {https://ui.adsabs.harvard.edu/abs/2012MNRAS.427..343M}
}

@ARTICLE{2014A&A...570A.113M,
       author = {{Mayer}, A. and {Jorissen}, A. and {Paladini}, C. and {Kerschbaum}, F. and {Pourbaix}, D. and {Siopis}, C. and {Ottensamer}, R. and {Me{\v{c}}ina}, M. and {Cox}, N.~L.~J. and {Groenewegen}, M.~A.~T. and {Klotz}, D. and {Sadowski}, G. and {Spang}, A. and {Cruzal{\`e}bes}, P. and {Waelkens}, C.},
        title = "{Large-scale environments of binary AGB stars probed by Herschel. II. Two companions interacting with the wind of {\ensuremath{\pi}}$^{1}$ Gruis}",
      journal = {\aap},
         year = 2014,
        month = oct,
       volume = {570},
          eid = {A113},
        pages = {A113},
          doi = {10.1051/0004-6361/201424465},
archivePrefix = {arXiv},
       eprint = {1408.3965},
 primaryClass = {astro-ph.SR},
       adsurl = {https://ui.adsabs.harvard.edu/abs/2014A&A...570A.113M}
}

@ARTICLE{2014A&A...566A.145R,
       author = {{Ramstedt}, S. and {Olofsson}, H.},
        title = "{The $^{12}$CO/$^{13}$CO ratio in AGB stars of different chemical type. Connection to the $^{12}$C/$^{13}$C ratio and the evolution along the AGB}",
      journal = {\aap},
         year = 2014,
        month = jun,
       volume = {566},
          eid = {A145},
        pages = {A145},
          doi = {10.1051/0004-6361/201423721},
archivePrefix = {arXiv},
       eprint = {1405.6404},
 primaryClass = {astro-ph.SR},
       adsurl = {https://ui.adsabs.harvard.edu/abs/2014A&A...566A.145R}
}

@ARTICLE{1986ApJ...311..335W,
       author = {{Wannier}, P.~G. and {Sahai}, R.},
        title = "{Mass Loss from Giant and Supergiant Stars}",
      journal = {\apj},
         year = 1986,
        month = dec,
       volume = {311},
        pages = {335},
          doi = {10.1086/164775},
       adsurl = {https://ui.adsabs.harvard.edu/abs/1986ApJ...311..335W}
}

@ARTICLE{2017A&A...606A.124D,
       author = {{Danilovich}, T. and {Van de Sande}, M. and {De Beck}, E. and {Decin}, L. and {Olofsson}, H. and {Ramstedt}, S. and {Millar}, T.~J.},
        title = "{Sulphur-bearing molecules in AGB stars. I. The occurrence of hydrogen sulphide}",
      journal = {\aap},
         year = 2017,
        month = oct,
       volume = {606},
          eid = {A124},
        pages = {A124},
          doi = {10.1051/0004-6361/201731203},
archivePrefix = {arXiv},
       eprint = {1707.06003},
 primaryClass = {astro-ph.SR},
       adsurl = {https://ui.adsabs.harvard.edu/abs/2017A&A...606A.124D}
}

@ARTICLE{2024A&A...685A..11K,
       author = {{Khouri}, T. and {Olofsson}, H. and {Vlemmings}, W.~H.~T. and {Schirmer}, T. and {Tafoya}, D. and {Maercker}, M. and {De Beck}, E. and {Nyman}, L.-{\r{A}}. and {Saberi}, M.},
        title = "{An empirical view of the extended atmosphere and inner envelope of the asymptotic giant branch star R Doradus. I. Physical model based on CO lines}",
      journal = {\aap},
         year = 2024,
        month = may,
       volume = {685},
          eid = {A11},
        pages = {A11},
          doi = {10.1051/0004-6361/202348382},
archivePrefix = {arXiv},
       eprint = {2402.13676},
 primaryClass = {astro-ph.SR},
       adsurl = {https://ui.adsabs.harvard.edu/abs/2024A&A...685A..11K}
}

@ARTICLE{2017NatAs...1..848V,
       author = {{Vlemmings}, Wouter and {Khouri}, Theo and {O'Gorman}, Eamon and {De Beck}, Elvire and {Humphreys}, Elizabeth and {Lankhaar}, Boy and {Maercker}, Matthias and {Olofsson}, Hans and {Ramstedt}, Sofia and {Tafoya}, Daniel and {Takigawa}, Aki},
        title = "{The shock-heated atmosphere of an asymptotic giant branch star resolved by ALMA}",
      journal = {Nature Astronomy},
         year = 2017,
        month = oct,
       volume = {1},
        pages = {848-853},
          doi = {10.1038/s41550-017-0288-9},
archivePrefix = {arXiv},
       eprint = {1711.01153},
 primaryClass = {astro-ph.SR},
       adsurl = {https://ui.adsabs.harvard.edu/abs/2017NatAs...1..848V}
}

@ARTICLE{2019A&A...626A..81V,
       author = {{Vlemmings}, W.~H.~T. and {Khouri}, T. and {Olofsson}, H.},
        title = "{Resolving the extended stellar atmospheres of asymptotic giant branch stars at (sub)millimetre wavelengths}",
      journal = {\aap},
         year = 2019,
        month = jun,
       volume = {626},
          eid = {A81},
        pages = {A81},
          doi = {10.1051/0004-6361/201935329},
archivePrefix = {arXiv},
       eprint = {1904.06374},
 primaryClass = {astro-ph.SR},
       adsurl = {https://ui.adsabs.harvard.edu/abs/2019A&A...626A..81V}
}

@ARTICLE{1999ApJ...517..209G,
       author = {{Goldsmith}, Paul F. and {Langer}, William D.},
        title = "{Population Diagram Analysis of Molecular Line Emission}",
      journal = {\apj},
         year = 1999,
        month = may,
       volume = {517},
       number = {1},
        pages = {209-225},
          doi = {10.1086/307195},
       adsurl = {https://ui.adsabs.harvard.edu/abs/1999ApJ...517..209G}
}

@ARTICLE{2021A&A...653A.141A,
       author = {{Asplund}, M. and {Amarsi}, A.~M. and {Grevesse}, N.},
        title = "{The chemical make-up of the Sun: A 2020 vision}",
      journal = {\aap},
         year = 2021,
        month = sep,
       volume = {653},
          eid = {A141},
        pages = {A141},
          doi = {10.1051/0004-6361/202140445},
archivePrefix = {arXiv},
       eprint = {2105.01661},
 primaryClass = {astro-ph.SR},
       adsurl = {https://ui.adsabs.harvard.edu/abs/2021A&A...653A.141A}
}

@ARTICLE{2019MNRAS.484..494D,
       author = {{Danilovich}, T. and {Richards}, A.~M.~S. and {Karakas}, A.~I. and {Van de Sande}, M. and {Decin}, L. and {De Ceuster}, F.},
        title = "{An ALMA view of CS and SiS around oxygen-rich AGB stars}",
      journal = {\mnras},
         year = 2019,
        month = mar,
       volume = {484},
       number = {1},
        pages = {494-509},
          doi = {10.1093/mnras/stz002},
archivePrefix = {arXiv},
       eprint = {1901.00070},
 primaryClass = {astro-ph.SR},
       adsurl = {https://ui.adsabs.harvard.edu/abs/2019MNRAS.484..494D}
}

@ARTICLE{2015ApJ...808...36M,
       author = {{Matthews}, L.~D. and {Reid}, M.~J. and {Menten}, K.~M.},
        title = "{New Measurements of the Radio Photosphere of Mira Based on Data from the JVLA and ALMA}",
      journal = {\apj},
         year = 2015,
        month = jul,
       volume = {808},
       number = {1},
          eid = {36},
        pages = {36},
          doi = {10.1088/0004-637X/808/1/36},
archivePrefix = {arXiv},
       eprint = {1506.03075},
 primaryClass = {astro-ph.SR},
       adsurl = {https://ui.adsabs.harvard.edu/abs/2015ApJ...808...36M}
}

@ARTICLE{1993AJ....105..595S,
       author = {{Sahai}, Raghvendra and {Bieging}, John H.},
        title = "{Interferometric Observations of Non-Maser SiO Emission From Circumstellar Envelopes of AGB Stars: Acceleration Regions and SIO Depletion}",
      journal = {\aj},
         year = 1993,
        month = feb,
       volume = {105},
        pages = {595},
          doi = {10.1086/116456},
       adsurl = {https://ui.adsabs.harvard.edu/abs/1993AJ....105..595S}
}

@ARTICLE{1998A&A...329..219P,
       author = {{Pardo}, J.~R. and {Cernicharo}, J. and {Gonzalez-Alfonso}, E. and {Bujarrabal}, V.},
        title = "{Simultaneous observations of maser lines of $^{28}$SiO in evolved stars}",
      journal = {\aap},
         year = 1998,
        month = jan,
       volume = {329},
        pages = {219-228},
       adsurl = {https://ui.adsabs.harvard.edu/abs/1998A&A...329..219P}
}

%%%%%%%%%%%%%%%%%%%%%%%%%%%%%%%%%%%%%%%%%%%%%%%%%%

%%%%%%%%%%%%%%%%% APPENDICES %%%%%%%%%%%%%%%%%%%%%

%\newpage
\appendix

\section{2D Gaussian component fitting}
\label{appendix:fitting}
In the current work (Section \ref{sec:dist:overview}), 2D Gaussian components were fitted to clumps of emission in the interferometric images, and the maser component maps of the SiO transitions listed in Table \ref{table:line-coverage} above were plotted. The extended-configuration data set of $\varv=0$ lines was neglected from the component fitting unless the line intensity ratios between maser and thermal emissions are sufficient, i.e., close to an order of magnitude difference, for the fitting to be systematically and accurately executed. The same technique was employed on the mid-configuration data in order to provide complementary information to the fitted component maps from the extended-configuration images.

We started by making cut-outs of the original data cubes, also called subimages in this work, centred on the SiO lines of interest. The cutouts were centred on the stellar positions, whose accuracy (a combination of astrometric uncertainty and noise errors) is 3 mas for all stars except SV Aqr, which is fainter, giving 5 mas. While the extended-configuration data were centred in the maps during imaging, the central position of the mid-configuration subimages needed to be shifted at this stage by determining whether the continuum peak is at a lower or higher pixel value than the pre-defined centre to draw the boundary for the cut-out.

The SiO subimage cubes were loaded into \textsc{aips} in order to fit 2D Gaussian models to the images by least-squares with {\small{\MakeUppercase{SAD}}} (Search And Destroy). The task sets an area of search based on a user-defined box, where it looks for pixels above the first flux cut-off level imposed by the user and merges these into contiguous islands. It then fits (multiple) 2D Gaussians in each island and repeats the process for descending flux cut-off levels. The noise cut-off was initially defined as 5$\sigma$ returned from the {\small{\MakeUppercase{imstat}}} task or $-$1.25 times the worst negative, whichever is greater. However, in some sources, this cut-off level proved to be too low as many (up to a few tens) spurious components were fitted by {\small{\MakeUppercase{SAD}}}. We chose 10$\sigma$ as the new baseline cut-off and modified it accordingly to the resulting component maps to focus on the brightest components detected in the image cubes. The fitted component size had the restoring beam deconvolved.

The position uncertainty in fitting a 2D Gaussian component to interferometry images with average uv coverage is (the size of the synthesised beam)/(S/N), where the synthesised beam is defined as the response of the interferometer to a point source whose size is determined by the maximum baseline. We also checked that the errors reported by {\small{\MakeUppercase{SAD}}} were consistent with this or greater (probably due to blending). We forced $\geq$ 1 mas for position error and $\geq$ 2 mas for component size since, at high S/N, deconvolution and other errors are also significant (\citealt{1996ApJS..103...81C}; \citealt{2011A&A...525A..56R}). Note that each component represents a line of sight through the cloud where maser activity is strongest within a specific velocity range, so the arrangement of components reflects the velocity gradient of the cloud. However, it may not fully depict the cloud’s shape, meaning that a linear or curved series of components can arise even from a spherical cloud, especially if the masers are tightly beamed. The fitted components were sorted into groups (features) consisting of series in adjacent velocity channels which overlapped by the greater of the position errors or the minimum likely size of a maser feature (10 mas). Any leftover isolated components were manually rejected.

\section{Additional S\MakeLowercase{i}O channel maps}
\label{appendix:additional-maps}
To complement the examples in Section \ref{sec:SiO-id:overview-maps}, this appendix presents channel maps from the extended-configuration data for the \ce{^{28}SiO} $\varv=1$, $J=6-5$ line towards $\pi^1$ Gru and the \ce{^{28}SiO} $\varv=3$, $J=6-5$ line towards U Her. Each figure displays a sequence of velocity-channel images at high angular resolution (beam sizes typically $\sim$20--30 mas), covering emission or absorption features across the relevant frequency range. The maps are centred on the stellar continuum peak, and angular offsets are shown in mas relative to this position. More channel maps are available in the online supplementary material.

\begin{figure*} 
   \includegraphics[width=0.87\textwidth, height=0.5\textwidth]{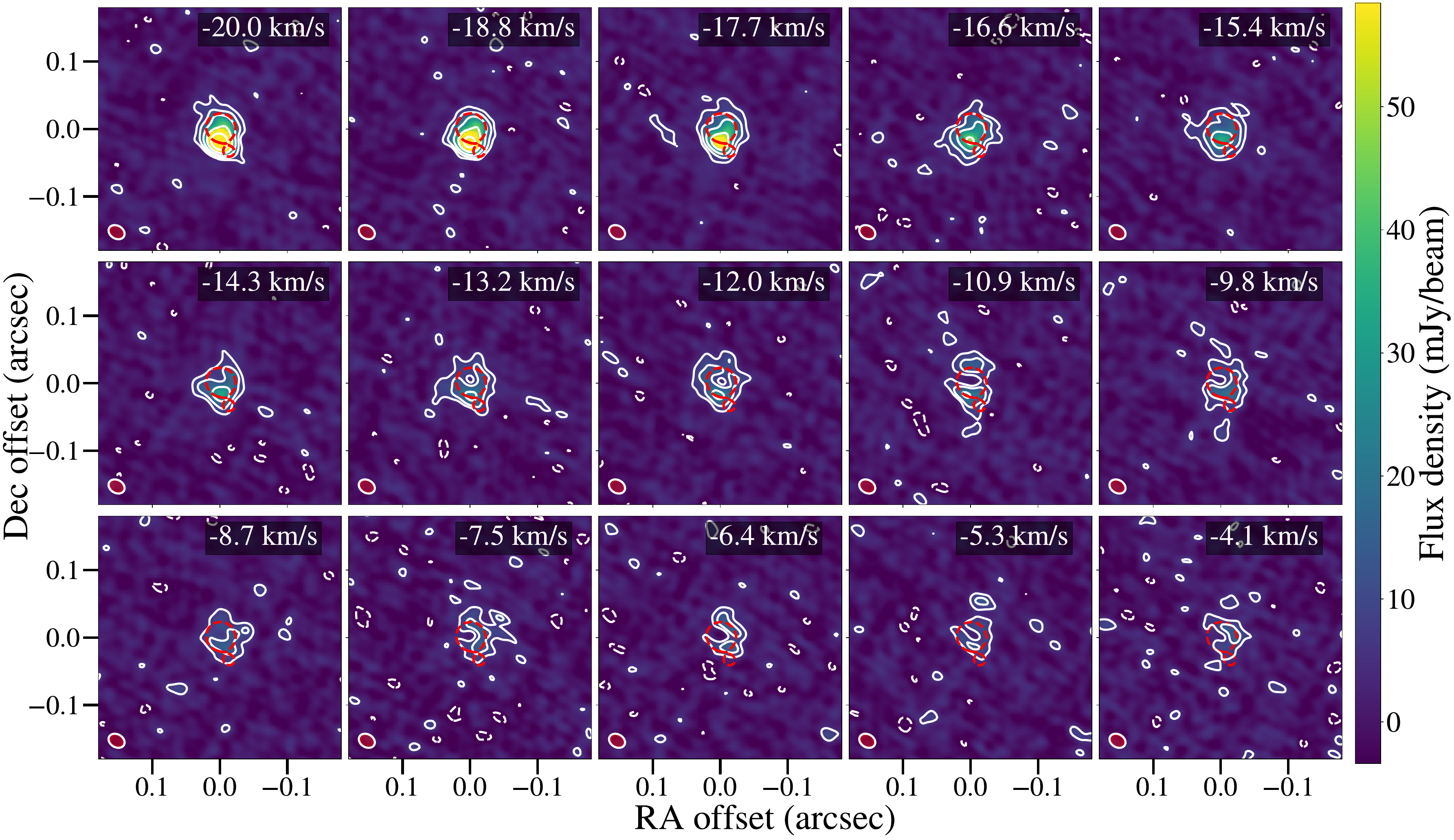} 
   \includegraphics[width=0.87\textwidth, height=0.5\textwidth]{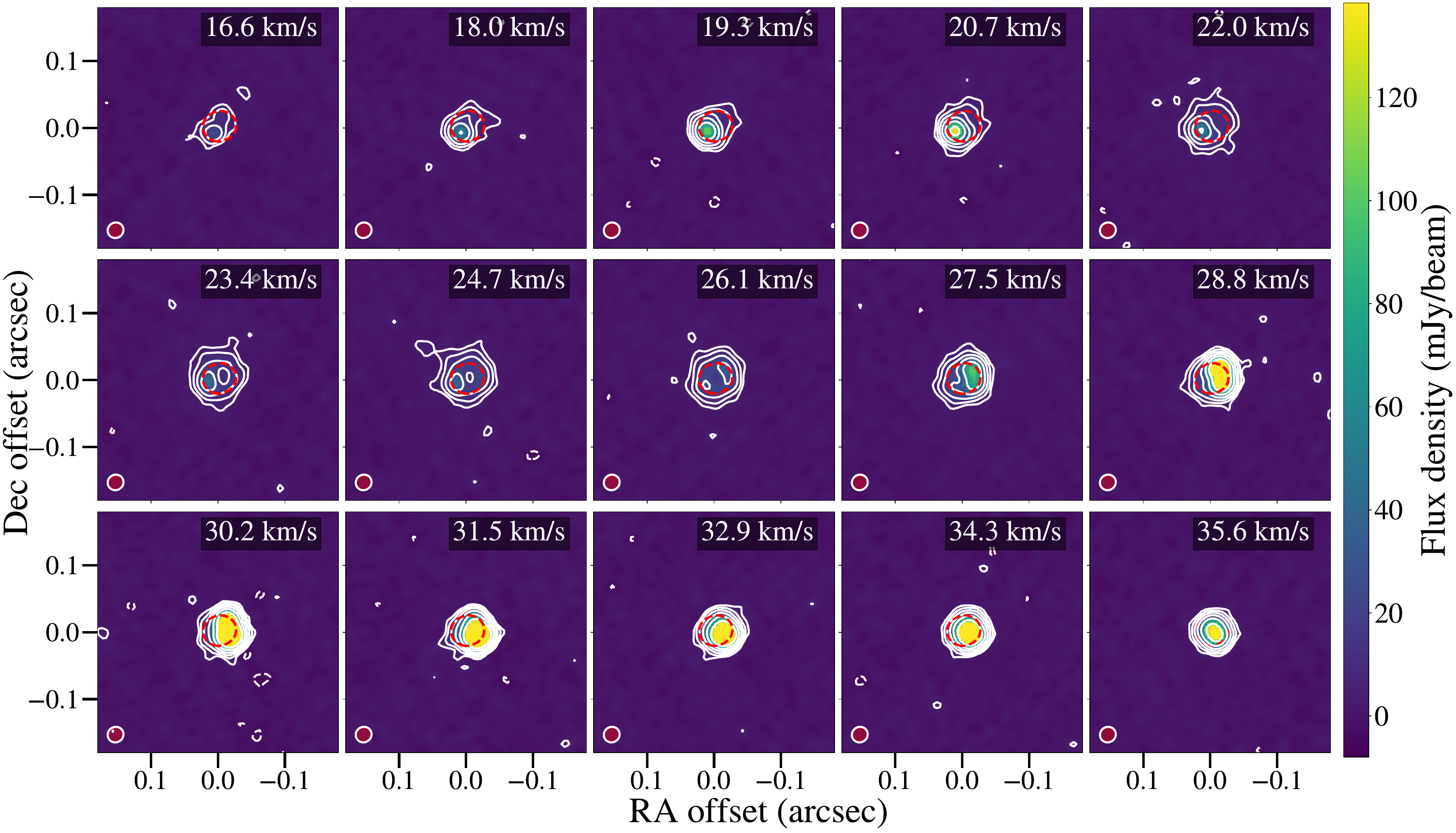} 
   \caption{High-resolution channel maps of the \ce{^{28}SiO} $\varv=1$, $J=6-5$ line at 258.707 GHz observed towards $\pi^1$ Gru (top) and the \ce{^{28}SiO} $\varv=1$, $J=5-4$ line at 215.596 GHz (bottom) towards T Mic. The spatial scales in the map are shown as RA and Dec offsets in arcsec with respect to the continuum position throughout this work. The corresponding angular offsets cover an area of 0.36 $\times$ 0.36 arcsec, centred on the continuum emission peak at the origin. The channel velocities are presented in the LSR frame based on $V_*$ from Table \ref{table:atomium-sources}. The scale for the line flux density per beam (in mJy beam$^{-1}$) is linear and is indicated by the vertical bar on the right of the figure. The white contours represent the $-$3$\sigma$, 3$\sigma$, 6$\sigma$, 12$\sigma$, 24$\sigma$, 48$\sigma$ and 96$\sigma$ levels. The line peak and typical rms noise are 278 and 2 mJy beam$^{-1}$ ($\pi^1$ Gru) and 658 and 1 mJy beam$^{-1}$ (T Mic). Negative contours, when present, are shown with dashed lines. The red dashed contour at the (0,0) position outlines the region where the continuum emission reaches half its peak intensity. The map also features an illustration of the clean line and continuum beams, depicted in white and solid red, located at the bottom left corner. For the line observations, these are (25 $\times$ 19) mas at PA 60$^\circ$ for $\pi^1$ Gru and (24 $\times$ 23) mas at PA $-$58$^\circ$ for T Mic, respectively. Similarly, the continuum parameters are (19 $\times$ 19) mas at PA 60$^\circ$ ($\pi^1$ Gru) and (24 $\times$ 21) mas at PA $-$73$^\circ$  (T Mic).} 
   \label{fig:add-maps-1}
\end{figure*}

\section{Absorption -- additional spectra and zeroth moment maps}
\label{appendix:absorption}
\begin{figure*}
   \begin{center}
   \includegraphics[width=0.35\textwidth, height=0.27\textwidth]{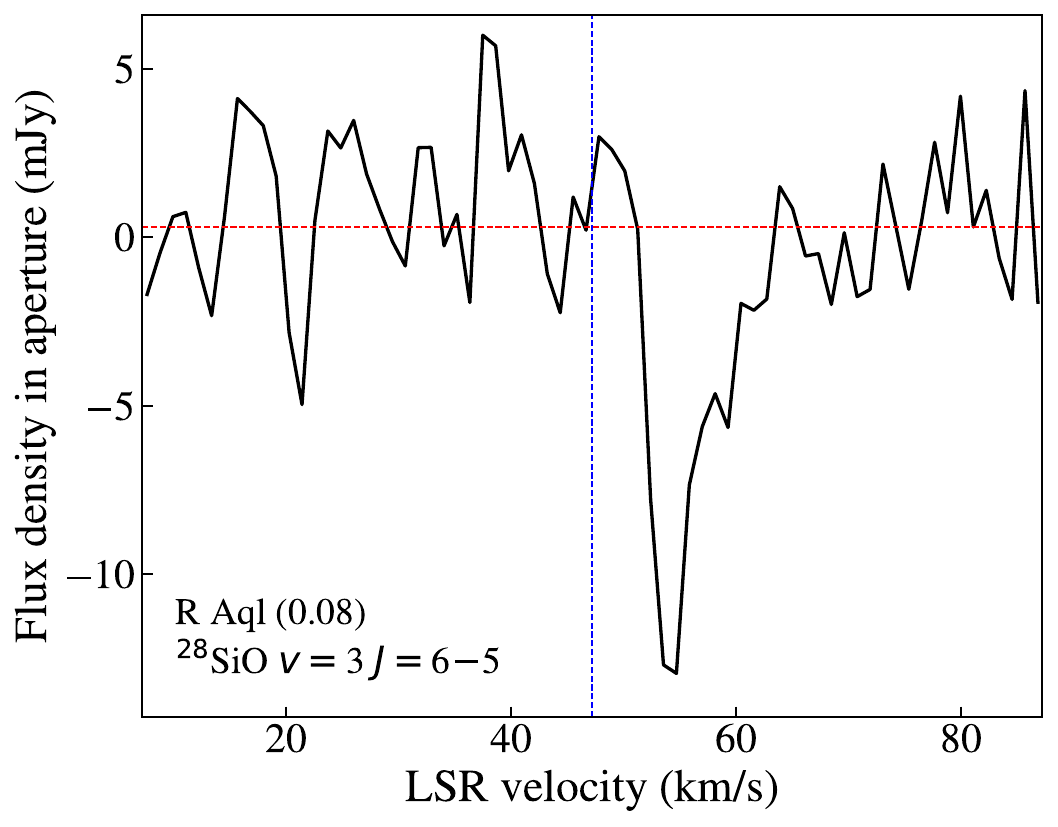}
   \includegraphics[width=0.35\textwidth, height=0.27\textwidth]{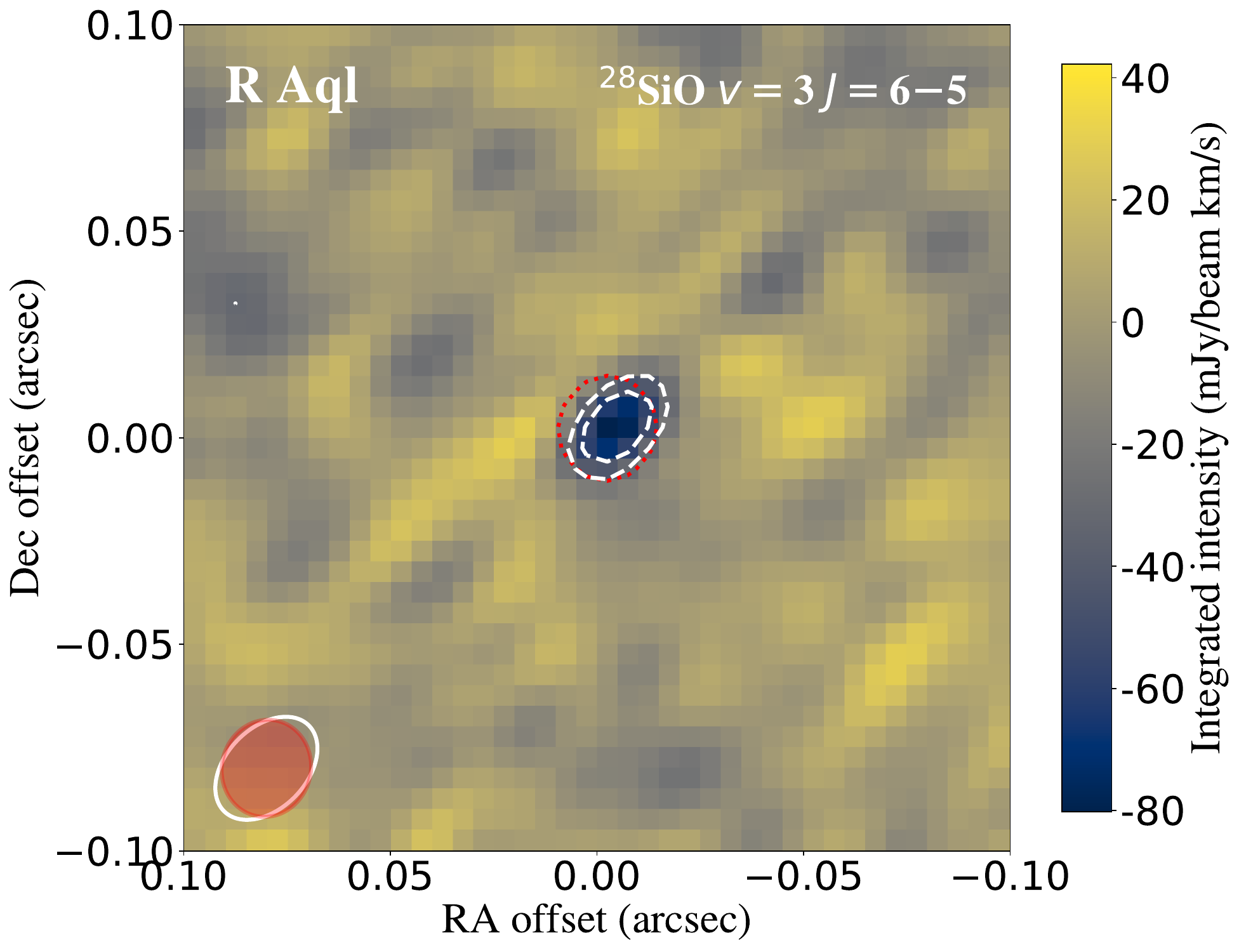} 
   \includegraphics[width=0.35\textwidth, height=0.27\textwidth]{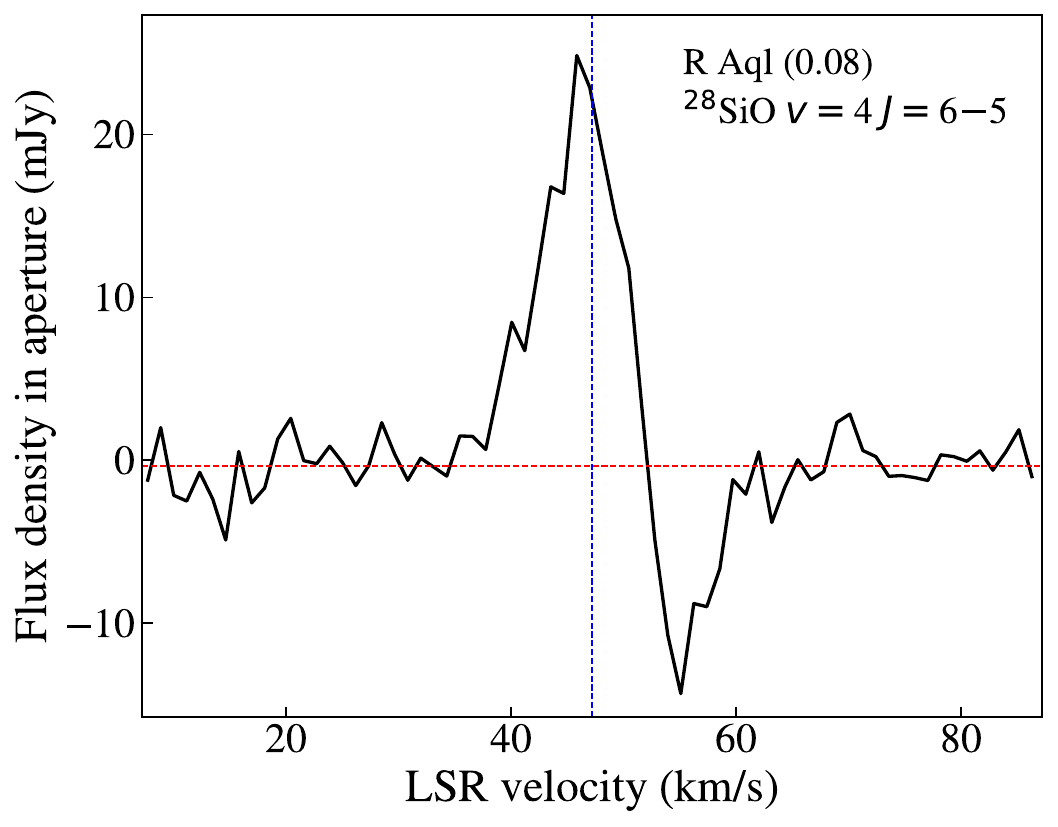}
   \includegraphics[width=0.35\textwidth, height=0.27\textwidth]{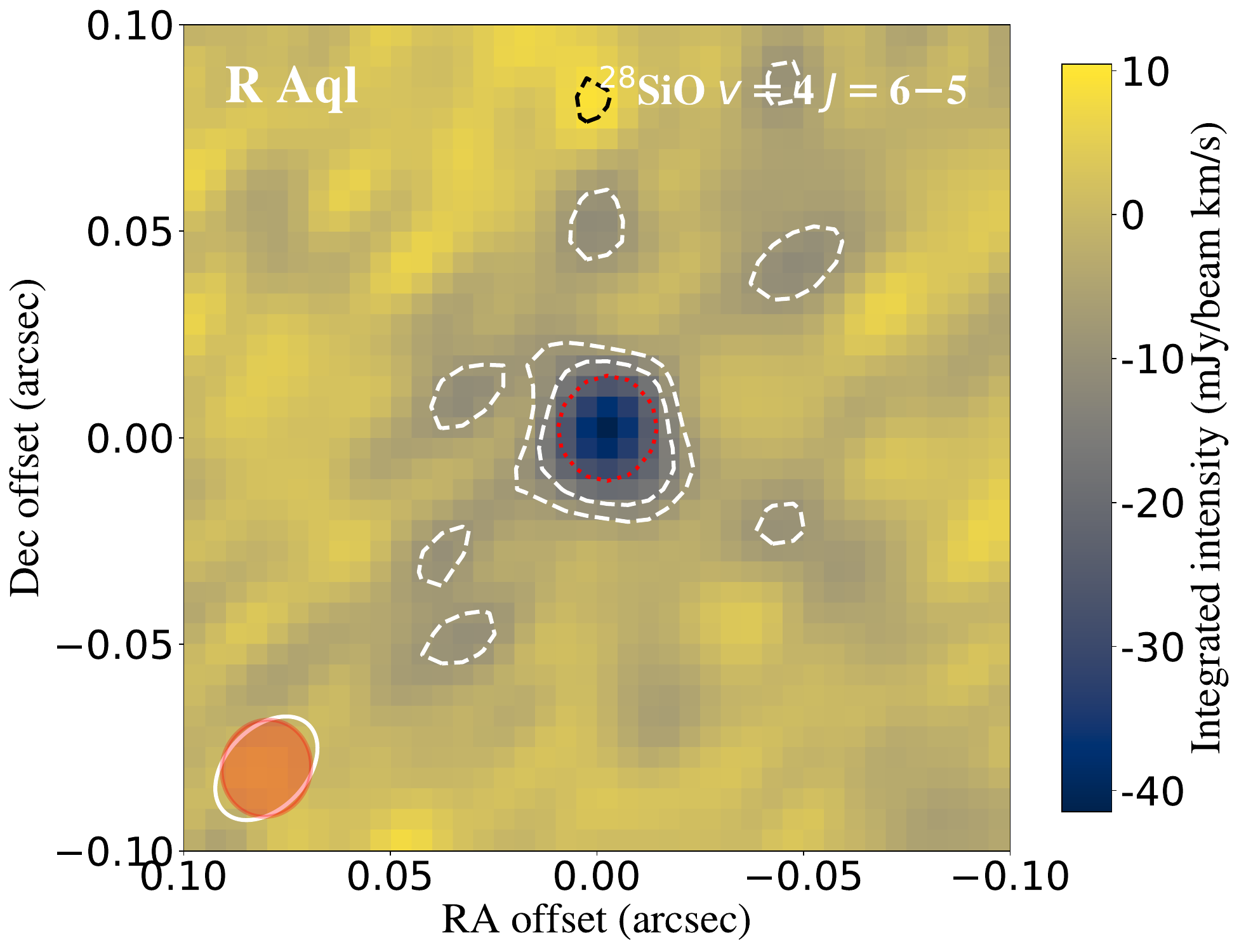}
   \includegraphics[width=0.35\textwidth, height=0.27\textwidth]{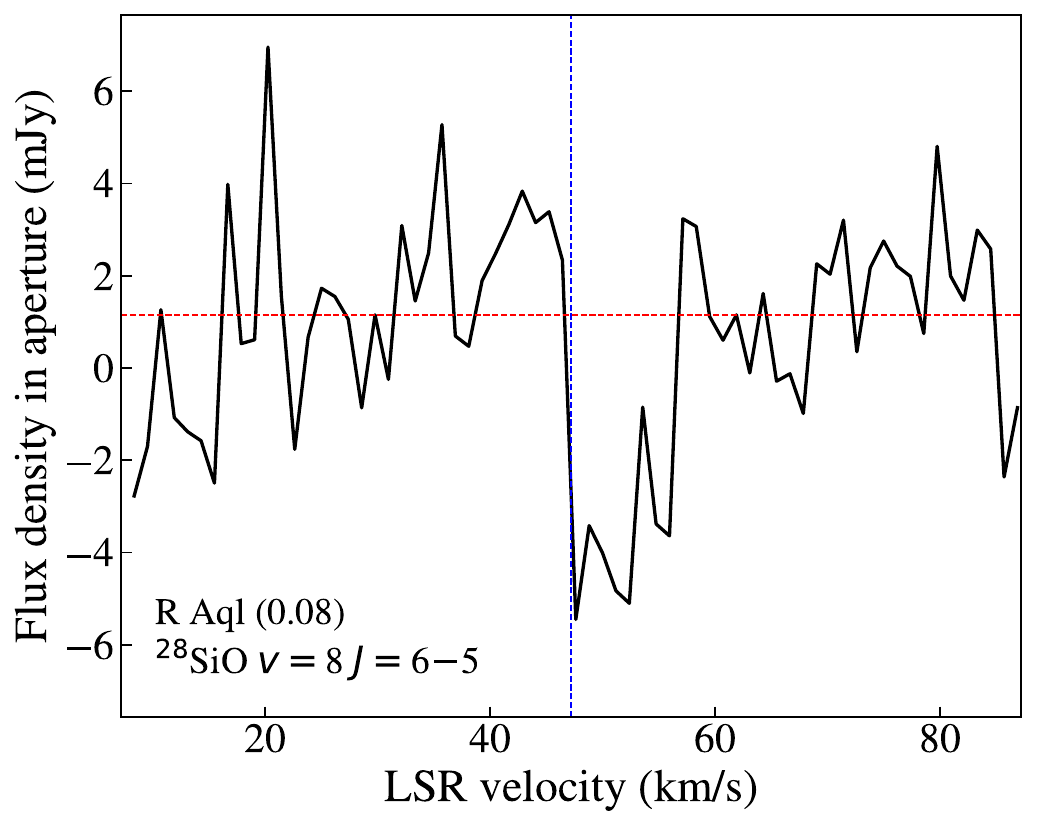}
   \caption{Absorption spectra and zeroth moment maps of the \ce{^{28}SiO} $\varv=3$, $J=6-5$ line at 255.091 GHz and the \ce{^{28}SiO} $\varv=4$, $J=6-5$ line at 253.286 GHz, as well as the absorption profile of the \ce{^{28}SiO} $\varv=8$, $J=6-5$ line at 246.078 GHz towards R Aql. The map field of view is 200$\times$200 mas. The white dashed lines show the $-$3\rm{$\sigma$} and $-$5\rm{$\sigma$} absorption contours. The red dotted line illustrates the half-intensity extent of the continuum emission. The white and red ellipses in the bottom left corner are the reference sizes of the HPBW beam and the continuum beam, respectively.}
   \label{fig:add-absorption-RAql}
   \end{center}
\end{figure*}

\begin{figure*}
   \begin{center}
   \includegraphics[width=0.35\textwidth, height=0.27\textwidth]{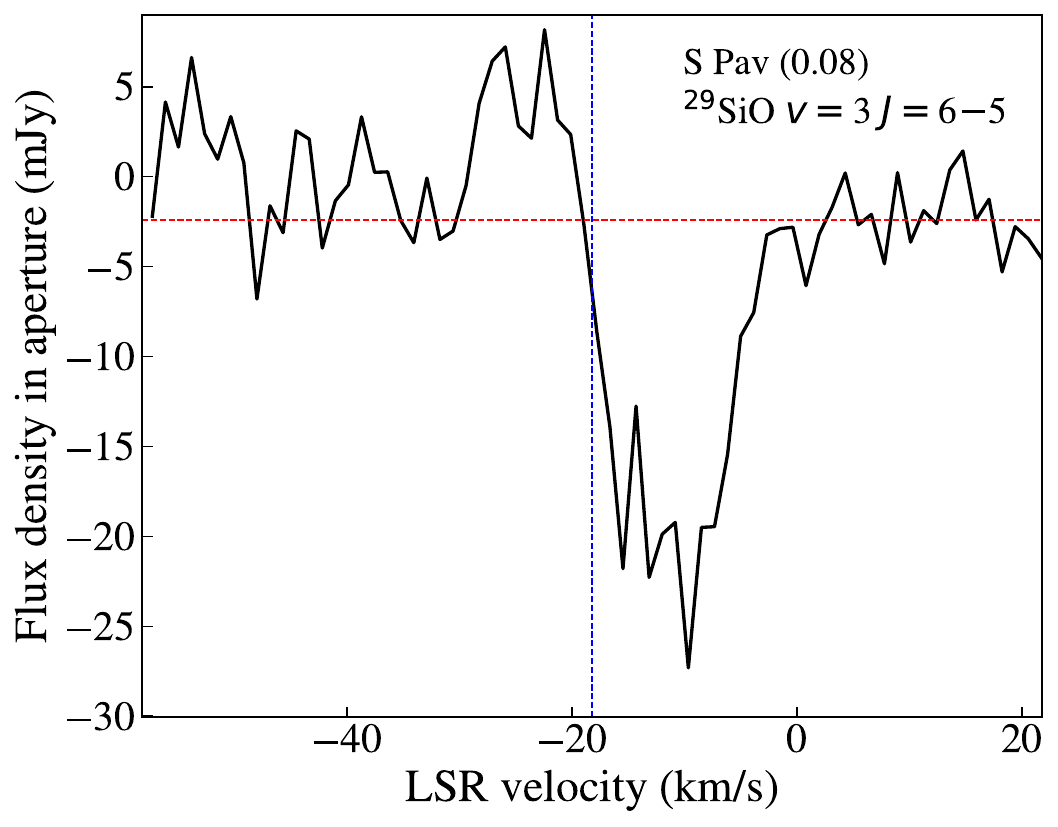}
   \includegraphics[width=0.35\textwidth, height=0.27\textwidth]{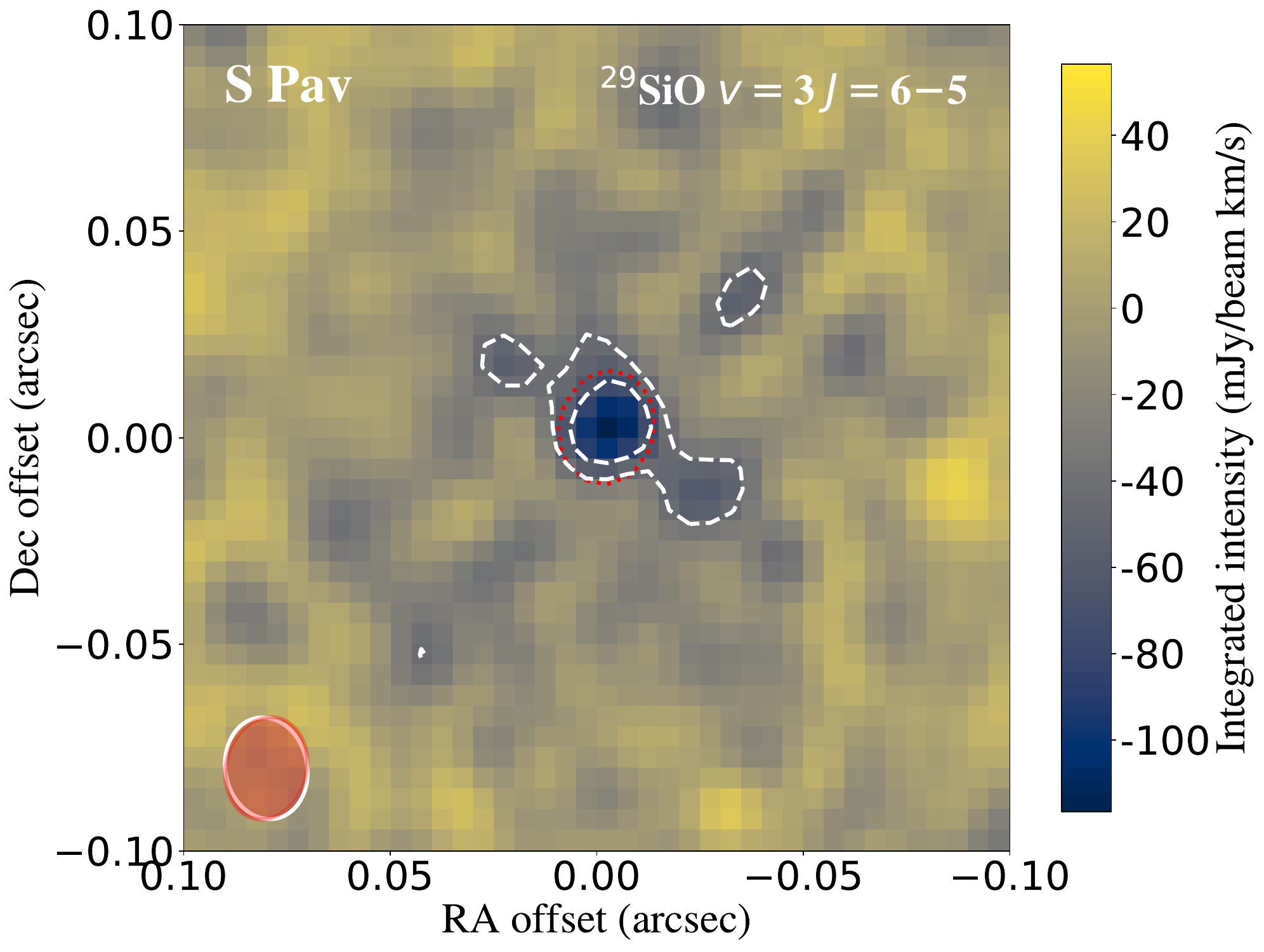}
   \includegraphics[width=0.35\textwidth, height=0.27\textwidth]{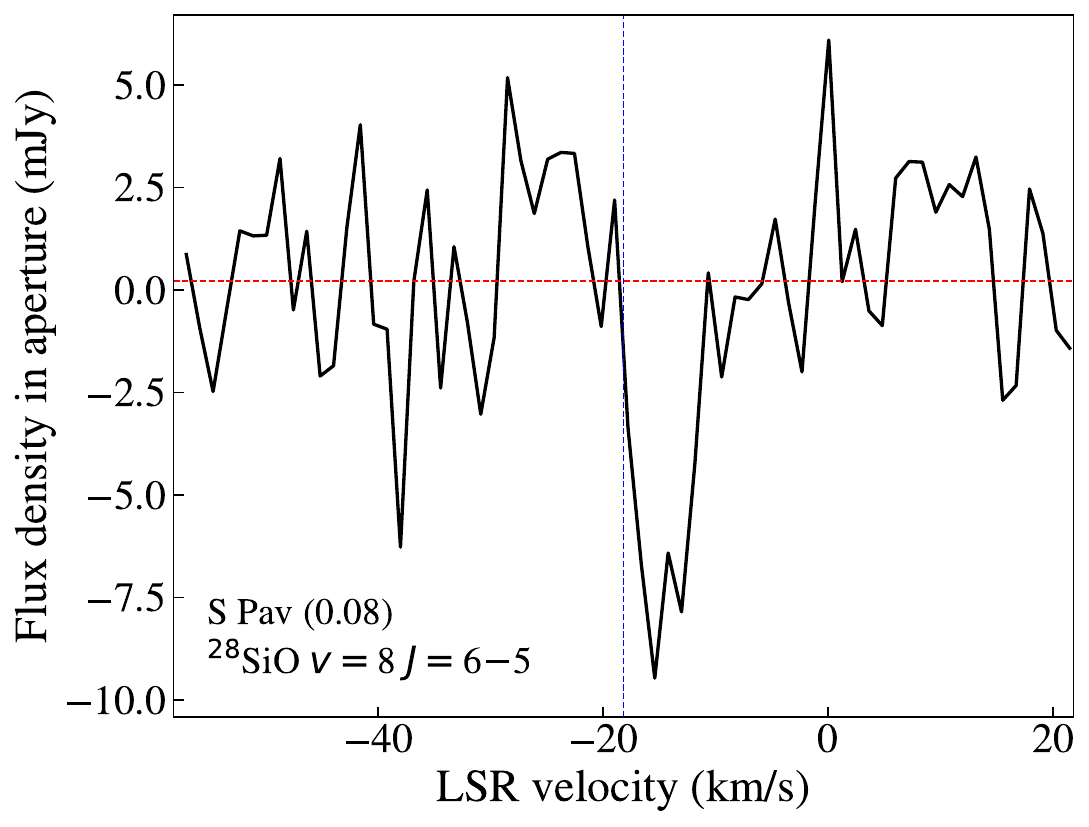}
   \caption{Absorption spectra of the \ce{^{29}SiO} $\varv=3$, $J=6-5$ line at 251.930 GHz and the \ce{^{28}SiO} $\varv=8$, $J=6-5$ line at 246.078 GHz as well as the zeroth moment map of the former towards S Pav. The rest of the caption follows that of Fig. \ref{fig:add-absorption-RAql}.}
   \label{fig:add-absorption-SPav}
   \end{center}
\end{figure*}

\begin{figure*}
   \begin{center}
   \includegraphics[width=0.35\textwidth, height=0.27\textwidth]{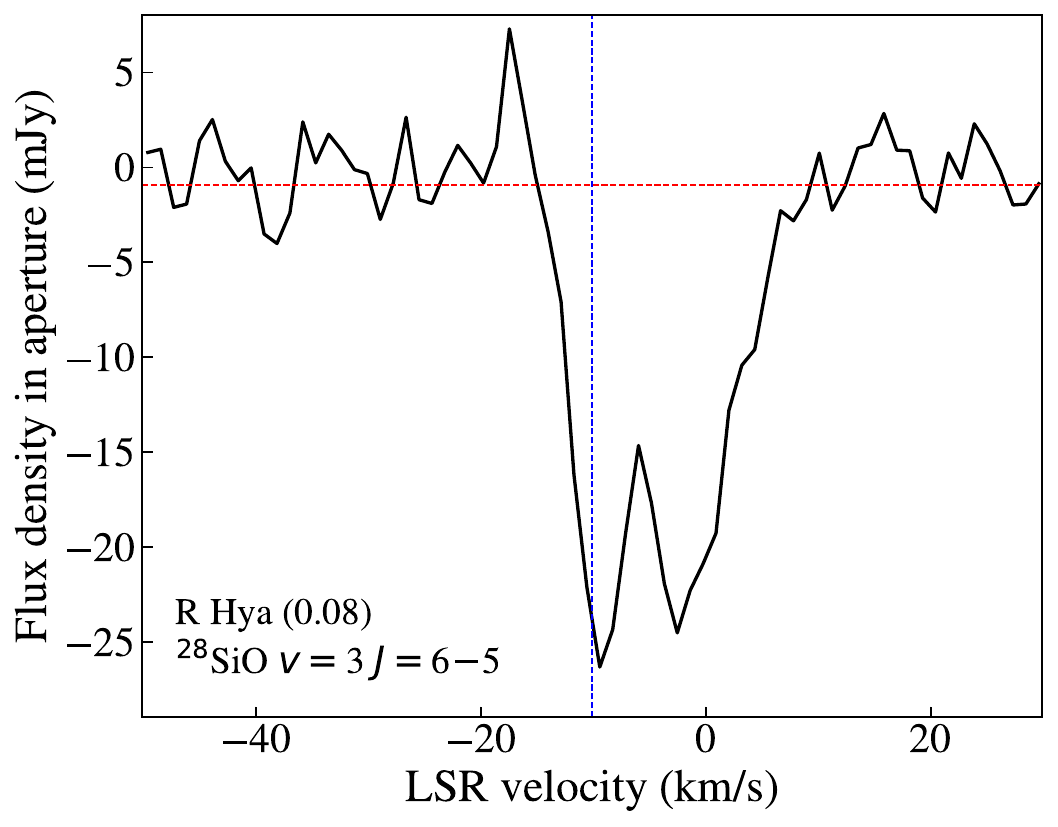}
   \includegraphics[width=0.35\textwidth, height=0.27\textwidth]{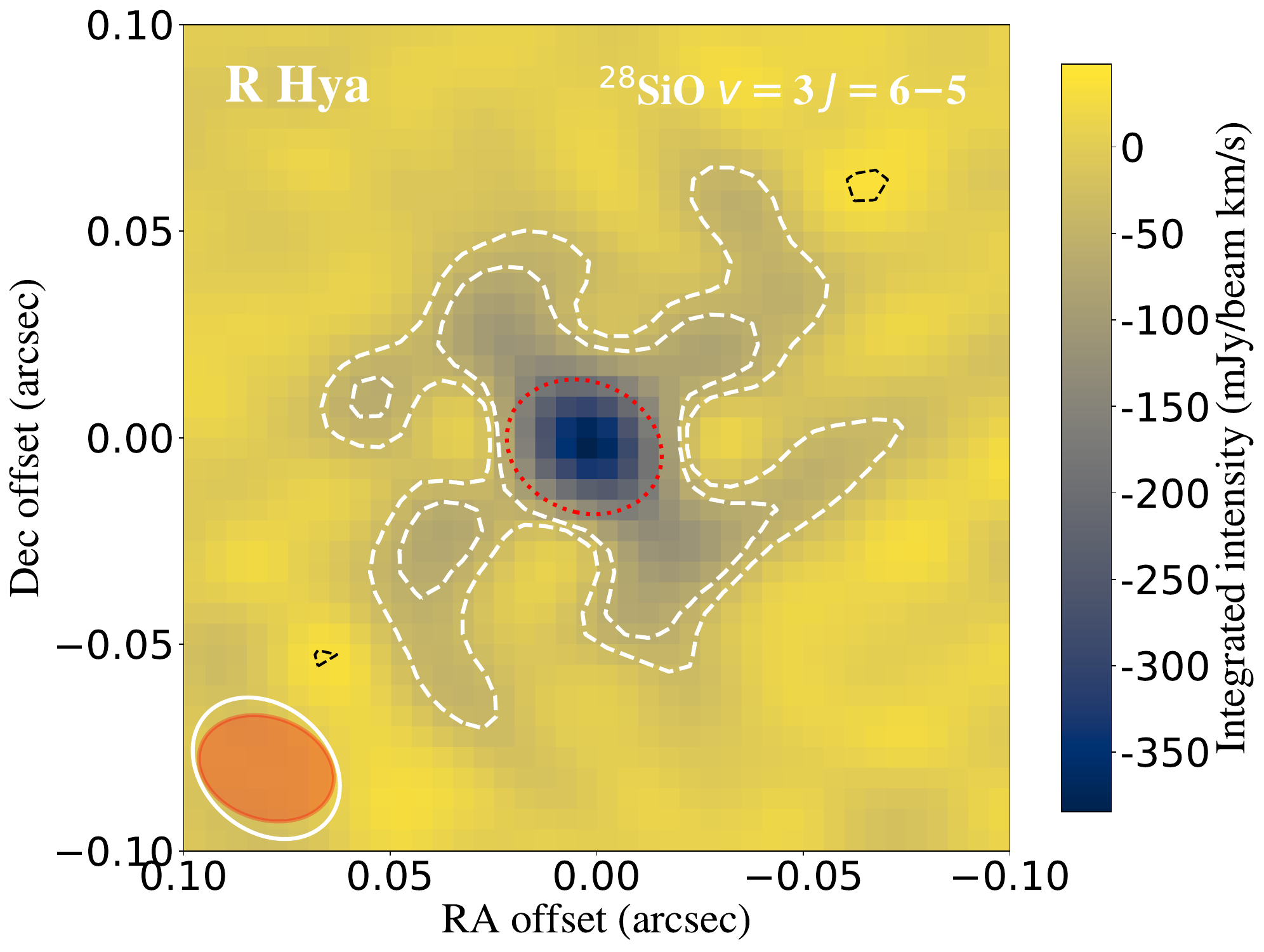} 
   \includegraphics[width=0.35\textwidth, height=0.27\textwidth]{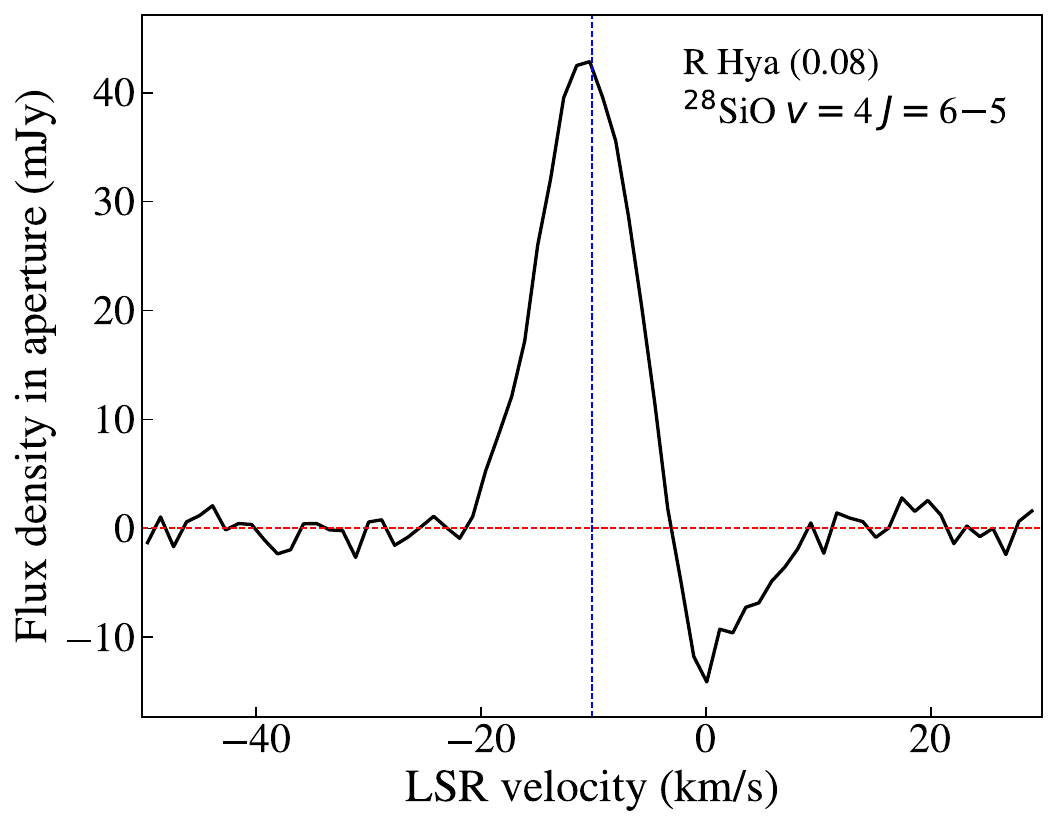}
   \includegraphics[width=0.35\textwidth, height=0.27\textwidth]{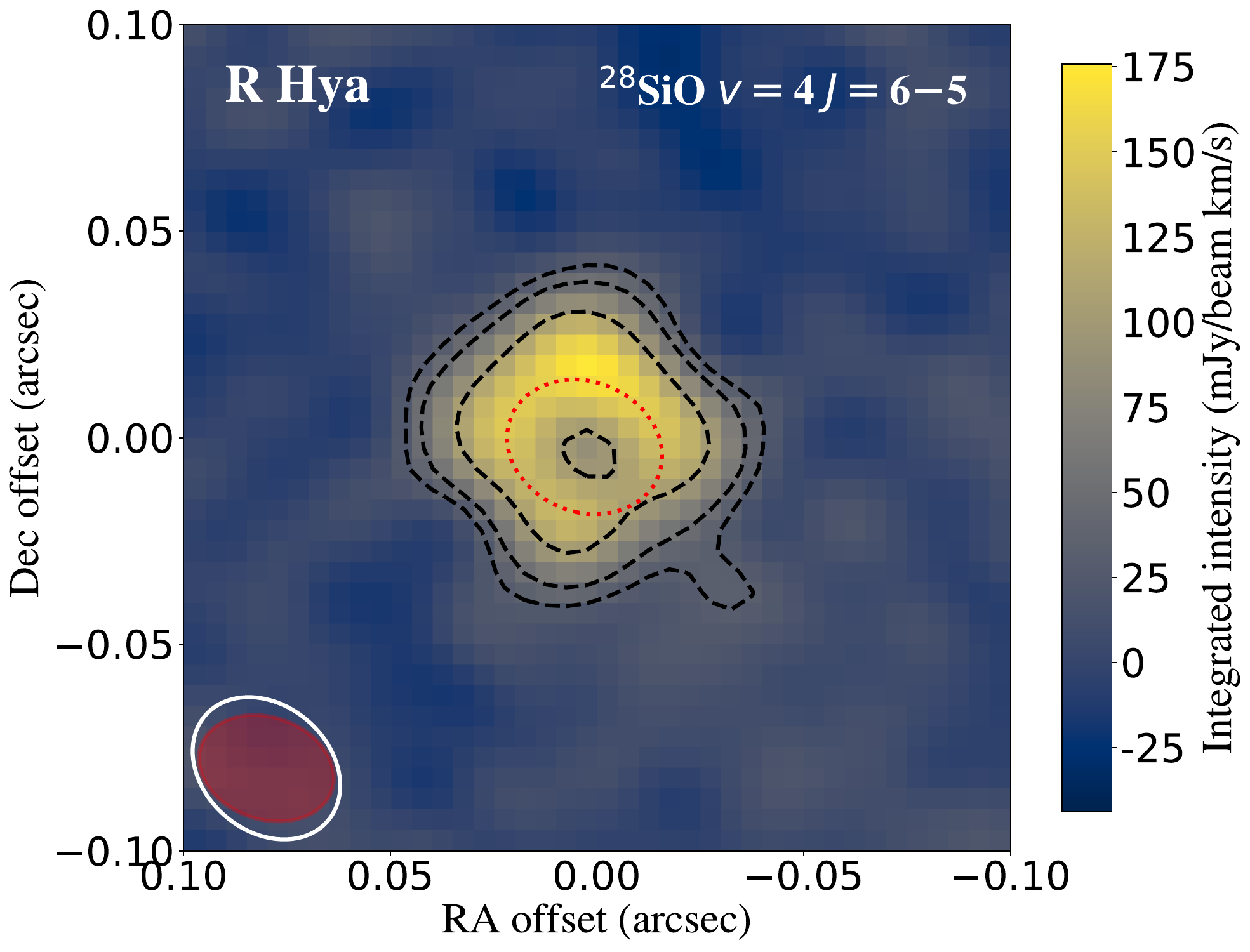}
   \caption{Absorption spectra and zeroth moment maps of the \ce{^{28}SiO} $\varv=3$, $J=6-5$ line at 255.091 GHz and the \ce{^{28}SiO} $\varv=4$, $J=6-5$ line at 253.286 GHz towards R Hya. The rest of the caption follows that of Fig. \ref{fig:add-absorption-RAql}.}
   \label{fig:add-absorption-RHya}
   \end{center}
\end{figure*}

The figures presented in this appendix (Figs. \ref{fig:add-absorption-RAql}--\ref{fig:add-absorption-RHya}) provide more examples of absorption spectra and mom0 maps observed in this study, namely the \ce{^{28}SiO} $\varv=3$, $J=6-5$ (R Aql, R Hya), \ce{^{28}SiO} $\varv=4$, $J=6-5$ (R Aql, R Hya), \ce{^{29}SiO} $\varv=3$, $J=6-5$ (S Pav) and \ce{^{28}SiO} $\varv=8$, $J=6-5$ (S Pav). The SiO line spectra from the extended-configuration data were extracted using a circular aperture of 0.08 arcsec in diameter (value shown in brackets) and were centred on the systemic velocity (blue dashed line) from Table \ref{table:atomium-sources}. Each panel is labelled with the corresponding source and transition. The red horizontal line indicates the fitted baseline used to estimate the rms and signal-to-noise ratio. The mom0 maps provide a field of view of 200$\times$200 mas, with the white dashed contours denoting $-$3$\sigma$ and $-$5$\sigma$ absorption features and the red dotted line outlining the half-intensity extent of the continuum emission. The intensity is integrated over the whole velocity range shown in the corresponding spectrum. The beam sizes for the line and continuum emission are depicted as white and red ellipses, respectively, in the bottom-left corner of each panel. These figures complement the analysis presented in Section \ref{sec:absorption}.

\section{Comparison between S\MakeLowercase{i}O lines}
\label{appendix:seven-spec}
\begin{figure*}
   \begin{center}
    \includegraphics[width=0.392\textwidth, height=0.308\textwidth]{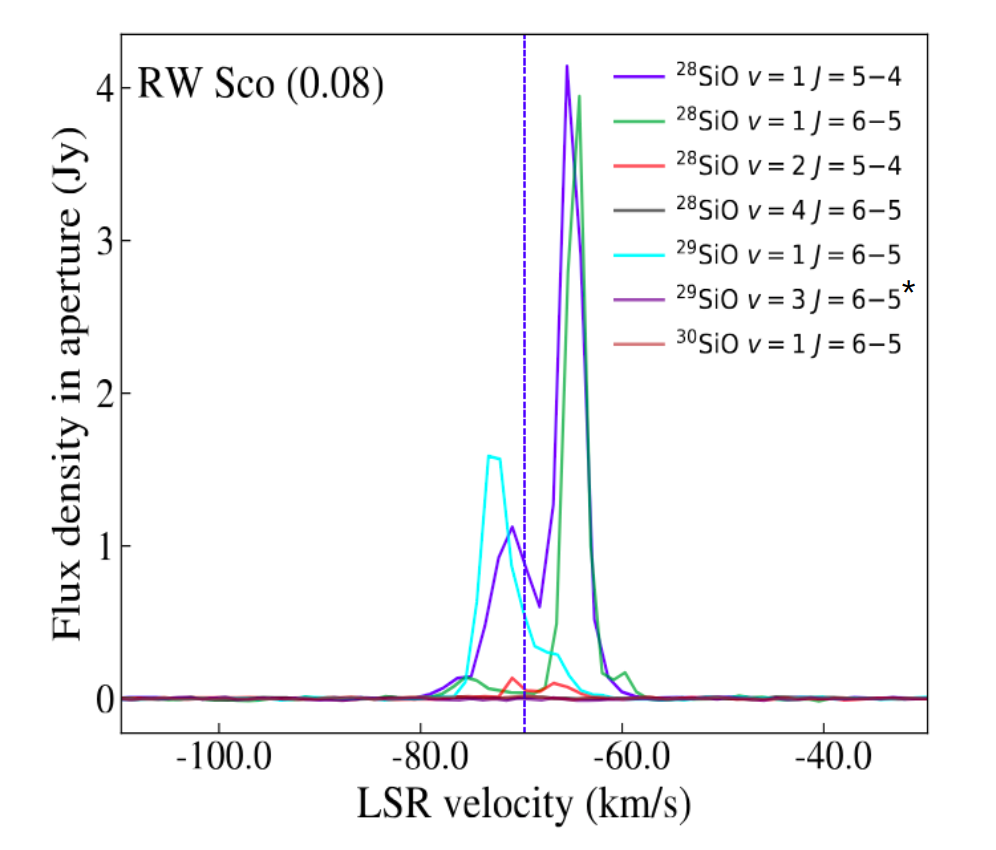}
    \includegraphics[width=0.392\textwidth, height=0.308\textwidth]{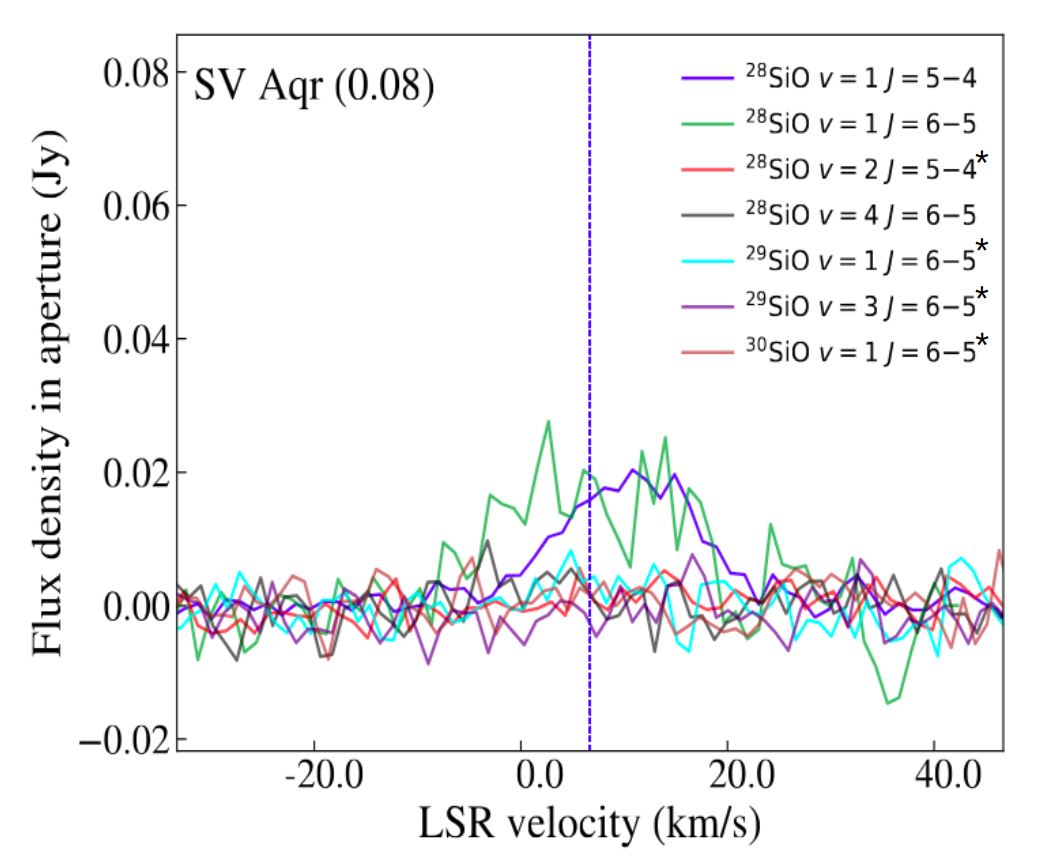}
    \includegraphics[width=0.392\textwidth, height=0.308\textwidth]{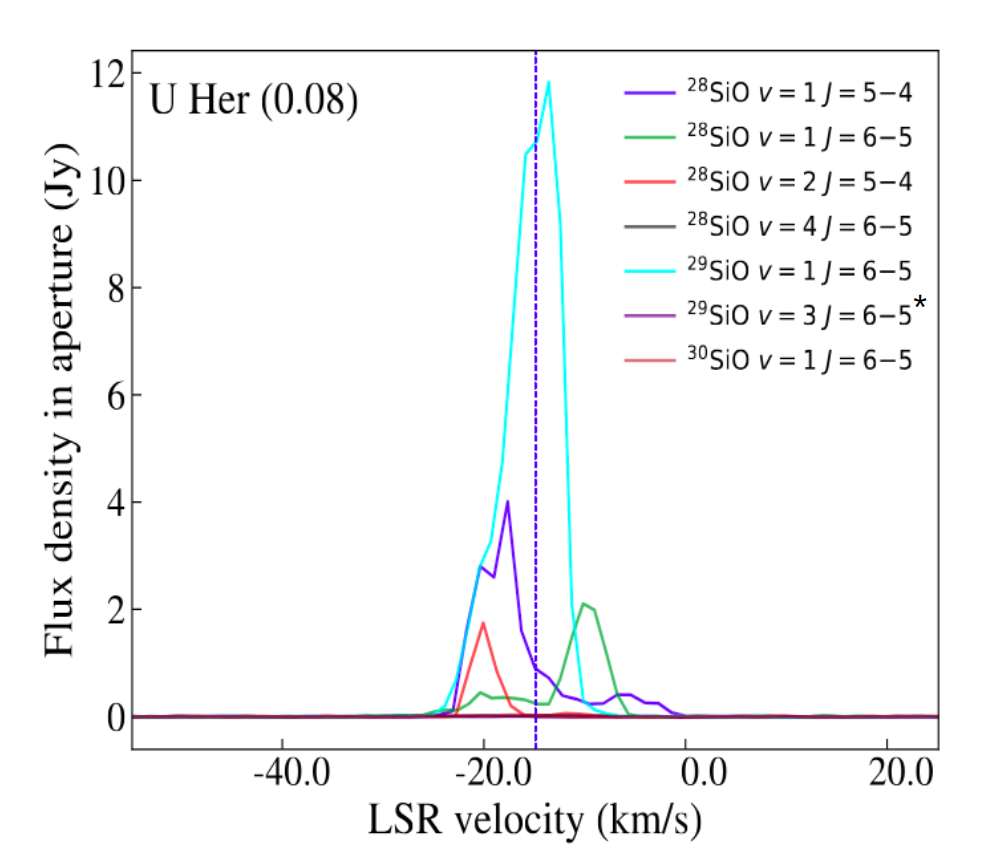}
    \includegraphics[width=0.392\textwidth, height=0.308\textwidth]{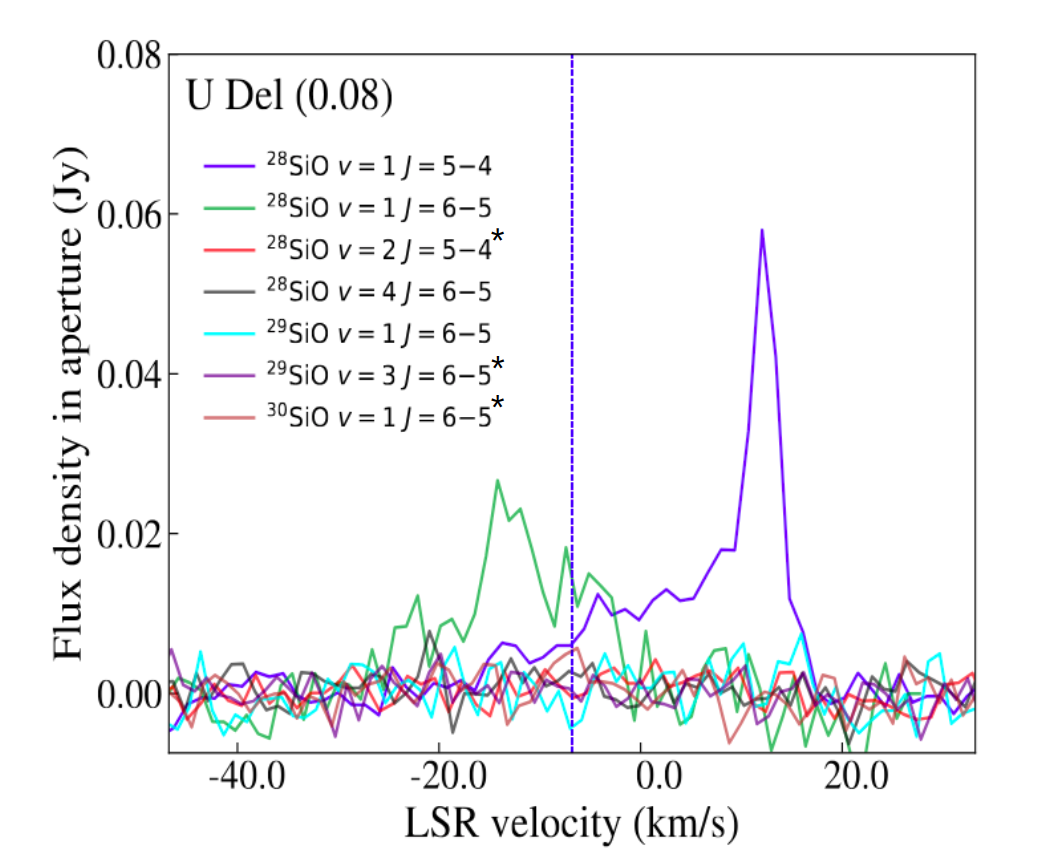}
    \includegraphics[width=0.392\textwidth, height=0.308\textwidth]{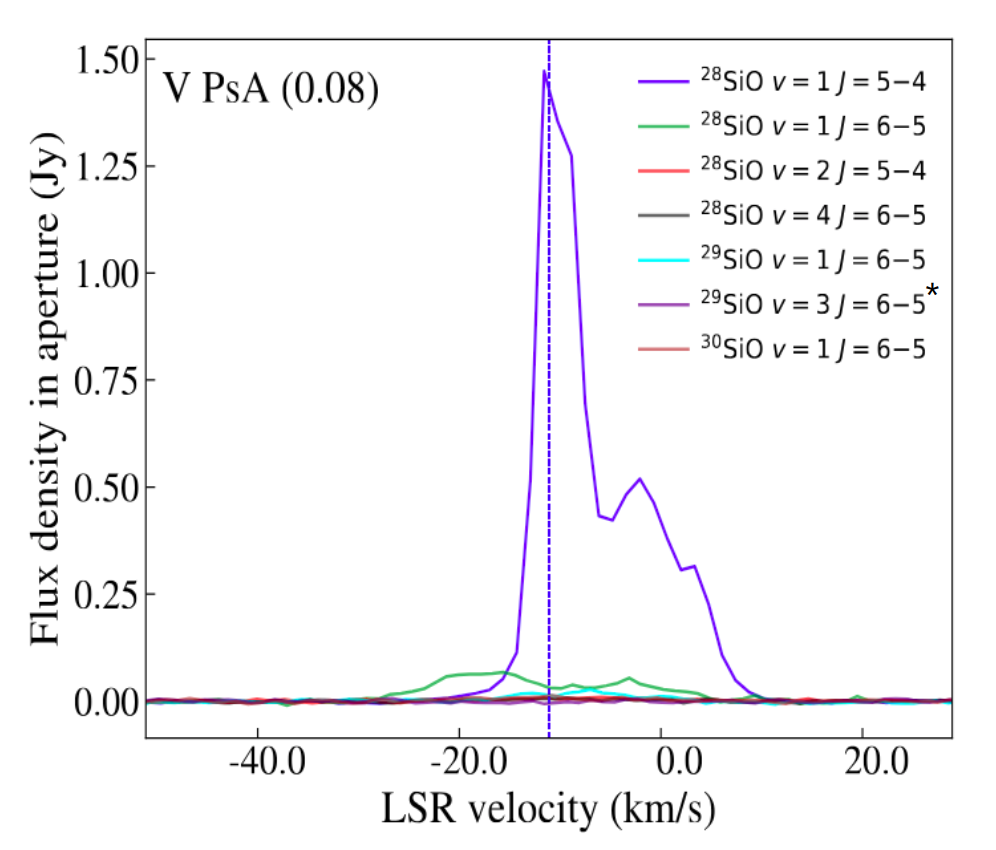}
    \includegraphics[width=0.392\textwidth, height=0.308\textwidth]{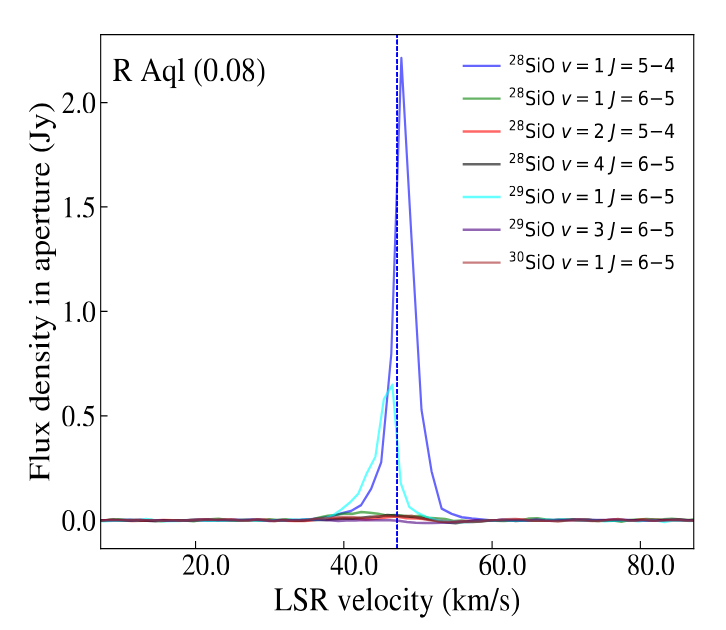}
    \includegraphics[width=0.392\textwidth, height=0.308\textwidth]{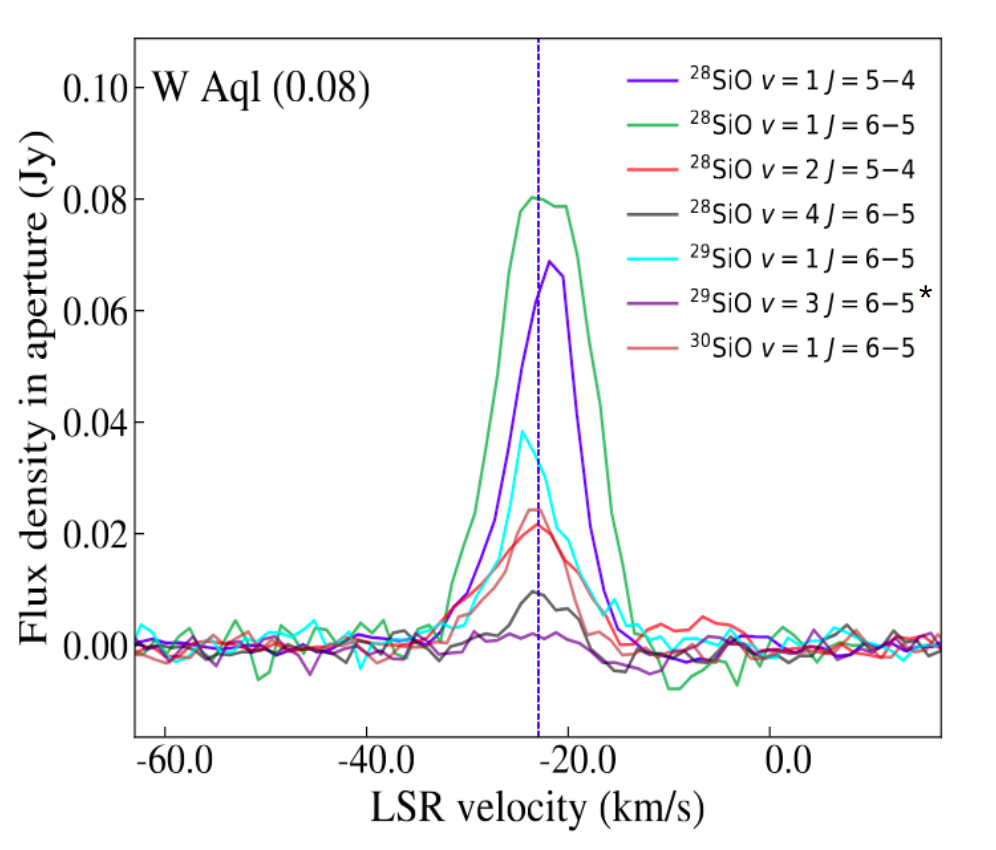}
    \includegraphics[width=0.392\textwidth, height=0.308\textwidth]{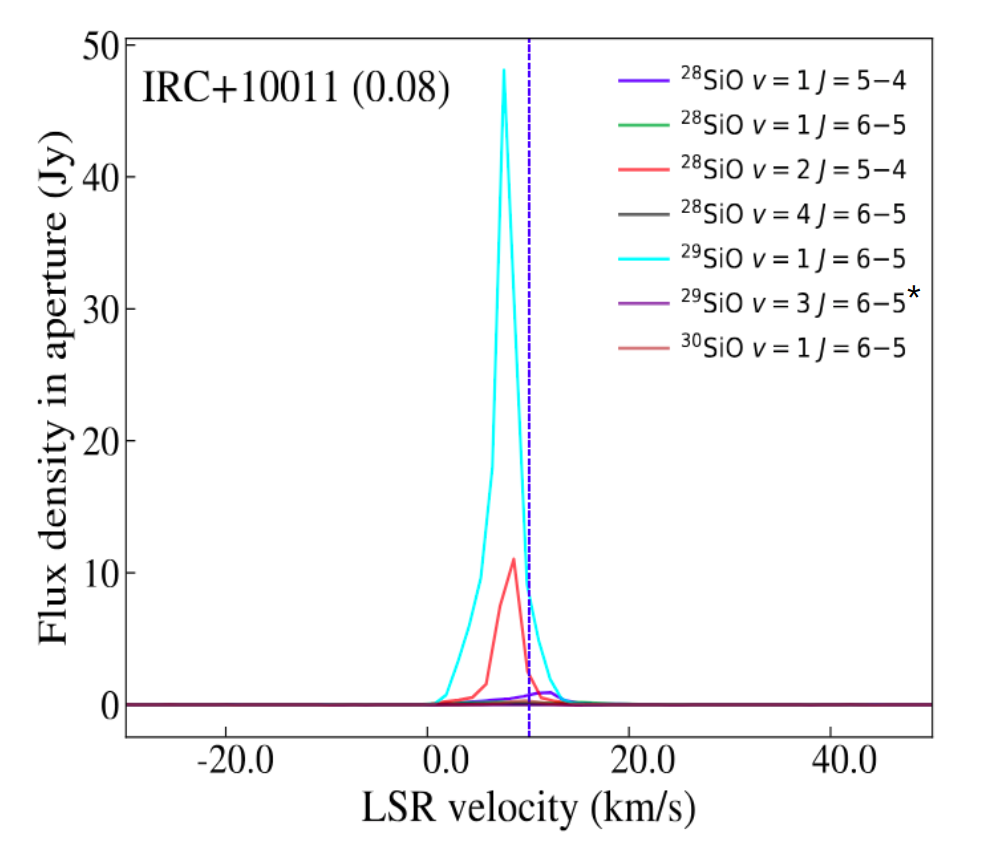}
   \caption{Spectra of seven SiO lines in the ATOMIUM data towards RW Sco, SV Aqr, U Her, U Del, V PsA, R Aql, W Aql and IRC+10011. The transitions along with their corresponding colours are given in the legend in the top right corner. The dashed blue line indicates $V_*$. Note that non-detection, marked by *, has been included in the plot for completeness and ease of direct comparison.} 
   \label{fig:add-spectra-compare-1}
   \end{center}
\end{figure*}

Here, we provide an overview of the spectra of the seven main high-$J$ \ce{SiO} transitions detected in the ATOMIUM survey, as shown in Fig. \ref{fig:spectra-compare}, for RW Sco, SV Aqr, U Her, U Del, V PsA, R Aql, W Aql and IRC+10011 (Fig. \ref{fig:add-spectra-compare-1}). Each panel features the transitions colour-coded as indicated in the legend in the top-right corner, allowing for a clear comparison of the maser profiles. The systemic velocity ($V_*$) is marked with a dashed blue line for reference. Non-detections are included in the plots for completeness, facilitating a direct comparison of spectral features and highlighting the variability or absence of maser activity in specific transitions or sources. These figures complement the discussion of \ce{SiO} maser line characteristics in Section \ref{sec:compare}.

%%%%%%%%%%%%%%%%%%%%%%%%%%%%%%%%%%%%%%%%%%%%%%%%%%

% Don't change these lines
\bsp	% typesetting comment
\label{lastpage}
\end{document}